\documentclass[11pt]{article}
\pdfoutput=1

\usepackage{amssymb}
\usepackage{amsmath}
\usepackage{bbold}
\usepackage{color}
\usepackage{colordvi}
\usepackage{colortbl}
\usepackage{fancybox}
\usepackage[footnotesize]{caption2}
\usepackage{graphicx}
\usepackage[center,footnotesize,hang]{subfigure}
\usepackage{bbm}
\usepackage{url}
\usepackage{multirow}
\usepackage{array}
\usepackage{arydshln}
\usepackage{pifont}
\usepackage{bm}
\usepackage{mathrsfs}
\usepackage{fancybox}
\usepackage{cite}
\usepackage[colorlinks]{hyperref}
\usepackage{enumitem}  
\usepackage{float} %
\usepackage{longtable}
\newcommand{\PreserveBackslash}[1]{\let\temp=\\#1\let\\=\temp}
\newcolumntype{C}[1]{>{\PreserveBackslash\centering}p{#1}}
\newcolumntype{R}[1]{>{\PreserveBackslash\raggedleft}p{#1}}
\newcolumntype{L}[1]{>{\PreserveBackslash\raggedright}p{#1}}
\allowdisplaybreaks \allowdisplaybreaks[2]

\definecolor{lightred}{rgb}{1,0.4,0.4}
\definecolor{lightgreen}{rgb}{0.4,1,0.4}

\makeatletter
\@addtoreset{equation}{section}
\makeatother

\begin{document}

\title{
\begin{flushright}
\hfill\mbox{\small USTC-ICTS-18-18} \\[5mm]
\begin{minipage}{0.2\linewidth}
\normalsize
\end{minipage}
\end{flushright}
{\Large \bf Lepton Mixing Predictions from $S_4$ in the Tri-Direct CP approach to Two Right-handed Neutrino Models
\\[2mm]}
}

\date{}

\author{
Gui-Jun~Ding$^{1}$\footnote{E-mail: {\tt
dinggj@ustc.edu.cn}},  \
Stephen~F.~King$^{2}$\footnote{E-mail: {\tt king@soton.ac.uk}}, \
Cai-Chang Li$^{1}$\footnote{E-mail: {\tt
lcc0915@mail.ustc.edu.cn}}  \
\\*[20pt]
\centerline{
\begin{minipage}{\linewidth}
\begin{center}
$^1${\it \small
Interdisciplinary Center for Theoretical Study and  Department of Modern Physics,\\
University of Science and Technology of China, Hefei, Anhui 230026, China}\\[2mm]
$^2${\it \small
Physics and Astronomy,
University of Southampton,
Southampton, SO17 1BJ, U.K.}\\
\end{center}
\end{minipage}}
\\[10mm]}
\maketitle
\thispagestyle{empty}

\begin{abstract}
\noindent

We perform an exhaustive analysis of all possible breaking patterns arising from $S_4\rtimes H_{CP}$ in a new {\it tri-direct CP approach} to the minimal seesaw model with two right-handed neutrinos, and construct a realistic flavour model along these lines.
According to this approach, separate residual flavour and CP symmetries persist in the charged lepton, ``atmospheric'' and ``solar'' right-handed neutrino sectors, i.e. we have {\it three} symmetry sectors rather than the usual two of the {\it semi-direct CP approach} (charged leptons and neutrinos).
Following the {\it tri-direct CP approach},
we find twenty-six kinds of independent phenomenologically interesting mixing patterns.
Eight of them predict a normal ordering (NO) neutrino mass spectrum and the other eighteen predict an inverted ordering (IO) neutrino mass spectrum. For each phenomenologically interesting mixing pattern, the corresponding predictions for the PMNS matrix, the lepton mixing parameters, the neutrino masses and the effective mass in neutrinoless double beta decay are given in a model independent way. One breaking pattern with NO spectrum and two breaking patterns with IO spectrum corresponds to form dominance. We find that the lepton mixing matrices of three kinds of breaking patterns with NO spectrum and one form dominance breaking pattern with IO spectrum preserve the first column of the  tri-bimaximal (TB) mixing matrix, i.e. yield a TM1 mixing matrix.

\end{abstract}
\newpage

\section{\label{sec:introduction}Introduction}
\indent

The discovery of neutrino oscillations imply that neutrinos have masses and there is mixing in the lepton sector. According to the neutrino oscillation experimental data, the $3\sigma$ ranges of the leptonic mixing angles and neutrino mass squared differences are~\cite{Esteban:2018azc}
\begin{equation}\label{eq:3sda}
\begin{array}{l}
0.272\leq\sin^2\theta_{12}\leq0.346,  \qquad  6.80\times10^{-5}\text{eV}^2\leq\Delta m^2_{21}\leq8.02\times10^{-5}\text{eV}^2\,,  \\
\left\{\begin{array}{lll}
0.01981\leq\sin^2\theta_{13}\leq0.02436, ~~~&~~~ 0.418\leq\sin^2\theta_{23}\leq0.613, ~~~&~~~ \text{(NO)}\,, \\
 0.02006\leq\sin^2\theta_{13}\leq0.02452, ~~~&~~~  0.435\leq\sin^2\theta_{23}\leq0.616, ~~~&~~~ \text{(IO)}\,,
\end{array}\right. \\
 \left\{\begin{array}{ll}
2.399\times10^{-3}\text{eV}^2\leq\Delta m^2_{31}\leq2.593\times10^{-3}\text{eV}^2, ~~~&~~~ \text{(NO)}\,, \\
-2.562\times10^{-3}\text{eV}^2\leq\Delta m^2_{32}\leq-2.369\times10^{-3}\text{eV}^2, ~~~&~~~ \text{(IO)}\,,
\end{array}\right.
\end{array}
\end{equation}
where the symbols ``NO'' and ``IO'' denotes  normal ordering and inverted ordering neutrino mass spectrums, respectively. We do not know the origin of neutrino mass and lepton mixing so far, although these results are consistent with some theories. The leading candidate for a framework of neutrino mass and lepton mixing is the type I seesaw  mechanism which involves additional heavy right-handed Majorana neutrinos~\cite{Minkowski:1977sc,Mohapatra:1979ia,Schechter:1980gr}. Since the masses of the right-handed Majorana neutrinos are typically far beyond reach of the LHC, the seesaw mechanism generally introduces many additional unconstrained parameters, making it very difficult to test experimentally. In order to obtain testable predictions, it is natural to follow the idea of minimality (as discussed in e.g.~\cite{King:2015sfk}), i.e. focusing on the seesaw theories with smaller numbers of parameters.

The most minimal version of the seesaw mechanism involves two additional right-handed neutrinos~\cite{King:1999mb,Frampton:2002qc}. In order to increase predictive power of the two right-handed neutrino seesaw model, various schemes to reduce the number of free parameters have been suggested, such as postulating one~\cite{King:2002nf} or two~\cite{Frampton:2002qc} texture zeros, however the latter models with two texture zero are now phenomenologically excluded for NO~\cite{Guo:2006qa,Harigaya:2012bw,Zhang:2015tea}. In the charged lepton diagonal basis, together with a diagonal right-handed neutrino mass matrix, the idea of constrained sequential dominance (CSD) has been proposed, involving a Dirac mass matrix with one texture zero and restricted form of the Yukawa couplings~\cite{King:2005bj}. The CSD($n$) scheme~\cite{King:2005bj,Antusch:2011ic,King:2013iva,King:2015dvf,King:2016yvg,King:2013xba,King:2013hoa,Bjorkeroth:2014vha}
assumes that the coupling of one right-handed neutrino (called ``atmospheric'') with $\nu_{L}$ is proportional to $(0,1,1)$, while the second right-handed neutrino (called ``solar'') has couplings to $\nu_{L}$ proportional to $(1,n,n-2)$ with positive integer $n$, where $\nu_{L}\equiv(\nu_e,\nu_{\mu},\nu_{\tau})^T_{L}$ denote the left-handed neutrino fields. The CSD($n$) models generally~\cite{King:2005bj,Antusch:2011ic,King:2013iva,King:2015dvf,King:2016yvg,King:2013xba,King:2013hoa,Bjorkeroth:2014vha}
predict a TM1 mixing matrix and normal mass hierarchy with a massless neutrino $m_{1}=0$~\cite{Harrison:2002er}. The predictions for lepton mixing parameters and neutrino mass have been studied for the cases of $n=1$~\cite{King:2005bj}, $n=2$~\cite{Antusch:2011ic}, $n=3$~\cite{King:2013iva,King:2015dvf,King:2016yvg}, $n=4$~\cite{King:2013xba,King:2013hoa} and $n\geq5$~\cite{Bjorkeroth:2014vha}. It turns out that the CSD(3) also called Littlest Seesaw (LS) model can successfully accommodate the experimental data on neutrino masses and mixing angles~\cite{King:2013iva,King:2015dvf,King:2016yvg}. The LS model can yield the baryon asymmetry of the Universe via leptogenesis~\cite{Bjorkeroth:2015tsa,Chianese:2018dsz,King:2018fqh}. The LS structure can also be incorporated into grand unified models~\cite{Bjorkeroth:2015ora,Bjorkeroth:2015uou,Bjorkeroth:2015tsa}. In practice the LS model can be achieved by introducing $S_{4}$ family symmetry, which is spontaneously broken by flavon fields with particular vacuum alignments governed by remnant subgroups of $S_{4}$~\cite{King:2015dvf,King:2016yvg}. Furthermore, from the breaking of $A_5$ flavor symmetry to different residual subgroups in the charged lepton, atmospheric neutrino and solar neutrino sectors, we can obtain the viable golden LS model which predicts the GR1 lepton mixing pattern~\cite{Ding:2017hdv}. Here GR1 mixing matrix  preserves the first column of the golden ratio mixing matrix.

The leptonic CP violation is one of the most urgent questions in neutrino oscillation physics. The indication of CP violation in neutrino sector has been reported by T2K~\cite{Abe:2017uxa} and NO$\nu$A~\cite{Adamson:2017gxd}, and the Dirac CP phase $\delta_{CP}$  will be intensively probed experimentally in the forthcoming years.
In order to address this question theoretically, non-Abelian discrete flavor symmetry combined with generalized CP symmetry have been widely exploited to explain the lepton mixing angles and to predict CP violating phases~\cite{Feruglio:2012cw,Holthausen:2012dk,Ding:2013hpa,Ding:2013bpa,Li:2013jya,Ding:2013nsa,Ding:2014ssa,Ding:2014hva,Li:2014eia,Ding:2014ora,Chen:2014wxa,Branco:2015hea,
Li:2015jxa,DiIura:2015kfa,Ballett:2015wia,Branco:2015gna,Chen:2015nha,Ding:2015rwa,Chen:2015siy,Li:2016ppt,Chen:2016ptr,
Yao:2016zev,Li:2016nap,Lu:2016jit,Li:2017zmk,Li:2017abz,Chen:2018lsv,Lu:2018oxc,Hagedorn:2016lva}. Both flavor symmetry $G_f$ and CP symmetry $H_{CP}$ are imposed at the high  energy scale, and the full symmetry is $G_f\rtimes H_{CP}$. In the successful {\it semi-direct CP approach},  the original symmetry $G_f\rtimes H_{CP}$ is spontaneously broken down to $G_{l}$ and $G_{\nu}\rtimes H^{\nu}_{CP}$ in the charged lepton sector and the neutrino sector at lower energies, respectively.

Recently we extended the above {\it semi-direct CP approach} to propose a so-called {\it tri-direct CP approach}~\cite{Ding:2018fyz} based on the two right-handed neutrino seesaw mechanism, and a new variant of the LS model is found. In the {\it tri-direct CP approach}, the common residual symmetry of the neutrino sector is
split into two branches: the residual symmetries $G_{\text{atm}}\rtimes H^{\text{atm}}_{CP}$ and $G_{\text{sol}}\rtimes H^{\text{sol}}_{CP}$ associated with the ``atmospheric'' and ``solar'' right-handed neutrino sectors respectively. An abelian subgroup $G_{l}$ is assumed to preserved by the charged lepton mass matrix and it allows the distinction of three generations. It is the combination of these {\it three} residual symmetries that provides a new way of fixing the lepton mixing parameters and neutrino masses in  {\it tri-direct CP approach}.

In the present work, we shall extend the analysis of the {\it tri-direct CP approach}  for two right-handed neutrino models considerably, beyond the few examples studied in~\cite{Ding:2018fyz}, to an exhaustive model independent analysis of {\it all} possible phenomenologically viable lepton flavor mixing patterns which arise from the breaking of the parent symmetry $S_{4}\rtimes H_{CP}$. The lepton mixing matrix is not restricted to TM1 mixing anymore and the mass ordering of the neutrino masses can be either NO or IO. We shall find eight independent phenomenologically interesting mixing patterns for the case of NO neutrino masses and eighteen independent phenomenologically interesting mixing patterns for the case of IO. The eight breaking patterns for NO are labeled as $\mathcal{N}_1\sim\mathcal{N}_{8}$ and the other eighteen for IO are labeled as $\mathcal{I}_1\sim\mathcal{I}_{18}$. For each possible breaking pattern, we numerically analyze the predictions of the mixing parameters, the three neutrino masses and the effective mass in neutrinoless double beta decay. We find that all the four breaking patterns $\mathcal{N}_{1}$, $\mathcal{N}_{2}$, $\mathcal{N}_3$ and $\mathcal{I}_5$ give rise to TM1 mixing. For the cases of $\mathcal{N}_5$, $\mathcal{I}_{4}$ and $\mathcal{I}_{5}$, the two columns of the Dirac neutrino mass matrix are orthogonal to each other and consequently the texture of form dominance~\cite{Chen:2009um,Choubey:2010vs,King:2010bk} is reproduced. Furthermore, we implement the case of $\mathcal{N}_4$ with $x=-4$ and $\eta=\pm3\pi/4$ in an explicit model based on $S_{4}\rtimes H_{CP}$, the required vacuum alignment needed to achieve the remnant symmetries is dynamically realized. In this model, the absolute value of the first column of PMNS matrix is fixed to be $\left(2\sqrt{\frac{6}{37}}, \sqrt{\frac{13}{74}}, \sqrt{\frac{13}{74}}\right)^T$.

The paper is organized as follows: in section~\ref{sec:framework}, we recall the framework of the {\it tri-direct CP approach} to two right-handed neutrino models, and we present the generic procedures of how to derive the lepton flavor mixing and neutrino masses from remnant symmetries in the {\it tri-direct CP approach} in a model independent way. In section~\ref{sec:TD_S4_GCP_NO}, we perform a model independent analysis of five kinds of phenomenologically viable breaking patterns achievable from the underlying symmetry $S_{4}\rtimes H_{CP}$ in the {\it tri-direct CP approach} with NO neutrino masses. In section~\ref{sec:TD_S4_GCP_IO}, a general analysis of five kinds of breaking patterns with IO neutrino masses are presented. In section~\ref{sec:model}, we present a new version of the LS model based on $S_{4}\rtimes H_{CP}$ from the {\it tri-direct CP approach}. The vacuum alignment, the LO structure and the NLO corrections of the model are discussed. Section~\ref{sec:Conclusion} is devoted to our conclusion. The group theory of $S_{4}$ and its all abelian subgroups are presented in appendix~\ref{sec:S4_group}. In appendix~\ref{sec:other_NO_mix}, we study the breaking patterns $\mathcal{N}_{6}\sim\mathcal{N}_8$ in a  model independent way. The analysis of the remaining thirteen kinds of breaking patterns with IO are given in appendix~\ref{sec:other_IO_mix}.

\section{\label{sec:framework} The tri-direct CP approach}

In the  scenario with a discrete flavor group $G_{f}$ and generalized CP symmetry $H_{CP}$, $G_{f}$ and $H_{CP}$ should be compatible with each other, and they fulfill the following consistency condition~\cite{Feruglio:2012cw,Holthausen:2012dk,Chen:2014tpa,Grimus:1995zi}
\begin{equation}
\label{eq:consistency}X_{\mathbf{r}}\rho^{*}_{\mathbf{r}}(g)X^{\dagger}_{\mathbf{r}}=\rho_{\mathbf{r}}(g^{\prime}),\quad
g,g^{\prime}\in G_{f}, \quad X_{\mathbf{r}}\in H_{CP}\,,
\end{equation}
where $\rho_{\mathbf{r}}(g)$ is the representation matrix of the element $g$ in the irreducible representation $\mathbf{r}$ of $G_{f}$, and $X_{\mathbf{r}}$ is the generalized CP transformation matrix of $H_{CP}$. Moreover, the physically well-defined generalized CP transformations should be class-inverting automorphisms of $G_f$~\cite{Chen:2014tpa}. It requires that the elements $g^{-1}$ and $g^{\prime}$ in Eq.~\eqref{eq:consistency} belong to the same conjugacy class of $G_{f}$.  The automorphism in Eq.~\eqref{eq:consistency} thus implies that the mathematical structure of the group comprising $G_f$ and CP is in general a semi-direct product $G_{f}\rtimes H_{CP}$~\cite{Feruglio:2012cw}.

\begin{figure}[t!]
\centering
\begin{tabular}{c}
\includegraphics[width=0.50\linewidth]{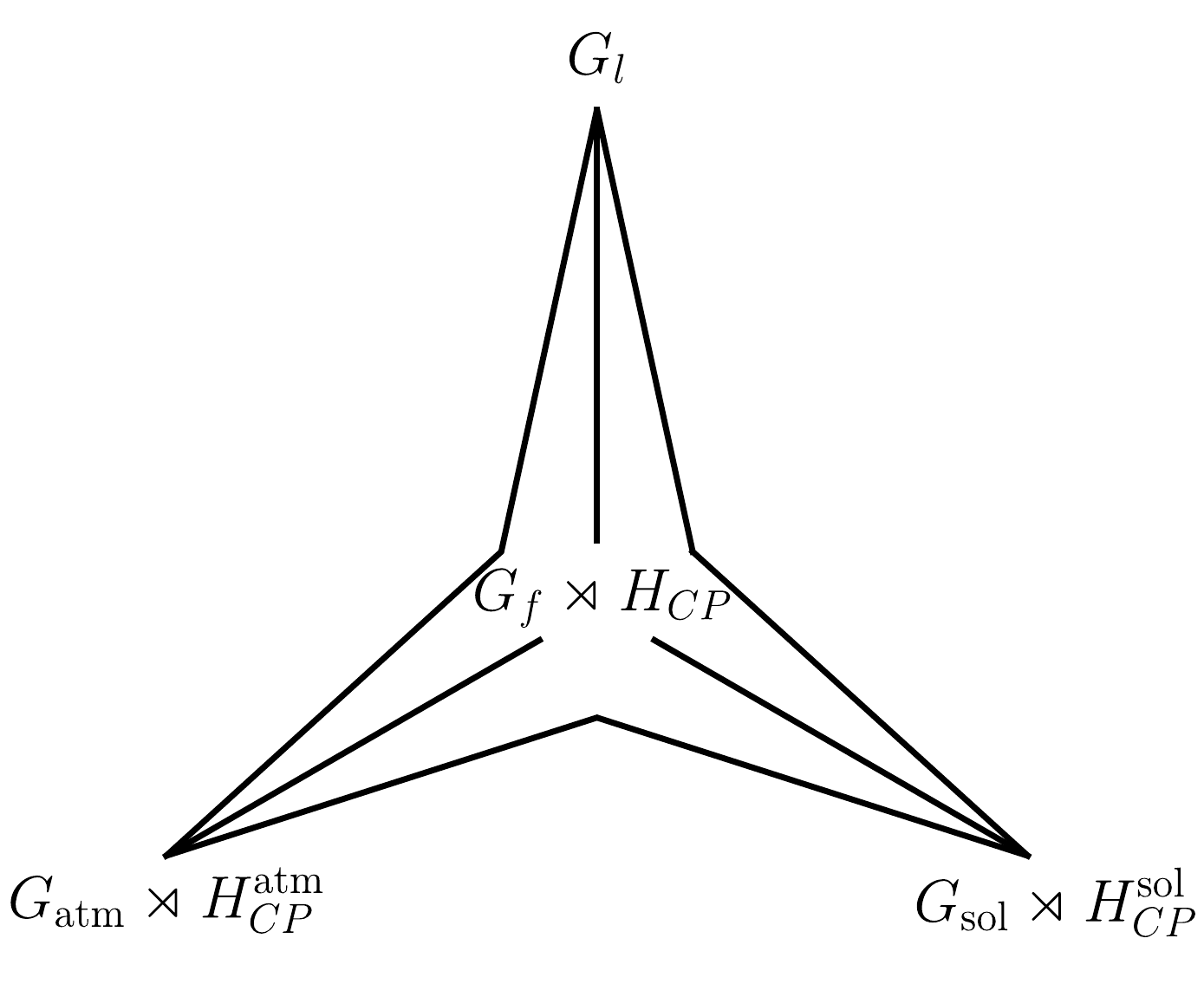}
\end{tabular}
\caption{\label{fig:tri_direct} A sketch of the  tri-direct CP approach for two right-handed neutrino models, where
the high energy family and CP symmetry $G_f\rtimes H_{CP}$
is spontaneously broken down to $G_{\text{atm}}\rtimes H_{CP}^{\text{atm}}$ in the sector of one of the right-handed neutrinos,
and $G_{\text{sol}}\rtimes H_{CP}^{\text{sol}}$ in the sector of the other right-handed neutrino, with
the charged lepton sector having a  different residual flavour symmetry $G_{l}$.}
\end{figure}

In the present work, we shall perform a comprehensive study of lepton mixing patterns which can be obtained from the flavor group $S_4$ and CP symmetry in the tri-direct CP approach~\cite{Ding:2018fyz}. In the following, we shall firstly review how the  tri-direct CP approach allows us to predict the lepton mixing and neutrino masses are predicted in terms of few parameters. In the tri-direct CP approach, the assumed family and CP symmetry $G_{f}\times H_{CP}$ at high energy scale is spontaneously broken down to an abelian subgroup $G_{l}$ which is capable of distinguishing the three generations in the charged lepton sector, and it is broken to $G_{\text{atm}}\rtimes H^{\text{atm}}_{CP}$ and $G_{\text{sol}}\rtimes H^{\text{sol}}_{CP}$ in the atmospheric and solar neutrino sectors respectively. A sketch of the tri-direct CP approach for two right-handed neutrino models is illustrated in figure~\ref{fig:tri_direct}. In the right-handed neutrino diagonal basis, the effective Lagrangian is given by
\begin{small}
\begin{equation}\label{eq:Lagrangian}
\mathcal{L}=-y_{l}L\phi_{l}E^{c}-y_{\mathrm{atm}}L\phi_{\mathrm{atm}}N^c_{\mathrm{atm}}-y_{\mathrm{sol}}L\phi_{\mathrm{sol}}N^c_{\mathrm{sol}}
-\frac{1}{2}x_{\mathrm{atm}}\xi_{\mathrm{atm}}N^c_{\mathrm{atm}}N^c_{\mathrm{atm}}-\frac{1}{2}x_{\mathrm{sol}}\xi_{\mathrm{sol}}N^{c}_{\mathrm{sol}}N^c_{\mathrm{sol}}
+\text{h.c.}\,,
\end{equation}
\end{small}
where $L$ stands for the left-handed lepton doublets and $E^c\equiv(e^{c}, \mu^{c}, \tau^{c})^T$ are the right-handed charged leptons, the flavons $\xi_{\text{atm}}$ and $\xi_{\text{sol}}$ are standard model singlets, the flavons $\phi_{l}$, $\phi_{\rm sol}$ and $\phi_{\rm atm}$ can be either Higgs fields or combinations of the electroweak Higgs doublet together with flavons. All the four coupling constants $y_{\mathrm{atm}}$, $y_{\mathrm{sol}}$, $x_{\mathrm{atm}}$ and $x_{\mathrm{sol}}$ would be constrained to be real if we impose CP as symmetry on the theory.

Without loss of generality, we assume that the three generations of left-handed leptons doublets transform as a faithful three-dimensional representation $\mathbf{3}$ under $G_{f}$. The residual symmetry $G_{l}$ in the charged lepton sector requires that the hermitian combination $m^{\dagger}_{l}m_{l}$ must be invariant under the action of $G_{l}$, i.e.
\begin{equation}\label{eq:ch_dia}
\rho^{\dagger}_{\mathbf{3}}(g_{l})m^{\dagger}_{l}m_{l}\rho_{\mathbf{3}}(g_{l})=m^{\dagger}_{l}m_{l}, \quad g_{l}\in G_{l}\,,
\end{equation}
where the charged lepton mass matrix $m_{l}$ is defined in the convention $l^{c}m_ll$. The diagonalization matrix of the hermitian combination $m^{\dagger}_{l}m_{l}$ is defined as $U_{l}$ with $U^{\dagger}_{l}m^{\dagger}_{l}m_{l}U_{l}=\text{diag}(m^2_{e},m^2_{\mu},m^2_{\tau})$. From Eq.~\eqref{eq:ch_dia}, we find that the unitary matrix $U_{l}$ can be derived from
\begin{equation}
\label{eq:Ul}U^{\dagger}_{l}\rho_{\mathbf{3}}(g_{l})U_{l}=\rho^{\text{diag}}_{\mathbf{3}}(g_{l})\,,
\end{equation}
where $\rho^{\text{diag}}_{\mathbf{3}}(g_{l})$ is a diagonal matrix with entries being three eigenvalues of $\rho_{\mathbf{3}}(g_{l})$. In the atmospheric neutrino sector and the solar neutrino sector, as the residual symmetries contain both flavor symmetry and CP symmetry, then the following restricted consistency conditions should be satisfied
\begin{subequations}
\begin{eqnarray}
\label{eq:nu_atm_consis}&&X^{\text{atm}}_{\mathbf{r}}\rho^{*}_{\mathbf{r}}(g^{\text{atm}}_{i})(X^{\text{atm}}_{\mathbf{r}})^{-1}
=\rho_{\mathbf{r}}(g^{\text{atm}}_{j}),\qquad g^{\text{atm}}_{i},g^{\text{atm}}_{j}\in G_{\text{atm}},\quad  X^{\text{atm}}_{\mathbf{r}}\in H^{\text{atm}}_{CP}\,,\\
\label{eq:nu_sol_consis}&&X^{\text{sol}}_{\mathbf{r}}\rho^{*}_{\mathbf{r}}(g^{\text{sol}}_{i})(X^{\text{sol}}_{\mathbf{r}})^{-1}
=\rho_{\mathbf{r}}(g^{\text{sol}}_{j}),\qquad g^{\text{sol}}_{i},g^{\text{sol}}_{j}\in G_{\text{sol}},\quad  X^{\text{sol}}_{\mathbf{r}}\in H^{\text{sol}}_{CP}\,.
\end{eqnarray}
\end{subequations}
The consistency conditions indicate that the mathematical structure of the residual flavor and CP symmetries is  a semi-direct product for $i\neq j$ and it reduces to a direct product for the case of $i=j$. The  consistency equations in Eqs.~\eqref{eq:nu_atm_consis} and~\eqref{eq:nu_sol_consis} can be used to find the residual CP consistent with the residual flavor symmetries of the atmospheric neutrino and the solar neutrino sectors, respectively. In the atmospheric and solar neutrino sectors, the residual symmetries imply that the vacuum alignments of flavons $\phi_{\text{atm}}$ and $\phi_{\text{sol}}$ should be invariant under the symmetries $G_{\text{atm}}\rtimes H^{\text{atm}}_{CP}$ and $G_{\text{sol}}\rtimes H^{\text{sol}}_{CP}$ respectively, i.e.
\begin{subequations}
\begin{eqnarray}
\label{eq:nu_atm_consis_vev}&&\rho_{\mathbf{r}}(g^{\text{atm}})\langle\phi_{\text{atm}}\rangle=\langle\phi_{\text{atm}}\rangle, \qquad X^{\text{atm}}_{\mathbf{r}}\langle\phi_{\text{atm}}\rangle^*=\langle\phi_{\text{atm}}\rangle\,,\\
\label{eq:nu_sol_consis_vev}&&\rho_{\mathbf{r}}(g^{\text{sol}})\langle\phi_{\text{sol}}\rangle=\langle\phi_{\text{sol}}\rangle, \qquad X^{\text{sol}}_{\mathbf{r}}\langle\phi_{\text{sol}}\rangle^*=\langle\phi_{\text{sol}}\rangle\,,
\end{eqnarray}
\end{subequations}
where $\langle\phi_{\text{atm}}\rangle$ and $\langle\phi_{\text{sol}}\rangle$ denote the vacuum alignments of flavons $\phi_{\text{atm}}$ and $\phi_{\text{sol}}$, respectively. After electroweak and flavor symmetry breaking, the flavons $\phi_{l}$, $\phi_{\text{atm}}$,  $\phi_{\text{sol}}$, $\xi_{\text{atm}}$ and $\xi_{\text{sol}}$ acquire nonvanishing vacuum expectation values (VEVs). From the Lagrangian in Eq.~\eqref{eq:Lagrangian}, one can read out the neutrino Dirac mass matrix and the heavy Majorana mass matrix,
\begin{equation}\label{eq:Dir_nu_mass}
m_{D}=\begin{pmatrix}
y_{\text{atm}}U_{a}\langle\phi_{\text{atm}}\rangle,  ~&~ y_{\text{sol}}U_{s}\langle\phi_{\text{sol}}\rangle
\end{pmatrix},\qquad m_{N}=\begin{pmatrix}
x_{\textrm{atm}}\langle\xi_{\text{atm}}\rangle ~&~ 0 \\
0  ~& ~ x_{\textrm{sol}}\langle\xi_{\text{sol}}\rangle
\end{pmatrix}\,,
\end{equation}
where $U_{a}$ and $U_{s}$ are two constants matrices and they are constituted by the Clebsch-Gordan (CG) coefficients which appear in the contractions $y_{\mathrm{atm}}L\phi_{\mathrm{atm}}N^c_{\mathrm{atm}}$ and $y_{\mathrm{sol}}L\phi_{\mathrm{sol}}N^c_{\mathrm{sol}}$, respectively. For the sake of convenience in the following, we shall parameterize the combinations $U_{a}\langle\phi_{\text{atm}}\rangle\equiv\bm{v}_{\text{atm}}v_{\phi_a}$ and $U_{s}\langle\phi_{\text{sol}}\rangle\equiv\bm{v}_{\text{sol}}v_{\phi_s}$, where $\bm{v}_{\text{atm}}$ and $\bm{v}_{\text{sol}}$ are three dimensional column vectors and they denote the directions of the vacuum alignment, $v_{\phi_a}$ and $v_{\phi_s}$ are the overall scale of corresponding flavons. The light effective Majorana neutrino mass matrix is given by the seesaw formula $m_{\nu}=-m_{D}m_{N}m^T_{D}$, then we find $m_{\nu}$ takes the from
\begin{eqnarray}
\nonumber m_{\nu}&=&-\frac{y^2_{\text{atm}}}{x_{\text{atm}}}\frac{U_{a}\langle\phi_{\text{atm}}\rangle\langle\phi_{\text{atm}}\rangle^{T}U^{T}_{a}}{\langle\xi_{\text{atm}}\rangle}
-\frac{y^2_{\text{sol}}}{x_{\text{sol}}}\frac{U_{s}\langle\phi_{\text{sol}}\rangle\langle\phi_{\text{sol}}\rangle^{T}U^{T}_{s}}{\langle\xi_{\text{sol}}\rangle}\,, \\
\label{eq:mnu} &\equiv& e^{i\varphi_{a}}\left[m_{a}\bm{v}_{\text{atm}}\bm{v}^T_{\text{atm}}+m_{s}e^{i\eta}\bm{v}_{\text{sol}}\bm{v}^T_{\text{sol}}\right]\,.
\end{eqnarray}
where the overall phase $\varphi_{a}$ is given by $\varphi_{a}=\text{arg}\left(-y^2_{\text{atm}}v^2_{\phi_a}/(x_{\text{atm}}\langle\xi_{\text{atm}}\rangle)\right)$,
$m_{a}=|y^2_{\text{atm}}v^2_{\phi_a}/(x_{\text{atm}}\langle\xi_{\text{atm}}\rangle)|$, $m_{s}=|y^2_{\text{sol}}v^2_{\phi_s}/(x_{\text{sol}}\langle\xi_{\text{sol}}\rangle)|$ and $\eta=\text{arg}\left(-y^2_{\text{sol}}v^2_{\phi_s}/(x_{\text{sol}}\langle\xi_{\text{sol}}\rangle)\right)-\varphi_{a}$. The overall phase $\varphi_{a}$ can be absorbed into the lepton field and it will always be omitted in the following. For convenience the notation $r\equiv m_s/m_a$ would be used throughout this paper. If the roles of $G_{\text{atm}}\rtimes H^{\text{atm}}_{CP}$ and $G_{\text{sol}}\rtimes H^{\text{sol}}_{CP}$ are switched, the two columns of the Dirac mass matrix $m_{D}$ would be exchanged. Thus the same neutrino mass matrix would be obtained if one interchanges $y_{\text{atm}}$ with $y_{\text{sol}}$ and $x_{\text{atm}}$ with $x_{\text{sol}}$.

In the following we shall give the detailed procedures for analyzing the phenomenological predictions of the tri-direct CP approach in a model independent way, and we shall present the generic expressions of lepton mixing matrix and neutrino masses. One can easily check that neutrino mass matrix $m_{\nu}$ of Eq.~\eqref{eq:mnu} satisfies
\begin{equation}
\label{eq:fixed_colm}m_{\nu}\bm{v}_{\text{fix}}=\left(
0,0,0\right)^T\,,
\end{equation}
with
\begin{equation}\label{eq:vdec}
 \bm{v}_{\text{fix}}\equiv \bm{v}_{\text{atm}}\times \bm{v}_{\text{sol}}\,,
\end{equation}
where $\bm{v}_{\text{atm}}\times \bm{v}_{\text{sol}}$ denotes the cross product of $\bm{v}_{\text{atm}}$ and $\bm{v}_{\text{sol}}$. The normalized vector of $\bm{v}_{\text{fix}}$ is defined as $\bm{\hat{v}}_{\text{fix}}\equiv \bm{v}_{\text{fix}}/\sqrt{\bm{v}^{\dagger}_{\text{fix}} \bm{v}_{\text{fix}}}$. Eq.~\eqref{eq:fixed_colm} implies that $\bm{\hat{v}}_{\text{fix}}$ is an eigenvector of $m_\nu$ with zero eigenvalue. As a result, the first (third) column of $U_{\nu}$ is determined to be $\bm{\hat{v}}_{\text{fix}}$ for NO (IO) mass spectrum, where $U_{\nu}$ is the diagonalization matrix of $m_{\nu}$ with $U^T_{\nu}m_{\nu}U_{\nu}=\text{diag}(0,m_{2},m_{3})$ for NO case and $U^T_{\nu}m_{\nu}U_{\nu}=\text{diag}(m_{1},m_{2},0)$ for IO case. In order to diagonalize the above neutrino mass matrix, we firstly perform a unitary transformation $U_{\nu1}$, where unitary matrix $U_{\nu1}$ can take the following form~\footnote{In general the unitary $U_{\nu1}$ is not unique. In the case $\bm{v}^\dagger_{\text{sol}}\bm{v}_{\text{atm}}\neq0$, $U_{\nu1}$ can multiplies a unitary rotation from the right-handed side in the $(23)$ and $(12)$ planes for NO and IO respectively.}
\begin{equation}\label{eq:Unu1}
U_{\nu1}=\left\{\begin{array}{l}
\left(\bm{\hat{v}}_{\text{fix}},\bm{\hat{v}}^*_{\text{atm}},\bm{\hat{v}}^\prime_{\text{sol}}\right)~~~\text{for}~~~\text{NO},\\
\left(\bm{\hat{v}}^*_{\text{atm}},\bm{\hat{v}}^\prime_{\text{sol}},\bm{\hat{v}}_{\text{fix}}\right)~~~\text{for}~~~\text{IO}\,,
\end{array}\right.
\end{equation}
with
\begin{equation}
\bm{\hat{v}}_{\text{atm}}=\frac{ \bm{v}_{\text{atm}}}{\sqrt{\bm{v}^{\dagger}_{\text{atm}} \bm{v}_{\text{atm}}}},\qquad
\bm{\hat{v}}^\prime_{\text{sol}}=\bm{\hat{v}}^*_{\text{fix}}\times \bm{\hat{v}}_{\text{atm}}\,.
\end{equation}
Then the neutrino mass matrix becomes
\begin{equation}\label{eq:mnup}
m^{\prime}_{\nu}=U^{T}_{\nu1}m_{\nu}U_{\nu1}=\left\{\begin{array}{l}
\begin{pmatrix}
0  &~   0  &~  0 \\
0  &~  y   &~ z  \\
0  &~  z  &~ w
\end{pmatrix}~~~\text{for}~~~\text{NO},\\
\begin{pmatrix}
y   &~ z  &~  0 \\
z  &~ w   &~  0   \\
0  &~  0 & ~0
\end{pmatrix}~~~\text{for}~~~\text{IO}\,,
\end{array}\right.
\end{equation}
where the expressions of the parameters $y$, $z$ and $w$ are
\begin{eqnarray}
\nonumber y&\equiv& |y|e^{i\phi_{y}}=m_{a}\bm{v}^{\dagger}_{\text{atm}} \bm{v}_{\text{atm}}+e^{i\eta}m_{s}\left(\bm{\hat{v}}^{\dagger}_{\text{atm}} \bm{v}_{\text{sol}}\right)^2\,, \\
\nonumber  z&\equiv& |z|e^{i\phi_{z}}=e^{i\eta}m_{s}\sqrt{\left(\bm{\hat{v}}_{\text{atm}}\times \bm{v}_{\text{sol}}\right)^\dagger \left(\bm{\hat{v}}_{\text{atm}}\times \bm{v}_{\text{sol}}\right)}\left(\bm{\hat{v}}^{\dagger}_{\text{atm}} \bm{v}_{\text{sol}}\right)\,, \\
  w&\equiv& |w|e^{i\phi_{w}}=e^{i\eta}m_{s}\left(\bm{\hat{v}}_{\text{atm}}\times \bm{v}_{\text{sol}}\right)^\dagger \left(\bm{\hat{v}}_{\text{atm}}\times \bm{v}_{\text{sol}}\right)\,.
\end{eqnarray}
The neutrino mass matrix $m'_{\nu}$ in Eq.~\eqref{eq:mnup} can be diagonalized through the standard procedure, as shown in Ref.~\cite{Ding:2013bpa,Ding:2018fyz},
\begin{equation}
\label{eq:mnup_Unu2D}U^{T}_{\nu2}m^{\prime}_{\nu}U_{\nu2}=\left\{\begin{array}{l}
\text{diag}(0, m_2, m_3)~~~\text{for}~~~\text{NO},\\
\text{diag}(m_1, m_2, 0)~~~\text{for}~~~\text{IO}\,,
\end{array}\right.
\end{equation}
where the unitary matrix $U_{\nu2}$ can be written as
\begin{equation}\label{eq:Unu2}
U_{\nu2}=\left\{\begin{array}{l}
\begin{pmatrix}
1  &~    0    &~   0 \\
0  &~  \cos\theta \,e^{i(\psi+\rho)/2}   &~  \sin\theta \,e^{i(\psi+\sigma)/2}   \\
0  &~   -\sin\theta\, e^{i(-\psi+\rho)/2}    &~~    \cos\theta\, e^{i(-\psi+\sigma)/2}
\end{pmatrix}~~~\text{for}~~~\text{NO},\\ [0.3in]
\begin{pmatrix}
\cos\theta \,e^{i(\psi+\rho)/2}   &~  \sin\theta \,e^{i(\psi+\sigma)/2}     &~   0 \\
 -\sin\theta\, e^{i(-\psi+\rho)/2}    &~~    \cos\theta\, e^{i(-\psi+\sigma)/2}  &~  0   \\
0  &~ 0 &~ 1
\end{pmatrix}~~~\text{for}~~~\text{IO}\,,
\end{array}\right.
\end{equation}
We find the light neutrino masses are
\begin{equation}\label{eq:nu_masses}
m^2_l=\frac{1}{2}\left[|y|^2+|w|^2+2|z|^2-\frac{|w|^2-|y|^2}{\cos2\theta}\right], \quad
m^2_h=\frac{1}{2}\left[|y|^2+|w|^2+2|z|^2+\frac{|w|^2-|y|^2}{\cos2\theta}\right]\,,
\end{equation}
with $m_1=0$, $m_2=m_l$, $m_3=m_{h}$ for NO case and $m_1=m_l$, $m_2=m_{h}$, $m_3=0$ for IO case. The rotation angle $\theta$ is determined by
\begin{eqnarray}
\nonumber&&\sin2\theta=\frac{2|z|\sqrt{|y|^2+|w|^2+2|y||w|\cos(\phi_{y}+\phi_{w}-2\phi_{z})}}
{\sqrt{(|w|^2-|y|^2)^2+4|z|^2\left[|y|^2+|w|^2+2|y||w|\cos(\phi_{y}+\phi_{w}-2\phi_{z})\right]}},\\
\label{eq:theta}&&\cos2\theta=\frac{|w|^2-|y|^2}{\sqrt{(|w|^2-|y|^2)^2+4|z|^2
\left[|y|^2+|w|^2+2|y||w|\cos(\phi_{y}+\phi_{w}-2\phi_{z})\right]}}\,.
\end{eqnarray}
It is obvious that $\sin2\theta$ is always non-negative. The expressions of the phases $\psi$, $\rho$ and $\sigma$ are given by
\begin{small}
\begin{eqnarray}
\nonumber&&\sin\psi=\frac{-|y|\sin(\phi_{y}-\phi_{z})+|w|\sin(\phi_{w}-\phi_{z})}
{\sqrt{|y|^2+|w|^2+2|y||w|\cos(\phi_{y}+\phi_{w}-2\phi_{z})}}, \\
\nonumber && \cos\psi=\frac{|y|\cos(\phi_{y}-\phi_{z})+|w|\cos(\phi_{w}-\phi_{z})}{\sqrt{|y|^2+|w|^2+2|y||w|\cos(\phi_{y}+\phi_{w}-2\phi_{z})}},\\
\nonumber&&\sin\rho=-\frac{(m^2_2-|z|^2)\sin\phi_{z}+|y||w|\sin(\phi_{y}+\phi_{w}-\phi_{z})}
{m_2\sqrt{|y|^2+|w|^2+2|y||w|\cos(\phi_{y}+\phi_{w}-2\phi_{z})}},\\
\nonumber && \cos\rho=\frac{(m^2_2-|z|^2)\cos\phi_{z}+|y||w|\cos(\phi_{y}+\phi_{w}-\phi_{z})}
{m_2\sqrt{|y|^2+|w|^2+2|y||w|\cos(\phi_{y}+\phi_{w}-2\phi_{z})}},\\
\nonumber &&\sin\sigma=-\frac{(m^2_3-|z|^2)\sin\phi_{z}+|y||w|\sin(\phi_{y}+\phi_{w}-\phi_{z})}
{m_3\sqrt{|y|^2+|w|^2+2|y||w|\cos(\phi_{y}+\phi_{w}-2\phi_{z})}},\\
\label{eq:par_prs} && \cos\sigma=\frac{(m^2_3-|z|^2)\cos\phi_{z}+|y||w|\cos(\phi_{y}+\phi_{w}-\phi_{z})}
{m_3\sqrt{|y|^2+|w|^2+2|y||w|\cos(\phi_{y}+\phi_{w}-2\phi_{z})}}\,.
\end{eqnarray}
\end{small}
Thus the lepton mixing matrix is determined to be
\begin{equation}\label{eq:PMNS}
\hskip-0.1in U_{PMNS}=P_{l}U^{\dagger}_{l}U_{\nu1}U_{\nu2}\,,
\end{equation}
where $P_{l}$ is a generic permutation matrix since the charged lepton masses are not constrained in this approach, and it can take the following six possible forms
\begin{equation}\label{eq:permutation_matrices}
\begin{array}{lll}
P_{123}=\begin{pmatrix}
1  &~ 0  ~&  0 \\
0  &~ 1  ~&  0\\
0  & ~0~  &  1
\end{pmatrix},~~&~~ P_{132}=\begin{pmatrix}
1  &  ~0~ &  0 \\
0  &  ~0~ &  1 \\
0  &  ~1~ &  0
\end{pmatrix},~~&~~ P_{213}=\begin{pmatrix}
0  &  ~1~  &  0 \\
1  &  ~0~  &  0 \\
0  &  ~0~  &  1
\end{pmatrix},\\
& & \\[-10pt]
P_{231}=\begin{pmatrix}
0   &  ~1~   &  0 \\
0   &  ~0~   &  1  \\
1   &  ~0~   &  0
\end{pmatrix},~~&~~ P_{312}=\begin{pmatrix}
0   &  ~0~  &   1  \\
1   &  ~0~  &   0 \\
0   &  ~1~  &  0
\end{pmatrix},~~&~~ P_{321}=\begin{pmatrix}
0    &   ~0~    &   1  \\
0    &   ~1~    &   0  \\
1    &   ~0~    &   0
\end{pmatrix}\,.
\end{array}
\end{equation}
If two mixing matrices are related by the exchange of the second and third rows, we shall only consider one of them. The reason is that the atmospheric mixing angle $\theta_{23}$ becomes $\pi/2-\theta_{23}$, the Dirac CP phases $\delta_{CP}$ becomes $\pi+\delta_{CP}$ and the other mixing parameters are unchanged after the second and third rows of a PMNS matrix are permuted. We notice that if both NO and IO neutrino mass spectrums can be achieved for a residual symmetry, the lepton mixing matrix of IO can be obtained from the corresponding one of NO by multiplying $P_{312}$ from the right side, and the expressions of the parameters $y$, $z$ and $w$ in $m^\prime_{\nu}$ are identical in NO and IO cases.

In the present work, we will adopt the standard parametrization of the lepton mixing matrix~\cite{Tanabashi:2018oca},
\begin{equation}\label{eq:PMNS_def}
U=\left(\begin{array}{ccc}
c_{12}c_{13}  &   s_{12}c_{13}   &   s_{13}e^{-i\delta_{CP}}  \\
-s_{12}c_{23}-c_{12}s_{13}s_{23}e^{i\delta_{CP}}   &  c_{12}c_{23}-s_{12}s_{13}s_{23}e^{i\delta_{CP}}  &  c_{13}s_{23}  \\
s_{12}s_{23}-c_{12}s_{13}c_{23}e^{i\delta_{CP}}   & -c_{12}s_{23}-s_{12}s_{13}c_{23}e^{i\delta_{CP}}  &  c_{13}c_{23}
\end{array}\right)\text{diag}(1,e^{i\frac{\beta}{2}},1)\,,
\end{equation}
where $c_{ij}\equiv \cos\theta_{ij}$, $s_{ij}\equiv \sin\theta_{ij}$, $\delta_{CP}$ is the Dirac CP violation phase and $\beta$ is the Majorana CP phase. There is a second Majorana phase if the lightest neutrino is not massless. As regards the CP violation, two weak basis invariants $J_{CP}$~\cite{Jarlskog:1985ht} and $I_1$~\cite{Branco:1986gr,Nieves:1987pp,Nieves:2001fc,Jenkins:2007ip,Branco:2011zb} associated with the CP phases $\delta_{CP}$ and $\beta$ respectively can be defined, \begin{eqnarray}
\nonumber && J_{CP}=\Im{(U_{11}U_{33}U^{*}_{13}U^{*}_{31})}=\frac{1}{8}\sin2\theta_{12}\sin2\theta_{13}\sin2\theta_{23}\cos\theta_{13}\sin\delta_{CP}\,, \\
\label{eq:CP_invariants}&& I_{1}=\left\{\begin{array}{l}
\Im{(U^{2}_{12}U^{*\,2}_{13})}=\frac{1}{4}\sin^2\theta_{12}\sin^22\theta_{13}\sin(\beta+2\delta_{CP})~~~\text{for~~~NO} \\[0.08in]
\Im{(U^{2}_{12}U^{*\,2}_{11})}=\frac{1}{4}\cos^4\theta_{13}\sin^22\theta_{12}\sin\beta ~~~\text{for~~~IO}
\end{array}\right.
\end{eqnarray}
Given a set of residual symmetry $\{G_{l},G_{\text{atm}}\rtimes H^{\text{atm}}_{CP}, G_{\text{sol}}\rtimes H^{\text{sol}}_{CP}\}$,
the explicit form of $U_{l}$, $\langle\phi_{\text{atm}}\rangle$ and $\langle\phi_{\text{sol}}\rangle$ can be straightforwardly determined. Using the general formulas of Eqs.~(\ref{eq:Ul}, \ref{eq:Unu1}, \ref{eq:Unu2}, \ref{eq:nu_masses}, \ref{eq:PMNS}), we can extract the predictions for lepton mixing matrix and neutrino masses.

\section{\label{sec:TD_S4_GCP_NO}Mixing patterns derived from $S_4$ with NO neutrino masses}

In this section, we shall consider all possible residual subgroups arising from the breaking of $S_4$ flavor symmetry and CP, the resulting predictions for lepton mixing parameters and neutrino masses are studied. The group theory of $S_{4}$ and all the CG coefficients in our basis are reported in appendix~\ref{sec:S4_group}. $S_{4}$ has 20 nontrivial abelian subgroups which contain nine $Z_2$ subgroups, four $Z_3$ subgroups, three $Z_4$ subgroups, four $K_4\cong Z_2\times Z_2$ subgroups. In our basis given in appendix~\ref{sec:S4_group}, the  generalized CP transformation compatible with the $S_{4}$ flavor symmetry is of the same form as the flavor symmetry transformation~\cite{Ding:2013hpa}, i.e.
\begin{equation}
X_{\bf{r}}=\rho_{\mathbf{r}}(g), \qquad g\in S_{4}\,,
\end{equation}
where $g$ can be any of the 24 group elements of $S_4$.

As discussed in section~\ref{sec:framework}, the flavor symmetry $S_4$ is broken to the abelian subgroup $G_{l}$ which is capable of distinguishing the three generations in the charged lepton sector. Then $G_{l}$ can take to be any one of the 11 subgroups $Z_{3}$, $Z_{4}$ and $K_{4}$ of $S_{4}$. The vacuum alignments of $\phi_{\text{atm}}$ and $\phi_{\text{sol}}$ preserve different residual symmetries $G_{\text{atm}}\rtimes H^{\text{atm}}_{CP}$ and $G_{\text{sol}}\rtimes H^{\text{sol}}_{CP}$, respectively. The residual flavor symmetries $G_{\text{atm}}$ and $G_{\text{sol}}$ can be any one of the 20 abelian subgroups of $S_{4}$. After including residual CP symmetry, we find there are altogether 4400 kinds of possible breaking patterns. But these breaking patterns are not all independent from each other. If a pair of residual flavor symmetries $\{G^{\prime}_{l},G^{\prime}_{\text{atm}}, G^{\prime}_{\text{sol}}\}$  is conjugated to the pair of groups $\{G_{l},G_{\text{atm}}, G_{\text{sol}}\}$ under an element of $S_4$, i.e.,
\begin{equation}\label{eq:conjugate}
G^{\prime}_{l}=hG_{l}h^{-1},\qquad G^{\prime}_{\text{atm}}=hG_{\text{atm}}h^{-1}, \qquad G^{\prime}_{\text{sol}}=hG_{\text{sol}}h^{-1}, \qquad h\in S_{4}\,,
\end{equation}
then these two breaking patterns will lead to the same predictions for mixing parameters~\cite{Ding:2013bpa,Li:2014eia,Ding:2014ssa}. As a result, it is sufficient to analyze the independent residual flavor symmetries not related by group conjugation and the compatible remnant CP. From appendix~\ref{sec:S4_group}, we find that all the $Z_3$ subgroups of $S_4$ are conjugate to each other, all the $Z_4$ subgroups  are related with each under group conjugation, $K^{(S, TST^{2})}_4$ is a normal subgroup of $S_4$, and the other three $K_4$ subgroups are conjugate to each other. As a consequence, it is sufficient to only consider four types of residual symmetries in the charged lepton sector, i.e. $G_{l}=Z^{T}_{3}$, $Z^{TSU}_{4}$, $K^{(S,TST^2)}_{4}$ and $K^{(S,U)}_{4}$, while both $G_{\text{atm}}$ and $G_{\text{sol}}$ can be any one of these 20 subgroups of $S_{4}$. In the present work, we assume that the three generations of left-handed leptons doublets are assigned to transform as $S_{4}$ triplet $\mathbf{3}$. For $G_{l}$ being above four kinds of subgroups, up to permutations and phases of the column vectors, the diagonalization matrix of the hermitian combination $m^{\dagger}_{l}m_{l}$ can be fixed to be
\begin{equation}\label{eq:ch_dia_matrix}
\begin{array}{lll}
 U_l=\begin{pmatrix}
 1  &~  0   ~&  0  \\
 0  &~  1  ~ &  0  \\
 0  & ~ 0  ~ &  1
\end{pmatrix}, & \qquad \text{for} & \qquad  G_{l}=Z^{T}_{3} \,, \\
& & \\[-0.12in]
  U_l=\frac{1}{2\sqrt{3}}
\begin{pmatrix}
 2 \omega  &~ -2 \omega  ~& 2 \omega  \\
 -\left(\sqrt{3}+1\right) \omega ^2 &~ \left(1-\sqrt{3}\right) \omega ^2 ~& 2 \omega ^2 \\
 \sqrt{3}-1 &~ \sqrt{3}+1 ~& 2 \\
\end{pmatrix}\,, & \qquad  \text{for} & \qquad  G_{l}=Z^{TSU}_{4} \,, \\
& & \\[-0.12in]
U_l=\frac{1}{\sqrt{3}}
\begin{pmatrix}
 \omega &~ 1~ & \omega^2 \\
 \omega^2 &~ 1 ~& \omega \\
 1 & ~1 ~& 1 \\
\end{pmatrix}, &\qquad  \text{for} &\qquad  G_{l}=K^{(S,TST^2)}_{4} \,, \\
& & \\[-0.12in]
U_l=
\begin{pmatrix}
 \sqrt{\frac{2}{3}} &~ \frac{1}{\sqrt{3}} ~& 0 \\
 -\frac{1}{\sqrt{6}} &~ \frac{1}{\sqrt{3}} ~& -\frac{1}{\sqrt{2}} \\
 -\frac{1}{\sqrt{6}} &~ \frac{1}{\sqrt{3}} ~& \frac{1}{\sqrt{2}} \\
\end{pmatrix}\,, & \qquad \text{for} & \qquad  G_{l}=K^{(S,U)}_{4} \,.
\end{array}
\end{equation}

The residual CP symmetries in the atmospheric neutrino sector and the solar neutrino sector have to be compatible with the residual flavor symmetries, and the restricted consistency conditions in Eqs.~\eqref{eq:nu_atm_consis} and~\eqref{eq:nu_sol_consis} must be fulfilled. For the residual flavor symmetries $G_{\text{atm}}$ and $G_{\text{sol}}$ being the 20 subgroups of $S_{4}$, the corresponding residual CP transformations consistent with these subgroups are listed in table~\ref{tab:res_CP}.
\begin{table}[t!]
\renewcommand{\tabcolsep}{1.1mm}
\renewcommand{\arraystretch}{1.25}
\centering
\footnotesize
\begin{tabular}{|c|c||c|c|}
\hline \hline
 $G_{\text{atm}}$ ($G_{\text{sol}}$) &  $X_{\text{atm}}$ ($X_{\text{sol}}$) & $G_{\text{atm}}$ ($G_{\text{sol}}$) &  $X_{\text{atm}}$ ($X_{\text{sol}}$)   \\ \hline

$Z_4^{TSU}$ & $\{SU,T,STS,TST^2U, U,ST,TS,T^2STU\}$  & $Z_4^{ST^2 U}$ & $\{SU,T^2,ST^2S,T^2STU, U,ST^2,T^2S,TST^2U \}$   \\ \hline

$Z_4^{TST^2 U}$ & $\{1,S,TST^2U,T^2STU,   TST^2,T^2ST,U,SU \}$   & $Z_3^{T}$ & $\{1,T, T^2,   U,TU, T^2U \}$   \\ \hline

$Z_3^{ST}$ & $\{S,STS,T^2,   U,STU,T^2SU \}$   & $Z_3^{TS}$ & $\{S,T,ST^2S,  U,TSU,ST^2U  \}$   \\ \hline

$Z_3^{STS}$ & $\{1,STS,ST^2S,  U,STSU,ST^2SU \}$   & $Z_2^{S}$ & $\{1,S,TST^2U,T^2STU,TST^2,T^2ST,U,SU  \}$   \\ \hline

$Z_2^{TST^2}$ & $\{SU,T^2STU, T^2,ST^2S, U,TST^2 U ,ST^2,T^2S  \}$   & $Z_2^{T^2ST }$ & $\{SU, TST^2U, T, STS, U,T^2STU, ST, TS \}$ \\ \hline

$Z_2^{U }$ & $\{1, U, S,  SU\}$   & $Z_2^{TU }$ & $\{U,T,STS,T^2STU \}$   \\ \hline

$Z_2^{SU }$ & $\{1,SU, S,U\}$   & $Z_2^{T^2 U}$ & $\{U, T^2,ST^2S,TST^2 U\}$   \\ \hline

$Z_2^{STS U}$ & $\{U, STS, T, T^2STU \}$   & $Z_2^{ST^2S U}$ & $\{U, ST^2S, T^2, TST^2U \}$   \\ \hline

$K_4^{(S,TST^2)}$ & all elements of $S_{4}$   & $K_4^{(S,U)}$ &  $\{1, S, U, SU, TST^2, T^2ST, TST^2U, T^2STU \}$   \\ \hline

$K_4^{(TST^2,T^2 U)}$ &  $\{U, T^2, ST^2S, TST^2 U,  SU, ST^2, T^2S, T^2STU \}$   & $K_4^{(T^2ST ,TU)}$ &  $\{U, T, STS,  T^2STU,  SU, ST, TS, TST^2 U\}$   \\ \hline\hline

\end{tabular}
\caption{\label{tab:res_CP}
The possible residual flavor subgroups and the compatible residual CP transformations.}
\end{table}
In this work, we assume that the flavon fields $\phi_{\text{atm}}$ and $\phi_{\text{sol}}$ are assigned to transform as $S_{4}$ triplet $\mathbf{3}$ and $\mathbf{3^\prime}$, respectively. In our working basis, the $S_4$ singlet contraction rules for $\mathbf{3}\otimes \mathbf{3}\rightarrow \mathbf{1}$ and $\mathbf{3}\otimes \mathbf{3}'\rightarrow \mathbf{1}'$ imply $\left(L\phi_{\text{atm}}\right)_{\mathbf{1}}=L_1\phi_{\text{atm}_1}+L_2\phi_{\text{atm}_3}+L_3\phi_{\text{atm}_2}$ and $\left(L\phi_{\text{sol}}\right)_{\mathbf{1}'}=L_1\phi_{\text{sol}_1}+L_2\phi_{\text{sol}_3}+L_3\phi_{\text{sol}_2}$. As a consequence, we can read out the matrices $U_{a}$ and $U_{s}$ as follow,
\begin{equation}
U_{a}=U_{s}=\begin{pmatrix}
1  &  ~0~ &  0 \\
0  &  ~0~ &  1 \\
0  &  ~1~ &  0
\end{pmatrix}\,.
\end{equation}
In other words, the column vectors $\bm{v}_{\text{atm}}$ and $\bm{v}_{\text{sol}}$ defined above Eq.~\eqref{eq:mnu} are $\bm{v}_{\text{atm}}=P_{132}\langle\phi_{\text{atm}}\rangle/v_{\phi_a}$ and $\bm{v}_{\text{sol}}=P_{132}\langle\phi_{\text{sol}}\rangle/v_{\phi_s}$. Hence the column vectors $\bm{v}_{\text{atm}}$ and $\bm{v}_{\text{sol}}$ can be obtained by exchanging the second and the third elements of the columns $\langle\phi_{\text{atm}}\rangle/v_{\phi_a}$ and $\langle\phi_{\text{sol}}\rangle/v_{\phi_s}$, respectively. The most general VEVs of the flavons $\phi_{\text{atm}}$ and $\phi_{\text{sol}}$ which preserve the possible residual symmetries in table~\ref{tab:res_CP} are summarized in table~\ref{tab:inv_VEV_CP}. For some residual flavor groups, not all the compatible residual CP transformations in table~\ref{tab:res_CP} are explicitly listed in table~\ref{tab:inv_VEV_CP}, this is because the invariant vacuum alignments for the shown residual CP and the unshown ones only differ in an overall factor $i$. The contribution of the overall $i$ can be absorbed into the couplings $x_{\text{atm}}$ and $x_{\text{sol}}$. Following the procedures presented in section~\ref{sec:framework},  we can straightforwardly obtain the expressions of the mixing parameters (three mixing angles, one Dirac CP phase and one Majorana CP phase) and the neutrino masses for each possible residual symmetry $\{G_{l},G_{\text{atm}}\rtimes H^{\text{atm}}_{CP}, G_{\text{sol}}\rtimes H^{\text{sol}}_{CP}\}$.

\begin{table}[t!]
\renewcommand{\tabcolsep}{0.1mm}
\renewcommand{\arraystretch}{1.5}
\centering
\tiny
\begin{tabular}{|c|c|c||c|c|c|c|c|c|c|c|}
\hline \hline
\multicolumn{6}{|c|}{VEVs of $\phi_{\text{atm}}$ }   \\ \hline
$G_{\text{atm}}$ & $X_{\text{atm}}$ & $\langle\phi_{\text{atm}}\rangle/v_{\phi_{a}}$ & $G_{\text{atm}}$ & $X_{\text{atm}}$ & $\langle\phi_{\text{atm}}\rangle/v_{\phi_{a}}$ \\ \hline

 $Z_2^S$ & $\{1,TST^2U,S,T^2STU\}$  & $\left( 1, 1, 1 \right)^T$ &  $Z_2^{TST^2}$ & $\{SU,T^2,ST^2S,T^2STU\}$ & $\left(1 , \omega^2 , \omega\right)^T$ \\ \hline

  $Z_2^{T^2ST }$ & $\{SU,T,TST^2U,STS\}$  &  $\left(1, \omega, \omega^2\right)^T$ & $Z_2^U$ & $\{1,U\}$ &  $\left(0, 1, -1\right)^T$ \\ \hline

  $Z_2^{TU}$ & $\{U,T\}$ & $\left( 0,-\omega, \omega^2\right)^T$ & $Z_2^{SU}$ & $\{1,SU\}$ &  $\left(2,-1,-1\right)^T$ \\ \hline

  $Z_2^{T^2 U}$ & $\{U,T^2\}$  &  $\left(0,-\omega^2 , \omega\right)^T$ & $Z_2^{STSU}$ & $\{T,T^2STU\}$ & $\left(2, -\omega ,-\omega^2\right)^T$ \\ \hline

 $Z_2^{ST^2SU}$ & $\{T^2,TST^2U\}$  & $\left(2, -\omega^2 ,-\omega\right)^T$ &  $Z_3^T$ & $\{1,T,T^2\}$ & $\left(1,0,0\right)^T$ \\ \hline

  $Z_3^{ST}$ & $\{S,STS,T^2\}$  & $\left(1, -2\omega^2,-2\omega\right)^T$ & $Z_3^{TS}$ & $\{S,T,ST^2S\}$ & $\left(1, -2\omega, -2\omega ^2\right)^T$ \\ \hline

    $Z_3^{STS}$ & $\{1,STS,ST^2S\}$ & $\left(1,-2,-2\right)^T$ & $Z_4^{TSU}$ & $\{SU,T,STS,TST^2U\}$ & $\left(1, \omega, \omega^2\right)^T$ \\ \hline

   $Z_4^{ST^2 U}$ & $\{SU,T^2,ST^2S,T^2STU\}$ & $\left(1, \omega^2, \omega \right)^T$  & $Z_4^{TST^2 U}$ & $\{1,S,TST^2U,T^2STU\}$ & $\left(1,1,1\right)^T$    \\  \hline \hline

   \multicolumn{6}{|c|}{VEVs of $\phi_{\text{sol}}$ }   \\ \hline
$G_{\text{sol}}$ & $X_{\text{sol}}$ & $\langle\phi_{\text{sol}}\rangle/v_{\phi_{s}}$ & $G_{\text{sol}}$ & $X_{\text{sol}}$ & $\langle\phi_{\text{sol}}\rangle/v_{\phi_{s}}$ \\ \hline

\multirow{2}{*}{$Z_2^U$} & $\{1,U\}$  &  $\left(1, x, x\right)^T$  & \multirow{2}{*}{$Z_2^{TU}$} & $\{U,T\}$ &  $\left(1,x\omega , x\omega^2 \right)^T$ \\ \cline{2-3} \cline{5-6}
 & $\{S,SU\}$ & $\left(1+2i x, 1-i x,1-i x\right)^T$ & & $\{STS,T^2STU\}$ & $\left(1+2 i x, \omega  (1-i x), \omega ^2 (1-i x)\right)^T$ \\ \hline

 \multirow{2}{*}{$Z_2^{SU}$} & $\{1,SU\}$  &   $\left(1, x, 2-x\right)^T$  & \multirow{2}{*}{$Z_2^{T^2 U}$} & $\{U,T^2\}$ &  $\left(1,x\omega^2, x\omega \right)^T$ \\ \cline{2-3} \cline{5-6}
 & $\{S,U\}$ &  $\left(1, 1+i x, 1-i x\right)^T$ & & $\{ST^2S,TST^2U\}$ & $\left(1+2 i x, \omega ^2 (1-i x),\omega  (1-i x)\right)^T$ \\ \hline

 \multirow{2}{*}{$Z_2^{STSU}$} & $\{U,STS\}$  &  $\left(\frac{\sqrt{3} x-1}{2} , 1+i x, 1-i x\right)^T$  & \multirow{2}{*}{$Z_2^{ST^2SU}$} & $\{U,ST^2S\}$ &  $\left(\frac{-1-\sqrt{3}x}{2}, 1+ix, 1-ix\right)^T$ \\ \cline{2-3} \cline{5-6}
 & $\{T,T^2STU\}$ & $\left(1, (2x+1)\omega , (1-2x)\omega^2\right)^T$   & & $\{T^2,TST^2U\}$ & $\left(1, (2x+1)\omega^2 , (1-2x)\omega \right)^T$  \\ \hline

$Z_2^S$ & $\{1,U,S,SU\}$  & $\left(1,1,1\right)^T$ & $Z_2^{TST^2}$ & $\{U,T^2,TST^2U,ST^2S\}$ & $\left(1, \omega^2, \omega\right)^T$ \\ \hline

$Z_2^{T^2ST }$  & $\{U,T,T^2STU,STS\}$ &  $\left(1, \omega, \omega^2\right)^T$ &  $Z_3^T$ & $\{1,U,T,TU,T^2,T^2U\}$ &  $\left(1, 0, 0\right)^T$ \\ \hline

 $Z_3^{ST}$ & $\{U,S,STU,STS,T^2SU,T^2\}$ &   $\left(1,-2\omega^2,-2\omega\right)^T$ & $Z_3^{TS}$ & $\{U,S,TSU,T,ST^2U,ST^2S\}$  & $\left(1,  -2\omega, -2\omega^2\right)^T$ \\ \hline

 $Z_3^{STS}$ & $\{1,U,STS,STSU,ST^2S,ST^2SU\}$ &  $\left(1,-2, -2\right)^T$  & $K_4^{(S, U)}$ & $\{1,U,S,SU\}$ &  $\left(1, 1, 1\right)^T$ \\ \hline

 $K_4^{\left(TST^2, T^2U\right)}$ & $\{U,T^2,TST^2U,ST^2S\}$ &  $\left(1, \omega^2, \omega \right)^T$  & $K_4^{\left(T^2ST, TU\right)}$  & $\{U,T,T^2STU,STS\}$ &  $\left(1, \omega, \omega ^2\right)^T$  \\ \hline \hline

\end{tabular}
\caption{\label{tab:inv_VEV_CP} The possible residual symmetries and the corresponding constraints on the vacuum configurations of the flavon fields $\phi_{\text{atm}}$ and $\phi_{\text{sol}}$ which transform as $\mathbf{3}$ and $\mathbf{3^\prime}$ respectively. The parameter $x$ is a generic real parameter. The VEVs of $\phi_{\text{atm}}$ invariant under the actions of the four $K_{4}$ subgroups are $(0,0,0)^T$. The VEVs of $\phi_{\text{sol}}$ invariant under the three $Z_{4}$ subgroups and the normal subgroup $K^{(S, TST^{2})}_4$ are also $(0,0,0)^T$. Comparing with table~\ref{tab:res_CP}, for some residual flavor subgroups we only show the invariant vacuum alignments for part of the CP transformations consistent with them. The reason is that the invariant VEVs for the remaining compatible CP transformations can be obtain by multiplying an overall factor $i$ from the above given VEVs, and the contribution of the overall $i$ can be be compensated by shifting the sign of the couplings $x_{\text{atm}}$ and $x_{\text{sol}}$.
}
\end{table}

In order to single out all independent viable breaking patterns from all possible breaking patterns in tri-direct CP approach. We will first find all possible independent pairs of $\{G_{l},G_{\text{atm}}\rtimes H^{\text{atm}}_{CP}, G_{\text{sol}}\rtimes H^{\text{sol}}_{CP}\}$ which are not related by group conjugation given in Eq.~\eqref{eq:conjugate}. In order to quantitatively assess how well a residual symmetry can describe the experimental data on mixing parameters and neutrino masses~\cite{Esteban:2018azc},  we define a $\chi^2$ function to estimate the goodness-of-fit of a chosen values of the input parameters,
\begin{equation}\label{eq:chisq}
\chi^2 = \sum_{i=1}^5 \left( \frac{P_i(x,\eta, m_a, r)-O_i}{\sigma_i}\right)^2\,,
\end{equation}
where the input parameters $m_{a}$, $r=m_{s}/m_{a}$ and $\eta$ are defined in Eq.~\eqref{eq:mnu}, the parameter $x$ parameterizes the vacuum of the flavon $\phi_{\text{sol}}$, $O_{i}$ denote the global best fit values of the observable quantities including the mixing angles $\sin^2\theta_{ij}$ and the mass splittings $\Delta m^2_{21}$ and $\Delta m^2_{3l}$ ($\Delta m^2_{3l}=\Delta m^2_{31}$ for NO and $\Delta m^2_{3l}=\Delta m^2_{32}$ for IO), and $\sigma_i$ refer to the $1\sigma$ deviations of the corresponding quantities. The values of $O_i$ and $\sigma_i$ are taken from the global data analysis~\cite{Esteban:2018azc}. $P_i\in\{\sin^2\theta_{12}, \sin^2\theta_{13}, \sin^2\theta_{23},  \Delta m_{21}^2,  \Delta m_{3l}^2\}$ are the theoretical predictions for the five physical observable quantities as functions of $x$, $\eta$, $m_a$, $r$. Here the contribution of the Dirac phase $\delta_{CP}$ is not included in the $\chi^2$ function. The reason is that the value of $\delta_{CP}$ is less constrained at present. For each set of the input parameters $x$, $\eta$, $m_{a}$ and $r$, we can extract the predictions for $P_i$ and the corresponding $\chi^2$. We have carried out the $\chi^2$ minimization. After performing the $\chi^2$ analysis for all possible breaking patterns in tri-direct CP approach, we find eight independent interesting mixing patterns with NO and eighteen interesting mixing patterns with IO. All the viable cases and the corresponding predictions for mixing parameters and neutrino masses are summarized in table~\ref{tab:bf_via_CP}. Then we proceed to study the eight NO viable cases (five cases in this section and three cases in appendix~\ref{sec:other_NO_mix}) one by one.

\begin{table}[hptb]
\renewcommand{\tabcolsep}{0.3mm}
\renewcommand{\arraystretch}{1.1}
\centering
\small
\begin{tabular}{|c|c|c| c c c c c c c c c|}  \hline \hline
\multicolumn{12}{|c|}{NO for $x$, $\eta$, $m_{a}$ and $r\equiv m_{s}/m_{a}$ being free parameters}   \\ \hline
& $(G_{l},G_{\text{atm}},G_{\text{sol}})$ & $X_{\text{sol}}$ &  $\chi^2_{\text{min}}$ &  $\sin^2\theta_{13}$  &$\sin^2\theta_{12}$  & $\sin^2\theta_{23}$  & $\delta_{CP}/\pi$ &  $\beta/\pi$  & $m_2(\text{meV})$  & $m_3(\text{meV})$ & $m_{ee}(\text{meV})$ \\   \hline
\multirow{2}{*}{$\mathcal{N}_{1}$} & \multirow{2}{*}{$(Z_3^T,Z_2^U,Z_2^{SU})$} & $1$ &  $0 .383$ & $0 .0224$ & $0 .318$ & $0 .580$ & $-0.386$ & $0 .335$ & $8 .597$ & $50 .249$ & $3 .100$ \\ \cline{3-12}
& & $U$ &  $0 .383$ & $0 .0224$ & $0 .318$ & $0 .580$ & $-0.386$ & $0 .910$ & $8 .597$ & $50 .249$ & $3 .725$ \\ \hline
\multirow{2}{*}{$\mathcal{N}_{2}$} & \multirow{2}{*}{$(Z_3^T,Z_3^{ST},Z_2^{SU})$} & $1$ &  $0 .383$ & $0 .0224$ & $0 .318$ & $0 .580$ & $-0.386$ & $0 .754$ & $8 .596$ & $50 .249$ & $3 .798$ \\ \cline{3-12}
& & $U$ &  $0 .383$ & $0 .0224$ & $0 .318$ & $0 .580$ & $-0.386$ & $0 .996$ & $8 .596$ & $50 .249$ & $3 .604$ \\ \hline
$\mathcal{N}_{3}$ & $(Z_3^T,Z_2^{S},Z_2^{SU})$ & $U$ &  $4 .321$ & $0 .0225$ & $0 .318$ & $0 .538$ & $-0.447$ & $0 .444$ & $8 .603$ & $50 .242$ & $3 .064$ \\ \hline
$\mathcal{N}_{4}$ & $(Z_3^T,Z_2^{TST^2},Z_2^{U})$ & $1$ &  $5 .081$ & $0 .0225$ & $0 .337$ & $0 .563$ & $-0.407$ & $0 .284$ & $8 .601$ & $50 .244$ & $2 .950$ \\ \hline
$\mathcal{N}_{5}$ &  $(K_4^{(S,U)},Z_2^{TU},Z_2^{TU})$ & $U$ & $20 .461$ & $0 .0225$ & $0 .256$ & $0 .582$ & $0$ & $-0.265$ & $8 .597$ & $50 .249$ & $3 .026$ \\ \hline
$\mathcal{N}_{6}$ & $(Z_4^{TSU},Z_3^T,Z_2^{SU})$ & $U$ &  $8 .698$ & $0 .0226$ & $0 .345$ & $0 .554$ & $-0.419$ & $0 .202$ & $8 .605$ & $50 .239$ & $2 .638$ \\ \hline
 \multirow{2}{*}{$\mathcal{N}_{7}$} & \multirow{2}{*}{$(K_4^{(S,TST^2)},Z_3^T,Z_2^{SU})$} & $1$ &  $12 .254$ & $0 .0224$ & $0 .328$ & $0 .513$ & $-0.482$ & $0 .502$ & $8 .600$ & $50 .245$ & $3 .099$ \\ \cline{3-12}
& & $U$ &  $11 .621$ & $0 .0224$ & $0 .327$ & $0 .514$ & $0$ & $0$ & $8 .601$ & $50 .244$ & $3 .877$ \\ \hline
$\mathcal{N}_{8}$ & $(K_4^{(S,TST^2)},Z_2^U,Z_2^{TU})$ & $U$ &  $5 .768$ & $0 .0228$ & $0 .298$ & $0 .537$ & $-0.451$ & $0 .365$ & $8 .539$ & $50 .326$ & $2 .615$ \\ \hline \hline

\multicolumn{12}{|c|}{IO for $x$, $\eta$, $m_{a}$ and $r\equiv m_{s}/m_{a}$ being free parameters}   \\ \hline
& $(G_{l},G_{\text{atm}},G_{\text{sol}})$ & $X_{\text{sol}}$ &  $\chi^2_{\text{min}}$ &  $\sin^2\theta_{13}$  &$\sin^2\theta_{12}$  & $\sin^2\theta_{23}$  & $\delta_{CP}/\pi$ &  $\beta/\pi$  & $m_1(\text{meV})$  & $m_2(\text{meV})$ & $m_{ee}(\text{meV})$ \\   \hline
$\mathcal{I}_{1}$ & $(Z_3^T,Z_3^{ST},Z_2^{U})$ & $1$ &  $17 .640$ & $0 .0226$ & $0 .310$ & $0 .5$ & $-0.928$ & $0 .306$ & $49 .377$ & $50 .120$ & $43 .792$ \\ \hline
$\mathcal{I}_{2}$ & $(Z_3^T,Z_2^{SU},Z_2^{TU})$ & $U$ &  $17 .640$ & $0 .0226$ & $0 .310$ & $0 .5$ & $-0.682$ & $0 .843$ & $49 .377$ & $50 .120$ & $21 .168$ \\ \hline
\multirow{2}{*}{$\mathcal{I}_{3}$} & \multirow{2}{*}{$(K_4^{(S,U)},Z_2^{TST^2},Z_2^{U})$} & $1$ & $17 .640$ & $0 .0226$ & $0 .310$ & $0 .5$ & $-0.495$ & $0 .102$ & $49 .377$ & $50 .120$ & $47 .946$ \\ \cline{3-12}
& & $S$ &  $17 .640$ & $0 .0226$ & $0 .310$ & $0 .5$ & $-0.495$ & $0 .102$ & $49 .377$ & $50 .120$ & $47 .946$ \\ \hline
$\mathcal{I}_{4}$ & $(K_4^{(S,U)},Z_2^{TU},Z_2^{TU})$ & $U$ &  $20 .419$ & $0 .0227$ & $0 .256$ & $0 .582$ & $0$ & $1$ & $49 .377$ & $50 .120$ & $23 .384$ \\ \hline
$\mathcal{I}_{5}$ & $(Z_3^T,Z_2^{SU},Z_2^{SU})$ & $U$ &  $18 .008$ & $0 .0227$ & $0 .318$ & $0 .5$ & $-0.5$ & $0 .743$ & $49 .377$ & $50 .120$ & $24 .840$  \\ \hline
$\mathcal{I}_{6}$ & $(Z_3^T,Z_2^{TST^2},Z_2^{U})$ & $1$ &  $17 .640$ & $0 .0226$ & $0 .310$ & $0 .5$ & $0 .913$ & $-0.389$ & $49 .377$ & $50 .120$ & $41 .048$ \\ \hline
$\mathcal{I}_{7}$ & $(Z_3^T,Z_2^{U},Z_2^{TU})$ & $U$ &  $17 .640$ & $0 .0226$ & $0 .310$ & $0 .5$ & $0 .975$ & $-0.175$ & $49 .377$ & $50 .120$ & $46 .918$ \\ \hline
$\mathcal{I}_{8}$ & $(Z_3^T,Z_2^{U},Z_2^{STSU})$ & $U$ & $17 .640$ & $0 .0226$ & $0 .310$ & $0 .5$ & $-0.761$ & $0 .759$ & $49 .377$ & $50 .119$ & $24 .569$ \\ \hline
$\mathcal{I}_{9}$ & $(Z_3^T,Z_2^{SU},Z_2^{STSU})$ & $U$ &  $17 .640$ & $0 .0226$ & $0 .310$ & $0 .5$ & $-0.954$ & $0 .249$ & $49 .377$ & $50 .120$ & $45 .347$ \\ \hline

\multirow{2}{*}{$\mathcal{I}_{10}$} & \multirow{2}{*}{$(Z_4^{TSU},Z_2^S,Z_2^{TU})$} & $U$ &  $17 .640$ & $0 .0226$ & $0 .310$ & $0 .5$ & $-0.00465$ & $-0.102$ & $49 .377$ & $50 .120$ & $47 .946$ \\ \cline{3-12}
& & $STS$ &  $17 .640$ & $0 .0226$ & $0 .310$ & $0 .5$ & $-0.00465$ & $-0.102$ & $49 .377$ & $50 .120$ & $47 .946$  \\ \hline
\multirow{2}{*}{$\mathcal{I}_{11}$} & \multirow{2}{*}{$(Z_4^{TSU},Z_2^S,Z_2^{T^2U})$} & $U$ & $17 .640$ & $0 .0226$ & $0 .310$ & $0 .5$ & $-0.128$ & $-0.548$ & $49 .377$ & $50 .120$ & $34 .480$  \\ \cline{3-12}
&& $ST^2S$ &  $17 .640$ & $0 .0226$ & $0 .310$ & $0 .5$ & $-0.372$ & $0 .548$ & $49 .377$ & $50 .120$ & $34 .480$ \\ \hline
$\mathcal{I}_{12}$ & $(Z_4^{TSU},Z_2^U,Z_2^{TU})$ & $U$ & $17 .640$ & $0 .0226$ & $0 .310$ & $0 .5$ & $-0.772$ & $0 .729$ & $49 .377$ & $50 .120$ & $25 .920$ \\ \hline
$\mathcal{I}_{13}$ & $(Z_4^{TSU},Z_2^{TU},Z_2^{U})$ & $1$ & $17 .640$ & $0 .0226$ & $0 .310$ & $0 .5$ & $0 .834$ & $-0.636$ & $49 .377$ & $50 .120$ & $30 .323$ \\ \hline
\multirow{2}{*}{$\mathcal{I}_{14}$} & \multirow{2}{*}{$(K_4^{(S,TST^2)},Z_3^T,Z_2^{SU})$} & $1$ &  $17 .640$ & $0 .0226$ & $0 .310$ & $0 .5$ & $-0.104$ & $-0.448$ & $49 .377$ & $50 .120$ & $38 .772$ \\ \cline{3-12}
& & $U$ &  $2 .046$ & $0 .0225$ & $0 .310$ & $0 .607$ & $-0.604$ & $-0.448$ & $49 .377$ & $50 .120$ & $38 .778$ \\ \hline
$\mathcal{I}_{15}$ & $(K_4^{(S,TST^2)},Z_2^U,Z_2^{TU})$ & $U$ &  $17 .640$ & $0 .0226$ & $0 .310$ & $0 .5$ & $-0.666$ & $-0.636$ & $49 .377$ & $50 .120$ & $30 .323$ \\ \hline
$\mathcal{I}_{16}$ & $(K_4^{(S,U)},Z_2^{TST^2},Z_2^{TU})$ & $STS$ & $17 .640$ & $0 .0226$ & $0 .310$ & $0 .5$ & $-0.872$ & $0 .548$ & $49 .377$ & $50 .120$ & $34 .480$ \\ \hline
$\mathcal{I}_{17}$ & $(K_4^{(S,U)},Z_2^{TU},Z_2^{U})$ & $S$ & $28 .676$ & $0 .0225$ & $0 .310$ & $0 .477$ & $0 .915$ & $-0.548$ & $49 .377$ & $50 .120$ & $34 .486$ \\ \hline
$\mathcal{I}_{18}$ & $(K_4^{(S,U)},Z_2^{TU},Z_2^{T^2U})$ & $ST^2S$ & $9 .241$ & $0 .0227$ & $0 .310$ & $0 .523$ & $-0.743$ & $0 .510$ & $49 .377$ & $50 .120$ & $36 .178$ \\ \hline \hline
\end{tabular}
\caption{\label{tab:bf_via_CP}The predictions for the lepton mixing angles, CP violation phases, neutrino masses and the effective Majorana mass $m_{ee}$ in neutrinoless double beta decay for all viable residual symmetries, where the parameters $x$, $\eta$, $m_{a}$ and $r\equiv m_{s}/m_{a}$ are treated as free parameters. The residual CP transformation associated with atmospheric neutrino can be read out from table~\ref{tab:inv_VEV_CP}. We only show one representative residual CP transformation of the solar neutrino sector since the other residual CP transformations can be obtained by multiplying the residual flavor symmetry $G_{\text{sol}}$ with the given CP transformation from the left-hand side.
}
\end{table}

\begin{description}[labelindent=-0.8em, leftmargin=0.3em]
\item[~~($\mathcal{N}_{1}$)]{$(G_{l},G_{\text{atm}},G_{\text{sol}})=(Z_3^T,Z_2^{U},Z_2^{SU})$, $X_{\text{atm}}=\{1,U\}$}

$\bullet$ $X_{\text{sol}}=\{1,SU\}$

For this breaking pattern, the charged lepton mass matrix $m^{\dagger}_lm_l$ is diagonal such that the unitary transformation $U_{l}$ is the identity matrix, as shown in Eq.~\eqref{eq:ch_dia_matrix}. From table~\ref{tab:res_CP}, we find that there are 4 possible residual CP transformations which are compatible with the residual family symmetry $Z_2^{U}$ in the atmospheric neutrino sector. For the residual CP transformations $X_{\text{atm}}=\{1,U\}$, the VEV alignment of flavon $\phi_{\text{atm}}$ is
\begin{equation}\label{eq:atm_VEV_N1_1}
\langle\phi_{\text{atm}}\rangle= v_{\phi_a}\left( 0, 1 , -1\right)^T\,,
\end{equation}
where $v_{\phi_a}$ is a real parameter with dimension of mass. For other two residual CP transformations $X_{\text{atm}}=\{S,SU\}$, the alignment of flavon $\phi_{\text{atm}}$ is
\begin{equation}\label{eq:atm_VEV_N1_2}
\langle\phi_{\text{atm}}\rangle= iv_{\phi_a}\left( 0, 1 , -1\right)^T\,,
\end{equation}
which differs from the vacuum configuration of Eq.~\eqref{eq:atm_VEV_N1_1} by an overall factor $i$. We see from Eq.~\eqref{eq:mnu} that this overall factor $i$ can be absorbed into the sign of the coupling constant $x_{\text{atm}}$. Hence the two alignments in Eq.~\eqref{eq:atm_VEV_N1_1} and Eq.~\eqref{eq:atm_VEV_N1_2} will give rise to the same light neutrino mass matrix, and it is sufficient to consider one of them. For other residual symmetries discussed in the following, if two alignments of $\phi_{\text{atm}}$ differ in an overall $i$ only one of them would be studied as well. Without loss of generality, here we shall choose the atmospheric vacuum in Eq.~\eqref{eq:atm_VEV_N1_1}, i.e. the residual CP is $X_{\text{atm}}=\{1,U\}$ in the atmospheric neutrino sector.

Firstly we consider the solar residual CP transformations $X_{\text{sol}}=\{1,SU\}$, then the VEV of flavon field  $\phi_{\text{sol}}$ reads as
\begin{equation}\label{eq:sol_VEV_N1}
\langle\phi_{\text{sol}}\rangle=v_{\phi_s}\left(1, x, 2-x\right)^T\,,
\end{equation}
where $x$ is a dimensionless real number and $v_{\phi_s}$ is a real parameter with dimension of mass. Consequently the Dirac neutrino mass matrix $m_D$ and the heavy right-handed neutrino Majorana mass matrix $m_{N}$ take the following form
\begin{equation}
m_{D}=\begin{pmatrix}
 0 &~ y_{\text{sol}}v_{\phi_s} \\
 -y_{\text{atm}}v_{\phi_a} &~ (2-x)y_{\text{sol}}v_{\phi_s} \\
 y_{\text{atm}}v_{\phi_a} &~  xy_{\text{sol}}v_{\phi_s} \\
\end{pmatrix}\,,
\qquad m_{N}=\begin{pmatrix}
x_{\textrm{atm}}\langle\xi_{\text{atm}}\rangle  &  0  \\
0  &   x_{\textrm{sol}}\langle\xi_{\text{sol}}\rangle
\end{pmatrix}
\end{equation}
where the couplings $y_{\text{atm}}$ and $y_{\text{sol}}$ are real since the theory is invariant under CP. Using the seesaw formula, we can obtain the low energy effective light neutrino mass matrix
\begin{equation}\label{eq:mnu1}
 m_{\nu} =m_{a}\begin{pmatrix}
 0 &~ 0 ~& 0 \\
 0 &~ 1 ~& -1 \\
 0 &~ -1 ~& 1 \\
\end{pmatrix}+m_{s}e^{i\eta}
\begin{pmatrix}
 1 &~ 2-x &~ x \\
 2-x &~ (x-2)^2 &~ (2-x) x \\
 x &~ (2-x) x &~ x^2 \\
\end{pmatrix}\,,
\end{equation}
where an overall unphysical phase has been omitted and it would be neglected hereinafter for the other cases, and the parameters $m_{a}$, $m_s$ and $\eta$ are defined in Eq.~\eqref{eq:mnu}. We find that the above neutrino mass matrix $m_{\nu}$ fulfills
\begin{equation}
m_{\nu}\begin{pmatrix}
2 \\
-1 \\
-1
\end{pmatrix}=\left(\begin{array}{c}
0\\
0\\
0
\end{array}\right)\,.
\end{equation}
It implies that the column vector $(2,-1, -1)^{T}$ is an eigenvector of $m_\nu$ with zero eigenvalue. Subsequently we follow the procedure given in section~\ref{sec:framework} to perform a unitary transformation $U_{\nu1}$,
\begin{equation}\label{eq:uni_tra_mu}
m^\prime_{\nu}=U^T_{\nu1}m_{\nu}U_{\nu1}\,,
\end{equation}
with
\begin{equation}\label{eq:Unu1_TB1}
U_{\nu1}=U_{TB}=
\begin{pmatrix}
 \sqrt{\frac{2}{3}} &~ \frac{1}{\sqrt{3}} &~ 0 \\
 -\frac{1}{\sqrt{6}} &~ \frac{1}{\sqrt{3}} &~ -\frac{1}{\sqrt{2}} \\
 -\frac{1}{\sqrt{6}} &~ \frac{1}{\sqrt{3}} &~ \frac{1}{\sqrt{2}} \\
\end{pmatrix}\,,
\end{equation}
where $U_{TB}$ is well-known tri-bimaximal (TB) mixing matrix. The neutrino mass matrix $m'_{\nu}$ is block diagonal, and its entries are given by
\begin{equation}
y=3m_{s}e^{i \eta }, \quad
z=\sqrt{6}  \left(x-1\right)m_{s}e^{i \eta },  \quad
w=2\left(m_{a}+(x-1)^2m_{s}e^{i \eta }\right)  \,.
\end{equation}
Furthermore, $m'_{\nu}$ be diagonalized by the unitary matrix $U_{\nu2}$ given in Eq.~\eqref{eq:Unu2}, i.e.
\begin{equation}
U^{T}_{\nu2}m^{\prime}_{\nu}U_{\nu2}=\text{diag}(0, m_2, m_3)\,.
\end{equation}
The expressions of $m_2$, $m_3$ and $U_{\nu2}$ can be straightforwardly obtained by inserting the parameters $y$, $z$ and $w$ into Eqs.~(\ref{eq:nu_masses}, \ref{eq:theta}, \ref{eq:par_prs}). As a result, we can read out the lepton mixing matrix as
\begin{equation}\label{eq:PMNS_UTM1_1}
U_{PMNS}=
\begin{pmatrix}
 \sqrt{\frac{2}{3}} &~ \frac{\cos \theta }{\sqrt{3}} &~ \frac{e^{i \psi } \sin \theta }{\sqrt{3}} \\
 -\frac{1}{\sqrt{6}} &~ \frac{\cos \theta }{\sqrt{3}}+\frac{e^{-i \psi } \sin \theta }{\sqrt{2}} &~ \frac{e^{i \psi } \sin \theta }{\sqrt{3}}-\frac{\cos \theta }{\sqrt{2}} \\
 -\frac{1}{\sqrt{6}} &~ \frac{\cos \theta }{\sqrt{3}}-\frac{e^{-i \psi } \sin \theta }{\sqrt{2}} &~ \frac{\cos \theta }{\sqrt{2}}+\frac{e^{i \psi } \sin \theta }{\sqrt{3}} \\
\end{pmatrix}P_{\nu}\,,
\end{equation}
with
\begin{equation}\label{eq:dia_Pnu}
P_{\nu}=\text{diag}(1, e^{i(\psi+\rho)/2}, e^{i(-\psi+\sigma)/2})\,.
\end{equation}
In the following, the Majorana phase matrix $P_{\nu}$ would be omitted for simplicity. We see that the neutrino mixing matrix is the so-called TM1 mixing pattern in which the first column of the tri-bimaximal mixing is preserved. The three lepton mixing angles for this mixing matrix are
\begin{equation}\label{eq:angles_TM1_1}
\sin^2\theta_{13}=\frac{\sin ^2\theta }{3}\,, \qquad \sin^2\theta_{12}=\frac{2 \cos ^2\theta }{5+\cos 2 \theta }\,, \qquad \sin^2\theta_{23}=\frac{1}{2}-\frac{ \sqrt{6} \sin 2 \theta  \cos \psi  }{5+\cos 2 \theta}\,.
\end{equation}
Eliminating the free parameter $\theta$, a sum rule between the solar mixing angle $\theta_{12}$ and the reactor mixing angle $\theta_{13}$ is found,
\begin{equation}\label{eq:sum_rule_TM1_1}
\cos^2\theta_{12}\cos^2\theta_{13}=\frac{2}{3}\,.
\end{equation}
Plugging in the best fit value of $\sin^2\theta_{13}=0.02241$~\cite{Esteban:2018azc}, we find the solar mixing angle is
\begin{equation}
\label{eq:sinSq12_TM1}\sin^2\theta_{12}\simeq0.318\,,
\end{equation}
which is within the $3\sigma$ region~\cite{Esteban:2018azc}. For the lepton mixing matrix in Eq.~\eqref{eq:PMNS_UTM1_1}, the two CP invariants are given by
\begin{equation}\label{eq:invariants_TM1_1}
J_{CP}=\frac{\sin 2\theta   \sin \psi }{6 \sqrt{6}}\,, \qquad I_{1}=\frac{1}{36} \sin ^22 \theta  \sin (\rho -\sigma )\,.
\end{equation}
The so-called TM1 mixing matrix indicates the following sum rule among the Dirac CP phase $\delta_{CP}$ and mixing angles
\begin{equation}
\label{eq:sum_rule2_TM1_1}\cos\delta_{CP}=\frac{(3-5 \cos2 \theta_{13}) \cot 2 \theta_{23}}{4\sin\theta_{13} \sqrt{3 \cos 2 \theta_{13}-1}}\,.
\end{equation}
It is easy to check that $\theta_{23}=\pi/4$ leads to $\cos\delta_{CP}=0$ which corresponds to maximal CP violation $\delta_{CP}=\pm\pi/2$. The neutrino masses $m_2$ and $m_3$ depend on all the four input parameters $x$, $\eta$, $m_{a}$ and $m_s$ while mixing parameters and mass ratio $m^2_2/m^2_3$ only depend on $x$, $\eta$ and $r\equiv m_{s}/m_{a}$. In the case $\eta$, $m_a$ and $r$ being free parameters, we find that the experimental data on the mixing angles and the neutrino masses can be achieved for some special $x$.

\begin{table}[t!]
\renewcommand{\tabcolsep}{0.5mm}
\renewcommand{\arraystretch}{1.3}
\footnotesize
\centering
\begin{tabular}{|c| c| c|c| c | c| c| c| c| c| c |c | c| c| c|}  \hline \hline
$\langle\phi_{\text{sol}}\rangle/v_{\phi_{s}}$ & $x$   & $\eta$ & $m_{a}$(meV) & $r$  	 & $\chi^2_{\text{min}}$ &  $\sin^2\theta_{13}$  &$\sin^2\theta_{12}$  & $\sin^2\theta_{23}$  & $\delta_{CP}/\pi$ &  $\beta/\pi$  & $m_2(\text{meV})$ & $m_3(\text{meV})$ & $m_{ee}(\text{meV})$\\   \hline
$\left(1,3,-1\right)^{T}$ & $3$ & $\pm\frac{2 \pi }{3}$ & $26 .843$ & $0 .0998$ & $19 .625$ & $0 .0222$ & $0 .318$ & $0 .488$ & $\mp0.516$ & $\mp0.403$ & $8 .586$ & $50 .263$ & $2 .680$ \\ \hline
$\left(1,-1,3\right)^{T}$ & $-1$ & $\pm\frac{2 \pi }{3}$ & $26 .798$ & $0 .101$ & $10 .716$ & $0 .0225$ & $0 .318$ & $0 .513$ & $\pm0 .482$ & $\mp0.401$ & $8 .628$ & $50 .212$ & $2 .694$ \\ \hline
\multirow{2}{*}{$\left(1,4,-2\right)^{T}$} & \multirow{2}{*}{$4$} & $\pm\frac{4 \pi }{5}$ & $35 .249$ & $0 .0565$ & $14 .196$ & $0 .0241$ & $0 .317$ & $0 .575$ & $\mp0.398$ & $\mp0.474$ & $8 .316$ & $50 .609$ & $1 .990$ \\ \cline{3-14}
&& $\pm\frac{5 \pi }{6}$ & $36 .720$ & $0 .0532$ & $3 .841$ & $0 .0227$ & $0 .318$ & $0 .610$ & $\mp0.338$ & $\mp0.554$ & $8 .560$ & $50 .297$ & $1 .954$ \\ \hline
$\left(1,-2,4\right)^{T}$ & $-2$ & $\pm\frac{4 \pi }{5}$ & $35 .242$ & $0 .0566$ & $68 .409$ & $0 .0243$ & $0 .317$ & $0 .425$ & $\pm0 .601$ & $\mp0.473$ & $8 .339$ & $50 .581$ & $1 .995$ \\ \hline
\multirow{2}{*}{$\left(1,\frac{7}{2},-\frac{3}{2}\right)^{T}$} & \multirow{2}{*}{$\frac{7}{2}$} & $\pm\frac{3 \pi }{4}$ & $31 .121$ & $0 .0734$ & $6 .567$ & $0 .0231$ & $0 .318$ & $0 .541$ & $\mp0.444$ & $\mp0.447$ & $8 .462$ & $50 .425$ & $2 .285$ \\ \cline{3-14}
&& $\pm\frac{4 \pi }{5}$ & $33 .006$ & $0 .0674$ & $9 .388$ & $0 .0210$ & $0 .319$ & $0 .589$ & $\mp0.366$ & $\mp0.544$ & $8 .806$ & $49 .994$ & $2 .223$ \\ \hline
$\left(1,\frac{8}{3},-\frac{2}{3}\right)^{T}$ & $\frac{8}{3}$ & $\pm\frac{3 \pi }{5}$ & $24 .618$ & $0 .121$ & $45 .788$ & $0 .0209$ & $0 .319$ & $0 .456$ & $\mp0.564$ & $\mp0.385$ & $8 .841$ & $49 .949$ & $2 .990$ \\ \hline
$\left(1,\frac{10}{3},-\frac{4}{3}\right)^{T}$ & $\frac{10}{3}$ & $\pm\frac{3 \pi }{4}$ & $30 .566$ & $0 .0777$ & $4 .332$ & $0 .0218$ & $0 .318$ & $0 .548$ & $\mp0.432$ & $\mp0.474$ & $8 .689$ & $50 .139$ & $2 .375$ \\ \hline
$\left(1,-\frac{1}{2},\frac{5}{2}\right)^{T}$ & $-\frac{1}{2}$ & $\pm\frac{\pi }{2}$ & $22 .359$ & $0 .145$ & $2 .475$ & $0 .0220$ & $0 .318$ & $0 .599$ & $\pm0 .354$ & $\mp0.316$ & $8 .672$ & $50 .158$ & $3 .242$ \\ \hline
$\left(1,-\frac{3}{2},\frac{7}{2}\right)^{T}$ & $-\frac{3}{2}$ & $\pm\frac{3 \pi }{4}$ & $31 .101$ & $0 .0737$ & $35 .893$ & $0 .0233$ & $0 .317$ & $0 .460$ & $\pm0 .555$ & $\mp0.445$ & $8 .493$ & $50 .386$ & $2 .293$ \\ \hline
$\left(1,-\frac{2}{3},\frac{8}{3}\right)^{T}$ & $-\frac{2}{3}$ & $\pm\frac{3 \pi }{5}$ & $24 .560$ & $0 .123$ & $13 .654$ & $0 .0212$ & $0 .319$ & $0 .545$ & $\pm0 .435$ & $\mp0.383$ & $8 .888$ & $49 .890$ & $3 .009$ \\ \hline
$\left(1,-\frac{4}{3},\frac{10}{3}\right)^{T}$ & $-\frac{4}{3}$ & $\pm\frac{3 \pi }{4}$ & $30 .550$ & $0 .0780$ & $38 .724$ & $0 .0220$ & $0 .318$ & $0 .453$ & $\pm0 .567$ & $\mp0.472$ & $8 .716$ & $50 .105$ & $2 .383$ \\  \hline
$\left(1,-\frac{3}{4},\frac{11}{4}\right)^{T}$ & $-\frac{3}{4}$ & $\pm\frac{3 \pi }{5}$ & $24 .578$ & $0 .120$ & $2 .837$ & $0 .0222$ & $0 .318$ & $0 .551$ & $\pm0 .429$ & $\mp0.367$ & $8 .668$ & $50 .164$ & $2 .948$  \\ \hline
$\left(1,-\frac{5}{4},\frac{13}{4}\right)^{T}$ & $-\frac{5}{4}$ & $\pm\frac{3 \pi }{4}$ & $30 .265$ & $0 .0802$ & $46 .446$ & $0 .0213$ & $0 .319$ & $0 .450$ & $\pm0 .573$ & $\mp0.486$ & $8 .824$ & $49 .971$ & $2 .428$ \\ \hline

$\left(1,-\frac{3}{5},\frac{13}{5}\right)^{T}$ & $-\frac{3}{5}$ & $\pm\frac{\pi }{2}$ & $22 .215$ & $0 .142$ & $11 .399$ & $0 .0232$ & $0 .317$ & $0 .606$ & $\pm0 .347$ & $\mp0.297$ & $8 .312$ & $50 .614$ & $3 .156$ \\ \hline
$\left(1,-\frac{4}{5},\frac{14}{5}\right)^{T}$ & $-\frac{4}{5}$ & $\pm\frac{3 \pi }{5}$ & $24 .587$ & $0 .118$ & $2 .595$ & $0 .0228$ & $0 .318$ & $0 .554$ & $\pm0 .425$ & $\mp0.357$ & $8 .532$ & $50 .335$ & $2 .911$ \\ \hline
$\left(1,-\frac{6}{5},\frac{16}{5}\right)^{T}$ & $-\frac{6}{5}$ & $\pm\frac{3 \pi }{4}$ & $30 .090$ & $0 .0816$ & $53 .229$ & $0 .0208$ & $0 .319$ & $0 .448$ & $\pm0 .577$ & $\mp0.494$ & $8 .887$ & $49 .891$ & $2 .456$ \\ \hline
$\left(1,-\frac{7}{5},\frac{17}{5}\right)^{T}$ & $-\frac{7}{5}$ & $\pm\frac{3 \pi }{4}$ & $30 .772$ & $0 .0762$ & $35 .632$ & $0 .0225$ & $0 .318$ & $0 .455$ & $\pm0 .562$ & $\mp0.461$ & $8 .627$ & $50 .213$ & $2 .346$ \\ \hline
$\left(1,-\frac{5}{6},\frac{17}{6}\right)^{T}$ & $-\frac{5}{6}$ & $\pm\frac{3 \pi }{5}$ & $24 .592$ & $0 .117$ & $4 .998$ & $0 .0231$ & $0 .318$ & $0 .556$ & $\pm0 .422$ & $\mp0.351$ & $8 .441$ & $50 .452$ & $2 .886$ \\ \hline \hline

\end{tabular}
\caption{\label{tab:bf_NO1_1}The predictions for the lepton mixing angles, CP violation phases, neutrino masses and the effective Majorana mass $m_{ee}$ for the breaking pattern $\mathcal{N}_{1}$ with $(G_{l},G_{\text{atm}},G_{\text{sol}})= (Z^{T}_{3},Z^{U}_{2},Z^{SU}_{2}) $ and $X_{\text{sol}}=\{1,SU\}$. Here we choose many benchmark values for the parameters $x$ and $\eta$. Notice that the lightest neutrino mass is vanishing $m_1=0$. }
\end{table}

In order to show concrete examples, some benchmark values of the parameter $x$ and $\eta$ are considered and the numerical results of the mixing parameters and neutrino masses are listed in table~\ref{tab:bf_NO1_1}. The solar flavon alignment $\phi_{\text{sol}}$ for these representative values of $x$ takes a relatively simple form, consequently we expect it should be not difficult to be realized dynamically in an explicit model. We show the predictions for the effective Majorana mass $m_{ee}$ in neutrinoless double beta decay in the last column of table~\ref{tab:bf_NO1_1}, where the effective mass $m_{ee}$ is defined as~\cite{Tanabashi:2018oca},
\begin{equation}
m_{ee}=|m_{1}U^2_{e1}+m_{2}U^2_{e2}+m_{3}U^2_{e3}|\,.
\end{equation}
From table~\ref{tab:bf_NO1_1}, we can see that the measured values of the lepton mixing angles and the mass splittings $\Delta m^2_{21}$ and $\Delta m^2_{31}$ can be accommodated for certain choices of $x$, $\eta$, $m_{a}$ and $r$. For the benchmark value $x=-1$, the solar flavon alignment is  $\langle\phi_{\text{sol}}\rangle=(1, -1, 3)^Tv_{\phi_s}$, it is exactly the Littlest seesaw model with CSD(3) which is originally proposed in~\cite{King:2015dvf}. The solar vacuum $\langle\phi_{\text{sol}}\rangle=(1, -3, 1)^Tv_{\phi_s}$ for $x=3$ corresponds to another version of Littlest seesaw~\cite{King:2016yvg}. Moreover, the value $x=4$ leads to the vacuum $\langle\phi_{\text{sol}}\rangle=(1, 4, -2)^Tv_{\phi_s}$, the CSD(4) scenario~\cite{King:2013xba} is reproduced. From table~\ref{tab:bf_NO1_1}, we see that a smaller $\chi^2$ than the original LS model~\cite{King:2015dvf,King:2016yvg,King:2013xba,Ballett:2016yod} can be achieved for the values of $x=-1/2$ and $\eta=\pm\pi/2$, the corresponding vacuum alignment $\langle\phi_{\text{sol}}\rangle\propto \left(2,-1,5\right)$ seems simple and it should be easy to realize in a concrete model.

Furthermore, we perform a comprehensive numerical analysis. The three input parameters $x$, $r$ and $\eta$ are randomly scanned over $x\in[-20, 20]$, $r\in[0, 20]$ and $\eta\in[-\pi, \pi]$. We only keep the points for which the resulting mixing angles $\sin^2\theta_{ij}$ and the mass ratio $m^2_2/m^2_3$ are in the experimentally preferred $3\sigma$ regions~\cite{Esteban:2018azc}. The parameter $m_a$ can be fixed by requiring that the individual squared mass differences $\Delta m^2_{21}$ and $\Delta m^2_{31}$ are reproduced. Then the predictions for the CP violating phase $\delta_{CP}$ and $\beta$ and the neutrino masses as well as $m_{ee}$ can be extracted. In the end we find the allowed regions of the parameter $x$ is $[-2.072,-0.287]\cup[2.463,4.683]$, of the parameter $|\eta|$ is $[0.414\pi,0.861\pi]$ and of the parameter $r$ is $[0.0400,0.166]$. As regards the predictions for the mixing angles, we find that any values of $\sin^2\theta_{13}$ and $\sin^2\theta_{23}$ in their $3\sigma$ ranges can be achieved. The the solar mixing angle is in a narrow region $0.317\leq\sin^2\theta_{12}\leq0.319$, which arise from the TM1 sum rule in Eq.~\eqref{eq:sum_rule_TM1_1}. The predicted ranges of Dirac CP phase $|\delta_{CP}|$ and Majorana CP phase $|\beta|$ are $[0.299\pi,0.624\pi]$ and $[0.273\pi,0.608\pi]$, respectively. These predictions may be tested at future long baseline experiments, as discussed in~\cite{Ballett:2016yod}. The allowed ranges of the mixing parameters for other breaking patterns are also obtained by randomly varying the parameters $x$, $r$ and $\eta$ in the ranges $x\in[-20, 20]$, $r\in[0, 20]$ and $\eta\in[-\pi, \pi]$. We will not explicitly mention this point in the following.

$\bullet$ $X_{\text{sol}}=\{S,U\}$

From table~\ref{tab:inv_VEV_CP}, we find that the VEV of $\phi_{\text{sol}}$ is proportional to $\left(1, 1+ix, 1-ix\right)^T$. Inserting the vacuum configuration of the flavons $\phi_{\text{atm}}$ and $\phi_{\text{sol}}$ into Eq.~\eqref{eq:mnu}, we obtain the light effective Majorana neutrino mass matrix is
\begin{equation}\label{eq:mnu_N12}
 m_{\nu}=m_{a}\begin{pmatrix}
 0 &~ 0 ~& 0 \\
 0 &~ 1 ~& -1 \\
 0 &~ -1 ~& 1 \\
\end{pmatrix}+m_{s}e^{i\eta}
\begin{pmatrix}
 1 &~ 1-i x &~ 1+i x \\
 1-i x &~ (1-ix)^2 &~ 1+x^2 \\
 1+i x &~ 1+x^2 &~ (1+i x)^2 \\
\end{pmatrix}\,,
\end{equation}
It is easy to check that the column vector $(2,-1, -1)^{T}$ is an eigenvector of $m_\nu$ with zero eigenvalue. In order to diagonalize light neutrino mass matrix $m_{\nu}$ in above equation, we first perform a unitary transformation $U_{\nu1}$, where $U_{\nu1}$ is taken to be TB mixing matrix $U_{TB}$. Then the neutrino mass matrix $m^\prime_{\nu}$ is of block diagonal form with nonzero elements $y$, $z$ and $w$,
\begin{equation}
y=3m_{s}  e^{i \eta }, \quad
z=i\sqrt{6}xm_{s}  e^{i \eta },  \quad
w=2 \left(m_{a}-x^2m_{s} e^{i \eta } \right)\,.
\end{equation}
The neutrino mass matrix $m^\prime_{\nu}$ can be exactly diagonalized by $U_{\nu2}$ shown in Eq.~\eqref{eq:Unu2}. It is easy to check that the PMNS matrix takes the same form as Eq.~\eqref{eq:PMNS_UTM1_1}, and it is also a TM1 mixing matrix. Therefore the expressions of mixing angles and CP invariants are still given by Eqs.~\eqref{eq:angles_TM1_1} and \eqref{eq:invariants_TM1_1}, respectively. However, the explicit dependence of the parameters $y$, $z$ and $w$ on $m_a$, $m_s$, $\eta$, $x$ differs from that of the above case with $X_{\text{sol}}=\{1,SU\}$. Hence distinct predictions for mixing  parameters are reached. We can check that the neutrino mass matrix $m_{\nu}$ in Eq.~\eqref{eq:mnu_N12} has the following symmetry properties
\begin{equation}\label{eq:sym_mnu2}
m_{\nu}(-x,r,\eta)=P^{T}_{132}m_{\nu}(x,r,\eta)P_{132},\qquad
m_{\nu}(-x,r,-\eta)=m^{*}_{\nu}(x,r,\eta)\,,
\end{equation}
The former implies that the reactor and solar mixing angles are invariant, the atmospheric angle changes from $\theta_{23}$ to $\pi/2-\theta_{23}$ and the Dirac phase changes from $\delta_{CP}$ to $\pi+\delta_{CP}$ under the transformation $x\rightarrow-x$. The later implies that all the lepton mixing angles are kept intact and the signs of all CP violation phases are reversed by changing $x\rightarrow -x$ and $\eta\rightarrow-\eta$. Once the values of $x$ and $\eta$ are fixed, the light neutrino mass matrix $m_{\nu}$ would only depend on two free parameters $m_a$ and $m_{s}$ whose values can be determined by the neutrino mass squared differences $\Delta m^2_{21}$ and  $\Delta m^2_{31}$. Then we can extract the predictions for three lepton mixing angles and CP violation phases $\delta_{CP}$ and $\beta$. The best fit values of the mixing parameters and neutrino masses for some benchmark values of $x$ and $\eta$ are shown in table~\ref{tab:bf_UTB1_2}. The most interesting points are $\eta=0$ and $\pi$ which predict maximal atmospheric mixing angle, maximal Dirac phase and trivial Majorana phase. The reason is because the general neutrino mass $m_\nu$ shown in Eq.~\eqref{eq:mnu_N12} has an accidental $\mu\tau$ reflection symmetry in the case of $\eta=0$ and $\pi$~\cite{King:2018kka}. The realistic values of mixing angles and mass ratio $m^2_2/m^2_3$ can be obtained for $x=\pm4$, $\pm7/2$, $\pm7/6$ in the case of $\eta=0$ or $\pi$. In order to describe the experimental data at $3\sigma$ level~\cite{Esteban:2018azc}, the three input parameters are constrained to be $|x|\in[1.045,1.346]\cup[2.952,4.754]$, $|\eta|\in[0,0.112\pi]\cup[0.674\pi,\pi]$ and $r\in[0.0250,0.0519]\cup[0.169,0.214]$. For this mixing pattern, the solar angle $\theta_{12}$ is predicted to be in the ranges of $[0.317,0.319]$ and the other two mixing angles $\theta_{13}$ and $\theta_{23}$ can take any values within their $3\sigma$ ranges. Furthermore, the absolute values of the two CP phases $|\delta_{CP}|$ and $|\beta|$ are predicted to lie in the regions $[0.298\pi,0.623\pi]$ and $[0,0.251\pi]\cup[0.830\pi,\pi]$, respectively.

\begin{table}[t!]
\renewcommand{\tabcolsep}{0.5mm}
\renewcommand{\arraystretch}{1.3}
\footnotesize
\centering
\begin{tabular}{|c| c| c| c | c| c| c| c| c| c |c |c|c|c|}  \hline \hline
$\langle\phi_{\text{sol}}\rangle/v_{\phi_{s}}$ & $x$   & $\eta$ & $m_{a}$(meV) & $r$  	 & $\chi^2_{\text{min}}$ &  $\sin^2\theta_{13}$  &$\sin^2\theta_{12}$  & $\sin^2\theta_{23}$  & $\delta_{CP}/\pi$ &  $\beta/\pi$  & $m_2$(meV) & $m_3$(meV) & $m_{ee}$(meV) \\ \hline
$\left(1,1\pm4 i,1\mp4 i\right)^{T}$ & $\pm4$ & $0$ & $47 .378$ & $0 .0320$ & $15 .257$ & $0 .0228$ & $0 .318$ & $0 .5$ & $\mp0.5$ & $0$ & $8 .562$ & $50 .295$ & $1 .515$ \\ \hline
$\left(1,1\pm\frac{7 i}{2},1\mp\frac{7 i}{2}\right)^{T}$ & $\pm\frac{7}{2}$ & $0$ & $43 .643$ & $0 .0383$ & $20 .623$ & $0 .0211$ & $0 .319$ & $0 .5$ & $\mp0.5$ & $0$ & $8 .734$ & $50 .083$ & $1 .670$ \\ \hline
$\left(1,1\pm\frac{5 i}{4},1\mp\frac{5 i}{4}\right)^T$ & $\pm\frac{5}{4}$ & $\pm\frac{3 \pi }{4}$ & $19 .427$ & $0 .187$ & $7 .914$ & $0 .0230$ & $0 .318$ & $0 .603$ & $\pm0 .352$ & $\mp0.888$ & $8 .361$ & $50 .554$ & $3 .626$ \\ \hline
\multirow{4}{*}{$\left(1,1\pm\frac{6 i}{5},1\mp\frac{6 i}{5}\right)^T$} & \multirow{4}{*}{$\pm\frac{6}{5}$} & $\pm\frac{3 \pi }{4}$ & $19 .634$ & $0 .187$ & $2 .177$ & $0 .0223$ & $0 .318$ & $0 .602$ & $\pm0 .350$ & $\mp0.881$ & $8 .619$ & $50 .222$ & $3 .675$ \\ \cline{3-14}
 & &$\pm\frac{4 \pi }{5}$ & $19 .242$ & $0 .192$ & $2 .540$ & $0 .0228$ & $0 .318$ & $0 .583$ & $\pm0 .382$ & $\mp0.909$ & $8 .455$ & $50 .434$ & $3 .694$ \\ \cline{3-14}
 & & $\pm\frac{5 \pi }{6}$ & $19 .030$ & $0 .195$ & $6 .400$ & $0 .023$ & $0 .318$ & $0 .570$ & $\pm0 .403$ & $\mp0.926$ & $8 .362$ & $50 .552$ & $3 .702$ \\ \cline{3-14}
 &  & $\mp\frac{5 \pi }{6}$ & $19 .049$ & $0 .194$ & $56 .898$ & $0 .0229$ & $0 .318$ & $0 .430$ & $\pm0 .597$ & $\pm0 .926$ & $8 .355$ & $50 .561$ & $3 .696$  \\ \hline
\multirow{5}{*}{$\left(1,1\pm\frac{7 i}{6},1\mp\frac{7 i}{6}\right)^T$} & \multirow{5}{*}{$\pm\frac{7}{6}$} & $\pi$ & $18 .720$ & $0 .201$ & $21 .977$ & $0 .0231$ & $0 .318$ & $0 .5$ & $\pm0 .5$ & $1$ & $8 .335$ & $50 .586$ & $3 .754$ \\ \cline{3-14}
&  & $\pm\frac{3 \pi }{4}$ & $19 .777$ & $0 .187$ & $6 .184$ & $0 .0218$ & $0 .318$ & $0 .602$ & $\pm0 .348$ & $\mp0.876$ & $8 .792$ & $50 .011$ & $3 .705$ \\ \cline{3-14}
 &  & $\pm\frac{4 \pi }{5}$ & $19 .391$ & $0 .192$ & $0 .557$ & $0 .0223$ & $0 .318$ & $0 .583$ & $\pm0 .381$ & $\mp0.905$ & $8 .634$ & $50 .204$ & $3 .726$ \\ \cline{3-14}
 & & $\pm\frac{5 \pi }{6}$ & $19 .184$ & $0 .195$ & $0 .917$ & $0 .0226$ & $0 .318$ & $0 .570$ & $\pm0 .402$ & $\mp0.923$ & $8 .546$ & $50 .317$ & $3 .736$ \\ \cline{3-14}
& & $\mp\frac{5 \pi }{6}$ & $19 .204$ & $0 .194$ & $51 .371$ & $0 .0225$ & $0 .318$ & $0 .431$ & $\pm0 .598$ & $\pm0 .923$ & $8 .537$ & $50 .328$ & $3 .729$ \\ \hline \hline
\end{tabular}
\caption{\label{tab:bf_UTB1_2}The predictions for the lepton mixing angles, CP violation phases, neutrino masses and the effective Majorana mass $m_{ee}$ for the breaking pattern $\mathcal{N}_{1}$ with $(G_{l},G_{\text{atm}},G_{\text{sol}})= (Z^{T}_{3},Z^{U}_{2},Z^{SU}_{2}) $ and $X_{\text{sol}}=\{S,U\}$. Here we choose many benchmark values for the parameters $x$ and $\eta$. Notice that the lightest neutrino mass is vanishing $m_1=0$.}
\end{table}

\item[~~($\mathcal{N}_{2}$)]{$(G_{l},G_{\text{atm}},G_{\text{sol}})=(Z_3^T,Z_3^{ST},Z_2^{SU})$, $X_{\text{atm}}=\{S,STS,T^2\}$}

$\bullet$ $X_{\text{sol}}=\{1,SU\}$

The unitary transformation $U_{l}$ is an identity matrix up to permutation of columns because the residual symmetry $G_{l}=Z^T_3$ is diagonal in our working basis. The possible residual CP transformations $X_{\text{sol}}$ and the corresponding VEVs of the flavon field $\phi_{\text{sol}}$ are the same as those of case $\mathcal{N}_{1}$. Here the VEVs of the flavon $\phi_{\text{atm}}$ is proportional to $\left(1, -2\omega^2,-2\omega\right)^T$, i.e.
\begin{equation}
\langle\phi_{\text{atm}}\rangle= v_{\phi_a}\left( 1, -2\omega^2 , -2\omega\right)^T\,,
\end{equation}
where $v_{\phi_a}$ is a real number with dimension of mass. For $X_{\text{sol}}=\{1,SU\}$, the alignment of $\phi_{\text{sol}}$ is given in Eq.~\eqref{eq:sol_VEV_N1}. Then the light neutrino mass matrix is
\begin{equation}\label{eq:nu_N21}
m_{\nu}=m_{a}\begin{pmatrix}
 1 &~ -2 \omega  &~ -2 \omega ^2 \\
 -2 \omega  &~ 4 \omega ^2 &~ 4 \\
 -2 \omega ^2 &~ 4 &~ 4 \omega  \\
\end{pmatrix}+m_{s}e^{i\eta}
\begin{pmatrix}
 1 &~ 2-x &~ x \\
 2-x &~ (x-2)^2 &~ (2-x) x \\
 x &~ (2-x) x &~ x^2 \\
\end{pmatrix}\,.
\end{equation}
It can be block diagonalized by the TB mixing matrix,
\begin{equation}\label{eq:mup_N21}
m^\prime_{\nu}=U^T_{TB}m_{\nu}U_{TB}=\begin{pmatrix}
0  &~   0  &~  0 \\
0  &~  y   &~ z  \\
0  &~  z  &~ w
\end{pmatrix}\,,
\end{equation}
with
\begin{eqnarray}
y= 3( m_{a}+m_{s}e^{i \eta }), \quad
z=3\sqrt{2} im_{a}+\sqrt{6}(x-1)m_{s}e^{i \eta }, \quad
 w= -6m_{a}+2(x-1)^2m_{s}e^{i \eta }\,.
\end{eqnarray}
Eq.~\eqref{eq:mup_N21} implies that the first column of $U_{TB}$ is an eigenvector of $m_\nu$ with zero eigenvalue. Hence the lepton mixing matrix is the TM1 mixing pattern. In order to achieve the Dirac CP phase $\delta_{CP}$ around $-\pi/2$ which is preferred by the present data~\cite{Esteban:2018azc}, we take $U_{l}=P_{132}$. Then the PMNS matrix can be obtained by exchanging the second and third rows of the mixing matrix in Eq.~\eqref{eq:PMNS_UTM1_1}. Comparing with the expressions of the three mixing angles in Eq.~\eqref{eq:angles_TM1_1}, $\sin^2\theta_{23}$ becomes $1-\sin^2\theta_{23}$ and the other two mixing angles are kept intact. The overall sign of the Jarlskog invariant $J_{CP}$ in Eq.~\eqref{eq:invariants_TM1_1} is reversed while the Majorana invariant $I_1$ is invariant. The sum rules among mixing angles and Dirac CP phase in Eqs.~(\ref{eq:sum_rule_TM1_1}, \ref{eq:sum_rule2_TM1_1}) are fulfilled as well.

Similar to previous cases, we perform a $\chi^2$ analysis for the neutrino mass matrix in Eq.~\eqref{eq:nu_N21}, and the numerical results for some benchmark values of $x$ and $\eta$ are reported in table~\ref{tab:bf_UTB1_3}. We can see that the measured values of the mixing angles and the neutrino masses can be well accommodated and the Dirac CP phase $\delta_{CP}$ is approximately maximal for all the typical values of $x$ and $\eta$. Furthermore, we notice that the vacuum $\langle\phi_{\text{sol}}\rangle\propto(2, 7, -3)^{T}$ for $x=7/2$, $\eta=-4\pi/5$ is relatively simple and it can describe the experimental data quite well. Similar to the LS model~\cite{King:2015dvf,King:2016yvg,King:2013xba,Ballett:2016yod}, we expect this alignment might provide an interesting opportunity for model building. Requiring all three mixing angles and the mass ratio $m^2_2/m^2_3$ lie in their $3\sigma$ ranges, we find the allowed regions of the parameter $x$ is $[-9.433,-1.314]\cup[3.189,7.120]$, of the parameter $\eta$ is $[-0.854\pi,-0.614\pi]\cup[0.555\pi,0.831\pi]$ and of the parameter $r$ is $[0.0485,0.756]$. The possible values of $\delta_{CP}$ lie in the interval $[-0.623\pi,-0.296\pi]$, and the allowed range of the Majorana phase is $[-0.773\pi,-0.527\pi]\cup[0.558\pi,0.810\pi]$. We see that this breaking pattern can accommodate a nearly maximal Dirac CP phase. The other mixing angles except $\theta_{12}$ can take any values within their $3\sigma$ ranges. The solar mixing angle $\sin^2\theta_{12}$ is close to 0.318 and this is generally true for TM1 mixing, as shown in Eqs.~(\ref{eq:sum_rule_TM1_1}, \ref{eq:sinSq12_TM1}).

\begin{table}[t!]
\renewcommand{\tabcolsep}{0.5mm}
\renewcommand{\arraystretch}{1.3}
\centering
\footnotesize
\begin{tabular}{|c| c| c| c | c| c| c| c| c| c |c |c |c|c|}  \hline \hline
$\langle\phi_{\text{sol}}\rangle/v_{\phi_{s}}$ & $x$   & $\eta$ & $m_{a}$(meV) & $r$  	 & $\chi^2_{\text{min}}$ &  $\sin^2\theta_{13}$  &$\sin^2\theta_{12}$  & $\sin^2\theta_{23}$  & $\delta_{CP}/\pi$ &  $\beta/\pi$  & $m_2$(meV) & $m_{3}$(meV) & $m_{ee}$(meV) \\   \hline
$\left(1 ,4,-2\right)^{T}$ & $4$ &$-\frac{3 \pi }{4}$ & $4 .030$ & $0 .369$ & $8 .810$ & $0 .0222$ & $0 .318$ & $0 .519$ & $-0.473$ & $-0.640$ & $8 .581$ & $50 .270$ & $3 .158$ \\ \hline
$\left(1 ,5,-3\right)^{T}$ & $5$ & $-\frac{2 \pi }{3}$ & $4 .120$ & $0 .226$ & $38 .945$ & $0 .0223$ & $0 .318$ & $0 .452$ & $-0.567$ & $-0.715$ & $8 .726$ & $50 .093$ & $3 .742$ \\ \hline
$\left(1 ,-2,4\right)^{T}$ & $-2$ & $\frac{3 \pi }{4}$ & $4 .050$ & $0 .365$ & $22 .417$ & $0 .0227$ & $0 .318$ & $0 .482$ & $-0.525$ & $0 .642$ & $8 .570$ & $50 .285$ & $3 .182$ \\ \hline
$\left(1 ,-3,5\right)^{T}$ & $-3$ & $\frac{2 \pi }{3}$ & $4 .132$ & $0 .224$ & $3 .930$ & $0 .0226$ & $0 .318$ & $0 .549$ & $-0.432$ & $0 .717$ & $8 .720$ & $50 .100$ & $3 .755$  \\ \hline
$\left(1 ,-5,7\right)^{T}$ & $-5$ & $\frac{3 \pi }{5}$ & $3 .955$ & $0 .120$ & $4 .755$ & $0 .0224$ & $0 .318$ & $0 .601$ & $-0.352$ & $0 .765$ & $8 .781$ & $50 .025$ & $3 .835$ \\ \hline
$\left(1 ,-6,8\right)^{T}$ & $-6$ & $\frac{3 \pi }{5}$ & $3 .832$ & $0 .0935$ & $3 .674$ & $0 .0224$ & $0 .318$ & $0 .607$ & $-0.342$ & $0 .778$ & $8 .502$ & $50 .374$ & $3 .737$ \\ \hline
\multirow{2}{*}{$\left(1 ,\frac{7}{2},-\frac{3}{2}\right)^{T}$} & \multirow{2}{*}{$\frac{7}{2}$} & $-\frac{4 \pi }{5}$ & $3 .862$ & $0 .531$ & $3 .408$ & $0 .0225$ & $0 .318$ & $0 .572$ & $-0.397$ & $-0.577$ & $8 .783$ & $50 .022$ & $2 .512$ \\ \cline{3-14}
&& $-\frac{5 \pi }{6}$ & $3 .574$ & $0 .600$ & $5 .012$ & $0 .0228$ & $0 .318$ & $0 .603$ & $-0.350$ & $-0.564$ & $8 .430$ & $50 .466$ & $2 .024$ \\ \hline
$\left(1 ,\frac{11}{2},-\frac{7}{2}\right)^{T}$ & $\frac{11}{2}$ & $-\frac{2 \pi }{3}$ & $4 .011$ & $0 .190$ & $43 .619$ & $0 .0221$ & $0 .318$ & $0 .446$ & $-0.576$ & $-0.733$ & $8 .434$ & $50 .462$ & $3 .690$ \\ \hline
$\left(1 ,-\frac{7}{2},\frac{11}{2}\right)^{T}$ & $-\frac{7}{2}$ & $\frac{2 \pi }{3}$ & $4 .022$ & $0 .189$ & $4 .322$ & $0 .0224$ & $0 .318$ & $0 .554$ & $-0.424$ & $0 .734$ & $8 .430$ & $50 .466$ & $3 .702$  \\ \hline
$\left(1 ,-\frac{11}{2},\frac{17}{2}\right)^{T}$ & $-\frac{11}{2}$ & $\frac{3 \pi }{5}$ & $3 .889$ & $0 .106$ & $2 .511$ & $0 .0224$ & $0 .318$ & $0 .604$ & $-0.347$ & $0 .772$ & $8 .630$ & $50 .210$ & $3 .782$ \\ \hline
$\left(1 ,-\frac{5}{3},\frac{11}{3}\right)^{T}$ & $-\frac{5}{3}$ & $\frac{4 \pi }{5}$ & $3 .858$ & $0 .474$ & $48 .139$ & $0 .0230$ & $0 .318$ & $0 .437$ & $-0.587$ & $0 .596$ & $8 .499$ & $50 .377$ & $2 .609$ \\ \hline
$\left(1,-\frac{10}{3},\frac{16}{3}\right)^{T}$ & $-\frac{10}{3}$ &$\frac{2 \pi }{3}$ & $4 .056$ & $0 .200$ & $2 .635$ & $0 .0224$ & $0 .318$ & $0 .553$ & $-0.426$ & $0 .729$ & $8 .519$ & $50 .352$ & $3 .718$ \\ \hline
\multirow{2}{*}{$\left(1 ,-\frac{7}{4},\frac{15}{4}\right)^{T}$} & \multirow{2}{*}{$-\frac{7}{4}$} & $\frac{3 \pi }{4}$ & $4 .137$ & $0 .409$ & $32 .646$ & $0 .0231$ & $0 .318$ & $0 .476$ & $-0.532$ & $0 .623$ & $8 .884$ & $49 .895$ & $3 .175$ \\ \cline{3-14}
&& $\frac{4 \pi }{5}$ & $3 .835$ & $0 .454$ & $49 .278$ & $0 .0228$ & $0 .318$ & $0 .440$ & $-0.583$ & $0 .604$ & $8 .380$ & $50 .529$ & $2 .633$ \\ \hline
$\left(1,-\frac{9}{4},\frac{17}{4}\right)^{T}$ & $-\frac{9}{4}$ & $\frac{3 \pi }{4}$ & $3 .979$ & $0 .326$ & $27 .092$ & $0 .0224$ & $0 .318$ & $0 .488$ & $-0.517$ & $0 .660$ & $8 .309$ & $50 .618$ & $3 .195$ \\ \hline
$\left(1,-\frac{11}{4},\frac{19}{4}\right)^{T}$ & $-\frac{11}{4}$ & $\frac{2 \pi }{3}$ & $4 .196$ & $0 .246$ & $10 .766$ & $0 .0227$ & $0 .318$ & $0 .545$ & $-0.437$ & $0 .707$ & $8 .896$ & $49 .879$ & $3 .787$ \\ \hline
$\left(1,-\frac{13}{4},\frac{21}{4}\right)^{T}$ & $-\frac{13}{4}$ & $\frac{2 \pi }{3}$ & $4 .074$ & $0 .205$ & $2 .306$ & $0 .0225$ & $0 .318$ & $0 .552$ & $-0.428$ & $0 .726$ & $8 .565$ & $50 .291$ & $3 .727$ \\ \hline
$\left(1,-\frac{8}{5},\frac{18}{5}\right)^{T}$ & $-\frac{8}{5}$ & $\frac{4 \pi }{5}$ & $3 .877$ & $0 .491$ & $49 .170$ & $0 .0231$ & $0 .318$ & $0 .435$ & $-0.591$ & $0 .590$ & $8 .600$ & $50 .245$ & $2 .591$ \\ \hline
\multirow{2}{*}{$\left(1,-\frac{9}{5},\frac{19}{5}\right)^{T}$} & \multirow{2}{*}{$-\frac{9}{5}$} & $\frac{3 \pi }{4}$ & $4 .118$ & $0 .399$ & $29 .007$ & $0 .0230$ & $0 .318$ & $0 .477$ & $-0.531$ & $0 .627$ & $8 .817$ & $49 .980$ & $3 .176$ \\ \cline{3-14}
&& $\frac{4 \pi }{5}$ & $3 .823$ & $0 .443$ & $51 .042$ & $0 .0228$ & $0 .318$ & $0 .442$ & $-0.581$ & $0 .609$ & $8 .313$ & $50 .614$ & $2 .648$ \\ \hline
$\left(1,-\frac{11}{5},\frac{21}{5}\right)^{T}$ & $-\frac{11}{5}$ & $\frac{3 \pi }{4}$ & $3 .992$ & $0 .334$ & $25 .304$ & $0 .0225$ & $0 .318$ & $0 .486$ & $-0.519$ & $0 .656$ & $8 .358$ & $50 .558$ & $3 .192$ \\ \hline\hline

\end{tabular}
\caption{\label{tab:bf_UTB1_3}The predictions for the lepton mixing angles, CP violation phases, neutrino masses and the effective Majorana mass $m_{ee}$ for the breaking pattern $\mathcal{N}_{2}$ with $(G_{l},G_{\text{atm}},G_{\text{sol}})= (Z^{T}_{3},Z^{ST}_{3},Z^{SU}_{2}) $ and $X_{\text{sol}}=\{1,SU\}$. Here we choose many benchmark values for the parameters $x$ and $\eta$. Notice that the lightest neutrino mass is vanishing $m_1=0$.
}
\end{table}

$\bullet$ $X_{\text{sol}}=\{S,U\}$

The explicit form of the vacuum of $\phi_{\text{atm}}$ and $\phi_{\text{sol}}$ invariant under the assumed residual symmetries can be found from table~\ref{tab:inv_VEV_CP}, i.e. $\langle\phi_{\text{atm}}\rangle\propto\left(1, -2\omega^2,-2\omega\right)^T$ and $\langle\phi_{\text{sol}}\rangle\propto\left(1, 1+ix, 1-ix\right)^T$. The most general neutrino mass matrix takes the following form
\begin{equation}\label{eq:mu_N22}
m_{\nu}=m_{a}\begin{pmatrix}
 1 &~ -2 \omega  &~ -2 \omega ^2 \\
 -2 \omega  &~ 4 \omega ^2 &~ 4 \\
 -2 \omega ^2 &~ 4 &~ 4 \omega  \\
\end{pmatrix}+m_{s}e^{i\eta}
\begin{pmatrix}
 1 &~ 1-i x &~ 1+i x \\
 1-i x &~ (1-ix)^2 &~ x^2+1 \\
 1+i x &~ x^2+1 &~ (1+i x)^2 \\
\end{pmatrix}\,.
\end{equation}
We can perform a TB transformation to obtain the block diagonal neutrino mass matrix $m^\prime_{\nu}$. The non-vanishing elements $y$, $z$ and $w$ of $m^\prime_{\nu}$ are given by
\begin{equation}
y=3\left( m_{a}+  m_{s} e^{i \eta }\right),\quad
 z=\sqrt{6}i \left(\sqrt{3}m_{a} +xm_{s}e^{i \eta }\right), \quad
w=-2  \left( 3m_{a} +x^2m_{s}e^{i \eta }\right)\,.
\end{equation}
We can further introduce the unitary transformation $U_{\nu2}$ to diagonalize the neutrino mass matrix $m^\prime_{\nu}$, as generally shown in Eqs.~(\ref{eq:mnup_Unu2D},\ref{eq:Unu2}). As a consequence, the lepton mixing matrix is also the TM1 pattern, and the sum rules in Eq.~\eqref{eq:sum_rule_TM1_1} and Eq.~\eqref{eq:sum_rule2_TM1_1} are satisfied as well. However, the dependence of the mixing parameters on the input parameters $m_{a}$, $m_{s}$, $\eta$ and $x$ are different, consequently the above two mixing patterns of $\mathcal{N}_{2}$ with $X_{\text{sol}}=\{1,SU\}$ and $X_{\text{sol}}=\{S,U\}$ lead to different predictions. In table~\ref{tab:bf_UTB1_4}, we present the results of our $\chi^2$ analysis for some simple values of $x$ and $\eta$. We find that accordance with experimental data can be achieved for certain values of $m_a$ and $r$. In the case of $\eta=\pi$, both the atmospheric mixing $\theta_{23}$ and the Dirac CP phase $\delta_{CP}$ are maximal, while the Majorana CP phase $\beta$ is trivial. We notice that realistic values of mixing angles and $m^2_2/m^2_3$ can be obtained for $x=3/2$, $-3$ in the case of $\eta=0$ or $\pi$. If requiring $\sin^2\theta_{ij}$ and $m^2_2/m^2_3$ lie in their $3\sigma$ regions~\cite{Esteban:2018azc}, we find the allowed regions of the parameter $x$ is $[-12.192,-2.744]\cup[1.350,1.534]$, of the parameter $\eta$ is $[-\pi,0.326\pi]\cup[0.804\pi,\pi]$ and of the parameter $r$ is $[0.0330,1.750]$. The predictions for CP phases are $|\delta_{CP}|\in[0.295\pi,0.624\pi]$ and  $\beta\in[-\pi,-0.749\pi]\cup[-0.248\pi,\pi]$.

\begin{table}[t!]
\renewcommand{\tabcolsep}{0.5mm}
\renewcommand{\arraystretch}{1.3}
\centering
\footnotesize
\begin{tabular}{|c| c| c| c | c| c| c| c| c| c |c |c |c |c |}  \hline \hline
$\langle\phi_{\text{sol}}\rangle/v_{\phi_{s}}$ & $x$   & $\eta$ & $m_{a}$(meV) & $r$  	 & $\chi^2_{\text{min}}$ &  $\sin^2\theta_{13}$  &$\sin^2\theta_{12}$  & $\sin^2\theta_{23}$  & $\delta_{CP}/\pi$ &  $\beta/\pi$  & $m_2$(meV) & $m_3$(meV) & $m_{ee}$(meV) \\   \hline
$\left(1,1+\frac{3 i}{2},1-\frac{3 i}{2}\right)^T$ & $\frac{3}{2}$ & $\pi$ & $35 .677$ & $1 .044$ & $16 .323$ & $0 .0218$ & $0 .318$ & $0 .5$ & $-0.5$ & $0$ & $8 .526$ & $50 .343$ & $1 .560$ \\ \hline
$\left(1,1-\frac{7 i}{2},1+\frac{7 i}{2}\right)^T$ & $-\frac{7}{2}$ & $-\frac{\pi }{3}$ & $1 .534$ & $1 .111$ & $2 .981$ & $0 .0215$ & $0 .319$ & $0 .590$ & $0 .367$ & $0 .745$ & $8 .535$ & $50 .330$ & $2 .806$ \\ \hline
\multirow{4}{*}{$\left(1,1-3 i,1+3 i\right)^T$} & \multirow{4}{*}{$-3$} &  $\frac{\pi }{5}$ & $1 .474$ & $1 .491$ & $44 .990$ & $0 .0234$ & $0 .317$ & $0 .444$ & $0 .576$ & $-0.848$ & $8 .674$ & $50 .157$ & $3 .498$ \\ \cline{3-14}
&  & $-\frac{ \pi }{5}$ & $1 .479$ & $1 .484$ & $4 .378$ & $0 .0233$ & $0 .317$ & $0 .556$ & $0 .423$ & $0 .849$ & $8 .699$ & $50 .126$ & $3 .501$ \\ \cline{3-14}
&  & $\frac{\pi }{6}$ & $1 .461$ & $1 .503$ & $36 .885$ & $0 .0224$ & $0 .318$ & $0 .453$ & $0 .565$ & $-0.874$ & $8 .568$ & $50 .288$ & $3 .537$ \\ \cline{3-14}
&  & $-\frac{ \pi }{6}$ & $1 .464$ & $1 .498$ & $2 .928$ & $0 .0223$ & $0 .318$ & $0 .547$ & $0 .434$ & $0 .874$ & $8 .587$ & $50 .262$ & $3 .538$ \\ \hline
\multirow{2}{*}{$\left(1,1-4 i,1+4 i\right)^T$} & \multirow{2}{*}{$-4$} & $-\frac{3 \pi }{7}$ & $1 .600$ & $0 .853$ & $4 .993$ & $0 .0217$ & $0 .319$ & $0 .61$ & $0 .333$ & $0 .667$ & $8 .557$ & $50 .303$ & $2 .322$ \\ \cline{3-14}
 & & $-\frac{4 \pi }{9}$ & $1 .596$ & $0 .859$ & $4 .271$ & $0 .0228$ & $0 .318$ & $0 .612$ & $0 .336$ & $0 .651$ & $8 .585$ & $50 .265$ & $2 .278$ \\ \hline \hline
\end{tabular}
\caption{\label{tab:bf_UTB1_4}The predictions for the lepton mixing angles, CP violation phases, neutrino masses and the effective Majorana mass $m_{ee}$ for the breaking pattern $\mathcal{N}_{2}$ with $(G_{l},G_{\text{atm}},G_{\text{sol}})= (Z^{T}_{3},Z^{ST}_{3},Z^{SU}_{2}) $ and $X_{\text{sol}}=\{S,U\}$. Here we choose many benchmark values for the parameters $x$ and $\eta$. Notice that the lightest neutrino mass is vanishing $m_1=0$.
}
\end{table}

\item[~~($\mathcal{N}_3$)]{$(G_{l},G_{\text{atm}}, G_{\text{sol}})=(Z_3^T,Z_2^{S},Z_2^{SU})$, $X_{\text{atm}}=\{1,S,TST^2U,T^2STU\}$, $X_{\text{sol}}=\{S,U\}$ }

For the concerned residual flavor symmetry $G_{\text{sol}}=Z_2^{SU}$ in the solar neutrino sector, the residual CP transformation $X_{\text{sol}}$ can only be $X_{\text{sol}}=\{S,U\}$ in order to achieve agreement with experimental data. In this case the charged lepton diagonalization matrix $U_{l}$ is also the identity matrix, and the vacuum expectation values of flavon fields $\phi_{\text{atm}}$ and $\phi_{\text{sol}}$ read as
\begin{equation}
\langle\phi_{\text{atm}}\rangle=v_{\phi_a}\left(1, 1 , 1\right)^T\,, \qquad
\langle\phi_{\text{sol}}\rangle=v_{\phi_s}\left(1, 1+i x , 1-i x\right)^T\,,
\end{equation}
The most general neutrino mass matrix  is determined to be
\begin{equation}
m_{\nu}=m_{a}\begin{pmatrix}
 1 &~ 1 &~ 1 \\
 1 &~ 1 &~ 1 \\
 1 &~ 1 &~ 1 \\
\end{pmatrix}+m_{s}e^{i\eta}
\begin{pmatrix}
 1 &~ 1-i x &~ 1+i x \\
 1-i x &~ (1-ix)^2 &~ x^2+1 \\
 1+i x &~ x^2+1 &~ (1+i x)^2 \\
\end{pmatrix}\,,
\end{equation}
We find that the column vector $(2,-1, -1)^{T}$ is an eigenvector of $m_\nu$ with zero eigenvalue. This neutrino mass matrix $m_{\nu}$ can be simplified into a block diagonal form by performing a $U_{TB}$ transformation. Then we can obtain the three nonzero elements  $y$, $z$ and $w$:
\begin{equation}
y=3 \left(m_{a}+m_{s} e^{i \eta } \right), \qquad
z=i\sqrt{6} x m_{s} e^{i \eta },  \qquad
w=-2 x^2  m_{s}e^{i \eta }\,.
\end{equation}
The neutrino mass matrix $m^\prime_{\nu}$ can be diagonalized by performing the unitary transformation $U_{\nu2}$. Thus the lepton mixing  matrix is the TM1 pattern shown in Eq.~\eqref{eq:PMNS_UTM1_1}. The
expressions of the lepton mixing angles are the same as those in Eq.~\eqref{eq:angles_TM1_1} and the two CP invariants $J_{CP}$ and $I_{1}$ are still given by Eq.~\eqref{eq:invariants_TM1_1}. The sum rules in Eq.~\eqref{eq:sum_rule_TM1_1} and Eq.~\eqref{eq:sum_rule2_TM1_1} are satisfied as well. Furthermore, we find that the neutrino mass matrix $m_{\nu}$ in Eq.~\eqref{eq:mu_N22} has the following symmetry properties
\begin{equation}\label{eq:sym_mnu2}
m_{\nu}(-x,r,\eta)=P^{T}_{132}m_{\nu}(x,r,\eta)P_{132},\qquad
m_{\nu}(-x,r,-\eta)=m^{*}_{\nu}(x,r,\eta)\,,
\end{equation}
\begin{table}[t!]
\renewcommand{\tabcolsep}{0.5mm}
\renewcommand{\arraystretch}{1.3}
\centering
\footnotesize
\begin{tabular}{|c| c| c| c | c| c| c| c| c| c |c |c |c |c |}  \hline \hline
$\langle\phi_{\text{sol}}\rangle/v_{\phi_{s}}$ & $x$   & $\eta$ & $m_{a}$(meV) & $r$  	 & $\chi^2_{\text{min}}$ &  $\sin^2\theta_{13}$  &$\sin^2\theta_{12}$  & $\sin^2\theta_{23}$  & $\delta_{CP}/\pi$ &  $\beta/\pi$  & $m_2$(meV) & $m_3$(meV) & $m_{ee}$(meV) \\   \hline
\multirow{7}{*}{$\left(1,1\pm4 i,1\mp4 i\right)^T$} & \multirow{7}{*}{$\pm4$} & $\pi$ & $3 .080$ & $0 .473$ & $20 .478$ & $0 .0209$ & $0 .319$ & $0 .5$ & $\pm0 .5$ & $0$ & $8 .562$ & $50 .296$ & $1 .624$ \\ \cline{3-14}
 &  & $\pm\frac{3 \pi }{4}$ & $3 .115$ & $0 .465$ & $10 .932$ & $0 .0239$ & $0 .317$ & $0 .532$ & $\pm0 .457$ & $\mp0.252$ & $8 .618$ & $50 .223$ & $2 .328$ \\ \cline{3-14}
 &  & $\mp\frac{3 \pi }{4}$ & $3 .107$ & $0 .466$ & $34 .132$ & $0 .0239$ & $0 .317$ & $0 .468$ & $\pm0 .543$ & $\pm0 .252$ & $8 .598$ & $50 .248$ & $2 .321$ \\ \cline{3-14}
 &  & $\pm\frac{4 \pi }{5}$ & $3 .106$ & $0 .467$ & $7 .318$ & $0 .0229$ & $0 .318$ & $0 .527$ & $\pm0 .464$ & $\mp0.201$ & $8 .609$ & $50 .235$ & $2 .113$  \\ \cline{3-14}
 &  & $\mp\frac{4 \pi }{5}$ & $3 .100$ & $0 .468$ & $26 .561$ & $0 .0229$ & $0 .318$ & $0 .474$ & $\pm0 .536$ & $\pm0 .201$ & $8 .592$ & $50 .255$ & $2 .107$ \\ \cline{3-14}
&  & $\pm\frac{5 \pi }{6}$ & $3 .100$ & $0 .468$ & $7 .928$ & $0 .0223$ & $0 .318$ & $0 .523$ & $\pm0 .469$ & $\mp0.168$ & $8 .599$ & $50 .246$ & $1 .980$ \\ \cline{3-14}
&  & $\mp\frac{5 \pi }{6}$ & $3 .095$ & $0 .469$ & $24 .266$ & $0 .0223$ & $0 .318$ & $0 .478$ & $\pm0 .531$ & $\pm0 .168$ & $8 .585$ & $50 .264$ & $1 .975$ \\ \hline
\multirow{6}{*}{$\left(1,1\pm5 i,1\mp5 i\right)^T$} & \multirow{6}{*}{$\pm5$} & $\pm\frac{\pi }{2}$ & $3 .055$ & $0 .310$ & $12 .152$ & $0 .0207$ & $0 .319$ & $0 .536$ & $\pm0 .447$ & $\mp0.503$ & $8 .630$ & $50 .210$ & $3 .198$ \\ \cline{3-14}
 &  & $\mp\frac{\pi }{2}$ & $3 .047$ & $0 .311$ & $38 .352$ & $0 .0207$ & $0 .319$ & $0 .464$ & $\pm0 .553$ & $\pm0 .503$ & $8 .606$ & $50 .238$ & $3 .190$ \\ \cline{3-14}
 &  & $\pm\frac{\pi }{3}$ & $3 .045$ & $0 .309$ & $12 .182$ & $0 .024$ & $0 .317$ & $0 .531$ & $\pm0 .459$ & $\mp0.670$ & $8 .556$ & $50 .303$ & $3 .609$ \\ \cline{3-14}
 &  & $\mp\frac{ \pi }{3}$ & $3 .038$ & $0 .310$ & $34 .705$ & $0 .024$ & $0 .317$ & $0 .469$ & $\pm0 .541$ & $\pm0 .670$ & $8 .537$ & $50 .328$ & $3 .603$ \\ \cline{3-14}
 &  & $\pm\frac{2 \pi }{5}$ & $3 .052$ & $0 .309$ & $5 .409$ & $0 .0228$ & $0 .318$ & $0 .534$ & $\pm0 .453$ & $\mp0.603$ & $8 .594$ & $50 .252$ & $3 .462$ \\ \cline{3-14}
 &  & $\mp\frac{2 \pi }{5}$ & $3 .045$ & $0 .310$ & $30 .253$ & $0 .0228$ & $0 .318$ & $0 .466$ & $\pm0 .547$ & $\pm0 .603$ & $8 .574$ & $50 .280$ & $3 .455$ \\ \hline
\multirow{4}{*}{$\left(1,1\pm\frac{9}{2} i,1\mp\frac{9}{2} i\right)^T$} & \multirow{4}{*}{$\pm\frac{9}{2}$} & $\pm\frac{2 \pi }{3}$ & $3 .067$ & $0 .378$ & $10 .006$ & $0 .021$ & $0 .319$ & $0 .535$ & $\pm0 .450$ & $\mp0.335$ & $8 .602$ & $50 .243$ & $2 .682$ \\ \cline{3-14}
 & & $\mp\frac{2\pi }{3}$ & $3 .059$ & $0 .379$ & $35 .123$ & $0 .021$ & $0 .319$ & $0 .465$ & $\pm0 .550$ & $\pm0 .335$ & $8 .580$ & $50 .271$ & $2 .675$ \\ \cline{3-14}
 &  & $\pm\frac{3\pi }{5}$ & $3 .074$ & $0 .376$ & $4 .416$ & $0 .0226$ & $0 .318$ & $0 .538$ & $\pm0 .447$ & $\mp0.403$ & $8 .602$ & $50 .243$ & $2 .931$ \\ \cline{3-14}
 &  & $\mp\frac{3\pi }{5}$ & $3 .066$ & $0 .378$ & $32 .040$ & $0 .0226$ & $0 .318$ & $0 .462$ & $\pm0 .553$ & $\pm0 .403$ & $8 .578$ & $50 .274$ & $2 .923$ \\ \hline
\multirow{9}{*}{$\left(1,1\pm\frac{11}{2} i,1\mp\frac{11}{2} i\right)^T$} &  \multirow{9}{*}{$\pm\frac{11}{2}$}  & $0$ & $3 .032$ & $0 .258$ & $15 .565$ & $0 .0229$ & $0 .318$ & $0 .5$ & $\pm0 .5$ & $1$ & $8 .578$ & $50 .274$ & $3 .816$ \\ \cline{3-14}
 &  & $\pm\frac{\pi }{4}$ & $3 .045$ & $0 .258$ & $10 .643$ & $0 .0213$ & $0 .319$ & $0 .523$ & $\pm0 .467$ & $\mp0.752$ & $8 .636$ & $50 .202$ & $3 .642$ \\ \cline{3-14}
 &  & $\mp\frac{\pi }{4}$ & $3 .040$ & $0 .258$ & $27 .523$ & $0 .0213$ & $0 .319$ & $0 .477$ & $\pm0 .533$ & $\pm0 .752$ & $8 .621$ & $50 .220$ & $3 .637$ \\ \cline{3-14}
 & & $\pm\frac{\pi }{5}$ & $3 .042$ & $0 .258$ & $9 .429$ & $0 .0219$ & $0 .318$ & $0 .519$ & $\pm0 .473$ & $\mp0.802$ & $8 .619$ & $50 .222$ & $3 .705$ \\ \cline{3-14}
&  & $\mp\frac{\pi }{5}$ & $3 .038$ & $0 .258$ & $23 .435$ & $0 .0219$ & $0 .318$ & $0 .481$ & $\pm0 .527$ & $\pm0 .802$ & $8 .607$ & $50 .237$ & $3 .701$ \\ \cline{3-14}
 &  & $\pm\frac{\pi }{6}$ & $3 .040$ & $0 .258$ & $9 .667$ & $0 .0222$ & $0 .318$ & $0 .516$ & $\pm0 .477$ & $\mp0.835$ & $8 .608$ & $50 .235$ & $3 .739$ \\ \cline{3-14}
 &  & $\mp\frac{\pi }{6}$ & $3 .036$ & $0 .258$ & $21 .569$ & $0 .0222$ & $0 .318$ & $0 .484$ & $\pm0 .523$ & $\pm0 .835$ & $8 .598$ & $50 .248$ & $3 .736$ \\ \hline \hline
\end{tabular}
\caption{\label{tab:bf_UTB1_5}The predictions for the lepton mixing angles, CP violation phases, neutrino masses and the effective Majorana mass $m_{ee}$ for the breaking pattern $\mathcal{N}_{3}$ with $(G_{l},G_{\text{atm}},G_{\text{sol}})= (Z^{T}_{3},Z^{S}_{2},Z^{SU}_{2}) $ and $X_{\text{sol}}=\{S,U\}$. Here we choose many benchmark values for the parameters $x$ and $\eta$. Notice that the lightest neutrino mass is vanishing $m_1=0$.}
\end{table}
These identities indicate that the mixing angles $\theta_{12}$ and $\theta_{13}$ keep invariant, $\theta_{23}$ becomes $\pi/2-\theta_{23}$ and the Dirac phase changes from $\delta_{CP}$ to $\pi+\delta_{CP}$ under the transformation $x\rightarrow-x$. Moreover, by changing $x$ to $-x$ and $\eta$ to $-\eta$ simultaneously, all the lepton mixing angles are unchanged and the signs of all CP violation phases are reversed. Detailed numerical analyses show that accordance with experimental data can be achieved for certain values of $x$, $m_a$, $r$ and $\eta$, and the corresponding benchmark numerical results are listed in table~\ref{tab:bf_UTB1_5}. We find that acceptable values of mixing angles and $m^2_2/m^2_3$ can be obtained for $x=\pm4$, $\eta=\pi$ and $x=\pm 11/2, \eta=0$. If all the three mixing angles and $m^2_2/m^2_3$ are restricted to their $3\sigma$ regions~\cite{Esteban:2018azc}. The viable ranges of the input parameters $|x|$ and $r$ are $[3.641,5.911]$ and $[0.213,0.568]$ respectively while any value of $\eta\in[-\pi, \pi]$ is viable. Then the atmospheric mixing angle $\sin^2\theta_{23}$ and the Dirac CP phase $\delta_{CP}$ are predicted to be $0.458\leq\sin^2\theta_{23}\leq0.542$ and  $|\delta_{CP}|\in[0.443\pi,0.557\pi]$, respectively. The Majorana CP phase $\beta$ can take any value between $-\pi$ and $\pi$.

In short summary, We find that all the above three breaking patterns $\mathcal{N}_1$, $\mathcal{N}_2$ and $\mathcal{N}_3$ predict TM1 lepton mixing matrix and the experimental data~\cite{Esteban:2018azc} can be described very well. All the three breaking patterns predict a normal mass hierarchy with $m_1=0$ and the sum rules in Eqs.~\eqref{eq:sum_rule_TM1_1} and~\eqref{eq:sum_rule2_TM1_1}. In fact these two sum rules are common to all TM1 mixing matrices. The prospects for testing the two sum rules in future neutrino facilities have been discussed~\cite{Ballett:2013wya}. Under the assumption of TM1 mixing, the structure of the Dirac mass matrix has been analyzed in Refs.~\cite{Shimizu:2017fgu,Shimizu:2017vwi} in the framework of two right-handed neutrino seesaw model, generally more parameters are involved than the tri-driect CP models.

\item[~~($\mathcal{N}_{4}$)]{$(G_{l},G_{\text{atm}},G_{\text{sol}})=(Z_3^T,Z_2^{TST^2},Z_2^U)$, $X_{\text{atm}}=\{SU,ST^2S,T^2,T^2STU\}$, $X_{\text{sol}}=\{1,U\}$}

This breaking pattern has been studied in great detail by us~\cite{Ding:2018fyz}. Hence we shall not repeat the analysis here. When all the three lepton mixing angles and the neutrino mass ratio $m^2_2/m^2_3$ are restricted in their $3\sigma$ regions~\cite{Esteban:2018azc}, we find that the parameters $x$, $|\eta|$ and $r$ should be in the ranges of $[-6.238,-3.365]$, $[0.347\pi,\pi]$ and $[0.154,0.607]$, respectively. Moreover, we show the results of $\chi^2$ analysis for some benchmark values of $x$ and $\eta$ in table~\ref{tab:bf_N4}. The highlighted case with red background in table~\ref{tab:bf_N4} has been realized in a concrete model~\cite{Ding:2018fyz}. In section~\ref{sec:model} of the present work, we shall construct a model to realize the breaking pattern with $x=-4$ and $\eta=\pm3\pi/4$. The corresponding best fit values of the input parameters, mixing angles, CP phases and neutrino masses are highlighted with green background in table~\ref{tab:bf_N4}.

\begin{table}[t!]
\renewcommand{\tabcolsep}{0.7mm}
\renewcommand{\arraystretch}{1.3}
\small
\centering
\begin{tabular}{|c|c| c| c| c | c| c| c| c| c| c|c |c|c|c|}  \hline \hline
 $x$   & $\eta$  & $m_{a}(\text{meV})$ & $r$ 	 & $\chi^2_{\text{min}}$ &  $\sin^2\theta_{13}$  &$\sin^2\theta_{12}$  & $\sin^2\theta_{23}$  & $\delta_{CP}/\pi$ &  $\beta/\pi$ & $m_2(\text{meV})$ & $m_3(\text{meV})$ & $m_{ee}(\text{meV})$ \\   \hline
 \multirow{4}{*}{$-4$} & \cellcolor{lightgreen}  $\frac{3 \pi }{4}$ &  \cellcolor{lightgreen} $3 .708$ & \cellcolor{lightgreen} $0 .423$ & \cellcolor{lightgreen} $48 .401$ & \cellcolor{lightgreen} $0 .0227$ & \cellcolor{lightgreen} $0 .336$ & \cellcolor{lightgreen} $0 .441$ & \cellcolor{lightgreen} $-0.587$ &  \cellcolor{lightgreen} $-0.264$ & \cellcolor{lightgreen} $8 .566$ & \cellcolor{lightgreen} $50 .289$ & \cellcolor{lightgreen} $2 .825$ \\ \cline{2-15}
 & \cellcolor{lightgreen} $-\frac{3\pi }{4}$ & \cellcolor{lightgreen} $3 .723$ & \cellcolor{lightgreen} $0 .421$ & \cellcolor{lightgreen} $5 .168$ & \cellcolor{lightgreen} $0 .0226$ & \cellcolor{lightgreen} $0 .336$ & \cellcolor{lightgreen} $0 .560$ & \cellcolor{lightgreen} $-0.412$ & \cellcolor{lightgreen} $0 .264$ & \cellcolor{lightgreen} $8 .603$ & \cellcolor{lightgreen} $50 .242$ & \cellcolor{lightgreen} $2 .840$  \\ \cline{2-15}
 & $\frac{4\pi }{5}$ & $3 .674$ & $0 .430$ & $51 .698$ & $0 .0205$ & $0 .338$ & $0 .451$ & $-0.576$ & $-0.211$ & $8 .535$ & $50 .331$ & $2 .569$  \\ \cline{2-15}
 &$-\frac{4\pi }{5}$ & $3 .686$ & $0 .428$ & $16 .028$ & $0 .0204$ & $0 .338$ & $0 .549$ & $-0.424$ & $0 .211$ & $8 .565$ & $50 .291$ & $2 .581$ \\ \hline
\multirow{2}{*}{$-5$} & $\frac{3 \pi }{5}$ & $3 .723$ & $0 .266$ & $67 .144$ & $0 .021$ & $0 .345$ & $0 .424$ & $-0.621$ & $-0.422$ & $8 .594$ & $50 .252$ & $3 .545$ \\ \cline{2-15}
 & $-\frac{3 \pi }{5}$ & $3 .745$ & $0 .264$ & $11 .786$ & $0 .0211$ & $0 .345$ & $0 .577$ & $-0.379$ & $0 .422$ & $8 .643$ & $50 .193$ & $3 .565$ \\ \hline

\multirow{2}{*}{$-6$} & $\frac{2 \pi }{5}$ & $3 .720$ & $0 .181$ & $69 .680$ & $0 .024$ & $0 .349$ & $0 .428$ & $-0.606$ & $-0.627$ & $8 .441$ & $50 .452$ & $3 .980$ \\ \cline{2-15}
 & $-\frac{2 \pi }{5}$ & $3 .738$ & $0 .180$ & $17 .443$ & $0 .0241$ & $0 .349$ & $0 .572$ & $-0.394$ & $0 .627$ & $8 .479$ & $50 .404$ & $3 .997$ \\ \hline

$-\frac{7}{2}$ & \cellcolor{lightred}  $\pi$ &  \cellcolor{lightred} $3 .716$ & \cellcolor{lightred} $0 .557$ & \cellcolor{lightred} $17 .524$ & \cellcolor{lightred} $0 .0227$ & \cellcolor{lightred} $0 .331$ & \cellcolor{lightred} $0 .5$ & \cellcolor{lightred} $-0.5$ & \cellcolor{lightred} $0$ & \cellcolor{lightred} $8 .611$ & \cellcolor{lightred} $50 .232$ & \cellcolor{lightred} $1 .647$ \\ \hline

 \multirow{2}{*}{$-\frac{9}{2}$} & $\frac{2 \pi }{3}$ & $3 .708$ & $0 .332$ & $58 .776$ & $0 .0216$ & $0 .341$ & $0 .429$ & $-0.609$ & $-0.352$ & $8 .567$ & $50 .289$ & $3 .271$ \\ \cline{2-15}
 & $-\frac{2\pi }{3}$ & $3 .727$ & $0 .330$ & $7 .352$ & $0 .0216$ & $0 .341$ & $0 .571$ & $-0.391$ & $0 .352$ & $8 .611$ & $50 .232$ & $3 .289$ \\ \hline \hline
\end{tabular}
\caption{\label{tab:bf_N4}The predictions for the lepton mixing angles, CP violation phases, neutrino masses and the effective Majorana mass $m_{ee}$ for the breaking pattern $\mathcal{N}_{4}$ with $(G_{l},G_{\text{atm}},G_{\text{sol}})= (Z_3^T,Z_2^{TST^2},Z_2^U)$ and $X_{\text{sol}}=\{1,U\}$. Here we choose many benchmark values for the parameters $x$ and $\eta$. Notice that the lightest neutrino mass is vanishing $m_1=0$.
}
\end{table}

\item[~~($\mathcal{N}_{5}$)]{$(G_{l},G_{\text{atm}},G_{\text{sol}})=(K_4^{(S,U)},Z_2^{TU},Z_2^{TU})$, $X_{\text{atm}}=\{U,T\}$, $X_{\text{sol}}=\{U,T\}$ }

In this case, the unitary transformation $U_{l}$ is TB the mixing matrix $U_{TB}$. From table~\ref{tab:inv_VEV_CP}, we find that the vacuum alignments of the flavons $\phi_{\text{atm}}\sim\mathbf{3}$ and $\phi_{\text{sol}}\sim\mathbf{3^\prime}$ are dictated by the residual symmetry to be
\begin{equation}
\langle\phi_{\text{atm}}\rangle=v_{\phi_a}\left(0, -\omega , \omega^2\right)^T\,, \qquad \langle\phi_{\text{sol}}\rangle=v_{\phi_s}\left(1,  \omega x , \omega^2x\right)^T\,.
\end{equation}
It is easy to check that the two  column vectors $\langle\phi_{\text{atm}}\rangle$ and $\langle\phi_{\text{sol}}\rangle$ are orthogonal to each other, i.e. $\langle\phi_{\text{atm}}\rangle^\dagger\langle\phi_{\text{sol}}\rangle=0$. This scenario is referred to as form dominance in the literature~\cite{Chen:2009um,Choubey:2010vs,King:2010bk}.
Applying the seesaw relation in Eq.~\eqref{eq:mnu} gives rise to a light neutrino mass matrix,
\begin{equation}
 m_{\nu}=m_{a}\begin{pmatrix}
 0 &~ 0 &~ 0 \\
 0 &~ \omega  &~ -1 \\
 0 &~ -1 &~ \omega ^2 \\
\end{pmatrix}+m_{s}e^{i\eta}
\begin{pmatrix}
 1 &~ x \omega ^2 &~ x \omega  \\
 x \omega ^2 &~ x^2 \omega  &~ x^2 \\
 x \omega  &~ x^2 &~ x^2 \omega ^2 \\
\end{pmatrix}\,.
\end{equation}
In this case, $\langle\phi_{\text{atm}}\rangle$ and $\langle\phi_{\text{sol}}\rangle$ are proportional to two columns of $U_{\nu}$ which is the diagonalization matrix of $ m_{\nu}$,
\begin{equation}
U_{\nu}=\begin{pmatrix}
 -\frac{2 x}{\sqrt{2+4 x^2}} &~ 0 &~ \frac{1}{\sqrt{1+2 x^2}} \\
 \frac{\omega  }{ \sqrt{2(1+2 x^2)}} &~ \frac{\omega }{\sqrt{2}} &~ \frac{\omega  x}{\sqrt{1+2 x^2}} \\
 \frac{\omega ^2 }{ \sqrt{2(1+2 x^2)}} &~ -\frac{\omega ^2}{\sqrt{2}} &~ \frac{\omega ^2 x}{\sqrt{1+2 x^2}} \\
\end{pmatrix}\text{diag}(1,1,e^{-\frac{i\eta}{2}})\,,
\end{equation}
with
\begin{equation}
U^T_{\nu}m_{\nu}U_{\nu}=\text{diag}(0,2m_{a},\left(1+2x^2\right)  m_{s} )\,.
\end{equation}
It implies that the three neutrino masses are $0$, $2m_{a}$ and $\left(1+2x^2\right) m_{s}$ which are independent of the phase $\eta$. Including the contribution $U_{l}=U_{TB}$ from the charged lepton sector, we find the lepton mixing matrix is given by
\begin{equation}
  U=
\begin{pmatrix}
 \frac{1-4 x}{2 \sqrt{3(1+2 x^2)}} &~ -\frac{i}{2} &~ \frac{2 +x}{ \sqrt{6(1+2 x^2)}} \\
 -\frac{i \sqrt{3} }{2 \sqrt{1+2 x^2}} &~ \frac{1}{2} &~ -\frac{ 3i x}{\sqrt{6(1+2 x^2)}} \\
 -\frac{1+2 x}{ \sqrt{6(1+2 x^2)}} &~ \frac{i}{\sqrt{2}} &~ \frac{1-x}{ \sqrt{3(1+2 x^2)}} \\
\end{pmatrix}\text{diag}(1,1,e^{-\frac{i\eta}{2}})\,.
\end{equation}
The three lepton mixing angles read as
\begin{equation}\label{eq:mix_angle_N5}
 \sin^2\theta_{13}=\frac{(2+x)^2}{6 \left(1+2 x^2\right)}, \qquad
  \sin^2\theta_{12}=\frac{3 \left(1+2 x^2\right)}{2(2 -4 x+11 x^2)}, \qquad
\sin^2\theta_{23}=\frac{9x^2}{2-4 x+11 x^2}\,,
\end{equation}
which are expressed in terms of one real parameter $x$. Furthermore, we can derive the following sum rules among the mixing angles
\begin{eqnarray}
\nonumber &&\sin^2\theta_{12}\cos^2\theta_{13}=\frac{1}{4}\,, \\
\label{eq:sum_BM}&& \sin^2\theta_{23}=\frac{6-7 \sin^ 2 \theta_{13}\pm2 \sin \theta_{13}\sqrt{2(3-4\sin^2 \theta_{13})}}{9\cos^2\theta_{13}}\simeq
\frac{6- \sin^ 2 \theta_{13}\pm2\sqrt{6} \sin \theta_{13}}{9}\,,
\end{eqnarray}
where the first sum rule in Eq.~\eqref{eq:sum_BM} has already appeared in the literature~\cite{Li:2014eia}, and the sign ``$\pm$'' in the second sum rule depends on the value of $x$. For the best fitting value of the reactor angle $\sin^2\theta_{13}=0.02241$~\cite{Esteban:2018azc}, the solar mixing angle is determined to be $\sin^2\theta_{12}=0.256$ and the atmospheric mixing angle is $\sin^2\theta_{23}=0.746$  or $\sin^2\theta_{23}=0.582$. We see that the later value of $\sin^2\theta_{23}$ is compatible with the preferred values from global data analysis~\cite{Esteban:2018azc}. The value of $\sin^2\theta_{12}$ is  rather close to its $3\sigma$ lower limit $0.275$. As a result, we suggest this mixing pattern is a good leading order approximation since accordance with experimental data should be easily achieved after subleading contributions are taken into account in a concrete model. Furthermore, we find that the two CP rephasing invariants $J_{CP}$ and $I_1$ are
\begin{equation}\label{eq:CP_invariants_N5}
 J_{CP}=0\,, \qquad I_{1}=-\frac{(2 +x)^2 \sin \eta }{24 \left(1+2 x^2\right)} \,.
\end{equation}
Hence the Dirac CP phase is trivial for any value of $x$. From the expressions of mixing angles in Eq.~\eqref{eq:mix_angle_N5} and Majorana invariant in Eq.~\eqref{eq:CP_invariants_N5}, we find that the Majorana CP phase $\beta$ is determined to be
\begin{equation}
\beta=\eta+\pi\,.
\end{equation}
It implies that  trivial Majorana CP phase is obtained for $\eta=0$ or $\pi$. For $\eta=\pm\pi/2$, the Majorana CP phase is maximal. As an example, we take the representative value $x=-5/4$. Thus the VEV of the flavon field $\phi_{\text{sol}}$ is proportional to $\left(1,-\frac{5}{4}\omega,\frac{5}{4}\omega^2\right)^T$ and the PMNS matrix is
\begin{equation}
 U=
\begin{pmatrix}
 -2 \sqrt{\frac{2}{11}} &~ \frac{i}{2} &~ -\frac{1}{2 \sqrt{11}} \\
 -i \sqrt{\frac{2}{11}} &~ \frac{1}{2} &~ \frac{5 i}{2 \sqrt{11}} \\
 \frac{1}{\sqrt{11}} &~ \frac{i}{\sqrt{2}} &~ \frac{3}{\sqrt{22}} \\
\end{pmatrix}\text{diag}(1,1,e^{-\frac{i\eta}{2}})\,,
\end{equation}
The three mixing angles are determined to be
\begin{equation}
 \sin^2\theta_{13}=\frac{1}{44}\simeq0.0227\,, \quad
  \sin^2\theta_{12}=\frac{11}{44}\simeq0.256\,, \quad
\sin^2\theta_{23}=\frac{25}{44}\simeq0.581\,.
\end{equation}
Wee see that the reactor and atmospheric mixing angles are compatible with the preferred values from global fit~\cite{Esteban:2018azc} at the $3\sigma$ level. Furthermore, in the case of $x=-5/4$ the two neutrino mass squared differences only depend on the values of $m_{a}$ and $r$, as shown in figure~\ref{fig:Delta_MSq_N5}. We find that the best fit values of the two neutrino mass squared differences $\Delta m^2_{21}$ and $\Delta m^2_{32}$
can be reproduced.

In order to increase the readability of the paper, the remaining three viable cases $\mathcal{N}_{6}$, $\mathcal{N}_{7}$ and $\mathcal{N}_{8}$ are moved to the appendix~\ref{sec:other_NO_mix}. The reason is that the diagonalization matrix of the charged lepton mass matrix and the phenomenologically interesting alignments of the flavon $\phi_{\text{sol}}$ may be not simple enough to be realized in a concrete model.

\begin{figure}[t!]
\centering
\begin{tabular}{c}
\includegraphics[width=0.50\linewidth]{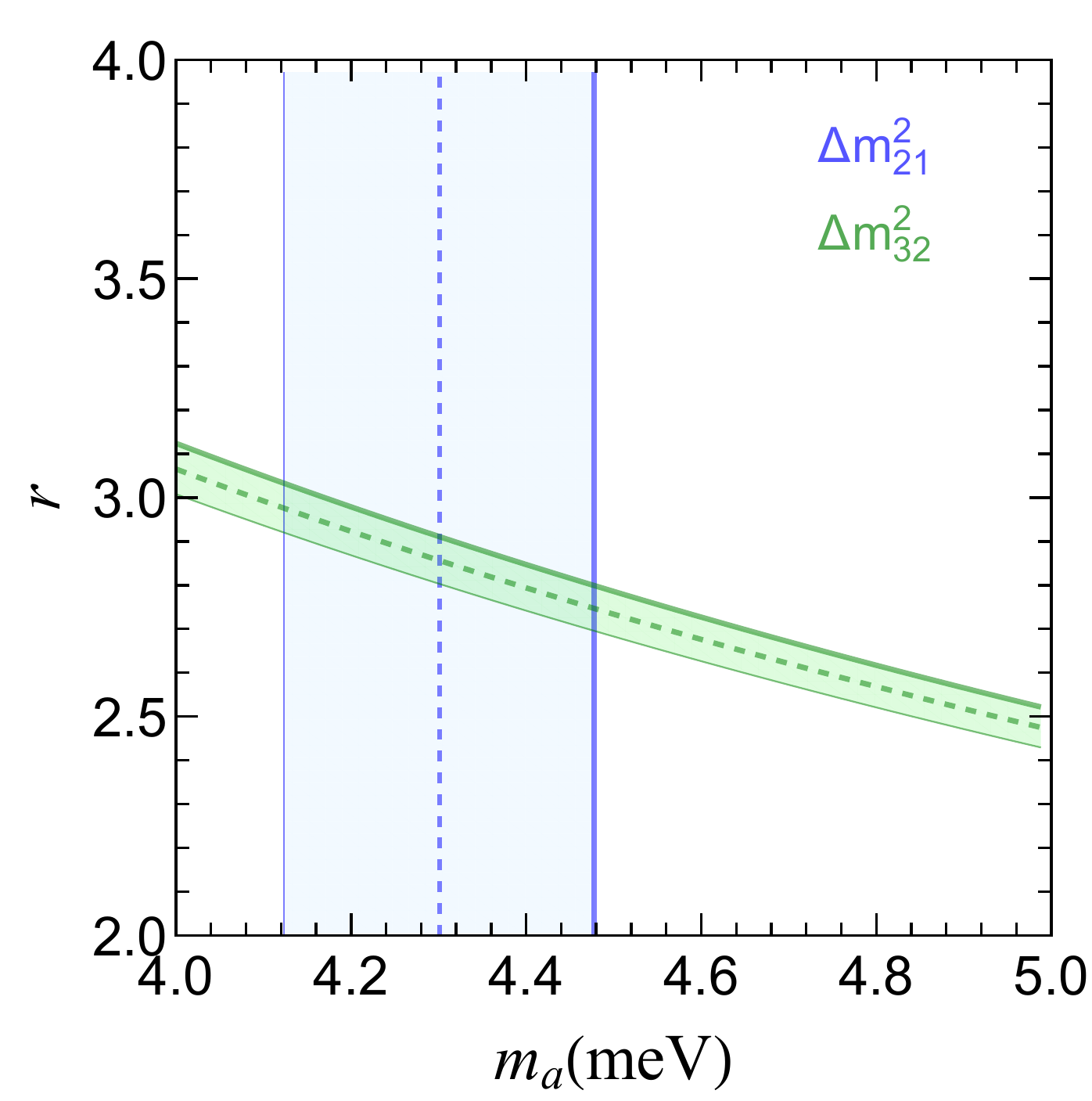}
\end{tabular}
\caption{\label{fig:Delta_MSq_N5} Contour plot of $\Delta m^2_{21}$ and $\Delta m^2_{32}$ in the $r-m_{a}$ plane for the residual symmetry $\mathcal{N}_{5}$ with the benchmark value $x=-5/4$. The $3\sigma$ lower (upper) bounds of the neutrino mass squared differences are labelled with thin (thick) solid curves, and the dashed contour lines represent the corresponding best fit values.}
\end{figure}

\end{description}

\section{\label{sec:TD_S4_GCP_IO}Mixing patterns derived from $S_4$ with IO neutrino masses}

From table~\ref{tab:bf_via_CP}, we see that the breaking of $S_4$ and CP symmetries in the tri-direct CP approach can lead to eighteen viable mixing patterns with IO neutrino masses. In the following, we proceed to study five viable cases among them and present their predictions for lepton mixing angles, CP violating phases and neutrino masses. The other viable breaking patterns are shown in appendix~\ref{sec:other_IO_mix}. The last two of the five breaking patterns in this section will lead to the form dominance texture.

\begin{description}[labelindent=-0.8em, leftmargin=0.3em]

\item[~~($\mathcal{I}_{1}$)]{$(G_{l},G_{\text{atm}},G_{\text{sol}})=(Z_3^T,Z_3^{ST},Z_2^{U})$, $X_{\text{atm}}=\{S,STS,T^2\}$, $X_{\text{sol}}=\{1,U\}$ }

The diagonalization matrix of the charged lepton mass matrix $m^\dagger_lm_l$ is an unity matrix because of the residual symmetry $G_{l}=Z_3^T$. The given residual symmetries fix the vacuum of the flavon fields $\phi_{\text{atm}}$ and $\phi_{\text{sol}}$ to be
\begin{equation}
\langle\phi_{\text{atm}}\rangle=v_{\phi_a}\left(1,-2 \omega ^2,-2 \omega\right)^T\,, \qquad \langle\phi_{\text{sol}}\rangle=v_{\phi_s}\left(1 ,  x ,  x\right)^T\,.
\end{equation}
Thus the light neutrino mass matrix is given by
\begin{equation}
 m_{\nu}=m_{a}\begin{pmatrix}
 1 &~ -2 \omega  &~ -2 \omega ^2 \\
 -2 \omega  &~ 4 \omega ^2 &~ 4 \\
 -2 \omega ^2 &~ 4 &~ 4 \omega  \\
\end{pmatrix}+m_{s}e^{i\eta}
\begin{pmatrix}
 1 &~ x &~ x \\
 x &~ x^2 &~ x^2 \\
 x &~ x^2 &~ x^2 \\
\end{pmatrix}\,,
\end{equation}
It can be simplified into a block diagonal form $m'_{\nu}$ by performing a unitary transformation $U_{\nu1}$, where the unitary matrix $U_{\nu1}$ is
\begin{equation}
U_{\nu1}=\begin{pmatrix}
0 &~ -i \sqrt{\frac{x^2-2 x+4}{7 x^2-2 x+4}} &~ \frac{i \sqrt{6} x}{\sqrt{7 x^2-2 x+4}} \\
\frac{-2 i \omega ^2-i x}{\sqrt{2 \left(x^2-2 x+4\right)}} &~ \frac{\sqrt{3} x \left(2 \omega ^2+x\right)}{\sqrt{\left(x^2-2 x+4\right) \left(7 x^2-2 x+4\right)}} &~ \frac{2 \omega ^2+x}{\sqrt{2 \left(7 x^2-2 x+4\right)}} \\
\frac{-i x-2 i \omega }{\sqrt{2 \left(x^2-2 x+4\right)}} &~ -\frac{\sqrt{3} x (x+2 \omega )}{\sqrt{\left(x^2-2 x+4\right) \left(7 x^2-2 x+4\right)}} &~ \frac{-x-2 \omega }{\sqrt{2 \left(7 x^2-2 x+4\right)}} \\
\end{pmatrix}\,,
\end{equation}
As shown in Eq.~\eqref{eq:mnup}, $m'_{\nu}$ can be parametrized by three parameters $y$, $z$ and $w$ with
\begin{eqnarray}
\nonumber &&y= -\frac{2 \left((x-4)^2m_{a} +x^2(x-1)^2m_{s}e^{i \eta }   \right)}{x^2-2 x+4}\,,  \\
\nonumber && z=-\frac{\sqrt{2 \left(7 x^2-2 x+4\right)} \left((x-4)m_{a} +x(x-1)m_{s}e^{i \eta }   \right)}{x^2-2 x+4}\,, \\
&& w= -\frac{\left(7 x^2-2 x+4\right) \left(m_{a}+m_{s}e^{i \eta } \right)}{x^2-2 x+4}\,.
\end{eqnarray}
The neutrino mass matrix $m'_{\nu}$ can be exactly diagonalized by a second unitary transformation $U_{\nu2}$ in Eq.~\eqref{eq:Unu2}. Then the lightest neutrino mass $m_3$ is vanishing and the other two neutrino masses $m_{1,2}$ can be obtained from Eq.~\eqref{eq:nu_masses}. The lepton mixing matrix is determined to be
\begin{equation}\label{eq:PMNS_IO1}
   U=
\begin{pmatrix}
 e^{-i \psi } \sqrt{\frac{x^2-2 x+4}{7 x^2-2 x+4}} \sin \theta  &~ -\sqrt{\frac{x^2-2 x+4}{7 x^2-2 x+4}} \cos \theta  &~ \frac{\sqrt{6} x}{\sqrt{7 x^2-2 x+4}} \\
 -\frac{i \cos \theta }{\sqrt{2}}-\frac{\sqrt{3}  xe^{-i \psi } \sin \theta }{\sqrt{7 x^2-2 x+4}} &~ \frac{\sqrt{3} x \cos \theta }{\sqrt{7 x^2-2 x+4}}-\frac{i e^{i \psi } \sin \theta }{\sqrt{2}} &~ \sqrt{\frac{x^2-2 x+4}{2(7 x^2-2 x+4)}} \\
 \frac{i \cos \theta }{\sqrt{2}}-\frac{\sqrt{3}  x e^{-i \psi }\sin \theta }{\sqrt{7 x^2-2 x+4}} &~ \frac{ie^{i \psi }  \sin \theta }{\sqrt{2}}+\frac{\sqrt{3}x \cos \theta  }{\sqrt{7 x^2-2 x+4}} &~ \sqrt{\frac{x^2-2 x+4}{2(7 x^2-2 x+4)}} \\
\end{pmatrix}P_{\nu}\,,
\end{equation}
with
\begin{equation}
\label{eq:P_nu_IO}P_{\nu}=\begin{pmatrix}
 e^{\frac{i }{2} (\rho +\psi )} &~ 0 &~ 0 \\
 0 &~ e^{\frac{ i}{2} (\sigma -\psi )} &~ 0 \\
 0 &~ 0 &~ 1 \\
\end{pmatrix}\,.
\end{equation}
In the IO case, $P_{\nu}$ will always be defined as Eq.~\eqref{eq:P_nu_IO} and it will be omitted for simplicity in the following. From the lepton mixing given in Eq.~\eqref{eq:PMNS_IO1}, we can extract the expressions of mixing angles and CP invariants as follow,
\begin{eqnarray}
 \nonumber \hskip-0.3in && \sin^2\theta_{13}=\frac{6 x^2}{4 -2 x+7 x^2}\,,\qquad \quad \sin^2\theta_{12}=\cos ^2\theta \,,\qquad  \quad \sin^2\theta_{23}=\frac{1}{2}\,, \\
\hskip-0.3in && J_{CP}=-\frac{\sqrt{3} x \left(x^2-2 x+4\right) \sin 2\theta  \cos \psi }{2\sqrt{2}\left(7 x^2-2 x+4\right)^{3/2}}\,, \qquad
 I_{1}=-\frac{\left(x^2-2 x+4\right)^2 \sin ^22 \theta  \sin (\rho -\sigma )}{4 \left(7 x^2-2 x+4\right)^2}\,.
\end{eqnarray}
where the definitions of CP invariants is given in Eq.~\eqref{eq:CP_invariants}. Note that the atmospheric mixing angle $\theta_{23}$ is maximal for any combination of input parameters and the reactor mixing angle $\theta_{13}$ only depend on input parameter $x$. Hence the viable range of $x$ can be obtained by varying $\theta_{13}$ over their $3\sigma$ range $0.02068\leq\sin^2\theta_{13}\leq0.02463$. Then input parameters $x$ is restricted in the range of $[-0.134,-0.122]\cup[0.115,0.126]$.

Subsequently we shall perform a comprehensive numerical analysis. The input parameters $x$, $r$ and $\eta$ are treated as random numbers in the intervals $[-20,20]$, $[0, 20]$ and $[-\pi, \pi]$ respectively, then we calculate the values of mixing parameters and mass ratio $m^2_1/m^2_2$. Imposing the $3\sigma$ allowed regions of the three mixing angles and $\Delta m^2_{21}/|\Delta m^2_{32}|$~\cite{Esteban:2018azc}, we find the allowed regions of the parameter $x$, $|\eta|$ and $r$ are $[-0.134,-0.122]\cup[0.115,0.126]$, $[0.9796\pi,0.9916\pi]$ and  $[8.630,8.713]$, respectively. As a consequence, $\theta_{13}$ and $\theta_{12}$ can take any values of their $3\sigma$ ranges~\cite{Esteban:2018azc}. The Dirac CP phase $\delta_{CP}$ and the absolute value of the Majorana CP phase $\beta$ are predicted to be in the range of $[-0.944\pi,-0.913\pi]\cup[-0.0873\pi,-0.0563\pi]\cup[0.0326\pi,0.0503\pi]\cup[0.950\pi,0.967\pi]$ and $[0.178\pi,0.193\pi]\cup[0.294\pi,0.321\pi]$, respectively. Here we find the Dirac CP phase is approximately trivial. Hence this breaking pattern would be ruled out if the signal of maximal $\delta_{CP}$ is confirmed by future neutrino facilities.

In order to show concrete examples, we give the predictions for mixing parameters, neutrino masses and the effective mass in neutrinoless double beta decay for $x=-1/8$ which gives a relatively simple VEV of $\phi_{\text{sol}}$.  In the case of $x=-1/8$, we find the third column of the PMNS matrix is $\left(\sqrt{\frac{2}{93}},\sqrt{\frac{91}{186}},\sqrt{\frac{91}{186}}\right)^T\simeq\left(0.147,0.699,0.699\right)^T$ which agrees with all measurements to date~\cite{Esteban:2018azc}. In order to evaluate how well the predicted mixing patterns agree with the experimental data on mixing angles and neutrino masses, we shall perform a $\chi^2$ analysis which uses the global fit results of Ref.~\cite{Esteban:2018azc}. The $\chi^2$ analysis results are
\begin{eqnarray}
\nonumber && \eta=-0.983\pi, \qquad m_{a}=5.721\,\text{meV},\qquad r=8.673, \qquad \chi^2_{\text{min}}=20.595, \qquad \sin^2\theta_{13}=0.0215\,, \\
\nonumber && \sin^2\theta_{12}=0.310, \qquad \sin^2\theta_{23}=0.5, \qquad \delta_{CP}=-0.0715\pi, \qquad \beta=-0.304\pi, \\
&&  m_1=49.377\,\text{meV}, \qquad m_2=50.120\,\text{meV}, \qquad  m_3=0\,\text{meV}, \qquad m_{ee}=43.904\,\text{meV}\,.
\end{eqnarray}

\item[~~($\mathcal{I}_{2}$)]{$(G_{l},G_{\text{atm}},G_{\text{sol}})=(Z_3^T,Z_2^{SU},Z_2^{TU})$, $X_{\text{atm}}=\{1,SU\}$, $X_{\text{sol}}=\{U,T\}$ }

For this breaking pattern, the VEV of flavon $\phi_{\text{atm}}$ is proportional to $\left(2,-1,-1\right)^T$.  Here only the residual CP symmetry $X_{\text{sol}}=\{U,T\}$ can give phenomenologically viable mixing patterns. The VEV of $\phi_{\text{sol}}$ is $\langle\phi_{\text{sol}}\rangle=v_{\phi_s}\left(1,x\omega , x\omega^2 \right)^T$. From the general VEVs of flavons $\phi_{\text{atm}}$ and $\phi_{\text{sol}}$, we find the general form of the neutrino mass matrix is given by
\begin{equation}\label{eq:mu_I2}
 m_{\nu}=m_{a}\begin{pmatrix}
 4 &~ -2 &~ -2 \\
 -2 &~ 1 &~ 1 \\
 -2 &~ 1 &~ 1 \\
\end{pmatrix}+m_{s}e^{i\eta}
\begin{pmatrix}
  1 &~ x \omega ^2 &~ x \omega  \\
 x \omega ^2 &~ x^2 \omega  &~ x^2 \\
 x \omega  &~ x^2 &~ x^2 \omega ^2 \\
\end{pmatrix}\,,
\end{equation}
After perform a unitary transformation $U_{\nu1}$ to $m_{\nu}$, where unitary matrix $U_{\nu1}$ takes the following form
\begin{equation}
U_{\nu1}=\begin{pmatrix}
 0 &~ i \sqrt{\frac{8 x^2-4 x+2}{11 x^2-4 x+2}} &~ \frac{i \sqrt{3} x}{\sqrt{11 x^2-4 x+2}} \\
 \frac{-2 x \omega -1}{\sqrt{2 \left(4 x^2-2 x+1\right)}} &~ -\frac{\sqrt{3} x (2 x \omega +1)}{\sqrt{2 \left(4 x^2-2 x+1\right) \left(11 x^2-4 x+2\right)}} &~ \frac{2 x \omega +1}{\sqrt{11 x^2-4 x+2}} \\
 \frac{-2 x \omega ^2-1}{\sqrt{2 \left(4 x^2-2 x+1\right)}} &~ \frac{\sqrt{3} x \left(2 x \omega ^2+1\right)}{\sqrt{2 \left(4 x^2-2 x+1\right) \left(11 x^2-4 x+2\right)}} &~ \frac{-2 x \omega ^2-1}{\sqrt{11 x^2-4 x+2}} \\
\end{pmatrix}\,,
\end{equation}
Then neutrino mass matrix $m^\prime_{\nu}$ is a block diagonal matrix with the nonzero elements $y$, $z$ and $w$ being
\begin{eqnarray}
\nonumber &&y= \frac{4(x-1)^2 m_{a} +x^2 (1-4 x)^2m_{s}e^{i \eta }  }{8 x^2-4 x+2}\,,  \\
\nonumber && z=\frac{i \sqrt{11 x^2-4 x+2} \left(4(x-1) m_{a} +x (4 x-1)m_{s}e^{i \eta }  \right)}{8 x^2-4 x+2}\,, \\
&& w= -\frac{(11 x^2-4 x+2) \left(4 m_{a}+m_{s}e^{i \eta } \right)}{8 x^2-4 x+2}\,.
\end{eqnarray}
The general form of the diagonalization matrix of $m^\prime_{\nu}$ is $U_{\nu2}$ which given in Eq.~\eqref{eq:Unu2} for IO case.  As the diagonalization matrix of the charged lepton mass matrix is the identity matrix. Then the PMNS matrix reads
\begin{equation}\label{eq:PMNS_I2}
\hskip-0.1in  U=
\begin{pmatrix}
 - \sqrt{\frac{8 x^2-4 x+2}{11 x^2-4 x+2}}e^{-i \psi } \sin \theta  &~ \sqrt{\frac{8 x^2-4 x+2}{11 x^2-4 x+2}} \cos \theta  &~ \frac{\sqrt{3} x}{\sqrt{11 x^2-4 x+2}} \\
 -\frac{\cos \theta }{\sqrt{2}}+\frac{\sqrt{3}x e^{-i \psi }  \sin \theta }{\sqrt{2(11 x^2-4 x+2)}} &~ \frac{-e^{i \psi } \sin \theta }{\sqrt{2}}-\frac{\sqrt{3} x \cos \theta }{\sqrt{2(11 x^2-4 x+2)}} &~ \sqrt{\frac{4 x^2-2 x+1}{11 x^2-4 x+2}} \\
 \frac{\cos \theta }{\sqrt{2}}+\frac{\sqrt{3}xe^{-i \psi } \sin \theta  }{\sqrt{2(11 x^2-4 x+2)}} &~ \frac{e^{i \psi } \sin \theta }{\sqrt{2}}-\frac{\sqrt{3} x \cos \theta }{\sqrt{2(11 x^2-4 x+2)}} &~ \sqrt{\frac{4 x^2-2 x+1}{11 x^2-4 x+2}} \\
\end{pmatrix}\,,
\end{equation}
 The lepton mixing parameters are predicted to be
\begin{eqnarray}
 \nonumber \hskip-0.3in && \sin^2\theta_{13}=\frac{3 x^2}{11 x^2-4 x+2}\,,\qquad
 \sin^2\theta_{12}=\cos^2\theta\,,\qquad
\sin^2\theta_{23}=\frac{1}{2}\,, \\
\hskip-0.3in && J_{CP}=\frac{\sqrt{3} x \left(4 x^2-2 x+1\right) \sin 2\theta   \sin \psi }{2(11 x^2-4 x+2)^{3/2}}\,, \qquad
 I_{1}=-\frac{\left(4 x^2-2 x+1\right)^2 \sin ^22 \theta \sin (\rho -\sigma )}{\left(11 x^2-4 x+2\right)^2}\,.
\end{eqnarray}
Note that the atmospheric mixing angle $\theta_{23}$ is predicted to be $45^\circ$. The admissible range of $x$ is $[-0.153,-0.138]\cup[0.108,0.118]$ which is obtained from the requirement that $\theta_{13}$ is in the experimentally preferred $3\sigma$ range. In order to see how well the lepton mixing angles and the neutrino masses can be described by this breaking pattern and its prediction for CP phases, we perform a $\chi^2$ analysis for $x=-1/7$ and $\eta=\pi$. From PMNS matrix in Eq.~\eqref{eq:PMNS_I2}, we see that the fixed column of PMNS matrix is $\left(\sqrt{\frac{3}{137}},\sqrt{\frac{67}{137}},\sqrt{\frac{67}{137}}\right)^T$ for $x=-1/7$. The $\chi^2$ results are
\begin{eqnarray}
\nonumber && m_{a}=12 .453\,\text{meV},\qquad r=5 .707, \qquad \chi^2_{\text{min}}=26 .114, \qquad \sin^2\theta_{13}=0 .0219\,, \\
\nonumber && \sin^2\theta_{12}=0 .278, \qquad \sin^2\theta_{23}=0.5, \qquad \delta_{CP}=-0.5\pi, \qquad \beta=\pi, \\
&&  m_1=49 .374\,\text{meV}, \qquad m_2=50 .117\,\text{meV}, \qquad m_3=0\,\text{meV}, \qquad m_{ee}=21 .261\,\text{meV}\,.
\end{eqnarray}
We see that the results of $\eta=\pi$ are viable and it leads to maximal Dirac CP phase and trivial Majorana CP phase. The reason is that the neutrino mass matrix in Eq.~\eqref{eq:mu_I2} has the symmetry property $m_{\nu}(x,r,\eta)=P^T_{132}m^{*}_{\nu}(x,r,\eta)P_{132}$. The reason why we chose $\eta=\pi$ is that it is easy to realize in an explicit model, please see the model in Ref.~\cite{Ding:2018fyz}. Furthermore, we perform a comprehensive numerical analysis. When the $3\sigma$ constraints on mixing angles and mass ratio $m^2_1/m^2_2$ are imposed. We find that the allowed regions of the parameter $x$, $|\eta|$ and $r$ are $[-0.153,-0.138]\cup[0.108,0.118]$, $[0.9965\pi,\pi]$ and $[5.666,5.817]$, respectively. The predictions for the two CP phases are $\delta_{CP}\in[-0.777\pi,-0.223\pi]\cup[0.145\pi,0.239\pi]\cup[0.761\pi,0.855\pi]$ and $|\beta|\in[0.636\pi,\pi]$. The mixing angles $\theta_{13}$ and $\theta_{12}$ can take any values in their $3\sigma$ ranges.

\item[~~($\mathcal{I}_{3}$)]{$(G_{l},G_{\text{atm}},G_{\text{sol}})=(K_4^{(S,U)},Z_2^{TST^2},Z_2^{U})$, $X_{\text{atm}}=\{SU,T^2,ST^2S,T^2STU\}$}

$\bullet$ $X_{\text{sol}}=\{1,U\}$

In this combination of residual flavor symmetries, both the residual CP transformations $X_{\text{sol}}=\{1,U\}$ and $X_{\text{sol}}=\{U,S\}$ will give results which agree with the experimental data. For the former case, the general VEVs invariant under the actions of the residual symmetries in the atmospheric neutrino and the solar neutrino sectors are
\begin{equation}
\langle\phi_{\text{atm}}\rangle=v_{\phi_a}\left(1,\omega ^2, \omega\right)^T\,, \qquad \langle\phi_{\text{sol}}\rangle=v_{\phi_s}\left(1 ,  x ,  x\right)^T\,,
\end{equation}
The light neutrino mass matrix is given by the seesaw formula, yielding
\begin{equation}
 m_{\nu}=m_{a}\begin{pmatrix}
 1 &~ \omega  &~ \omega ^2 \\
 \omega  &~ \omega ^2 &~ 1 \\
 \omega ^2 &~ 1 &~ \omega  \\
\end{pmatrix}+m_{s}e^{i\eta}
\begin{pmatrix}
 1 &~ x &~ x \\
 x &~ x^2 &~ x^2 \\
 x &~ x^2 &~ x^2 \\
\end{pmatrix}\,,
\end{equation}
The diagonalization matrix of above neutrino mass matrix can be taken to be $U_{\nu}=U_{\nu1}U_{\nu2}$ with
\begin{equation}
U^T_{\nu}m_{\nu}U_{\nu}=\text{diag}(m_1,m_2,0)\,,
\end{equation}
where
\begin{equation}
U_{\nu1}=\begin{pmatrix}
 \frac{1}{\sqrt{3}} &~ -\frac{2 +x}{ \sqrt{3(2+2 x+5 x^2)}} &~ -\frac{i \sqrt{3} x}{\sqrt{2+2 x+5 x^2}} \\
 \frac{\omega ^2}{\sqrt{3}} &~ \frac{\omega ^2-(2-\omega ) x }{ \sqrt{3(2+2 x+5 x^2)}} &~ \frac{x-\omega ^2 }{\sqrt{2+2 x+5 x^2}} \\
 \frac{\omega }{\sqrt{3}} &~ \frac{\omega -\left(2-\omega ^2\right) x }{ \sqrt{3(2+2 x+5 x^2)}} &~ \frac{\omega  -x}{\sqrt{2+2 x+5 x^2}} \\
\end{pmatrix}\,,
\end{equation}
and $U_{\nu2}$ is shown in Eq.~\eqref{eq:Unu2} for IO case. Here if we only perform a unitary transformation $U_{\nu1}$ on $ m_{\nu}$, we obtain a block diagonal neutrino mass matrix $m^\prime_{\nu}$ with nonzero elements
\begin{eqnarray}
\nonumber &&y=3 m_{a}+\frac{1}{3}   (1-x)^2m_{s}e^{i \eta }\,,  \\
\nonumber && z=\frac{1}{3}   (x-1) \sqrt{2+2 x+5 x^2}m_{s}e^{i \eta } \,, \\
&& w= \frac{1}{3}   \left(2+2 x+5 x^2\right)m_{s}e^{i \eta }\,.
\end{eqnarray}
Form the expression of $U_{\nu}$, we find that the lepton mixing matrix takes the following form
\begin{equation}\label{eq:PMNS_I3_1}
\hskip-0.1in  U=
\begin{pmatrix}
 \frac{(1+2x) e^{-i \psi } \sin \theta }{\sqrt{2 +2 x+5 x^2}} &~ \frac{(1+2 x) \cos \theta }{\sqrt{2 +2 x+5 x^2}} &~ \frac{i (1-x)}{\sqrt{2 +2 x+5 x^2}} \\
 -\frac{\cos \theta }{\sqrt{2}}+\frac{(x-1) e^{-i \psi } \sin \theta }{\sqrt{2(2 +2 x+5 x^2)}} &~ \frac{(x-1) \cos \theta }{\sqrt{2(2 +2 x+5 x^2)}}+\frac{e^{i \psi } \sin \theta }{\sqrt{2}} &~ \frac{i (1+2 x)}{\sqrt{2(2 +2 x+5 x^2)}} \\
 \frac{\cos \theta }{\sqrt{2}}+\frac{(x-1) e^{-i \psi } \sin \theta }{\sqrt{2(2 +2 x+5 x^2)}} &~ \frac{(x-1) \cos \theta }{\sqrt{2(2 +2 x+5 x^2)}}-\frac{e^{i \psi } \sin \theta }{\sqrt{2}} &~ \frac{i (1+2 x)}{\sqrt{2(2 +2 x+5 x^2)}} \\
\end{pmatrix}\,,
\end{equation}
Accordingly we find the expressions of the mixing angles and CP invariants are
\begin{eqnarray}
\nonumber &&  \sin^2\theta_{13}=\frac{(x-1)^2}{2+2 x+5 x^2}, \qquad
  \sin^2\theta_{12}=\cos ^2\theta , \qquad
\sin^2\theta_{23}=\frac{1}{2}\,, \\
&& J_{CP}=\frac{(x-1) (2x+1)^2 \sin 2 \theta  \sin \psi }{4 \left(2+2 x+5 x^2\right)^{3/2}}, \qquad
I_{1}=-\frac{(2x+1)^4 \sin ^22 \theta  \sin (\rho -\sigma )}{4 \left(2+2 x+5 x^2\right)^2}\,.
\end{eqnarray}
This mixing matrix gives a maximal $\theta_{23}$ for any value of $x$. The admissible range $[0.638, 0.662]\cup[1.615 , 1.699]$ of $x$ is obtained by varying $\theta_{13}$ over its $3\sigma$ range. It the three mixing angles are required within their $3\sigma$ ranges. We find the other two input parameters $|\eta|$ and $r$ are restricted in the range of $[0.965\pi,0.975\pi]$ and $[0.441,0.480]\cup[1.594,1.630]$, respectively. The two CP phases are predicted to be $0.487\pi\leq|\delta_{CP}|\leq0.503\pi$ and $0.0962\pi\leq|\beta|\leq0.108\pi$.

Similar to previous cases, we also give an example which could be easily achieved in a model. The alignment parameter is $x=2/3$, thus the VEVs of $\phi_{\text{sol}}$ is $\left(1,\frac{2}{3},\frac{2}{3}\right)^Tv_{\phi_s}$. The fixed column of PMNS matrix takes the form $\left(\frac{1}{5 \sqrt{2}},\frac{7}{10},\frac{7}{10}\right)^T\simeq\left(0.141,0.7,0.7\right)^T$. The best fit values of mixing parameters and neutrino masses for this example are
\begin{eqnarray}
\nonumber && \eta=-0.969\pi, \qquad m_{a}=16 .794\,\text{meV},\qquad r=1 .579, \qquad \chi^2_{\text{min}}=33 .640, \qquad \sin^2\theta_{13}=0.02\,, \\
\nonumber && \sin^2\theta_{12}=0.310, \qquad \sin^2\theta_{23}= 0.5, \qquad \delta_{CP}=-0.497\pi, \qquad \beta=0.0964\pi, \\
&&  m_1=49 .377\,\text{meV}, \qquad m_2=50 .120\,\text{meV}, \qquad m_3=0\,\text{meV}, \qquad m_{ee}=48 .137\,\text{meV}\,.
\end{eqnarray}

We see that $\theta_{13}$ is rather close to its $3\sigma$ lower limit $0.2068$~\cite{Esteban:2018azc}. Hence this results should be considered as a good leading order approximation. The reason is that accordance with experimental data can be easily achieved after subleading contributions are taken into account.

$\bullet$ $X_{\text{sol}}=\{U,S\}$

For this kind of residual CP symmetries in the atmospheric neutrino sector and the solar neutrino sector, the neutrino mass matrix takes the form
\begin{equation}
\hskip-0.1in m_{\nu}=m_{a}\begin{pmatrix}
 0 &~ 0 &~ 0 \\
 0 &~ 1 &~ -1 \\
 0 &~ -1 &~ 1 \\
\end{pmatrix}+m_{s}e^{i\eta}
\begin{pmatrix}
 (1+2 x i)^2 &~ (1-i x) (1+2 x i) &~ (1-i x) (1+2 x i) \\
 (1-i x) (1+2 x i) &~ (1-i x)^2 &~ (1-i x)^2 \\
 (1-i x) (1+2 x i) &~ (1-i x)^2 &~ (1-i x)^2 \\
\end{pmatrix}\,,
\end{equation}
In order to diagonalize it, we first perform a unitary transformation $U_{\nu1}$, where unitary matrix $U_{\nu1}$ takes as the form
\begin{equation}
U_{\nu1}=\begin{pmatrix}
 \frac{1}{\sqrt{3}} &~ \frac{i x-1}{ \sqrt{3(1+x^2)}} &~ \frac{-i-x }{\sqrt{3(1+x^2)}} \\
 \frac{\omega ^2}{\sqrt{3}} &~ \frac{i \omega  x-1}{\sqrt{3(1+x^2)}} &~ \frac{-i  \omega-x }{\sqrt{3(1+x^2)}} \\
 \frac{\omega }{\sqrt{3}} &~ \frac{i \omega ^2 x-1}{\sqrt{3(1+x^2)}} &~ \frac{-i \omega ^2-x}{\sqrt{3(1+x^2)}} \\
\end{pmatrix}\,,
\end{equation}
Then we find that the neutrino mass matrix $m^\prime_{\nu}$ is a block diagonal matrix with nonzero elements
 \begin{equation}
y=3 m_{a}-3 x^2  m_{s}e^{i \eta }, \qquad
z=-3 i x   \sqrt{1+x^2}m_{s}e^{i \eta },  \qquad
w=3   \left(1+x^2\right)m_{s}e^{i \eta }\,.
\end{equation}
The neutrino mass matrix $m^\prime_{\nu}$ can be further diagonalized by the unitary matrix $U_{\nu2}$. Then the lepton mixing matrix is determined to be of the form
\begin{equation}\label{eq:PMNS_I3_2}
\hskip-0.1in  U
=\begin{pmatrix}
 \frac{ e^{-i \psi } \sin \theta}{\sqrt{1+x^2}} &~ \frac{ \cos \theta }{\sqrt{1+x^2}} &~ -\frac{x}{\sqrt{1+x^2}} \\
 \frac{i \cos \theta }{\sqrt{2}}+\frac{x e^{-i \psi } \sin \theta }{ \sqrt{2(1+x^2)}} &~ \frac{x \cos \theta }{\sqrt{2(1+x^2)}}-\frac{ie^{i \psi }  \sin \theta }{\sqrt{2}} &~ \frac{1}{\sqrt{2(1+x^2)}} \\
 \frac{i \cos \theta }{\sqrt{2}}-\frac{x e^{-i \psi } \sin \theta }{\sqrt{2(1+x^2)}} &~ -\frac{x \cos \theta }{\sqrt{2(1+x^2)}}-\frac{i e^{i \psi } \sin \theta }{\sqrt{2}} &~ -\frac{1}{\sqrt{2(1+x^2)}} \\
\end{pmatrix}\,,
\end{equation}
The lepton mixing angles and CP invariants turn out to take the form
\begin{eqnarray}
\nonumber && \sin^2\theta_{13}=\frac{x^2}{1+x^2}\,, \qquad
  \sin^2\theta_{12}=\cos^2\theta \,, \qquad
\sin^2\theta_{23}=\frac{1}{2}\,, \\
&& J_{CP}=-\frac{x \sin 2 \theta  \cos \psi }{4 \left(1+x^2\right)^{3/2}}\,, \qquad I_{1}=-\frac{ \sin^22\theta \sin (\rho -\sigma )}{4\left(1+x^2\right)^2}\,.
\end{eqnarray}
The atmospheric mixing angle is maximal and the value of $\theta_{13}$ relies on the value of $x$. If $\theta_{13}$ is required  in its $3\sigma$ range, we find that $|x|$ has to lie in the range of $[0.145,0.159]$. When all the three mixing angles and mass ratio $m^2_1/m^2_2$ are restricted in their $3\sigma$ ranges~\cite{Esteban:2018azc}. The other two input parameters  $|\eta|$ and $r$ should be constrained in regions $[0.0245\pi,0.0345\pi]$ and $[0.945,0.954]$, respectively. Then the absolute values of the two CP phases would generically be constrained in small regions $[0.488\pi,0.503\pi]$ and $[0.0971\pi,0.110\pi]$. Furthermore, we shall perform a comprehensive numerical analysis for $x=1/4\sqrt{3}$. Then the first column of PMNS matrix is fixed to be $\frac{1}{7}\left(1,2\sqrt{6},2\sqrt{6}\right)^T\simeq\left(0.143,0.700,0.700\right)^T$, and the best fit values of mixing parameters are
\begin{eqnarray}
\nonumber && \eta=0.0307\pi, \qquad m_{a}=16 .798\,\text{meV},\qquad r=0 .955, \qquad \chi^2_{\text{min}}=29 .075, \qquad \sin^2\theta_{13}=0.0204\,, \\
\nonumber && \sin^2\theta_{12}=0.310, \qquad \sin^2\theta_{23}= 0.5, \qquad \delta_{CP}=-0.497\pi, \qquad \beta=0.0973\pi, \\
&&  m_1=49.377\,\text{meV}, \qquad m_2=50 .120\,\text{meV}, \qquad m_3=0\,\text{meV}, \qquad m_{ee}=48 .108\,\text{meV}\,.
\end{eqnarray}

\item[~~($\mathcal{I}_{4}$)]{$(G_{l},G_{\text{atm}},G_{\text{sol}})=(K_4^{(S,U)},Z_2^{TU},Z_2^{TU})$, $X_{\text{atm}}=\{U,T\}$, $X_{\text{sol}}=\{U,T\}$}

In this case, only the residual CP transformation $X_{\text{sol}}=\{U,T\}$ will give results which agree with the experimental data. This mixing pattern is discussed in the case $\mathcal{N}_{5}$ with NO. Then the results of this case can be easy obtained from the predictions of case $\mathcal{N}_{5}$. The PMNS matrix in this case is
\begin{equation}
  U=
\begin{pmatrix}
 \frac{\sqrt{\frac{3}{2}}x}{\sqrt{1+2 x^2}} &~  \frac{i}{2} &~ \frac{\sqrt{3} }{2 \sqrt{1+2 x^2}} \\
  \frac{2 +x}{ \sqrt{6(1+2 x^2)}} &~ -\frac{i}{2} &~ \frac{1-4 x}{2 \sqrt{3(1+2 x^2)}} \\
  \frac{x-1}{\sqrt{3(1+2 x^2)}} &~ -\frac{i}{\sqrt{2}} &~ \frac{1+2 x}{ \sqrt{6(1+2 x^2)}} \\
\end{pmatrix}\text{diag}(e^{-\frac{i\eta}{2}},1,1)\,,
\end{equation}
One can straightforwardly extract the lepton mixing angles as follows,
\begin{equation}
  \sin^2\theta_{13}=\frac{3 }{4 +8 x^2}, \qquad
  \sin^2\theta_{12}=\frac{1+2 x^2}{1+8x^2}, \qquad
\sin^2\theta_{23}=\frac{(1-4 x)^2}{3 \left(1+8 x^2\right)}\,.
\end{equation}
The three mixing angles only depend on one input parameter $x$. Then two sum rules among the three mixing angles can be obtained. We find that the two sum rules here are the same as the sum rules in Eq.~\eqref{eq:sum_BM}. As same as $\mathcal{N}_{5}$, the two CP phases are predicted to be
\begin{equation}
\sin\delta_{CP}=0, \qquad \beta=\eta+\pi\,.
\end{equation}
It implies that the Dirac CP phase is trivial for input parameters taking any values. The Majorana CP phase relies only on input parameter $\eta$. In the case $\eta=0$ or $\pi$, the Majorana CP phase is also trivial. While the Majorana CP phase is maximal for $\eta=\pm\pi/2$. As an example, we give the predictions for $x=4$. Here the VEVs of the flavonr fields $\phi_{\text{sol}}$ is proportional to column vector $\left(1,4\omega,4\omega^2\right)^T$ and the PMNS matrix is
\begin{equation}
 U=
\begin{pmatrix}
 2 \sqrt{\frac{2}{11}} &~ \frac{i}{2} &~ \frac{1}{2 \sqrt{11}} \\
 \sqrt{\frac{2}{11}} &~ -\frac{i}{2} &~ -\frac{5}{2 \sqrt{11}} \\
 \frac{1}{\sqrt{11}} &~ -\frac{i}{\sqrt{2}} &~ \frac{3}{\sqrt{22}} \\
\end{pmatrix}\text{diag}(e^{-\frac{i\eta}{2}},1,1)\,,
\end{equation}
The three mixing angles are predicted to be
\begin{equation}
 \sin^2\theta_{13}=\frac{1}{44}\simeq0.0227\,, \quad
  \sin^2\theta_{12}=\frac{11}{44}\simeq0.256\,, \quad
\sin^2\theta_{23}=\frac{25}{44}\simeq0.581\,.
\end{equation}
We find that the solar mixing angles is rather close to its $3\sigma$ lower bound~\cite{Esteban:2018azc}.  Hence this mixing pattern can be regarded as a good leading order approximation.

\item[~~($\mathcal{I}_{5}$)]{$(G_{l},G_{\text{atm}},G_{\text{sol}})=(Z_3^{T},Z_2^{SU},Z_2^{SU})$, $X_{\text{atm}}=\{1,SU\}$, $X_{\text{sol}}=\{U,S\}$ }

When the residual CP $X_{\text{sol}}$ takes the transformations $\{U,S\}$, the flavons $\phi_{\text{atm}}\sim\mathbf{3}$ and $\phi_{\text{sol}}\sim\mathbf{3^\prime}$ are required to have the vacuum alignments
\begin{equation}
\langle\phi_{\text{atm}}\rangle=v_{\phi_a}\left(2, -1 , -1\right)^T\,, \qquad \langle\phi_{\text{sol}}\rangle=v_{\phi_s}\left(1,  1+i x , 1-ix\right)^T\,,
\end{equation}
It is easy to check that the column vectors $\langle\phi_{\text{atm}}\rangle$ and $\langle\phi_{\text{sol}}\rangle$ are orthogonal to each other. It leads to a form dominance breaking pattern~\cite{Chen:2009um,Choubey:2010vs,King:2010bk}. The light neutrino mass matrix is given by the seesaw relation with
\begin{equation}
 m_{\nu}=m_{a}\begin{pmatrix}
 4 &~ -2 &~ -2 \\
 -2 &~ 1 &~ 1 \\
 -2 &~ 1 &~ 1 \\
\end{pmatrix}+m_{s}e^{i\eta}
\begin{pmatrix}
 1 &~ 1-i x &~ 1+i x \\
  1-i x &~ (1-i x)^2 &~ 1+x^2 \\
  1+i x &~ 1+x^2 &~ (1+ ix)^2 \\
\end{pmatrix}\,,
\end{equation}
The diagonalization matrix of $m_{\nu}$ with  non-negative  eigenvalues can be taken to be
\begin{equation}
U_{\nu}=\begin{pmatrix}
 \sqrt{\frac{2}{3}} &~ \frac{1}{\sqrt{3+2 x^2}} &~ \frac{i \sqrt{2} x}{\sqrt{3(3+2 x^2)}} \\
 -\frac{1}{\sqrt{6}} &~ \frac{1+ix}{\sqrt{3+2 x^2}} &~ \frac{3+2 ix}{ \sqrt{6(3+2 x^2)}} \\
 -\frac{1}{\sqrt{6}} &~ \frac{1-i x}{\sqrt{3+2 x^2}} &~ \frac{2 i x-3 }{\sqrt{6(3+2 x^2)}} \\
\end{pmatrix}\text{diag}(1,e^{-\frac{i\eta}{2}},1)\,,
\end{equation}
where
\begin{equation}
U^T_{\nu}m_{\nu}U_{\nu}=\text{diag}(6 m_{a}, m_{s} \left(3 +2 x^2\right),0 )\,,
\end{equation}
Only input parameters $m_a$, $m_s$ and $x$ are involved in the two nonzero neutrino masses.  Since the diagonalization matrix of the charged lepton mass matrix is a $3\times3$ identity matrix up to the permutation of columns. Then lepton mixing matrix can be taken to be $U_{\nu}$, i.e. this form dominance breaking pattern predicts a TM1 mixing matrix. Then the three lepton mixing angles and two CP invariants are determined to be
\begin{eqnarray}
\nonumber &&  \sin^2\theta_{13}=\frac{2 x^2}{9 +6 x^2}, \qquad
  \sin^2\theta_{12}=\frac{3 }{9 +4 x^2} , \qquad
\sin^2\theta_{23}=\frac{1}{2}\,, \\
\label{eq:mix_par_I5}&& J_{CP}=-\frac{x}{9 +6 x^2}, \qquad
I_{1}=-\frac{2  \sin \eta }{9 +6x^2}\,.
\end{eqnarray}
We note that the atmospheric mixing angle $\theta_{23}$ is maximal and the other two mixing angles depend on a single real parameter $x$. Hence the following sum rule between the reactor mixing angle and the solar mixing angle is found to be satisfied
\begin{equation}
\cos^2\theta_{12}\cos^2\theta_{13}=\frac{2}{3}\,.
\end{equation}
This sum rule comes from the so called TM1 mixing matrix. From the CP invariants defined in Eq.~\eqref{eq:CP_invariants} and the expressions of mixing parameters in~\eqref{eq:mix_par_I5}, we find
\begin{equation}
\sin\delta_{CP}=-\text{sign}(x), \qquad \beta=-\eta\,.
\end{equation}
It predicts a maximal Dirac CP phase and the Majorana CP phase $\beta$ equals the negative value of input parameter $\eta$. As an example, we give the predictions for $x=1/3$. Here the VEVs of the flavon $\phi_{\text{sol}}$ is proportional to column vector $\left(1,1+\frac{i}{3},1-\frac{i}{3}\right)^T$ and the PMNS matrix for $x=1/3$  is
\begin{equation}
 U=
\begin{pmatrix}
 \sqrt{\frac{2}{3}} &~ \frac{3}{\sqrt{29}} &~ i \sqrt{\frac{2}{87}} \\
 -\frac{1}{\sqrt{6}} &~ \frac{3+i}{\sqrt{29}} &~ \frac{9+2 i}{\sqrt{174}} \\
 -\frac{1}{\sqrt{6}} &~ \frac{3-i}{\sqrt{29}} &~ -\frac{9-2 i}{\sqrt{174}} \\
\end{pmatrix}\text{diag}(1,1,e^{-\frac{i\eta}{2}})\,,
\end{equation}
The three mixing angles are predicted to be
\begin{equation}
 \sin^2\theta_{13}=\frac{2}{87}\simeq0.0230\,, \quad
  \sin^2\theta_{12}=\frac{27}{87}\simeq0.318\,, \quad
\sin^2\theta_{23}=\frac{1}{2}, \quad \delta_{CP}=-0.5\pi.
\end{equation}
These results agrees with all measurements to date~\cite{Esteban:2018azc}. In above we have given five kinds of viable breaking patterns with IO. From table~\ref{tab:bf_via_CP}, we know that other thirteen kinds of breaking patterns with IO are left. In order to increase an article's readability, we shall discuss them in appendix~\ref{sec:other_IO_mix}.

\end{description}

\section{\label{sec:model}Model construction}

In section~\ref{sec:TD_S4_GCP_NO}, section~\ref{sec:TD_S4_GCP_IO}, appendix~\ref{sec:other_NO_mix} and appendix~\ref{sec:other_IO_mix}, we have performed a model-independent analysis for the lepton mixing patterns which can be derived from $S_{4}\rtimes H_{CP}$ in the tri-direct CP approach. In the model-independent analysis, we have assumed that the flavons $\phi_{\text{atm}}$ and $\phi_{\text{sol}}$ transform as $S_{4}$ triplet $\mathbf{3}$ and $\mathbf{3^\prime}$, respectively. From model-independent analysis, we find 8 kinds of breaking patterns which are compatible with current experimental data for NO neutrino masses. In this section, we shall construct a supersymmetric model with the flavor symmetry $S_4$ and a CP symmetry, and the symmetry breaking pattern $\mathcal{N}_4$ is realized due to on-vanishing vacuum expectation values of some flavons. The phenomenological predictions of $\mathcal{N}_4$ for lepton mixing parameters and neutrino masses has been studied in detail in Ref.~\cite{Ding:2018fyz}. The reason why we construct a model to realize the breaking pattern $\mathcal{N}_4$ are: firstly the TM1 mixing matrix obtained in $\mathcal{N}_{1}\sim \mathcal{N}_{3}$ has been widely discussed. Secondly the vacuum alignment of flavon $\phi_{\text{sol}}$ and the fixed column of PMNS matrix are relatively simpler than other breaking patterns which don't give TM1 mixing. Thirdly the minimum $\chi^2$ in $\mathcal{N}_4$ is small for simple benchmark values of $x$ and $\eta$. We have tabulated many simple admissible values of $x$ and $\eta$ in table~\ref{tab:bf_N4}. We find that phenomenologically viable lepton mixing matrix can be obtained for the atmospheric flavon vacuum alignment $(1,\omega^2,\omega)^T$ and the solar flavon vacuum alignment $(1,-4,-4)^T$ with $\eta=\pm\frac{3\pi}{4}$. In the present work, the model based on above vacuum alignments will be realized, i.e. we shall present a supersymmetric model which contains two right-handed neutrinos and realizes the breaking pattern $\mathcal{N}_4$ with $x=-4$ and $\eta=\pm\frac{3\pi}{4}$. Here we assume that the three generations of left-handed lepton doublets $L$ furnish triplet representation $\mathbf{3}$ under the family symmetry $S_{4}$, while the right-handed charged leptons $e^{c}$, $\mu^{c}$ and $\tau^{c}$  are singlet $\mathbf{1}$ under the family symmetry $S_{4}$. The two right-handed neutrinos $\nu^{c}_{\text{atm}}$ and $\nu^{c}_{\text{sol}}$ are assumed to be the singlet representations $\mathbf{1}$ and $\mathbf{1^\prime}$, respectively. In order to ensure the needed vacuum alignment and to forbid unwanted couplings, the auxiliary symmetry $Z_{5}\times Z_8 \times Z^\prime_8$ is imposed. In this model, the original symmetry $S_{4}\rtimes H_{CP}$ will spontaneous break to $Z^T_3$, $Z^{TST^2}_2\times X_{\text{atm}}$ and $Z_2^U\times X_{\text{sol}}$ in the charged lepton, atmospheric neutrino and solar neutrino sectors, where the residual CP transformations $X_{\text{atm}}=SU$ and $X_{\text{sol}}=U$. As a consequence, the desired vacua $\langle\phi_{\text{atm}}\rangle\propto\left(1, \omega^2, \omega\right)^{T}$ and $\langle\phi_{\text{sol}}\rangle\propto\left(1,-4,-4\right)^{T}$ can be really accomplished. The relevant flavon fields and how they transform under the imposed flavor symmetry $S_4\times Z_{5}\times Z_8 \times Z^\prime_8$ are collected in table~\ref{tab:field}.

\begin{table}[t!]
\renewcommand{\tabcolsep}{0.5mm}
\begin{center}
\begin{tabular}{|c|c|c|c|c|c|c|c||c|c||c|c||c|c|c|c|c|c|c||c|c||c|c|c|c|c|c|c|c|c|c|c|c|}\hline\hline
 & $L$ & $e^c$  &  $\mu^c$ &  $\tau^c$  & $\nu^{c}_{\text{atm}}$ & $\nu^{c}_{\text{sol}}$  &$H_{u,d}$ & $\eta_{l}$ &  $\phi_{l}$   & $\xi_{a}$ & $\phi_{a}$  & $\xi_{s}$ & $\eta_{s}$ & $\chi_{s}$ &  $\varphi_{s}$ & $\Delta_{s}$  &  $\phi_{s}$ &  $\psi_{s}$ &  $\xi^0_{l}$  &  $\phi^0_{l}$ &  $\phi^0_{a}$ & $\sigma^0$ & $\rho^0$ &  $\eta^0$ & $\chi^0$ & $\varphi^0$  &   $\Delta^0$ &   $\kappa^0$   \\ \hline

$S_4$  & $\mathbf{3}$ &  $\mathbf{1}$  &  $\mathbf{1}$ & $\mathbf{1}$  & $\mathbf{1}$ & $\mathbf{1^\prime}$ & $\mathbf{1}$  &  $\mathbf{2}$ & $\mathbf{3}$  &  $\mathbf{1}$ & $\mathbf{3}$  & $\mathbf{1}$  & $\mathbf{2}$  &  $\mathbf{3}^{\prime}$ & $\mathbf{3}^{\prime}$  & $\mathbf{3}^{\prime}$  &  $\mathbf{3}^{\prime}$  &$\mathbf{3}^{\prime}$  &  $\mathbf{1}$ &  $\mathbf{3}^{\prime}$ &  $\mathbf{3^\prime}$ &  $\mathbf{2}$ &  $\mathbf{2}$ &  $\mathbf{2}$ & $\mathbf{3}^{\prime}$   & $\mathbf{3}^{\prime}$ &  $\mathbf{3}^{\prime}$ & $\mathbf{1}$   \\

$Z_5$ & $\omega^4_5$  & $\omega^3_5$   & $\omega^4_5$   & $1$  & $1$  & $1$  & $1$  & $\omega_5$  &  $\omega_5$  &  $1$  & $\omega_5$  & $1$   & $\omega^3_5$  &   $\omega^2_5$   & $1$ & $\omega^3_5$  & $\omega_5$ &  $\omega^4_5$  & $\omega^3_5$   & $\omega^3_5$   &$\omega^3_5$  &  $\omega^2_5$   & $\omega_5$ & $\omega^4_5$   & $1$  & $1$  & $\omega^2_5$  & $\omega^3_5$       \\

$Z_8$ & $\omega^7_8$ &  $\omega^6_8$  &  $\omega^7_8$  & $1$  & $\omega^5_8$ & $\omega_8$ & $1$ & $\omega_8$ & $\omega_8$ & $\omega^6_8$ &  $-1$ &  $\omega^6_8$  & $1$  & $1$ & $1$ &   $1$ &  $1$ &  $1$  & $\omega^6_8$ &  $\omega^6_8$ &   $1$  & $1$ & $1$ &  $1$ &  $1$ &   $1$  &  $1$  &   $1$     \\

$Z^\prime_8$ & $\omega_8$ & $\omega^7_8$  & $\omega^7_8$ & $\omega^7_8$ & $\omega^5_8$  & $1$  & $1$  &  $1$ & $1$ & $\omega^6_8$ & $\omega^2_8$  & $1$  & $\omega^5_8$  & $\omega^5_8$  & $\omega_8$ & $\omega^4_8$ & $\omega^7_8$ & $\omega^5_8$  & $1$ &   $1$ &   $\omega^4_8$  & $\omega^6_8$    & $\omega^6_8$ & $\omega^6_8$ & $\omega^6_8$ &  $\omega^7_8$ & $\omega^4_8$  &   $\omega^2_8$   \\ \hline\hline

\end{tabular}
\caption{\label{tab:field} The lepton, Higgs and flavon superfields and their transformation properties under the flavor symmetry $S_4\times Z_5 \times Z_8\times Z^\prime_8$, where $\omega_5\equiv e^{2\pi i/5}$ and $\omega_8\equiv e^{\pi i/4}$. In addition, we assume a standard $U(1)_{R}$ symmetry under which all lepton fields carry a unit charge while the Higgs and flavons have zero charge. }
\end{center}
\end{table}

\subsection{\label{subsec:alignment}Vacuum alignment}

We adopt the now-standard $F-$term alignment mechanism to generate the appropriate vacuum alignments of the flavor symmetry breaking flavons. The leading order (LO) driving superpotential $w_d$ which is invariant under the imposed $S_4\times Z_{5}\times Z_8 \times Z^\prime_8$ takes the following form
\begin{equation}
w_{d}=w_{d}^l+w_{d}^{\text{atm}}+w_{d}^{\text{sol}}\,,
\end{equation}
where $w_{d}^l$, $w_{d}^{\text{atm}}$ and $w_{d}^{\text{sol}}$ are used to  realize the LO vacuum alignments of the flavons in the charged lepton sector, the atmospheric neutrino sector and the solar neutrino sector, respectively. They can be expressed as
\begin{eqnarray}
\nonumber&&\hskip-0.15in w_{d}^l=g_{1}\xi^0_{l}\left(\eta_{l}\eta_{l}\right)_{\mathbf{1}}+g_2\xi^0_{l}\left(\phi_{l}\phi_{l}\right)_{\mathbf{1}}
+g_3\left(\phi^{0}_{l}\left(\eta_{l}\phi_{l}\right)_{\mathbf{3^\prime}}\right)_{\mathbf{1}}
+g_4\left(\phi^{0}_{l}\left(\phi_{l}\phi_{l}\right)_{\mathbf{3^\prime}}\right)_{\mathbf{1}}\,,\\
\nonumber&&\hskip-0.15in w_{d}^{\text{atm}}=h_1\left(\phi^{0}_{a}\left(\phi_{a}\phi_{a}\right)_{\mathbf{3^\prime}}\right)_{\mathbf{1}}\,,\\
\nonumber&&\hskip-0.15in w_{d}^{\text{sol}}=f_{1}\left(\rho^0\left(\chi_{s}\chi_{s}\right)_{\mathbf{2}}\right)_{\mathbf{1}}
+f_{2}\left(\sigma^0\left(\psi_{s}\psi_{s}\right)_{\mathbf{2}}\right)_{\mathbf{1}}
+f_{3}\left(\eta^0\left(\eta_{s}\eta_{s}\right)_{\mathbf{2}}\right)_{\mathbf{1}}
+f_{4}\left(\eta^0\left(\chi_{s}\psi_{s}\right)_{\mathbf{2}}\right)_{\mathbf{1}} \\
\nonumber &&\qquad+f_{5}\left(\chi^0\left(\eta_{s}\chi_{s}\right)_{\mathbf{3^\prime}}\right)_{\mathbf{1}}
+f_{6}\left(\chi^0\left(\varphi_{s}\varphi_{s}\right)_{\mathbf{3^\prime}}\right)_{\mathbf{1}}
+M_{\varphi}\left(\varphi^0\varphi_{s}\right)_{\mathbf{1}}
+f_{7}\left(\varphi^0\left(\chi_{s}\psi_{s}\right)_{\mathbf{3^\prime}}\right)_{\mathbf{1}} \\
 \label{eq:wd_exp}&&\qquad+M_{\Delta}\left(\Delta^0_{s}\Delta_{s}\right)_{\mathbf{1}}+f_{8}\left(\Delta^0_{s}\left(\chi_{s}\phi_{s}\right)_{\mathbf{3^\prime}}\right)_{\mathbf{1}}
+f_{9}\kappa^0_{s}\left(\chi_{s}\varphi_{s}\right)_{\mathbf{1}}
+f_{10}\kappa^0_{s}\left(\phi_{s}\phi_{s}\right)_{\mathbf{1}}\,,
\end{eqnarray}
where the subscript of $(\cdots)_{\mathbf{r}}$ denotes a contraction of the $S_4$ indices into the representation $\mathbf{r}$.  All the coupling $g_{i}$ $h_{1}$, $f_{i}$ and  mass parameters $M_{\varphi}$, $M_{\Delta}$ in Eq.~\eqref{eq:wd_exp} are real. The reason is that we have required that the theory is invariant under the imposed generalised CP symmetry. In the SUSY limit, the vacuum alignment is achieved via the requirement of vanishing $F-$terms of the driving fields.  In the charged lepton sector, the $F-$term conditions obtained from the driving fields $\xi^{0}_{l}$ and $\phi^{0}_{l}$ are given by
\begin{eqnarray}
\label{eq:align_ch}
\nonumber&&\frac{\partial w_{d}^l}{\partial\xi^{0}_{l}}=2 g_{1} \eta_{l_{1}} \eta_{l_{2}}+g_{2} \left(\phi_{l_{1}}^2+2 \phi_{l_{2}} \phi_{l_{3}}\right)=0\,,\\
\nonumber&&\frac{\partial w_{d}^l}{\partial\phi^{0}_{l_{1}}}=g_{3} (\eta_{l_{1}} \phi_{l_{2}}-\eta_{l_{2}} \phi_{l_{3}})+2g_{4} \left( \phi_{l_{1}}^2- \phi_{l_{2}} \phi_{l_{3}}\right)=0\,,\\
\nonumber&&\frac{\partial w_{d}^l}{\partial\phi^{0}_{l_{2}}}=g_{3} (\eta_{l_{1}} \phi_{l_{1}}-\eta_{l_{2}} \phi_{l_{2}})+2g_{4} \left( \phi_{l_{2}}^2- \phi_{l_{1}} \phi_{l_{3}}\right)=0\,,\\
&&\frac{\partial w_{d}^l}{\partial\phi^{0}_{l_{3}}}=g_{3} (\eta_{l_{1}} \phi_{l_{3}}-\eta_{l_{2}} \phi_{l_{1}})+2g_{4} \left( \phi_{l_{3}}^2- \phi_{l_{1}} \phi_{l_{2}}\right)=0\,.
\end{eqnarray}
By straightforward calculations, we find that these equations are satisfied by the alignment
\begin{equation}
\label{eq:vacuum_ch}\langle\eta_{l}\rangle=\left(0, 1\right)^Tv_{\eta_{l}},\qquad \langle\phi_{l}\rangle=\left(0, 1,0\right)^Tv_{\phi_{l}}, \quad \text{with} \quad v_{\phi_{l}}=\frac{g_{3}}{2g_{4}}v_{\eta_{l}}\,,
\end{equation}
where $v_{\eta_{l}}$ is undetermined.  In the atmospheric neutrino sector, the vacuum is determined by $F-$term conditions associated with the driving field $\phi^{0}_{a}$
\begin{equation}
\begin{aligned}
&\frac{\partial w_{d}^{\text{atm}}}{\partial\phi^{0}_{a_{1}}}=2h_{1} \left( \phi_{a_{1}}^2- \phi_{a_{2}} \phi_{a_{3}}\right)=0\,,\\
&\frac{\partial w_{d}^{\text{atm}}}{\partial\phi^{0}_{a_{2}}}=2h_{1} \left( \phi_{a_{2}}^2- \phi_{a_{1}} \phi_{a_{3}}\right)=0\,,\\
&\frac{\partial w_{d}^{\text{atm}}}{\partial\phi^{0}_{a_{3}}}=2h_{1} \left( \phi_{a_{3}}^2- \phi_{a_{1}} \phi_{a_{2}}\right)=0\,,
\end{aligned}
\end{equation}
These equations lead to the vacuum alignment of $\phi_{a}$ being
\begin{equation}
\label{eq:vacuum_atm} \langle\phi_{a}\rangle=\left(1, \omega^2, \omega\right)^T v_{\phi_{a}}\,.
\end{equation}
Now we turn to the $F-$term conditions of the solar neutrino sector. The resulting $F$-term conditions depend on the $S_{4}$ representation of $\rho^0$ and $\sigma^0$ are given as
\begin{equation}
\begin{aligned}
&\frac{\partial w_{d}^{\text{sol}}}{\partial\rho^{0}_1}=f_{1} \left(2 \chi_{s_{1}} \chi_{s_{2}}+\chi_{s_{3}}^2\right)=0\,,\\
&\frac{\partial w_{d}^{\text{sol}}}{\partial\rho^{0}_2}=f_{1} \left(2 \chi_{s_{1}} \chi_{s_{3}}+\chi_{s_{2}}^2\right)=0\,,\\
&\frac{\partial w_{d}^{\text{sol}}}{\partial\sigma^{0}_1}=f_{2} \left(2 \psi_{s_{1}} \psi_{s_{2}}+\psi_{s_{3}}^2\right)=0\,,\\
&\frac{\partial w_{d}^{\text{sol}}}{\partial\sigma^{0}_2}=f_{2} \left(2 \psi_{s_{1}} \psi_{s_{3}}+\psi_{s_{2}}^2\right)=0\,.\\
\end{aligned}
\end{equation}
The above equations lead to the vacuum alignments of $\chi_{s}$ and $\psi_{s}$ as follow
\begin{equation}\label{eq:vev_sol_1}
 \langle\chi_{s}\rangle=v_{\chi_{s}}(1,0,0)^T, \qquad
\langle\psi_{s}\rangle=v_{\psi_{s}}\left(1, -2,  -2\right)^T \,,
\end{equation}
The alignment of the doublet flavon $\eta$ originates in the $F$-term
 \begin{equation}
\begin{aligned}
&\frac{\partial w_{d}^{\text{sol}}}{\partial\eta^{0}_1}=f_{3} \eta_{s_{1}}^2+f_{4} (\chi_{s_{1}} \psi_{s_{2}}+\chi_{s_{2}} \psi_{s_{1}}+\chi_{s_{3}} \psi_{s_{3}})=0\,,\\
&\frac{\partial w_{d}^{\text{sol}}}{\partial\eta^{0}_2}=f_{3} \eta_{s_{2}}^2+f_{4} (\chi_{s_{1}} \psi_{s_{3}}+\chi_{s_{2}} \psi_{s_{2}}+\chi_{s_{3}} \psi_{s_{1}})=0\,,\\
\end{aligned}
\end{equation}
A straightforward calculation shows that the $F$-term conditions resulting from above determine the doublet alignment uniquely to
\begin{equation}\label{eq:vev_sol_2}
\langle\eta_{s}\rangle=v_{\eta_{s}}(1,1)^T, \qquad \text{with} \qquad v^2_{\eta_{s}}=\frac{2f_{4}}{f_{3}}v_{\chi_{s}}v_{\psi_{s}} \,.
\end{equation}
From these results, we immediately see the $F$-term conditions
\begin{equation}
\begin{aligned}
&\frac{\partial w_{d}^{\text{sol}}}{\partial\chi^{0}_1}=f_{5} (\eta_{s_{1}} \chi_{s_{2}}+\eta_{s_{2}} \chi_{s_{3}})+2f_{6} \left( \varphi_{s_{1}}^2- \varphi_{s_{2}} \varphi_{s_{3}}\right)=0\,,\\
&\frac{\partial w_{d}^{\text{sol}}}{\partial\chi^{0}_2}=f_{5} (\eta_{s_{1}} \chi_{s_{1}}+\eta_{s_{2}} \chi_{s_{2}})+2f_{6} \left( \varphi_{s_{2}}^2- \varphi_{s_{1}} \varphi_{s_{3}}\right)=0\,,\\
&\frac{\partial w_{d}^{\text{sol}}}{\partial\chi^{0}_3}=f_{5} (\eta_{s_{1}} \chi_{s_{3}}+\eta_{s_{2}} \chi_{s_{1}})+2f_{6} \left( \varphi_{s_{3}}^2- \varphi_{s_{1}} \varphi_{s_{2}}\right)=0\,, \\
\end{aligned}
\end{equation}
generate the alignment
\begin{equation}\label{eq:vev_sol_3}
\langle\varphi_{s}\rangle=v_{\varphi_{s}}(1,-1,-1)^T, \qquad \text{with} \qquad v^2_{\varphi_{s}}=-\frac{f_{5}}{4f_{6}}v_{\eta_{s}}v_{\chi_{s}} \,.
\end{equation}
Similarly, we can consider the $F$-term of the driving field $\varphi^0$
\begin{equation}
\begin{aligned}
&\frac{\partial w_{d}^{\text{sol}}}{\partial\varphi^{0}_1}=M_{\varphi} \varphi_{s_{1}}+f_{7} (2 \chi_{s_{1}} \Delta_{s_{1}}-\chi_{s_{2}} \Delta_{s_{3}}-\chi_{s_{3}} \Delta_{s_{2}})=0\,,\\
&\frac{\partial w_{d}^{\text{sol}}}{\partial\varphi^{0}_2}=M_{\varphi} \varphi_{s_{3}}+f_{7} (2 \chi_{s_{2}} \Delta_{s_{2}}-\chi_{s_{1}} \Delta_{s_{3}}-\chi_{s_{3}} \Delta_{s_{1}})=0\,,\\
&\frac{\partial w_{d}^{\text{sol}}}{\partial\varphi^{0}_3}=M_{\varphi} \varphi_{s_{2}}+f_{7} (2 \chi_{s_{3}} \Delta_{s_{3}}-\chi_{s_{1}} \Delta_{s_{2}}-\chi_{s_{2}} \Delta_{s_{1}})=0\,, \\
\end{aligned}
\end{equation}
These conditions determine the alignments of $\Delta_{s}$ being
\begin{equation}\label{eq:vev_sol_4}
\langle\Delta_{s}\rangle=v_{\Delta_{s}}(1,2,2)^T, \qquad \text{with} \qquad v_{\Delta_{s}}=-\frac{M_{\varphi}v_{\varphi_{s}}}{2f_{7}v_{\chi_{s}}} \,.
\end{equation}
In order to realise the alignment $\langle\phi_{s}\rangle\propto(1,-4,-4)^T$, we consider the $F$-term of driving field $\Delta^0$:
\begin{equation}
\begin{aligned}
&\frac{\partial w_{d}^{\text{sol}}}{\partial\Delta^{0}_{1}}=M_{\Delta} \Delta_{s_{1}}+f_{8} (2 \chi_{s_{1}} \phi_{s_{1}}-\chi_{s_{2}} \phi_{s_{3}}-\chi_{s_{3}} \phi_{s_{2}})=0\,,\\
&\frac{\partial w_{d}^{\text{sol}}}{\partial\Delta^{0}_{2}}=M_{\Delta} \Delta_{s_{3}}+f_{8} (2 \chi_{s_{2}} \phi_{s_{2}}-\chi_{s_{3}} \phi_{s_{1}}-\chi_{s_{1}} \phi_{s_{3}})=0\,,\\
&\frac{\partial w_{d}^{\text{sol}}}{\partial\Delta^{0}_{3}}=M_{\Delta} \Delta_{s_{2}}+f_{8} (2 \chi_{s_{3}} \phi_{s_{3}}-\chi_{s_{1}} \phi_{s_{2}}-\chi_{s_{2}} \phi_{s_{1}})=0\,, \\
\end{aligned}
\end{equation}
which uniquely specify the alignment
\begin{equation}\label{eq:vve_sol_5}
\langle\phi_{s}\rangle=v_{\phi_{s}}(1,-4,-4)^T, \qquad \text{with} \qquad v_{\phi_{s}}=-\frac{M_{\Delta}v_{\Delta_{s}}}{2f_{8}v_{\chi_{s}}} \,.
\end{equation}
For driving field $\kappa^0$, the $F-$flatness gives rise to
\begin{equation}
\frac{\partial w_{d}^{\text{sol}}}{\partial\kappa^{0}}=f_{9} (\varphi_{s_{1}} \chi_{s_{1}}+\varphi_{s_{2}} \chi_{s_{3}}+\varphi_{s_{3}} \chi_{s_{2}})+f_{10} \left(\phi_{s_{1}}^2+2 \phi_{s_{2}} \phi_{s_{3}}\right)=0
\end{equation}
It is easy to check that
\begin{equation}\label{eq:vev_sol_6}
v^2_{\phi_{s}}=-\frac{f_{9}v_{\chi_{s}}v_{\varphi_{s}}}{33f_{10}}\,.
\end{equation}

Now we have obtained the vacuum alignments of all flavons $\eta_s$, $\psi_s$, $\chi_s$, $\varphi_s$, $\Delta_s$ and $\phi_s$ in the solar neutrino sector by adopting the standard $F-$term alignment mechanism. In other words, the needed vacuum alignment $\langle\phi_s\rangle\propto(1,-4,-4)^T$ is realized. Next we shall fix the overall phases of all VEVs of flavons in the atmospheric and solar neutrino sectors. From the alignments of flavons $\xi_{s}$, $\eta_{s}$, $\chi_{s}$, $\psi_{s}$, $\varphi_{s}$ and $\phi_{s}$ shown above, we find the VEVs of these fields are invariant under the subgroup $Z^{U}_2$. In order to obtain the phase with $\eta=\pm\frac{3\pi}{4}$, we introduce the $S_{4}$ singlet fields in table~\ref{tab:field_2}. Then the driving superpotential which is used to obtain the phases of all the VEVs of flavons in the neutrino sector is
\begin{eqnarray}
\nonumber w_{d}^{\text{phase}}&=&M^2_{1}\zeta^0_{1}+x_{1}\zeta^0_{1}\Omega^2_{1}
+M^2_{2}\zeta^0_{2}+x_{2}\zeta^0_{2}\Omega^2_{2}+M^2_{3}\zeta^0_{3}+x_{3}\zeta^0_{3}\Omega_{3}\Omega_{4} +M^2_{4}\zeta^0_{4}+x_{4}\zeta^0_{4}\Omega_{5}\Omega_{6}+M_{\Omega_1}\Omega^0_1\Omega_1\\
\nonumber &&+x_{5}\Omega^0_1\xi^2_{s}+M_{\Omega_2}\Omega^0_2\Omega_2+x_{6}\Omega^0_2\xi^2_{a}
+M_{\Omega_3}\Omega^0_3\Omega_3+x_{7}\Omega^0_3\left(\eta_s\eta_s\right)_{\mathbf{1}} +x_{8}\Omega^0_3\left(\chi_s\psi_s\right)_{\mathbf{1}}
+M_{\Omega_4}\Omega^0_4\Omega_4\\
&& +x_{9}\Omega^0_4\left(\varphi_s\psi_s\right)_{\mathbf{1}} +M_{\Omega_5}\Omega^0_5\Omega_5+x_{10}\Omega^0_5\left(\phi_a\chi_s\right)_{\mathbf{1^\prime}}
  +M_{\Omega_6}\Omega^0_6\Omega_6+x_{11}\Omega^0_6\left(\phi_a\phi_s\right)_{\mathbf{1^\prime}}\,,
\end{eqnarray}
where couplings $x_{i}$ and mass parameters $M^2_{i}$ and $M_{\Omega_{i}}$ are all real. The F-term conditions from above superpotential are
\begin{equation}
\begin{array}{ll}
\frac{\partial w_{d}^{\text{phase}}}{\partial\zeta^{0}_{1}}=M^2_{1}+x_{1}\Omega^2_{1}=0\,,  & \qquad \frac{\partial w_{d}^{\text{phase}}}{\partial\zeta^{0}_{2}}=M^2_{2}+x_{2}\Omega^2_{2}=0\,,\\
\frac{\partial w_{d}^{\text{phase}}}{\partial\zeta^{0}_{3}}=M^2_{3}+x_{3}\Omega_{3}\Omega_{4}=0\,,  & \qquad \frac{\partial w_{d}^{\text{phase}}}{\partial\zeta^{0}_{4}}=M^2_{4}+x_{4}\Omega_{5}\Omega_{6}=0\,,\\
\frac{\partial w_{d}^{\text{phase}}}{\partial\Omega^{0}_{1}}=M_{\Omega_1}\Omega_{1}+x_{5}\xi^2_{s} =0\,,
 & \qquad\frac{\partial w_{d}^{\text{phase}}}{\partial\Omega^{0}_{2}}=M_{\Omega_2}\Omega_{2}+x_{6}\xi^2_{a} =0\,,\\
\frac{\partial w_{d}^{\text{phase}}}{\partial\Omega^{0}_{3}}=M_{\Omega_3}\Omega_{3}+2x_{7}v^2_{\eta_{s}}+x_{8}v_{\chi_{s}}v_{\psi_{s}} =0\,,
 & \qquad \frac{\partial w_{d}^{\text{phase}}}{\partial\Omega^{0}_{4}}=M_{\Omega_4}\Omega_{4}+5x_{9}v_{\varphi_{s}}v_{\psi_{s}} =0\,, \\
 \frac{\partial w_{d}^{\text{phase}}}{\partial\Omega^{0}_{5}}=M_{\Omega_5}\Omega_{5}+x_{10}v_{\chi_{s}}v_{\phi_{a}} =0\,,
 & \qquad \frac{\partial w_{d}^{\text{phase}}}{\partial\Omega^{0}_{6}}=M_{\Omega_6}\Omega_{6}+5x_{11}v_{\phi_{s}}v_{\phi_{a}} =0\,, \\
\end{array}
\end{equation}
From above equations and the relations among VEVs of flavons in the solar neutrino sector, we can achieve the phases of the VEVs of all the flavons in the solar and atmospheric neutrino sectors. Since the expressions of the flavon VEVs are a little redundant and there are not used in the following discussing. Hence we will not show them here. In order to obtain the phase $\eta=\pm\frac{3\pi}{4}$, the only valid ratio is $\frac{v^2_{\phi_{s}}v_{\xi_{a}}}{v^2_{\phi_{a}}v_{\xi_{s}}}$, where $v_{\xi_{s}}$ and $v_{\xi_{a}}$ are the VEVs of flavon fields $\xi_{s}$ and $\xi_{a}$, respectively. The expression of this ratio is
\begin{table}[t!]
\renewcommand{\tabcolsep}{2.3mm}
\begin{center}
\begin{tabular}{|c|c|c|c|c|c|c||c|c|c|c|c|c|c|}\hline\hline
 & $\Omega_1$ & $\Omega_2$  &  $\Omega_3$ &  $\Omega_4$  & $\Omega_5$ & $\Omega_6$  & $\Omega^0_1$ & $\Omega^0_2$  &  $\Omega^0_3$ &  $\Omega^0_4$  & $\Omega^0_5$ & $\Omega^0_6$  & $\zeta^0_{1,2,3,4}$ \\ \hline

$S_4$  & $\mathbf{1}$ &  $\mathbf{1}$  &  $\mathbf{1}$ & $\mathbf{1}$  & $\mathbf{1^\prime}$ & $\mathbf{1^\prime}$ & $\mathbf{1}$ &  $\mathbf{1}$  &  $\mathbf{1}$ & $\mathbf{1}$  & $\mathbf{1^\prime}$ & $\mathbf{1^\prime}$ & $\mathbf{1}$    \\

$Z_5$ & $1$  & $1$   & $\omega_5$   & $\omega^4_5$  & $\omega^3_5$ & $\omega^2_5$  & $1$  & $1$  &  $\omega^4_5$  &  $\omega_5$  & $\omega^2_5$  & $\omega^3_5$   & $1$     \\

$Z_8$ & $-1$ &  $-1$  &  $1$  & $1$  & $-1$ & $-1$ & $-1$ & $-1$ & $1$ & $1$ &  $-1$ &  $-1$  & $1$    \\

$Z^\prime_8$ & $1$ & $-1$  & $\omega^2_8$ & $\omega^6_8$ & $\omega^7_8$  & $\omega_8$  & $1$  &  $-1$ & $\omega^6_8$ & $\omega^2_8$ & $\omega_8$  & $\omega^7_8$  & $1$   \\ \hline\hline

\end{tabular}
\caption{\label{tab:field_2} The transformation rules of the singlet flavon and driving superfields which are used to determine the phase of the flavon VEVs.}
\end{center}
\end{table}
\begin{equation}\label{eq:VEV_ratio}
\frac{v^2_{\phi_{s}}v_{\xi_{a}}}{v^2_{\phi_{a}}v_{\xi_{s}}}=\frac{\left(\frac{5}{33}\right)^{3/4}x_{4}x_{10} x_{11} f_{5}M_{\Delta_{s}}M_{\varphi_{s}}}{16 \sqrt{2}f_{6} f_{7} f_{8}M_{\Omega_5} M_{\Omega_6}M_{4}^2}\left(\frac{x_1x^2_5f_{4}^2  f_{9}^3 M_{2}^2M_{3}^2 M_{\Delta_{s}}^2 M_{\varphi_{s}}^2  M^2_{\Omega_2}M_{\Omega_3} M_{\Omega_4} }{ (f_{3} x_{8}+4 f_{4} x_{7})x_2x_{3}x^2_6 x_{9} f_{3}  f_{7}^2 f_{8}^2 f_{10}^3 M_{1}^2 M^2_{\Omega_1}}\right)^{\frac{1}{4}},
\end{equation}
Since all the couplings $x_i$, $f_i$ and mass parameters $M_i$ in Eq.~\eqref{eq:VEV_ratio} are real. Then we see that the phases of the ratio $\frac{v^2_{\phi_{s}}v_{\xi_{a}}}{v^2_{\phi_{a}}v_{\xi_{s}}}$ is $e^{\frac{ik\pi}{4}}(i=0,1,...,7)$. In the present work we shall concentrate on the solution with
\begin{equation}\label{eq:VEV_ratio_arg}
\text{arg}\left(\frac{v^2_{\phi_{s}}v_{\xi_{a}}}{v^2_{\phi_{a}}v_{\xi_{s}}}\right)=\pm\frac{3\pi}{4}\,.
\end{equation}
In order to obtain the observed hierarchy among the charged lepton masses, we assume
\begin{equation}
\frac{v_{\eta_l}}{\Lambda}\sim\frac{v_{\phi_l}}{\Lambda}\sim\lambda^2\,,
\end{equation}
where $\lambda$ is the Cabibbo angle with $\lambda\simeq0.23$. Moreover, the VEVs of flavons in the neutrino sector are expected to be of the same order of magnitude and we will take them to be of the same order as the VEVs of flavons in the charged lepton sector, i.e.
\begin{equation}
\frac{v_{\xi_a}}{\Lambda}\sim\frac{v_{\phi_a}}{\Lambda}\sim\frac{v_{\xi_s}}{\Lambda}\sim\frac{v_{\phi_s}}{\Lambda}\sim
\frac{v_{\eta_s}}{\Lambda}\sim\frac{v_{\chi_s}}{\Lambda}\sim\frac{v_{\psi_s}}{\Lambda}\sim\frac{v_{\varphi_s}}{\Lambda}\sim
\frac{v_{\Delta_s}}{\Lambda}\sim\frac{v_{\Omega_i}}{\Lambda}\sim\lambda^2\,,
\end{equation}
where $v_{\Omega_i}(i=1,\cdots,6)$ are the VEVs of flavons $\Omega_i$. Now we will briefly touch on the subleading corrections to the driving superpotential given above. We first start with the corrections to the driving superpotential $w^l_d$ which contains the driving fields $\xi^0_l$ and $\phi^0_l$. We find that the NLO corrections of it is suppressed by $1/\Lambda^2$ with respect to the renormalizable terms in Eq.~\eqref{eq:wd_exp}. The subleading contributions to driving superpotential $w^\text{atm}_d$ and $w^\text{sol}_d$ involve three flavon fields. The corresponding corrections to the leading order terms in $w^\text{atm}_d$ and $w^\text{sol}_d$  are of relative order $\lambda^2$.

\subsection{\label{subsec:model}The structure of the model}

The lowest dimensional Yukawa operators of the charged lepton mass terms,  which are invariant under the imposed flavor symmetry $S_4\times Z_{5}\times Z_8 \times Z^\prime_8$,  can be written as
\begin{eqnarray}
\nonumber  w_l&&=\frac{y_{\tau}}{\Lambda}\left(L\phi_{l}\right)_{\mathbf{1}}\tau^{c}H_d
+\frac{y_{\mu_1}}{\Lambda^2}\left(L\left(\eta_{l}\phi_{l}\right)_{\mathbf{3}}\right)_{\mathbf{1}}\mu^{c}H_d
+\frac{y_{\mu_2}}{\Lambda^2} \left(L\left(\phi_{l}\phi_{l}\right)_{\mathbf{3}}\right)_{\mathbf{1}}\mu^{c}H_d\\
\nonumber&&+\frac{y_{e_1}}{\Lambda^3}\left(L\phi_{l}\right)_{\mathbf{1}}\left(\eta_{l}\eta_{l}\right)_{\mathbf{1}}e^cH_d
+\frac{y_{e_2}}{\Lambda^3}\left(\left(L\phi_{l}\right)_{\mathbf{2}}\left(\eta_{l}\eta_{l}\right)_{\mathbf{2}}\right)_{\mathbf{1}}e^cH_d
+\frac{y_{e_3}}{\Lambda^3}\left(\left(L\eta_{l}\right)_{\mathbf{3}}\left(\phi_{l}\phi_{l}\right)_{\mathbf{3}}\right)_{\mathbf{1}}e^cH_d \\
\nonumber&&
+\frac{y_{e_4}}{\Lambda^3}\left(\left(L\eta_{l}\right)_{\mathbf{3}^{\prime}}\left(\phi_{l}\phi_{l}\right)_{\mathbf{3}^{\prime}}\right)_{\mathbf{1}}e^cH_d
 +\frac{y_{e_5}}{\Lambda^3}\left(L\phi_{l}\right)_{\mathbf{1}}\left(\phi_{l}\phi_{l}\right)_{\mathbf{1}}e^{c}H_d
+\frac{y_{e_6}}{\Lambda^3}\left(\left(L\phi_{l}\right)_{\mathbf{2}}\left(\phi_{l}\phi_{l}\right)_{\mathbf{2}}\right)_{\mathbf{1}}e^cH_d\\
\label{eq:ch_Yukawa}&&+\frac{y_{e_7}}{\Lambda^3}\left(\left(L\phi_{l}\right)_{\mathbf{3}}\left(\phi_{l}\phi_{l}\right)_{\mathbf{3}}\right)_{\mathbf{1}}e^cH_d
+\frac{y_{e8}}{\Lambda^3}\left(\left(L\phi_{l}\right)_{\mathbf{3}^{\prime}}\left(\phi_{l}\phi_{l}\right)_{\mathbf{3}^{\prime}}\right)_{\mathbf{1}}e^cH_d\,,
\end{eqnarray}
where all couplings are real  due to the generalized CP symmetry. Substituting the flavon VEVs of $\eta_{l}$ and $\phi_{l}$ in Eq.~\eqref{eq:vacuum_ch}, we find the charged lepton mass matrix is diagonal with the three charged lepton  masses being
\begin{eqnarray}
\nonumber&& m_e=\left|\left(y_{e_6}-2y_{e_8}-2y_{e_4}v_{\eta_{l}}/v_{\phi_{l}}+y_{e_2}v^2_{\eta_{l}}/v^2_{\phi_{l}}\right)\frac{v^3_{\phi_{l}}}{\Lambda^3}\right|v_d,\\
&&m_{\mu}=\left|y_{\mu_1}\frac{v_{\eta_{l}}v_{\phi_{l}}}{\Lambda^2}\right|v_d,\qquad m_{\tau}=\left|y_{\tau}\frac{v_{\phi_{l}}}{\Lambda}\right|v_d\,,
\end{eqnarray}
where $v_{d}=\langle H_{d}\rangle$. Note, in order to obtain the mass hierarchies of the charged leptons $m_{e}:m_{\mu}:m_{\tau}\simeq \lambda^4:\lambda^2:1$, the auxiliary symmetry $Z_{8}$ is imposed, where $\lambda\simeq0.23$ is the Cabibbo angle. The auxiliary symmetry $Z_{8}$ imposes different powers of $\eta_{l}$ and $\phi_{l}$ to couple with the electron, muon and tau lepton mass terms. From Eq.~\eqref{eq:ch_Yukawa}, we find that the electron, muon and tau masses arise at order $(\langle\Phi_{l}\rangle/\Lambda)^3$, $(\langle\Phi_{l}\rangle/\Lambda)^2$ and $\langle\Phi_{l}\rangle/\Lambda$ respectively, where $\Phi_{l}$ refer to either $\eta_{l}$ or $\phi_{l}$. If we assume that $\langle\Phi_{l}\rangle/\Lambda$ is of order $\lambda^2$, then the  mass hierarchy of the charged leptons can be reproduced. Moreover, the subleading operators related to $e^{c}$, $\mu^{c}$ and $\tau^{c}$ comprise four flavons and consequently are suppressed by $1/\Lambda^4$. Such corrections for the charged lepton masses and lepton mixing parameters can be neglected.

Now we come to the neutrino sector. The light neutrino masses are  given by the famous type-I seesaw mechanism with two right-handed neutrinos. The most general LO superpotential for the neutrino masses is
\begin{equation}
w_{\nu}=\frac{y_{a}}{\Lambda}\left(L\phi_{a}\right)_{\mathbf{1}}H_{u}\nu^{c}_{\text{atm}}+
\frac{y_{s}}{\Lambda}\left(L\phi_{s}\right)_{\mathbf{1^\prime}}H_{u}\nu^{c}_{\text{sol}}
+x_{a}\nu^{c}_{\text{atm}}\nu^{c}_{\text{atm}}\xi_{a}
+x_{s}\nu^{c}_{\text{sol}}\nu^{c}_{\text{sol}}\xi_{s}\,,
\end{equation}
where the four coupling constants $y_{a}$, $y_{s}$, $x_{a}$ and $x_{s}$ are real because the theory is required to be invariant under the generalised CP transformation. From the vacuum alignments of flavons $\phi_{a}$ and $\phi_{s}$, we can read out the Dirac and Majorana mass matrices as follows
\begin{equation}
M_{D}=\begin{pmatrix}
y_{a}v_{\phi_{a}} ~&~ y_{s}v_{\phi_{s}} \\
\omega y_{a}v_{\phi_{a}}  ~&~ -4y_{s}v_{\phi_{s}} \\
\omega^2y_{a}v_{\phi_{a}}  ~&~ -4y_{s}v_{\phi_{s}}
\end{pmatrix}\frac{v_{u}}{\Lambda},\qquad
M_{N}=\begin{pmatrix}
x_{a}v_{\xi_{a}}  &  0  \\
0  &  x_{s}v_{\xi_{s}}
\end{pmatrix}\,,
\end{equation}
where $v_{u}=\langle H_{u}\rangle$ and the clarity of expressions of VEVs $v_{\xi_{a}}$, $v_{\xi_{s}}$, $v_{\phi_{a}}$, $v_{\phi_{s}}$ are shown in section~\ref{subsec:alignment}. After applying the seesaw formula, the effective light neutrino mass matrix can be written as
\begin{equation}\label{eq:nu_mass_LO}
m_{\nu}=-\frac{v^2_{u}}{\Lambda^2}\left[\frac{y^2_{a}v^2_{\phi_{a}}}{x_{a}v_{\xi_{a}}}\begin{pmatrix}
 1 &~ \omega  &~ \omega ^2 \\
 \omega  &~ \omega ^2 &~ 1 \\
 \omega ^2 &~ 1 &~ \omega  \\
\end{pmatrix}+\frac{y^2_{s}v^2_{\phi_{s}}}{x_{s}v_{\xi_{s}}}\begin{pmatrix}
 1 &~  -4 &~  -4 \\
 -4 &~ 16 &~ 16 \\
  -4 &~ 16 &~ 16 \\
\end{pmatrix}
\right]\,.
\end{equation}
In section~\ref{subsec:alignment}, we have taken the solution with the phase of the ratio $\frac{v^2_{\phi_{s}}v_{\xi_{a}}}{v^2_{\phi_{a}}v_{\xi_{s}}}$ being $\pm\frac{3\pi}{4}$. Up to the over all phase of the neutrino mass matrix in Eq.~\eqref{eq:nu_mass_LO},  we see that this neutrino mass matrix is of the same form as the general mass matrix in breaking pattern $\mathcal{N}_4$~\cite{Ding:2018fyz} but with
\begin{equation}
x=-4, \qquad m_a=\left|\frac{y^2_{a}v^2_{\phi_{a}}}{x_{a}v_{\xi_{a}}}\frac{v^2_{u}}{\Lambda^2}\right|,\qquad
m_se^{i\eta}=\pm\left|\frac{y^2_{s}v^2_{\phi_{s}}}{x_{s}v_{\xi_{s}}}\frac{v^2_{u}}{\Lambda^2}\right|e^{\mp\frac{3\pi i}{4}}\,.
\end{equation}
It is easy check that we can obtain the results with $\eta=\pm\frac{3\pi}{4}$ in the case of $x_ax_s>0$. In the following, we will briefly touch on the subleading corrections to the superpotential given in sections~\ref{subsec:alignment} and \ref{subsec:model}. Furthermore,we find that the next-to-leading operators of $w_{\nu}$ are suppressed by $1/\Lambda^2$ with respect to the LO contributions and therefore can be negligible.

From the standard procedure shown in section~\ref{sec:framework}, we find that the above model predicts the following LO lepton mixing matrix
\begin{equation}
U_{PMNS}=\frac{1}{\sqrt{74}}
\begin{pmatrix}
 4 \sqrt{3} &~ -i \sqrt{26} \cos \theta  &~ -i\sqrt{26}e^{i \psi } \sin \theta  \\
 \sqrt{13} &~ 2\sqrt{6} i  \cos \theta -\sqrt{37} e^{-i \psi } \sin \theta  &~ 2\sqrt{6}i e^{i \psi }  \sin \theta  +\sqrt{37}\cos \theta   \\
 \sqrt{13} &~ 2\sqrt{6} i \cos \theta  +\sqrt{37}e^{-i \psi } \sin \theta   &~ 2\sqrt{6} i  e^{i \psi } \sin \theta -\sqrt{37} \cos \theta  \\
\end{pmatrix}P_{\nu}\,,
\end{equation}
where the diagonal phase $P_{\nu}$ is given in Eq.~\eqref{eq:dia_Pnu}. All the parameters $\theta$, $\psi$, $\sigma$ and $\rho$ only depend on one input parameter $r$. In the case of $\eta=-\frac{3\pi}{4}$, the three mixing angles and the two CP invariants can be expressed in terms of $r=m_s/m_a$ as
\begin{eqnarray}
\nonumber &&\sin^2\theta_{13}=\frac{13}{74} \left(1-\frac{15 \left(781 r^2+3 \sqrt{2} r-7\right)}{13 C_{r}}\right)\,, \\
\nonumber &&\sin^2\theta_{12}=1-\frac{48 C_{r}}{15 \left(781 r^2+3 \sqrt{2} r-7\right)+61 C_{r}}\,, \\
\nonumber &&\sin^2\theta_{23}=\frac{1}{2}+\frac{740 \sqrt{6} r}{15 \left(781 r^2+3 \sqrt{2} r-7\right)+61 C_{r}}\,, \\
&&J_{CP}=\frac{3 \sqrt{3} \left(-308 r^2+25 \sqrt{2} r-2\right)}{37 C_{r}}, \qquad I_{1}=\frac{\sqrt{2} \left(1-196 r^2\right)}{37  C_{r}}\,,
\end{eqnarray}
where parameter $C_{r}$ is defined as
\begin{equation}
C_{r}=\sqrt{\left(1089 r^2-25 \sqrt{2} r+9\right)^2-21904 r^2}\,.
\end{equation}
A sum rule between the reactor mixing angle and the solar mixing angle is easy to obtain
\begin{equation}
\cos^2\theta_{12}\cos^2\theta_{13}=\frac{48}{74}\,.
\end{equation}
Inputting the experimentally preferred $3\sigma$ range $0.01981\leq\sin^2\theta_{13}\leq0.02436$~\cite{Esteban:2018azc}, we obtain the prediction for the solar mixing angle
\begin{equation}
0.3352\leq \sin^2\theta_{12}\leq0.3365 \,.
\end{equation}
Furthermore, we find that the two nonzero neutrino masses are determined to be
\begin{equation}
m^2_2=\frac{m^2_a}{2}  \left(1089 r^2-25 \sqrt{2} r+9- C_{r}\right)\,, \quad
m^2_3=\frac{m^2_a}{2}  \left(1089 r^2-25 \sqrt{2} r+9+ C_{r}\right)\,,
\end{equation}
The neutrino masses $m^2_2$ and $m^2_3$ are dependent on free parameters $m_a$ and $r$. If we require that all the three lepton mixing angles and two mass squared differences lie in their corresponding experimentally $3\sigma$ intervals~\cite{Esteban:2018azc}. Then the lepton mixing parameters and the neutrino masses are predicted to be
\begin{eqnarray}
\nonumber &&0.3362\leq \sin^2\theta_{12}\leq0.3364, \quad 0.02254\leq \sin^2\theta_{13}\leq0.02280, \quad 0.556\leq \sin^2\theta_{23}\leq0.564, \\
\nonumber &&-0.418\leq\delta_{CP}/\pi\leq-0.406, \quad  0.263\leq\beta/\pi\leq0.264, \quad  2.690\,\text{meV}\leq m_{ee}\leq2.985\,\text{meV}, \\
\label{eq:model_prediction} &&8.240\,\text{meV}\leq m_2\leq8.950\,\text{meV}, \quad 49.265\,\text{meV}\leq m_3\leq51.235\,\text{meV}\,.
\end{eqnarray}
We see that all mixing parameters and neutrino masses are restricted in rather narrow regions. It is straightforward to show that the model above is a powerful model to predicted lepton mixing parameters and neutrino masses, especially for mixing angle $\theta_{12}$. The next generation reactor neutrino oscillation experiments JUNO~\cite{An:2015jdp} and RENO-50~\cite{Kim:2014rfa} expect to reduce the error of $\theta_{12}$ to about $0.1^{\circ}$ or around $0.3\%$.  The oscillation parameters $\theta_{12}$, $\theta_{23}$ and $\delta_{CP}$ would be precisely measured by the future long baseline experiments DUNE~\cite{Acciarri:2016crz,Acciarri:2015uup,Acciarri:2016ooe}, T2HK~\cite{Abe:2014oxa}, T2HKK~\cite{Abe:2016ero}. Hence this breaking pattern can be checked by future neutrino facilities. Furthermore, we expect that a more ambitious facility such as the neutrino factory~\cite{Geer:1997iz,DeRujula:1998umv,Bandyopadhyay:2007kx} could provide a more stringent tests of our approach. We see that the light neutrino mass matrix in Eq.~\eqref{eq:nu_mass_LO} has the following symmetry property
\begin{equation}
m_{\nu}(m_a,r,-\eta)=P^{T}_{132}m^*_{\nu}(m_a,r,\eta)P_{132}\,.
\end{equation}
Therefore the atmospheric mixing angle changes from $\theta_{23}$ to $\pi/2-\theta_{23}$, the Dirac CP phase changes from $\delta_{CP}$ to $\pi-\delta_{CP}$, the Majorana CP phase will become the opposite and other observable quantities remain unchanged under the transformation $\eta\rightarrow-\eta$. The predictions for $\eta=\frac{3\pi}{4}$ can be easily be obtained from the results of $\eta=-\frac{3\pi}{4}$. Hence we shall not show the predictions for $\eta=\frac{3\pi}{4}$.

\section{\label{sec:Conclusion}Conclusion}

In the present paper, guided by the principles of symmetry and minimality, we have analyzed the possible symmetry breaking patterns of $S_{4}\rtimes H_{CP}$ in the tri-direct CP approach~\cite{Ding:2018fyz} based on the two right-handed neutrino seesaw mechanism. In the tri-direct CP approach, the high energy flavor and generalized CP symmetry $S_4\rtimes H_{CP}$ is spontaneously broken down to an abelian subgroup $G_{l}$ (non $Z_2$ subgroups) in the charged lepton sector, to $G_{\text{atm}}\rtimes H^{\text{atm}}_{CP}$ in one right-handed neutrino sector and to $G_{\text{sol}}\rtimes H^{\text{sol}}_{CP}$  in other right-handed neutrino sector, as illustrated in figure~\ref{fig:tri_direct}. In this work, we assume that the flavon field $\phi_{\text{atm}}$ which couples to right-handed $N_{\text{atm}}$ and the left-handed lepton doublets $L$ is assigned to transform as $S_{4}$ triplet $\mathbf{3}$,  and the flavon $\phi_{\text{sol}}$ which couples to right-handed $N_{\text{sol}}$ and the left-handed lepton doublets $L$ transforms as the three-dimensional representation $\mathbf{3^\prime}$ under $S_{4}$. Then the two columns of the neutrino Dirac mass matrix are determined by the vacuum alignments of $\phi_{\text{atm}}$ and $\phi_{\text{sol}}$, respectively. Furthermore, we have given the basic procedure of predicting lepton flavor mixing and neutrino mass from residual symmetries in the tri-direct CP approach in a model independent way and we find that first (third) column of PMNS matrix is fixed by the diagonalization matrix $U_{l}$ of the charged lepton mass matrix and the vacuum alignments of $\phi_{\text{atm}}$ and $\phi_{\text{sol}}$ for NO (IO) spectrum.

After considering all possible breaking patterns arising from $S_{4}$ flavor symmetry combined with the corresponding generalized CP symmetry in a model independent way,  we find eight phenomenologically interesting mixing patterns with NO spectrum labeled as $\mathcal{N}_1\sim\mathcal{N}_8$ and eighteen phenomenologically interesting mixing patterns with IO spectrum labeled as $\mathcal{I}_1\sim\mathcal{I}_{18}$, please see table~\ref{tab:bf_via_CP}. For each phenomenologically interesting mixing pattern, we have analyzed the corresponding predictions for the PMNS matrix, the lepton mixing parameters, the neutrino masses and the effective mass in neutrinoless double beta decay in a model independent way in the tri-direct CP approach. There are one form dominance breaking pattern with NO spectrum ($\mathcal{N}_5$) and two form dominance breaking patterns with IO spectrum ($\mathcal{I}_4$ and $\mathcal{I}_5$). We find that three kinds of breaking patterns with NO spectrum ($\mathcal{N}_1\sim\mathcal{N}_3$) and one form dominance breaking pattern with IO spectrum ($\mathcal{I}_5$) yield TM1 mixing matrix. For each of these four kinds of breaking patterns with TM1 mixing matrix, two sum rules among mixing angles and Dirac CP phase corresponding to TM1 mixing are obtained. Furthermore, we perform a numerical analysis for each breaking pattern which is able to give a successful description of the lepton mixing parameters and the neutrino masses in terms of four real input parameters $x$, $\eta$, $m_a$ and $r$. In the breaking patterns with NO spectrum, we also give the $\chi^2$ results for some benchmark values of $x$ and $\eta$, where the parameter $x$ comes from the VEV of flavon $\phi_{\text{sol}}$. The simple values of $x$ and $\eta$ are very useful in model building. Once the values of $x$ and $\eta$ are fixed, we obtain a highly predictive theory of neutrino mass and lepton mixing, in which all lepton mixing parameters and the neutrino masses are determined by only two real input parameters $m_a$ and $r$. In the breaking pattern $\mathcal{N}_1$, for the benchmark value $x=-1$ which leads to $\langle\phi_{\text{sol}}\rangle=(1, -1, 3)^Tv_{\phi_s}$, it is exactly the Littlest seesaw model with CSD(3) which is originally proposed in~\cite{King:2015dvf}. The solar vacuum $\langle\phi_{\text{sol}}\rangle=(1, -3, 1)^Tv_{\phi_s}$ for $x=3$, it corresponds to another version of Littlest seesaw~\cite{King:2016yvg}. Moreover, for the vacuum $\langle\phi_{\text{sol}}\rangle=(1, 4, -2)^Tv_{\phi_s}$ with $x=4$, the CSD(4) scenario~\cite{King:2013xba} is reproduced. Furthermore, we show the best fit values of the  neutrino masses and the mixing parameters for a simple value of $x$ for each of the eighteen breaking patterns with IO spectrum.

Guided by above model independent analysis, we construct a successful flavor model involving two right-handed neutrinos based on $S_{4}$ and generalized CP symmetry to realise the breaking pattern $\mathcal{N}_4$ with $x=-4$ and $\eta=\pm\frac{3\pi}{4}$, in which the original symmetry $S_{4}\rtimes H_{CP}$ is spontaneously broken down to $Z^T_3$ in the charged lepton sector, to $Z^{TST^2}_2\times X_{\text{atm}}$ in the atmospheric neutrino sector and to $Z_2^U\times X_{\text{sol}}$ in the solar neutrino sector, where the residual CP transformations $X_{\text{atm}}=SU$ and $X_{\text{sol}}=U$. In this model, the first column of PMNS matrix is fixed to be $\left(2 \sqrt{\frac{6}{37}},\sqrt{\frac{13}{74}},\sqrt{\frac{13}{74}}\right)^T$. This model has not so far appeared in the literature. We find that this model is a powerful model to predicted lepton mixing parameters and neutrino masses. In particular, all the lepton mixing parameters and the neutrino masses are restricted in rather narrow regions in this model as in Eq.~\eqref{eq:model_prediction}.

In summary, we have performed an exhaustive analysis of all possible breaking patterns arising from $S_4\rtimes H_{CP}$ in a new tri-direct CP approach to the minimal seesaw model with two right-handed neutrinos and have constructed a realistic flavour model along these lines. According to this approach, separate residual flavour and CP symmetries persist in the charged lepton, ``atmospheric'' and ``solar'' right-handed neutrino sectors, resulting in {\it three} symmetry sectors rather than the usual two of the semi-direct CP approach. Following the tri-direct CP approach, we have found twenty-six kinds of independent phenomenologically interesting mixing patterns. Eight of them predict a normal ordering (NO) neutrino mass spectrum and the other eighteen predict an inverted ordering (IO) neutrino mass spectrum. For each phenomenologically interesting mixing pattern, the corresponding predictions for the PMNS matrix, the lepton mixing parameters, the neutrino masses and the effective mass in neutrinoless double beta decay are given in a model independent way. One breaking pattern with NO spectrum and two breaking patterns with IO spectrum corresponds to form dominance. We have found that the lepton mixing matrices of three kinds of breaking patterns with NO spectrum and one form dominance breaking pattern with IO spectrum preserve the first column of the  tri-bimaximal (TB) mixing matrix, corresponding to the TM1 mixing matrix.

\subsection*{Acknowledgements}
G.-J.\,D. acknowledges the support of the National Natural Science Foundation of China under Grant Nos 11522546 and 11835013.
S.\,F.\,K. acknowledges the STFC Consolidated Grant ST/L000296/1
and the European Union's Horizon 2020 research and innovation programme under the Marie Sk\l{}odowska-Curie grant agreements
Elusives ITN No.\ 674896 and InvisiblesPlus RISE No.\ 690575. C.-C.\, L. is supported by China Postdoctoral Science Foundation  Grant Nos. 2017M620258 and 2018T110617, CPSF-CAS Joint Foundation for Excellent Postdoctoral Fellows No. 2017LH0003,   the Fundamental Research Funds for the Central Universities under Grant No. WK2030040090 and the CAS Center for Excellence in Particle Physics (CCEPP).

\newpage

\section*{\label{sec:appendix}Appendix}

\begin{appendix}

\section{\label{sec:S4_group}Group Theory of $S_{4}$}

$S_4$ is the permutation group of four objects, and it has $24$ elements. In the present work, we shall adopt the same convention as~\cite{Ding:2013hpa}. The $S_{4}$ group can be generated by three  generators $S$, $T$ and $U$ with  the multiplication rules
\begin{eqnarray}
S^2=T^3=U^2=(ST)^3=(SU)^2=(TU)^2=(STU)^4=1\,.
\end{eqnarray}
$S_{4}$ group has twenty  abelian subgroups which contain nine $Z_2$ subgroups, four $Z_3$ subgroups, three $Z_4$ subgroups and four $K_4\cong Z_2\times Z_2$ subgroups. These abelian subgroups can be expressed in terms of the generators $S$, $T$ and $U$ as follows.

\begin{itemize}[leftmargin=1.5em]

\item{$Z_2$ subgroups}
\begin{equation}
\label{eq:Z2_subgroups}
\begin{array}{lll}
Z_2^{ST^{2}SU}=\{1,ST^{2}SU\},& ~~~~ Z_2^{TU}=\{1,TU\},& ~~~~ Z_2^{STSU}=\{1,STSU\},\\
Z_2^{T^2U}=\{1,T^2U\},&~~~~ Z_2^{U}=\{1,U\}, &~~~~ Z_2^{SU}=\{1,SU\}, \\
Z_2^{S}=\{1,S\},&~~~~ Z_2^{T^2ST}=\{1,T^2ST\}, &~~~~ Z_2^{TST^{2}}=\{1,TST^{2}\}\,.
\end{array}
\end{equation}
The former six $Z_{2}$ subgroups are conjugate to each other, and the latter three subgroups are related to each other by group conjugation as well.
\item{$Z_3$ subgroups}
\begin{equation}
\label{eq:Z3_subgroups}
\begin{array}{ll}
Z_3^{ST}=\{1,ST,T^{2}S\},&\qquad Z_3^{T}=\{1,T,T^{2}\},\\
Z_3^{STS}=\{1,STS,ST^2S\},&\qquad Z_3^{TS}=\{1,TS,ST^{2}\}\,.
\end{array}
\end{equation}
which are related with each other under group conjugation.
\item{$Z_4$ subgroups}
\begin{eqnarray}\label{eq:Z4_subgroups}
\nonumber&&Z_4^{TST^{2}U}=\{1,TST^{2}U,S,T^{2}STU\},\qquad Z_4^{ST^2U}=\{1,ST^2U,TST^{2},T^2SU\},\\
&&Z_4^{TSU}=\{1,TSU,T^2ST,STU\}\,,
\end{eqnarray}
All the above $Z_4$ subgroups are conjugate to each other.
\item{$K_4$ subgroups}
\begin{equation}\label{eq:K4_subgroups}
\begin{array}{l}
K^{(S, TST^{2})}_4\equiv Z_2^{S}\times Z_2^{TST^{2}}=\{1,S,TST^{2},T^{2}ST\},\\
K^{(S,U)}_4\equiv Z_2^{S}\times Z_2^{U}=\{1,S,U,SU\}, \\
K^{(TST^{2}, T^{2}U)}_4\equiv Z_2^{TST^{2}}\times Z_2^{T^{2}U}\equiv \{1,TST^{2},T^2U,ST^{2}SU\}, \\
K^{(T^{2}ST, TU)}_4\equiv Z_2^{T^{2}ST}\times Z_2^{TU}=\{1,T^{2}ST,TU,STSU\}\,,
\end{array}
\end{equation}
where $K^{(S, TST^{2})}_4$ is a normal subgroup of $S_4$, and the other three $K_4$ subgroups are conjugate to each other.

\end{itemize}

\begin{table}[t!]
\begin{center}
\begin{tabular}{|c|c|c|c|}\hline\hline
 ~~  &  $S$  &   $T$    &  $U$  \\ \hline
~~~${\bf 1}$, ${\bf 1^\prime}$ ~~~ & 1   &  1  & $\pm1$  \\ \hline
   &   &    &    \\ [-0.16in]
${\bf 2}$ &  $\left( \begin{array}{cc}
    1&~0 \\
    0&~1
    \end{array} \right) $
    & $\left( \begin{array}{cc}
    \omega&~0 \\
    0&~\omega^2
    \end{array} \right) $
    & $\left( \begin{array}{cc}
    0&~1 \\
    1&~0
    \end{array} \right)$\\ [0.12in]\hline
   &   &    &    \\ [-0.16in]
${\bf 3}$, ${\bf 3^\prime}$ & $\frac{1}{3} \left(\begin{array}{ccc}
    -1&~ 2  ~& 2  \\
    2  &~ -1  ~& 2 \\
    2 &~ 2 ~& -1
    \end{array}\right)$
    & $\left( \begin{array}{ccc}
    1 &~ 0 ~& 0 \\
    0 &~ \omega^{2} ~& 0 \\
    0 &~ 0 ~& \omega
    \end{array}\right) $
    & $\mp\left( \begin{array}{ccc}
    1 &~ 0 ~& 0 \\
    0 &~ 0 ~& 1 \\
    0 &~ 1 ~& 0
    \end{array}\right)$
\\[0.22in] \hline\hline
\end{tabular}
\caption{\label{tab:S4_rep}The representation matrices of the generators $S$, $T$ and $U$ for the five irreducible representations of $S_4$ in the chosen basis, where $\omega=e^{2\pi i/3}$. }
\end{center}
\end{table}

$S_4$ has five irreducible representations which contain two singlet irreducible representations $\mathbf{1}$ and $\mathbf{1^{\prime}}$, one two-dimensional representation $\mathbf{2}$ and two three-dimensional irreducible representations $\mathbf{3}$ and $\mathbf{3^{\prime}}$. In this work, we choose the same basis as that of~\cite{Ding:2013hpa}, i.e. the representation matrix of the generator $T$ is diagonal. The representation matrices for the three generators are listed in table~\ref{tab:S4_rep}. Moreover, the Kronecker products of two irreducible representations of $S_4$ group are
\begin{eqnarray}
\nonumber && \bf{1}\otimes \mathbf{R}=\mathbf{R},\qquad \bf{1^\prime}\otimes \bf{1^\prime}=\bf{1},\qquad \bf{1^\prime}\otimes\bf{2}=\bf{2},\qquad \bf{1^\prime}\otimes\bf{3}=\bf{3^\prime},\qquad \bf{1^\prime}\otimes\bf{3^\prime}=\bf{3},  \\
\nonumber && \bf{2}\otimes\bf{2}=\bf{1}\oplus\bf{1^\prime}\oplus\bf{2},\qquad \bf{2}\otimes\bf{3}=\bf{2}\otimes\bf{3^\prime}=\bf{3}\otimes\bf{3^\prime},\\
&& \bf{3}\otimes\bf{3}=\bf{3^\prime}\otimes\bf{3^\prime}=\bf{1}\oplus\bf{2}\oplus\bf{3}\oplus\bf{3^\prime},\qquad \bf{3}\otimes\bf{3^\prime}=\bf{1^\prime}\oplus\bf{2}\oplus\bf{3}\oplus\bf{3^\prime}\,,
\end{eqnarray}
where $\mathbf{R}$ stands for any irreducible representation of $S_4$.

We now list the CG coefficients for our basis. All the CG coefficients can be reported in the form of $\mathbf{R_1}\otimes \mathbf{R_2}$, where $\mathbf{R_1}$ and $\mathbf{R_2}$ are two irreducible representations of $S_4$. We shall use $\alpha_i$ to denote the elements of first representation and $\beta_i$ stands for the elements of the second representation of the tensor product. For the product of the singlet $\mathbf{1^{\prime}}$ with a doublet or a triplet, we have
\begin{eqnarray*}
\begin{array}{|c|c|c|}\hline\hline
~~\mathbf{1^\prime}\otimes\mathbf{2}~~ &~ ~\mathbf{1^\prime}\otimes\mathbf{3}=\mathbf{3^\prime}~ ~&~  ~\mathbf{1^\prime}\otimes\mathbf{3^\prime} \\ \hline
 & & \\[-0.18in]
~~\mathbf{2}\sim
\left(\begin{array}{c}\alpha\beta_1 \\-\alpha\beta_2\end{array}\right)
~ ~ &~
~  \mathbf{3^\prime} \sim
\left(\begin{array}{c}\alpha\beta_1  \\ \alpha\beta_2  \\ \alpha\beta_3 \end{array}\right)
~~ &~
~\mathbf{3}\sim
\left(\begin{array}{c} \alpha\beta_1  \\ \alpha\beta_2  \\ \alpha\beta_3 \end{array}\right)~~
 \\ \hline \hline
\end{array}
\end{eqnarray*}
The CG coefficients for the products involving the doublet   representation $\mathbf{2}$ are found to be
\begin{eqnarray*}
\begin{array}{|c|c|c|}\hline\hline
~\mathbf{2}\otimes\mathbf{2}=\mathbf{1}\oplus\mathbf{1^\prime}\oplus\mathbf{2}~ & ~\mathbf{2}\otimes\mathbf{3}=\mathbf{3}\oplus\mathbf{3^\prime}~ &  ~\mathbf{2}\otimes\mathbf{3^\prime}=\mathbf{3}\oplus\mathbf{3^\prime} \\ \hline
 & & \\[-0.18in]
~~ {\mathbf1}\sim\alpha_1\beta_2+\alpha_2\beta_1~~  & ~~ & ~~  \\ & & \\[-0.35in]
~~{\mathbf 1^\prime}\sim\alpha_1\beta_2-\alpha_2\beta_1
~ ~&~
~{\mathbf3}\sim
\left(\begin{array}{c} \alpha_1\beta_2+\alpha_2\beta_3 \\ \alpha_1\beta_3+\alpha_2\beta_1  \\ \alpha_1\beta_1+\alpha_2\beta_2 \end{array}\right)
~~ &~
~\mathbf{3}\sim
\left(\begin{array}{c} \alpha_1\beta_2-\alpha_2\beta_3 \\ \alpha_1\beta_3-\alpha_2\beta_1  \\ \alpha_1\beta_1-\alpha_2\beta_2 \end{array}\right) ~~ \\
& & \\[-0.18in]
~~{\mathbf2}\sim
\left(\begin{array}{c}\alpha_2\beta_2  \\ \alpha_1\beta_1 \end{array}\right)
~ ~ &~
~  {\mathbf3^\prime}\sim
\left(\begin{array}{c}\alpha_1\beta_2-\alpha_2\beta_3 \\ \alpha_1\beta_3-\alpha_2\beta_1  \\ \alpha_1\beta_1-\alpha_2\beta_2 \end{array}\right)
~~ &~
~\mathbf{3^\prime}\sim
\left(\begin{array}{c}\alpha_1\beta_2+\alpha_2\beta_3 \\ \alpha_1\beta_3+\alpha_2\beta_1  \\ \alpha_1\beta_1+\alpha_2\beta_2\end{array}\right) ~~
 \\ \hline \hline
\end{array}
\end{eqnarray*}
Finally, for the products among the triplet representations $\mathbf{3}$ and $\mathbf{3^{\prime}}$, we have
\begin{eqnarray*}
\begin{array}{|c|c|} \hline\hline
~ ~ \mathbf{3}\otimes\mathbf{3}=\mathbf{3^\prime}\otimes\mathbf{3^\prime}=\mathbf{1}\oplus\mathbf{2}\oplus\mathbf{3}\oplus\mathbf{3^\prime}~~ & ~ ~\mathbf{3}\otimes\mathbf{3^\prime}=\mathbf{1^\prime}\oplus\mathbf{2}\oplus\mathbf{3}\oplus\mathbf{3^\prime}~~ \\ \hline
  & \\[-0.16in]
 ~~ {\mathbf1}\sim\alpha_1\beta_1+\alpha_2\beta_3+\alpha_3\beta_2 ~~& ~~ {\mathbf1^\prime}\sim\alpha_1\beta_1+\alpha_2\beta_3+\alpha_3\beta_2 ~~ \\
  & \\[-0.16in]
~~{\mathbf2}\sim
\left(\begin{array}{c}\alpha_2\beta_2+\alpha_1\beta_3+\alpha_3\beta_1  \\ \alpha_3\beta_3+\alpha_1\beta_2+\alpha_2\beta_1 \end{array}\right)
~~ & ~   ~ {\mathbf2}\sim
\left(\begin{array}{c}\alpha_2\beta_2+\alpha_1\beta_3+\alpha_3\beta_1  \\ -(\alpha_3\beta_3+\alpha_1\beta_2+\alpha_2\beta_1) \end{array}\right)~~\\
  & \\[-0.16in]
~ ~{\mathbf3}\sim
\left(\begin{array}{c}\alpha_2\beta_3-\alpha_3\beta_2  \\ \alpha_1\beta_2-\alpha_2\beta_1  \\ \alpha_3\beta_1-\alpha_1\beta_3 \end{array}\right)
~~ &~
~  {\mathbf3}\sim
\left(\begin{array}{c}2\alpha_1\beta_1- \alpha_2\beta_3-\alpha_3\beta_2  \\ 2\alpha_3\beta_3- \alpha_1\beta_2-\alpha_2\beta_1  \\
 2\alpha_2\beta_2-\alpha_3\beta_1-\alpha_1\beta_3 \end{array}\right)
~~  \\
 &  \\[-0.16in]
~~ {\mathbf3^\prime}\sim
\left(\begin{array}{c} 2\alpha_1\beta_1-\alpha_2\beta_3-\alpha_3\beta_2  \\  2\alpha_3\beta_3-\alpha_1\beta_2-\alpha_2\beta_1  \\
2\alpha_2\beta_2-\alpha_3\beta_1-\alpha_1\beta_3 \end{array}\right)
~~ &~
~{\mathbf3^\prime}\sim
\left(\begin{array}{c}\alpha_2\beta_3-\alpha_3\beta_2  \\ \alpha_1\beta_2-\alpha_2\beta_1  \\ \alpha_3\beta_1-\alpha_1\beta_3 \end{array}\right)
~~ \\ \hline\hline
\end{array}
\end{eqnarray*}

\section{\label{sec:other_NO_mix}Other mixing patterns with NO}

\begin{description}[labelindent=-0.8em, leftmargin=0.3em]

\item[~~($\mathcal{N}_6$)]{$(G_{l},G_{\text{atm}},G_{\text{sol}})=(Z_4^{TSU},Z_3^T,Z_2^{SU})$, $X_{\text{atm}}=\{1,T,T^2\}$, $X_{\text{sol}}=\{U,S\}$ }

Here the diagonalization matrix of charged lepton mass matrix $U_{l}$ is given in Eq.~\eqref{eq:ch_dia_matrix}. Here only residual CP transformations $X_{\text{sol}}=\{U,S\}$ can accommodate the present experimental data on lepton mixing. In this kind of breaking pattern the VEVs of flavon fields $\phi_{\text{atm}}$ and  $\phi_{\text{sol}}$ are
\begin{equation}
 \langle\phi_{\text{atm}}\rangle=v_{\phi_a} \left(1, 0, 0\right)^T, \qquad \langle\phi_{\text{sol}}\rangle= v_{\phi_s}\left(1, 1+i x, 1-i x\right)^T\,.
\end{equation}
Straightforward calculations demonstrate that the neutrino mass matrix is of the following form
\begin{equation}\label{eq:mu_N6}
 m_{\nu}=m_{a}\begin{pmatrix}
 1 &~ 0 ~& 0 \\
 0 &~ 0 ~& 0 \\
 0 &~ 0 ~& 0 \\
\end{pmatrix}+m_{s}e^{i\eta}
\begin{pmatrix}
 1 &~ 1-i x &~ 1+i x \\
 1-i x &~ (1-ix)^2 &~ x^2+1 \\
 1+i x &~ x^2+1 &~ (1+i x)^2 \\
\end{pmatrix}\,,
\end{equation}
The neutrino mass matrix $m_{\nu}$ has an eigenvalue 0, and the corresponding eigenvector is $(0, 1+ix, -1+ix)^{T}$. It is convenient to firstly perform a constant unitary transformation $U_{\nu1}$. The unitary matrix $U_{\nu1}$ can take the form
\begin{equation}\label{eq:Unu1_N6}
U_{\nu1}=\begin{pmatrix}
0  &~ i \sqrt{\frac{2 \left(x^2+1\right)}{5 x^2-2 \sqrt{3} x+3}} &~ \frac{1-\sqrt{3} x}{\sqrt{5 x^2-2 \sqrt{3} x+3}} \\
\frac{i x+1}{\sqrt{2 \left(x^2+1\right)}}  &~ \frac{(i-x) \left(\sqrt{3} x-1\right)}{\sqrt{2 \left(x^2+1\right) \left(5 x^2-2 \sqrt{3} x+3\right)}} &~ \frac{i x+1}{\sqrt{5 x^2-2 \sqrt{3} x+3}}  \\
 \frac{-1+ix}{\sqrt{2 \left(x^2+1\right)}}  &~ \frac{(i+x) \left(\sqrt{3} x-1\right)}{\sqrt{2 \left(x^2+1\right) \left(5 x^2-2 \sqrt{3} x+3\right)}} &~ \frac{1-i x}{\sqrt{5 x^2-2 \sqrt{3} x+3}}  \\
\end{pmatrix}\,.
\end{equation}
Then the neutrino mass matrix $m^\prime_{\nu}$ is a block diagonal matrix and it  can be diagonalized by a unitary matrix $U_{\nu2}$ in the (2,3) sector. The nonzero parameters $y$, $z$ and $w$ in the $m^\prime_{\nu}$ are given by
\begin{eqnarray}
\nonumber &&y=\frac{2 \left(x^2+1\right) \left(m_{a}+3x^2m_{s} e^{i \eta }  \right)}{5 x^2-2 \sqrt{3} x+3}\,,  \\
\nonumber && z= i\frac{ \sqrt{2(x^2+1)} \left((1-\sqrt{3}x) m_{a} + \sqrt{3}x \left(2 x^2-\sqrt{3}x+3 \right)m_{s}e^{i \eta }\right)}{5 x^2-2 \sqrt{3} x+3}\,,  \\
&& w=\frac{ \left(\sqrt{3} x-1\right)^2m_{a}+  \left(4 x^4-4 \sqrt{3} x^3+15 x^2-6 \sqrt{3} x+9\right)m_{s}e^{i \eta }}{5 x^2-2 \sqrt{3} x+3}\,.
\end{eqnarray}
Then the lepton mixing matrix is determined to be
\begin{equation}\label{eq:PMNS_N6}
\hskip-0.1in  U=\frac{1}{2\sqrt{3}}\begin{pmatrix}
 \frac{\sqrt{2} \left(x+\sqrt{3}\right)}{\sqrt{x^2+1}} &~ -\sqrt{\frac{10 x^2-4 \sqrt{3} x+6}{x^2+1}} \cos \theta  &~  -\sqrt{\frac{10 x^2-4 \sqrt{3} x+6}{x^2+1}} e^{i \psi }\sin \theta \\
 \sqrt{\frac{5 x^2-2 \sqrt{3} x+3}{x^2+1}} &~ \frac{\left(x+\sqrt{3}\right) \cos \theta }{\sqrt{x^2+1}}-\sqrt{6} e^{-i \psi } \sin \theta  &~ \sqrt{6}\cos \theta+\frac{ \left(\sqrt{3}+x \right)e^{i \psi } \sin \theta }{ \sqrt{x^2+1}} \\
 \sqrt{\frac{5 x^2-2 \sqrt{3} x+3}{x^2+1}} &~ \frac{\left(x+\sqrt{3}\right) \cos \theta }{\sqrt{x^2+1}}+\sqrt{6} e^{-i \psi } \sin \theta  &~ -\sqrt{6}\cos \theta+\frac{ \left(\sqrt{3}+x \right)e^{i \psi } \sin \theta }{ \sqrt{x^2+1}} \\
\end{pmatrix}\,,
\end{equation}
One can straightforwardly extract the lepton mixing angles and CP phases as follows
\begin{eqnarray}
\nonumber &&\sin^2\theta_{13}=\frac{\left(5 x^2-2 \sqrt{3} x+3\right) \sin ^2\theta }{6 \left(x^2+1\right)}\,,\\
 \nonumber &&\sin^2\theta_{12}=\frac{\left(5 x^2-2 \sqrt{3} x+3\right) \cos ^2\theta }{6 \left(x^2+1\right)-\left(5 x^2-2 \sqrt{3} x+3\right) \sin ^2\theta }\,,\\
\nonumber &&\sin^2\theta_{23}=\frac{1}{2}+\frac{\left(\sqrt{3} x+3\right) \sqrt{x^2+1} \sin 2 \theta  \cos \psi }{\sqrt{2} \left(6 \left(x^2+1\right)-\left(5 x^2-2 \sqrt{3} x+3\right) \sin ^2\theta \right)}\,, \\
\nonumber &&J_{CP}=-\frac{\left(5 \sqrt{3} x^3+9 x^2-3 \sqrt{3} x+9\right) \sin 2 \theta  \sin \psi }{72 \sqrt{2} \left(x^2+1\right)^{3/2}}\,, \\
&&I_{1}=\frac{\left(5 x^2-2 \sqrt{3} x+3\right)^2 \sin ^22 \theta  \sin (\rho -\sigma )}{144 \left(x^2+1\right)^2}\,.
\end{eqnarray}
A sum rule between the solar mixing angle $\theta_{12}$ and the reactor mixing angle $\theta_{13}$ is satisfied,
\begin{equation}\label{eq:sum_rule_N6_1}
\cos^2\theta_{12}\cos^2\theta_{13}=\frac{1}{6}+\frac{ 1+\sqrt{3} x}{3 \left(1+x^2\right)}\,.
\end{equation}
For the fixed value of $x$, mixing angles $\theta_{13}$ and $\theta_{12}$ are strongly correlated with each other. On the other hand, the $3\sigma$ ranges of $\theta_{13}$ and $\theta_{12}$~\cite{Esteban:2018azc} will restrict the possible value of input parameter $x$ ($0.310\leq x\leq0.925$). Furthermore, we can derive the following sum rule among the Dirac CP phase $\delta_{CP}$ and mixing angles
\begin{equation}\label{eq:sum_rule_N6_2}
\cos\delta_{CP}=\frac{\cos 2 \theta_{23} \left((9+2\sqrt{3}x+7x^2)\sin^2 \theta_{13}-3+2\sqrt{3}x-5x^2\right)}{2\sin \theta_{13} \sin 2 \theta_{23} \sqrt{\left(\sqrt{3}+x\right)^2 \left(3 \left(1+x^2\right) \cos 2 \theta_{13}+2 x \left(x-\sqrt{3}\right)\right)}}\,.
\end{equation}
We note that maximal $\theta_{23}$ leads to maximal Dirac CP phase $\delta_{CP}$.  It is easy to check that the neutrino mass matrix $m_{\nu}$ in Eq.~\eqref{eq:mu_N6} has the symmetry property
\begin{equation}
m_{\nu}(x,r,-\eta)=P^{T}_{132}m^*_{\nu}(x,r,\eta)P_{132}\,,
\end{equation}
It implies that the atmospheric angle changes from $\theta_{23}$ to $\pi/2-\theta_{23}$, the Dirac phase $\delta_{CP}$ becomes $\pi-\delta_{CP}$, the Majorana CP phase $\beta$ will add a negative factor and the other output parameters are kept intact under a change of the sign of the parameter $\eta$. For the fixed value of $x$ and $\eta$, all the mixing angles, CP phases and neutrino masses are fully determined by $m_a$ and  $r$. As an example, we shall give the predictions for $x=\frac{1}{3}$ and $\eta=-\frac{4\pi}{5}$. For this case, the fixed column of PMNS matrix is $\frac{1}{2 \sqrt{15}}\left(3 \sqrt{3}+1,\sqrt{16-3 \sqrt{3} }, \sqrt{16-3 \sqrt{3} }\right)^T$. Furthermore, we shall perform a conventional $\chi^2$ analysis, the  best fit values of the input and output parameters read
\begin{eqnarray}
\nonumber &&  m_{a}=11.910\,\text{meV},\qquad r=1.372, \qquad \chi^2_{\text{min}}=8.753, \qquad \sin^2\theta_{13}=0.0227\,, \\
\nonumber && \sin^2\theta_{12}=0.345, \qquad \sin^2\theta_{23}=0.557, \qquad \delta_{CP}/\pi=-0.415, \quad \beta/\pi=0.215, \\
&& m_1=0, \qquad  m_2=8.606\,\text{meV}, \qquad m_3=50.238\,\text{meV}, \qquad m_{ee}=2.720\,\text{meV}\,.
\end{eqnarray}
By comprehensively scanning over the parameter space of $x$, $\eta$ and $r$, if we require the three mixing angles and mass ratio $m^2_2/m^2_3$ in their $3\sigma$ regions~\cite{Esteban:2018azc}, we find that all the three input parameters are  restricted in narrow intervals of $0.311\leq x\leq0.381$, $0.730\pi\leq |\eta|\leq\pi$ and $1.270\leq r\leq1.487$. Furthermore, $\theta_{12}$ is found to lie in a narrow interval around its $3\sigma$ upper bound $0.334\leq\sin^2\theta_{12}\leq0.350$, and $\theta_{23}$ can only take the value in the range $[0.425,0.575]$. The predictions for the two CP phases $\delta_{CP}$ and $\beta$ are $-0.611\pi\leq\delta_{CP}\leq-0.389\pi$ and $-0.281\pi\leq\beta\leq0.281\pi$, respectively.

\item[~~($\mathcal{N}_7$)]{$(G_{l},G_{\text{atm}},G_{\text{sol}})=(K_4^{(S,TST^2)},Z_3^T,Z_2^{SU})$, $X_{\text{atm}}=\{1,T,T^2\}$}

$\bullet$ $X_{\text{sol}}=\{1,SU\}$

In this breaking pattern, both the two types of the residual CP transformations ($X_{\text{sol}}=\{1,SU\}$ and $X_{\text{sol}}=\{S,U\}$) corresponding to residual flavor $G_{\text{sol}}=Z_2^{SU}$ are viable. For these two types of residual CP transformations, the vacuum alignments  $\langle\phi_{\text{sol}}\rangle$ which are invariant under the residual flavor and CP symmetries are proportional to $\left(1,x,2-x\right)^T$ and $\left(1,1+ix,1-ix\right)^T$, respectively. When the original symmetry $S_{4}\rtimes H_{CP}$ is broken into $K_4^{(S,TST^2)}$ in the charged lepton sector, the diagonalization matrix of the hermitian combination of the charged lepton mass matrix $m^{\dagger}_{l}m_{l}$ is given in Eq.~\eqref{eq:ch_dia_matrix}. For $X_{\text{sol}}=\{1,SU\}$, the most general form of the neutrino mass matrix  read as
\begin{equation}
 m_{\nu}=m_{a}\begin{pmatrix}
 1 &~ 0 ~& 0 \\
 0 &~ 0 ~& 0 \\
 0 &~ 0 ~& 0 \\
\end{pmatrix}+m_{s}e^{i\eta}
\begin{pmatrix}
 1 &~ 2-x &~ x \\
 2-x &~ (x-2)^2 &~ (2-x) x \\
 x &~ (2-x) x &~ x^2 \\
\end{pmatrix}\,,
\end{equation}
It is easy to check that the column vector $(0,x, x-2)^{T}$ is an eigenvector of the above neutrino mass matrix and the corresponding eigenvalue is 0. Before diagonalizing the neutrino mass matrix $ m_{\nu}$, it is useful to perform a unitary transformation $U_{\nu1}$, and this unitary transformation will make $m^\prime_{\nu}$ to be a block diagonal matrix. Here the unitary matrix $U_{\nu1}$ takes the following form
\begin{equation}
U_{\nu1}=\begin{pmatrix}
 0 &~ -\sqrt{\frac{x^2-2 x+2}{x^2-2 x+4}} &~ -\frac{ \sqrt{2}}{\sqrt{x^2-2 x+4}} \\
 \frac{x}{\sqrt{2 \left(x^2-2 x+2\right)}} &~ \frac{x-2}{\sqrt{\left(x^2-2 x+2\right) \left(x^2-2 x+4\right)}} &~\frac{ 2-x}{\sqrt{2 \left(x^2-2 x+4\right)}} \\
 \frac{x-2}{\sqrt{2 \left(x^2-2 x+2\right)}} &~ -\frac{x}{\sqrt{\left(x^2-2 x+2\right) \left(x^2-2 x+4\right)}} &~ \frac{ x}{\sqrt{2 \left(x^2-2 x+4\right)}} \\
\end{pmatrix}\,,
\end{equation}
Then the nonzero parameters $y$, $z$ and $w$ of $m^\prime_{\nu}$ are given by
\begin{eqnarray}
\nonumber &&y=\frac{(x^2-2 x+2) \left(m_{a}+9m_{s} e^{i \eta } \right)}{x^2-2 x+4}\,,  \\
\nonumber && z=\frac{  \sqrt{2(x^2-2 x+2)} \left(m_{a}-3(x-1)^2m_{s} e^{i \eta }  \right)}{x^2-2 x+4}\,,  \\
&& w=\frac{2 \left(m_{a}+(x-1)^4m_{s}e^{i \eta }  \right)}{x^2-2 x+4}\,.
\end{eqnarray}
Then we need to put $m^\prime_{\nu}$ into diagonal form with real non-negative masses, which can be done exactly by using the standard procedure shown in section~\ref{sec:framework}, i.e. $U^T_{\nu2}m^\prime_{\nu}U_{\nu2}=\text{diag}(0,m_2,m_3)$.  Hence the lepton mixing matrix is determined to be
\begin{equation}\label{eq:PMNS_N7_1}
\hskip-0.1in U=
\begin{pmatrix}
 \frac{\sqrt{2} (x-1)}{\sqrt{3(x^2-2 x+2)}} &~ -\sqrt{\frac{x^2-2 x+4}{3(x^2-2 x+2)}} \cos \theta  &~  -\sqrt{\frac{x^2-2 x+4}{3(x^2-2 x+2)}}e^{i \psi } \sin \theta  \\
\sqrt{\frac{x^2-2 x+4}{6(x^2-2 x+2)}} &~ \frac{(x-1) \cos \theta }{ \sqrt{3(x^2-2 x+2)}}+\frac{ie^{-i \psi } \sin \theta }{\sqrt{2}} &~\frac{ (x-1)e^{i \psi } \sin \theta }{\sqrt{3(x^2-2 x+2)}}- \frac{i\cos \theta }{\sqrt{2}} \\
\sqrt{\frac{x^2-2 x+4}{6(x^2-2 x+2)}} &~ \frac{(x-1) \cos \theta }{ \sqrt{3(x^2-2 x+2)}}-\frac{ie^{-i \psi } \sin \theta }{\sqrt{2}} &~ \frac{ (x-1)e^{i \psi } \sin \theta }{ \sqrt{3(x^2-2 x+2)}}+\frac{i\cos \theta }{\sqrt{2}} \\
\end{pmatrix}\,,
\end{equation}
The lepton mixing angles and CP invariants can be read out as
\begin{eqnarray}
\nonumber &&\sin^2\theta_{13}=\frac{\left(x^2-2 x+4\right) \sin ^2\theta }{3 \left(x^2-2 x+2\right)}\,,\\
 \nonumber &&\sin^2\theta_{12}=1-\frac{2 (x-1)^2}{3 \left(x^2-2 x+2\right)-\left(x^2-2 x+4\right) \sin ^2\theta }\,,\\
\nonumber &&\sin^2\theta_{23}=\frac{1}{2}-\frac{\sqrt{6} (x-1) \sqrt{x^2-2 x+2} \sin 2 \theta  \sin \psi }{2 \left(3 \left(x^2-2 x+2\right)-\left(x^2-2 x+4\right) \sin ^2\theta \right)}\,, \\
\nonumber &&J_{CP}=\frac{\left(x^3-3 x^2+6 x-4\right) \sin 2 \theta  \cos \psi }{6 \sqrt{6} \left(x^2-2 x+2\right)^{3/2}}\,, \\
 &&I_{1}=\frac{\left(x^2-2 x+4\right)^2 \sin ^22 \theta  \sin (\rho -\sigma )}{36 \left(x^2-2 x+2\right)^2}\,.
\end{eqnarray}
As a consequence, sum rules among the Dirac CP phase $\delta_{CP}$ and the mixing angles can be found as follows
\begin{eqnarray}
\nonumber && \cos^2\theta_{12}\cos^2\theta_{13}=\frac{2(1-x)^2}{3( x^2-2 x+2)}\,, \\
&& \cos\delta_{CP}=\frac{ \cot 2 \theta_{23} \left(3 x(x-2)-\left(5 x^2-10x+8\right) \cos 2 \theta_{13}\right)}{ 4\sin \theta_{13}\sqrt{(1-x)^2 \left(2+2 x-x^2+3 \left(x^2-2x+2\right) \cos 2 \theta_{13}\right)}}\,.
\end{eqnarray}
For a given value of $x$,  the possible ranges of $\sin^2\theta_{12}$ can be obtained from the above correlations by varying $\theta_{13}$ over its $3\sigma$ ranges and we also can obtain the prediction for $\cos\delta_{CP}$ from the $3\sigma$ ranges of mixing angles $\theta_{13}$ and $\theta_{23}$. For fixed $x$ and $\eta$, all mixing parameters and neutrino masses depend on two input parameters $m_{a}$ and $r$. The results of the $\chi^2$ analysis for some benchmark values of $x$ and $\eta$  are reported in table~\ref{tab:bf_N7_1}. From table~\ref{tab:bf_N7_1} we find that the results for $\eta=0$ are viable for both $x=-8$ and $10$. Furthermore,  maximal atmospheric mixing angle, maximal Dirac CP phase and trivial Majorana CP phase are obtained for $\eta=0$.

\begin{table}[t!]
\renewcommand{\tabcolsep}{0.4mm}
\renewcommand{\arraystretch}{1.3}
\small
\centering
\begin{tabular}{|c|c| c| c| c | c| c| c| c| c| c |c |c |c |c |}  \hline \hline
$\langle\phi_{\text{sol}}\rangle/v_{\phi_{s}}$ &  $x$   &  $\eta$ & $m_{a}$(meV) & $r$  	 & $\chi^2_{\text{min}}$ &  $\sin^2\theta_{13}$  &$\sin^2\theta_{12}$  & $\sin^2\theta_{23}$  & $\delta_{CP}/\pi$ &  $\beta/\pi$   & $m_{2}$(meV)  & $m_{3}$(meV)  & $m_{ee}$(meV)  \\   \hline
\multirow{3}{*}{$\left(1,10,-8\right)^T$}
& \multirow{3}{*}{$10$} & $0$ & $8 .682$ & $0 .0350$ & $22 .926$ & $0 .0207$ & $0 .328$ & $0 .5$ & $0 .5$ & $1$ & $8 .618$ & $50 .223$ & $3 .806$  \\ \cline{3-14}
&&  $\frac{\pi }{6}$ & $8 .684$ & $0 .0350$ & $23 .417$ & $0 .0205$ & $0 .328$ & $0 .506$ & $0 .491$ & $-0.833$ & $8 .621$ & $50 .220$ & $3 .713$ \\ \cline{3-14}
&&   $-\frac{ \pi }{6}$ & $8 .680$ & $0 .0350$ & $27 .567$ & $0 .0205$ & $0 .328$ & $0 .494$ & $0 .509$ & $0 .833$ & $8 .618$ & $50 .224$ & $3 .711$ \\ \hline
\multirow{3}{*}{$\left(1,-8,10\right)^T$} & \multirow{3}{*}{$-8$} & $0$ & $8 .682$ & $0 .035$ & $22 .926$ & $0 .0207$ & $0 .328$ & $0 .5$ & $-0.5$ & $1$ & $8 .618$ & $50 .223$ & $3 .806$ \\ \cline{3-14}
&&  $\frac{\pi }{6}$ & $8 .680$ & $0 .0350$ & $27 .567$ & $0 .0205$ & $0 .328$ & $0 .494$ & $-0.509$ & $-0.833$ & $8 .618$ & $50 .224$ & $3 .711$ \\ \cline{3-14}
&& $-\frac{ \pi }{6}$ & $8 .684$ & $0 .0350$ & $23 .417$ & $0 .0205$ & $0 .328$ & $0 .506$ & $-0.491$ & $0 .833$ & $8 .621$ & $50 .220$ & $3 .713$ \\ \hline \hline
\end{tabular}
\caption{\label{tab:bf_N7_1}
The predictions for the lepton mixing angles, CP violation phases, neutrino masses and the effective Majorana mass $m_{ee}$ for the breaking pattern $(G_{l},G_{\text{atm}},G_{\text{sol}})= (K_4^{(S,TST^2)},Z_3^T,Z_2^{SU}) $ and $X_{\text{sol}}=\{1,SU\}$. Here we choose many benchmark values for the parameters $x$ and $\eta$. The fixed column of the PMNS mixing matrix is $\left(3 \sqrt{\frac{3}{41}},\sqrt{\frac{7}{41}},\sqrt{\frac{7}{41}}\right)^T$ for both $x=10$ and $x=-8$. Notice that the lightest neutrino mass is vanishing $m_1=0$.
}
\end{table}

The admissible range of $x$, $r$ and $\eta$ can be obtained from the requirement that the three mixing angles and mass ratio $m^2_2/m^2_3$ are in the experimentally preferred $3\sigma$ ranges~\cite{Esteban:2018azc}.   When all three mixing angles and mass ratio $m^2_2/m^2_3$ lie in their $3\sigma$ ranges, we  find the allowed regions of the input parameters $x$, $\eta$ and $r$ are in the intervals $[-8.094,-6.351]\cup[8.351,10.094]$, $[-\pi,\pi]$ and $[0.0324,0.0550]$, respectively. Furthermore, the mixing angles $\sin^2\theta_{12}$ and $\sin^2\theta_{23}$ are predicted to be in the ranges of $[0.326,0.330]$ and $[0.486,0.514]$, respectively. The Dirac CP phase $\delta_{CP}$ is found to lie in rather narrow region around maximal, i.e.  $|\delta_{CP}|\in[0.481\pi,0.519\pi]$. While the Majorana CP phase $\beta$ can take any value from $-\pi$ to $\pi$.

$\bullet$ $X_{\text{sol}}=\{U,S\}$

For $X_{\text{sol}}=\{U,S\}$, the most general neutrino mass matrix is of the form
\begin{equation}
 m_{\nu}=m_{a}\begin{pmatrix}
 1 &~ 0 ~& 0 \\
 0 &~ 0 ~& 0 \\
 0 &~ 0 ~& 0 \\
\end{pmatrix}+m_{s}e^{i\eta}
\begin{pmatrix}
 1 &~ 1-i x &~ 1+i x \\
 1-i x &~ (1-ix)^2 &~ x^2+1 \\
 1+i x &~ x^2+1 &~ (1+i x)^2 \\
\end{pmatrix}\,.
\end{equation}
Subsequently a unitary transformation $U_{\nu1}$ is performed on the light neutrino fields, then $m^\prime_{\nu}=U^T_{\nu1}m_{\nu}U_{\nu1}$ becomes a block diagonal matrix with a block of $2\times2$ matrix. The unitary transformation matrix $U_{\nu1}$ can take the following form
\begin{equation}
U_{\nu1}=\begin{pmatrix}
 0 &~ -\frac{i \left(x^2+1\right)}{\sqrt{\left(x^2+1\right) \left(x^2+3\right)}} &~ -\frac{i \sqrt{2}}{\sqrt{x^2+3}} \\
 \frac{i x+1}{\sqrt{2 \left(x^2+1\right)}} &~ \frac{-i+x}{\sqrt{\left(x^2+1\right) \left(x^2+3\right)}} &~ -\frac{-i+x}{\sqrt{2 \left(x^2+3\right)}} \\
 \frac{i (i+x)}{\sqrt{2 \left(x^2+1\right)}} &~ -\frac{i+x}{\sqrt{\left(x^2+1\right) \left(x^2+3\right)}} &~ \frac{i+x}{\sqrt{2 \left(x^2+3\right)}} \\
\end{pmatrix}\,,
\end{equation}
The parameters $y$, $z$ and $w$ in neutrino mass matrix $m^\prime_{\nu}$ are
\begin{equation}\label{eq:yzw_N7_2}
y=-\frac{\left(x^2+1\right) \left(m_{a}+9 m_{s}e^{i \eta } \right)}{x^2+3},  \quad
 z=\frac{ \sqrt{2(x^2+1)} \left(-m_{a}+3x^2 m_{s}e^{i \eta }  \right)}{x^2+3},  \quad
 w=-\frac{2 \left(m_{a}+x^4m_{s}e^{i \eta }  \right)}{x^2+3}\,.
\end{equation}
Then the block diagonal neutrino mass matrix $m^\prime_{\nu}$  can be diagonalized by a unitary rotation matrix $U_{\nu2}$ for NO case given in Eq.~\eqref{eq:Unu2}. From the expressions of the matrices $U_{l}$ in Eq.~\eqref{eq:ch_dia_matrix}, $U_{\nu1}$ and $U_{\nu2}$, we find the lepton mixing matrix is
\begin{equation}\label{eq:PMNS_N7_2}
  U=\begin{pmatrix}
 \frac{\sqrt{2} x}{\sqrt{3(x^2+1)}} &~ -\sqrt{\frac{x^2+3}{3(x^2+1)}} \cos \theta  &~ - \sqrt{\frac{x^2+3}{3(x^2+1)}} e^{i \psi }\sin \theta  \\
 \frac{x+\sqrt{3}}{ \sqrt{6(x^2+1)}} &~ \frac{x \left(x +\sqrt{3}\right) \cos \theta }{ \sqrt{3(x^2+1)( x^2+3)}}+\frac{ \left(x-\sqrt{3}\right) e^{-i \psi }\sin \theta }{ \sqrt{2(x^2+3)}} &~ \frac{\left(\sqrt{3}-x\right) \cos \theta }{ \sqrt{2(x^2+3)}}+\frac{ x \left(x +\sqrt{3}\right)e^{i \psi } \sin \theta }{\sqrt{3(x^2+1)( x^2+3)}} \\
 \frac{x-\sqrt{3}}{ \sqrt{6(x^2+1)}} &~ \frac{x \left( x-\sqrt{3}\right) \cos \theta }{\sqrt{3(x^2+1)( x^2+3)}}-\frac{ \left(x+\sqrt{3}\right) e^{-i \psi }\sin \theta }{ \sqrt{2(x^2+3)}} & ~ \frac{\left(x+\sqrt{3}\right) \cos \theta }{ \sqrt{2(x^2+3)}}+\frac{ x \left( x-\sqrt{3}\right) e^{i \psi }\sin \theta }{\sqrt{3(x^2+1)( x^2+3)}} \\
\end{pmatrix}\,,
\end{equation}
The expressions for the lepton mixing angles are as follows,
\begin{eqnarray}
\nonumber \sin^2\theta_{13}&=&\frac{\left(x^2+3\right) \sin ^2\theta }{3 \left(x^2+1\right)}\,, \qquad
  \sin^2\theta_{12}=1-\frac{2 x^2}{2 x^2+\left(x^2+3\right) \cos ^2\theta }\,,\\
\sin^2\theta_{23}&=&\frac{1}{2}-\frac{\sqrt{3} x}{x^2+3}+\frac{8 \sqrt{3} x^3 \sin ^2\theta -\sqrt{6} x \left(x^2-3\right) \sqrt{x^2+1} \sin 2 \theta  \cos \psi }{2 \left(x^2+3\right) \left(2 x^2+\left(x^2+3\right) \cos ^2\theta \right)}\,.
\end{eqnarray}
These give a sum rule between mixing angles $\theta_{12}$ and $\theta_{13}$ with
\begin{equation}
\cos^2\theta_{12}\cos^2\theta_{13}=\frac{2 x^2}{3 \left(1+x^2\right)}\,.
\end{equation}
On the one hand, for the fixed value of $x$, the possible values of $\theta_{12}$ will always be limited to a narrow range by varying the mixing angle $\theta_{13}$ over its $3\sigma$ range. On the other hand, from the $3\sigma$ ranges of mixing angles $\theta_{13}$ and $\theta_{12}$, we find that $x\leq-4.997$ or $x>5.392$ should be satisfied. From the PMNS matrix, we find that the two CP rephasing invariants $J_{CP}$ and $I_1$ are predicted to be
\begin{equation}
 J_{CP}=\frac{x \left(x^2-3\right) \sin 2 \theta  \sin \psi }{6 \sqrt{6} \left(x^2+1\right)^{3/2}}\,, \qquad
I_{1}=\frac{\left(x^2+3\right)^2 \sin ^22 \theta  \sin (\rho -\sigma )}{36 \left(x^2+1\right)^2}\,.
\end{equation}
We can derive the following sum rule among the Dirac CP phase $\delta_{CP}$ and mixing angles
\begin{equation}
\cos\delta_{CP}=\frac{2  \cos ^2\theta_{13} \left(2 \sqrt{3} x-3  \cos 2 \theta_{23}\right)+x^2 (3-5 \cos 2 \theta_{13}) \cos 2 \theta_{23}}{4\sin \theta_{13} \sin 2 \theta_{23} \sqrt{x^2 \left(3 -x^2+3 \left(1+x^2\right) \cos 2 \theta_{13}\right)}}\,.
\end{equation}
For a given value of $x$, the possible ranges of $\cos\delta_{CP}$ can be obtained from the above sum rule by varying $\theta_{13}$ and $\theta_{23}$ over their $3\sigma$ regions. Detailed numerical analyses show that all the three mixing angles and mass ratio $m^2_2/m^2_3$ can simultaneously lie in their respective $3\sigma$ ranges for input parameters $|x|$, $\eta$ and $r$  lying in the ranges $[7.347,9.104]$, $[-\pi,\pi]$ and  $[0.0324,0.0549]$, respectively. Then $\theta_{12}$ is found to lie in a narrow interval $0.326\leq\sin^2\theta_{12}\leq0.330$, the atmospheric mixing angle is constrained to lie in the interval $0.485\leq\sin^2\theta_{23}\leq0.515$ and
$|\delta_{CP}|$ is predicted to be in the range of $[0,0.0190\pi]\cup[0.981\pi,\pi]$. Any value between $-\pi$ and $\pi$ is permitted for Majorana CP phase $\beta$. Here we find the Dirac CP phase is approximate trivial. Hence this breaking pattern would be ruled out if the signal of maximal $\delta_{CP}$ is confirmed by future neutrino facilities. In table~\ref{tab:bf_N7_2} we present the predictions for mixing angles, CP violating phases, light neutrino masses and the effective mass in neutrinoless double beta decay for some benchmark values of the parameters $x$ and $\eta$. We find that the results of $\eta=0$ are viable. It is useful in model building. However, $\eta=0$ leads to trivial Dirac CP phase and Majorana CP phase. The reason is that parameters $y$, $z$ and $w$ in Eq.~\eqref{eq:yzw_N7_2} are all real. From the expressions of parameters $\psi$, $\rho$ and $\sigma$ given in Eq.~\eqref{eq:par_prs}, we find that the three parameters can only take $0$ or $\pi$. Then up to the diagonal phase matrix $P_{\nu}$ with  entries $\pm1$ or $\pm i$, it is easy to check that PMNS matrix in Eq.~\eqref{eq:PMNS_N7_2} is a real matrix. This mixing matrix gives trivial Dirac CP phase and Majorana CP phase.

\begin{table}[t!]
\renewcommand{\tabcolsep}{0.2mm}
\renewcommand{\arraystretch}{1.3}
\small
\centering
\begin{tabular}{|c|c| c| c| c | c| c| c| c| c| c |c |c |c |c |}  \hline \hline
$\langle\phi_{\text{sol}}\rangle/v_{\phi_{s}}$ &  $x$  & $\eta$ & $m_{a}$(meV)  & $r$  	 & $\chi^2_{\text{min}}$ &  $\sin^2\theta_{13}$  &$\sin^2\theta_{12}$  & $\sin^2\theta_{23}$  & $\delta_{CP}/\pi$ &  $\beta/\pi$   & $m_{2}$(meV)  & $m_{3}$(meV)  & $m_{ee}$(meV) \\   \hline
\multirow{8}{*}{$\left(1,1+8 i,1-8 i\right)^T$} &  \multirow{8}{*}{$8$} & $\pi$ & $8 .639$ & $0 .0445$ & $27 .361$ & $0 .0208$ & $0 .330$ & $0 .489$ & $0$ & $1$ & $8 .582$ & $50 .268$ & $1 .727$ \\ \cline{3-14}
& &   $\pm\frac{\pi }{2}$ & $8 .662$ & $0 .0443$ & $18 .077$ & $0 .0234$ & $0 .328$ & $0 .502$ & $\mp0 .0178$ & $\pm0.501$ & $8 .594$ & $50 .253$ & $3 .108$ \\ \cline{3-14}
& &  $\pm\frac{2 \pi }{3}$ & $8 .658$ & $0 .0443$ & $18 .400$ & $0 .0221$ & $0 .329$ & $0 .496$ & $\mp0.0158$ & $\pm0 .667$ & $8 .595$ & $50 .251$ & $2 .516$ \\ \cline{3-14}
& &  $\pm\frac{3 \pi }{4}$ & $8 .652$ & $0 .0444$ & $21 .067$ & $0 .0215$ & $0 .329$ & $0 .493$ & $\mp0.0131$ & $\pm0 .751$ & $8 .592$ & $50 .256$ & $2 .224$ \\ \cline{3-14}
& &  $\pm\frac{2 \pi }{5}$ & $8 .657$ & $0 .0443$ & $22 .196$ & $0 .0243$ & $0 .327$ & $0 .506$ & $\mp0.0166$ & $\pm0 .401$ & $8 .585$ & $50 .264$ & $3 .421$ \\ \cline{3-14}
& &  $\pm\frac{3 \pi }{5}$ & $8 .661$ & $0 .0443$ & $17 .255$ & $0 .0226$ & $0 .328$ & $0 .498$ & $\mp0.0172$ & $\pm0 .601$ & $8 .596$ & $50 .249$ & $2 .758$ \\ \cline{3-14}
& &  $\pm\frac{4 \pi }{5}$ & $8 .648$ & $0 .0444$ & $22 .923$ & $0 .0213$ & $0 .329$ & $0 .492$ & $\mp0.0109$ & $\pm0 .800$ & $8 .589$ & $50 .259$ & $2 .065$  \\ \cline{3-14}
& &  $\pm\frac{5 \pi }{6}$ &  $8 .645$ & $0 .0444$ & $24 .117$ & $0 .0211$ & $0 .329$ & $0 .491$ & $\mp0.00932$ & $\pm0 .834$ & $8 .587$ & $50 .262$ & $1 .970$ \\ \hline
\multirow{2}{*}{$\left(1,1+9 i,1-9 i\right)^T$}  & \multirow{2}{*}{$9$} & $0$ & $8 .686$ & $0 .035$ & $18 .468$ & $0 .0207$ & $0 .328$ & $0 .513$ & $0$ & $0$ & $8 .623$ & $50 .218$ & $3 .807$ \\ \cline{3-14}
&& $\pm\frac{\pi }{6}$ & $8 .686$ & $0 .035$ & $21 .429$ & $0 .0205$ & $0 .328$ & $0 .512$ & $\mp0.00861$ & $\pm0 .167$ & $8 .623$ & $50 .217$ & $3 .713$ \\ \hline
\multirow{8}{*}{$\left(1,1-8 i,1+8 i\right)^T$} & \multirow{8}{*}{$-8$} & $\pi$ &  $8 .644$ & $0 .0444$ & $19 .680$ & $0 .0208$ & $0 .330$ & $0 .511$ & $1$ & $1$ & $8 .587$ & $50 .262$ & $1 .729$ \\ \cline{3-14}
&&  $\pm\frac{\pi }{2}$ & $8 .659$ & $0 .0443$ & $19 .667$ & $0 .0234$ & $0 .328$ & $0 .498$ & $\pm0.982$ & $\pm0.501$ & $8 .591$ & $50 .257$ & $3 .107$ \\ \cline{3-14}
&&  $\pm\frac{2 \pi }{3}$ & $8 .659$ & $0 .0443$ & $15 .344$ & $0 .0221$ & $0 .329$ & $0 .504$ & $\pm0 .984$ & $\pm0 .667$ & $8 .597$ & $50 .249$ & $2 .517$ \\ \cline{3-14}
&&  $\pm\frac{3 \pi }{4}$ & $8 .654$ & $0 .0444$ & $16 .091$ & $0 .0215$ & $0 .329$ & $0 .507$ & $\pm0 .987$ & $\pm0 .751$ & $8 .595$ & $50 .252$ & $2 .225$ \\ \cline{3-14}
&&  $\pm\frac{2 \pi }{5}$ & $8 .652$ & $0 .0443$ & $26 .655$ & $0 .0243$ & $0 .327$ & $0 .494$ & $\pm0 .983$ & $\pm0 .401$ & $8 .580$ & $50 .271$ & $3 .419$ \\ \cline{3-14}
&&  $\pm\frac{3 \pi }{5}$ & $8 .661$ & $0 .0443$ & $15 .973$ & $0 .0226$ & $0 .328$ & $0 .502$ & $\pm0 .983$ & $\pm0 .601$ & $8 .596$ & $50 .250$ & $2 .758$ \\ \cline{3-14}
&&  $\pm\frac{4 \pi }{5}$ & $8 .651$ & $0 .0444$ & $17 .004$ & $0 .0213$ & $0 .329$ & $0 .508$ & $\pm0 .989$ & $\pm0 .800$ & $8 .593$ & $50 .254$ & $2 .066$  \\ \cline{3-14}
&&  $\pm\frac{5 \pi }{6}$ & $8 .649$ & $0 .0444$ & $17 .671$ & $0 .0211$ & $0 .329$ & $0 .509$ & $\pm0 .991$ & $\pm0 .834$ & $8 .591$ & $50 .256$ & $1 .971$ \\ \hline
\multirow{2}{*}{$\left(1,1-9 i,1+9 i\right)^T$}  & \multirow{2}{*}{$-9$} & $0$ & $8 .676$ & $0 .035$ & $28 .194$ & $0 .0207$ & $0 .328$ & $0 .487$ & $1$ & $0$ & $8 .613$ & $50 .230$ & $3 .804$ \\ \cline{3-14}
&&   $\pm\frac{\pi }{6}$ & $8 .677$ & $0 .035$ & $30 .042$ & $0 .0205$ & $0 .328$ & $0 .488$ & $\pm0 .991$ & $\pm0 .167$ & $8 .614$ & $50 .228$ & $3 .710$ \\ \hline \hline
\end{tabular}
\caption{\label{tab:bf_N7_2}
The predictions for the lepton mixing angles, CP violation phases, neutrino masses and the effective Majorana mass $m_{ee}$ for the breaking pattern $(G_{l},G_{\text{atm}},G_{\text{sol}})= (K_4^{(S,TST^2)},Z_3^T,Z_2^{SU}) $ and $X_{\text{sol}}=\{U,S\}$. Here we choose many benchmark values for the parameters $x$ and $\eta$. The first column of PMNS matrix are fixed to be $\left(\frac{16}{\sqrt{390}},\frac{\sqrt{3}\pm8}{\sqrt{390}},\frac{8\mp\sqrt{3}}{\sqrt{390}}\right)^T$ and $\left(\frac{6 \sqrt{3}}{2\sqrt{41}},\frac{3 \sqrt{3}\pm1}{2\sqrt{41}},\frac{3 \sqrt{3}\mp1}{2\sqrt{41}}\right)^T$ for $x=\pm8$ and $x=\pm9$ respectively. Notice that the lightest neutrino mass is vanishing $m_1=0$.
}
\end{table}

\item[~~($\mathcal{N}_8$)]{$(G_{l},G_{\text{atm}},G_{\text{sol}})=(K_4^{(S,TST^2)},Z_2^U,Z_2^{TU})$, $X_{\text{atm}}=\{1,U\}$, $X_{\text{sol}}=\{U,T\}$ }

For this lepton mixing pattern, only the residual CP transformation $X_{\text{sol}}$ taking the values $\{U,T\}$ is viable. Then the VEV of flavon $\phi_{\text{sol}}$ is proportional to $\left(1,\omega x, \omega^2 x\right)^T$. The most general neutrino mass matrix is
\begin{equation}
 m_{\nu}=m_{a}\begin{pmatrix}
 0 &~ 0 &~ 0 \\
 0 &~ 1 &~ -1 \\
 0 &~ -1 &~ 1 \\
\end{pmatrix}+m_{s}e^{i\eta}
\begin{pmatrix}
 1 &~ x \omega ^2 &~ x \omega  \\
 x \omega ^2 &~ x^2 \omega  &~ x^2 \\
 x \omega  &~ x^2 &~ x^2 \omega ^2 \\
\end{pmatrix}\,.
\end{equation}
It is easy to check that the above neutrino mass matrix has an eigenvalue 0 with eigenvector $(x,1, 1)^{T}$. The neutrino mass matrix $ m_{\nu}$ can be block diagonalised by the unitary matrix $U_{\nu1}$, where $U_{\nu1}$ is
\begin{equation}
U_{\nu1}=\begin{pmatrix}
 -\frac{x}{\sqrt{2+x^2}} &~ 0 &~ -\frac{\sqrt{2} }{\sqrt{2+x^2}} \\
 -\frac{1}{\sqrt{2+x^2}} &~ -\frac{1}{\sqrt{2}} &~ \frac{x}{ \sqrt{2(2+x^2)}} \\
 -\frac{1}{\sqrt{2+x^2}} &~ \frac{1}{\sqrt{2}} &~ \frac{x}{ \sqrt{2(2+x^2)}} \\
\end{pmatrix}\,,
\end{equation}
Then the three nonzero parameters $y$, $z$ and $w$ of $m^\prime_{\nu}$ are
\begin{equation}
y=2 m_{a}-\frac{3}{2} x^2 m_{s}e^{i \eta } , \qquad
z=-\frac{1}{2} i x   \sqrt{3(2+x^2)}m_{s}e^{i \eta },  \qquad
w=\frac{1}{2}   \left(2 +x^2\right)m_{s}e^{i \eta }\,.
\end{equation}
The neutrino mass matrix $m^\prime_{\nu}$ can be diagonalized by unitary matrix $U_{\nu2}$ which show in Eq.~\eqref{eq:Unu2}. Then the lepton mixing matrix is determined to be
\begin{equation}
 U=\begin{pmatrix}
 -\frac{2+x}{ \sqrt{3(2+x^2)}} &~ -\frac{\sqrt{2} (x-1) e^{-i \psi } \sin \theta }{\sqrt{3(2+x^2)}} &~ \frac{\sqrt{2} (1-x) \cos \theta }{\sqrt{3(2+x^2)}} \\
 \frac{1-x}{\sqrt{3(2+x^2)}} &~ \frac{i \cos \theta }{\sqrt{2}}+\frac{(2 +x) e^{-i \psi } \sin \theta }{ \sqrt{6(2+x^2)}} &~ \frac{(2+x) \cos \theta }{\sqrt{6(2+x^2)}}-\frac{ie^{i \psi }  \sin \theta }{\sqrt{2}} \\
 \frac{1-x}{\sqrt{3(2+x^2)}} &~ -\frac{i \cos \theta }{\sqrt{2}}+\frac{(2 +x) e^{-i \psi } \sin \theta }{\sqrt{6(2+x^2)}} &~ \frac{(2 +x) \cos \theta }{\sqrt{6(2+x^2)}}+\frac{i e^{i \psi } \sin \theta }{\sqrt{2}} \\
\end{pmatrix}\,,
\end{equation}
One can straightforwardly extract the lepton mixing angles and the two CP rephasing invariants $J_{CP}$ and $I_1$  as follows,
\begin{eqnarray}
\nonumber &&\sin^2\theta_{13}=\frac{2 (1-x)^2 \cos ^2\theta }{3 \left(2+x^2\right)}\,,
\qquad \sin^2\theta_{12}=\frac{2 (1-x)^2 \sin ^2\theta }{3 \left(2+x^2\right)-2 (1-x)^2 \cos ^2\theta }\,,\\
\nonumber &&\sin^2\theta_{23}=\frac{1}{2}+\frac{\sqrt{3} (2+x) \sqrt{2 +x^2} \sin 2 \theta \sin \psi }{2 \left(3 \left(2+x^2\right)-2 (1-x)^2 \cos ^2\theta \right)}\,, \\
 && J_{CP}=-\frac{(1-x)^2 (2+x) \sin 2 \theta  \cos \psi }{6 \sqrt{3} \left(2 +x^2\right)^{3/2}}\,, \qquad I_{1}=\frac{(1-x)^4 \sin ^22 \theta  \sin (\rho -\sigma )}{9 \left(2+x^2\right)^2}\,.
\end{eqnarray}
 We see that the three mixing angles and Dirac CP phase only depend on two free parameters $\theta$ and $\psi$. Then mixing parameters are strongly correlated such that the following sum rules among the mixing angles and Dirac CP phase are found to be satisfied
\begin{equation}
 \hskip-0.12in \cos^2\theta_{12}\cos^2\theta_{13}=\frac{(2+ x)^2}{3 \left(2+x^2\right)}\,, \quad
 \cos\delta_{CP}=\frac{ \left[3+6x-\left(5 +2 x+2 x^2\right) \cos 2 \theta_{13}\right]\cot 2 \theta_{23}}{2\sin\theta_{13}\sqrt{3 (2+x)^2 \left(2 +x^2\right) \cos ^2\theta_{13}-(2+x)^4}}\,.
\end{equation}
The former correlation implies that the solar mixing angle $\theta_{12}$ is restricted by the observed value of $\theta_{13}$ for a given $x$. From the sum rule among $\delta_{CP}$ and mixing angles, we find that maximal atmospheric mixing angle $\theta_{23}=45^\circ$ leads to maximal Dirac CP phase, i.e. $\cos\delta_{CP}=0$. For fixed value of $x$, the possible value of $\delta_{CP}$ is determined by the $3\sigma$ ranges of mixing angles $\theta_{13}$ and $\theta_{23}$.

\begin{table}[t!]
\renewcommand{\tabcolsep}{0.3mm}
\renewcommand{\arraystretch}{1.2}
\small
\centering
\begin{tabular}{|c|c| c| c| c | c| c| c| c| c| c |c |c |c |c |}  \hline \hline
$\langle\phi_{\text{sol}}\rangle/v_{\phi_{s}}$ & $x$   & $\eta$ & $m_{a}$(meV) & $r$  	 & $\chi^2_{\text{min}}$ &  $\sin^2\theta_{13}$  &$\sin^2\theta_{12}$  & $\sin^2\theta_{23}$  & $\delta_{CP}/\pi$ &  $\beta/\pi$  & $m_{2}$(meV) & $m_{3}$(meV) & $m_{ee}$(meV) \\   \hline
$\left(1,4\omega,4\omega^2\right)^T$ &  $4$ & $\pm\frac{\pi }{3}$ & $26 .798$ & $0 .0335$ & $10 .716$ & $0 .0225$ & $0 .318$ & $0 .513$ & $\mp0.482$ & $\pm0 .401$ & $8 .628$ & $50 .212$ & $2 .694$ \\ \hline
\multirow{2}{*}{$\left(1,\frac{7}{2}\omega,\frac{7}{2}\omega^2\right)^T$} &  \multirow{2}{*}{$\frac{7}{2}$} & $\pm\frac{\pi }{3}$ & $26 .502$ & $0 .0444$ & $30 .264$ & $0 .0210$ & $0 .277$ & $0 .517$ & $\mp0.478$ & $\pm0 .397$ & $8 .922$ & $49 .846$ & $2 .453$ \\ \cline{3-14}
&&  $\pm\frac{2\pi }{5}$ & $24 .191$ & $0 .0503$ & $19 .851$ & $0 .0235$ & $0 .275$ & $0 .577$ & $\mp0.405$ & $\pm0 .313$ & $8 .278$ & $50 .656$ & $2 .534$ \\ \hline
$\left(1,\frac{15}{4}\omega,\frac{15}{4}\omega^2\right)^T$ &  $\frac{15}{4}$ & $\pm\frac{\pi }{3}$ & $26 .661$ & $0 .0384$ & $13 .801$ & $0 .0218$ & $0 .299$ & $0 .515$ & $\mp0.480$ & $\pm0 .399$ & $8 .766$ & $50 .044$ & $2 .582$ \\ \hline
$\left(1,\frac{19}{5}\omega,\frac{19}{5}\omega^2\right)^T$ & $\frac{19}{5}$ & $\pm\frac{\pi }{3}$ & $26 .690$ & $0 .0374$ & $12 .276$ & $0 .0220$ & $0 .303$ & $0 .514$ & $\mp0.481$ & $\pm0 .399$ & $8 .737$ & $50 .079$ & $2 .606$ \\ \hline \hline
\end{tabular}
\caption{\label{tab:bf_N8}
The predictions for the lepton mixing angles, CP violation phases, neutrino masses and the effective Majorana mass $m_{ee}$ for the breaking pattern $\mathcal{N}_{8}$ with $(G_{l},G_{\text{atm}},G_{\text{sol}})= (K_4^{(S,TST^2)},Z_2^U,Z_2^{TU}) $ and $X_{\text{sol}}=\{U,T\}$. Here we choose many benchmark values for the parameters $x$ and $\eta$. The TM1 mixing matrix is reproduced in the case of $x=4$, and the first column of the PMNS matrix are fixed to be $\left(\frac{11}{3 \sqrt{19}},\frac{5}{3 \sqrt{19}},\frac{5}{3 \sqrt{19}}\right)^T$, $\left(\frac{23}{\sqrt{771}},\frac{11}{\sqrt{771}},\frac{11}{\sqrt{771}}\right)^T$ and $\left(\frac{29}{3 \sqrt{137}},\frac{14}{3 \sqrt{137}},\frac{14}{3 \sqrt{137}}\right)^T$ for $x=7/2$, $15/4$ and $19/5$ respectively. Notice that the lightest neutrino mass is vanishing $m_1=0$.
}
\end{table}

In order to see how well the lepton mixing angles can be described by this breaking pattern and its prediction for CP phases, we perform a $\chi^2$ analysis defined in Eq.~\eqref{eq:chisq} for some benchmark values of $x$ and $\eta$. The results are listed in table~\ref{tab:bf_N8}. We can see that the minimum value of $\chi^2$ can be quite small for $x=4$, $15/4$ and $19/5$. Furthermore, we also find that the mixing pattern with $x=4$ is equivalent to the breaking pattern $\mathcal{N}_{1}$ with $X_{\text{sol}}=\{1,SU\}$ and $x=-1$. In order to obtain all possible values of mixing angles and CP phases, we consider input parameters $x$, $\eta$ and $r$ being free parameters and require all three mixing angles and mass ratio $m^2_2/m^2_3$ lying in their $3\sigma$ ranges. Then we find the allowed values of input parameters $x$, $|\eta|$ and $r$ are $[3.472,4.481]$, $[0.253\pi,0.412\pi]$ and $[0.0240,0.0516]$, respectively. Moreover, any value of $\theta_{12}$ within its $3\sigma$ range can be achieved and $\theta_{23}$ is restricted in the range of $0.450\leq\sin^2\theta_{23}\leq0.588$. The absolute values of the Dirac CP phase and Majorana CP phase are predicted in the ranges of $[0.393\pi,0.579\pi]$ and $[0.298\pi,0.512\pi]$, respectively.

\end{description}

\section{\label{sec:other_IO_mix}Other mixing patterns with IO}

In this appendix, we shall list the other possible choices for the residual symmetries $(G_{l}$, $G_{\text{atm}}$, $G_{\text{sol}}$ and the resulting predictions for lepton mixing parameters and neutrino masses.

\begin{description}[labelindent=-0.8em, leftmargin=0.3em]

\item[~~($\mathcal{I}_{6}$)]{$(G_{l},G_{\text{atm}},G_{\text{sol}})=(Z_3^T,Z_2^{TST^2},Z_2^U)$, $X_{\text{atm}}=\{SU,ST^2S,T^2,T^2STU\}$, $X_{\text{sol}}=\{1,U\}$}

Here the residual symmetries in the charged lepton sector, atmospheric neutrino sector and solar neutrino sector are the same as $\mathcal{N}_4$ case which is discussed in section~\ref{sec:TD_S4_GCP_NO}. Then the light neutrino mass matrix $m_{\nu}$ takes the same form as it in Ref.~\cite{Ding:2018fyz}. From the discussing below Eq.~\eqref{eq:permutation_matrices}, the PMNS matrix of this case can be directly obtained from the PMNS matrix in $\mathcal{N}_4$:
\begin{eqnarray}
\hskip-0.1in \nonumber  U=\frac{1}{\sqrt{2}}\begin{pmatrix}
  2i\sqrt{\frac{x^2+x+1}{5x^2+2x+2}} \cos \theta  &~ 2i\sqrt{\frac{x^2+x+1}{5x^2+2x+2}} e^{i \psi }\sin \theta  &~ \frac{ \sqrt{6} x}{\sqrt{5x^2+2x+2}} \\
  -e^{-i \psi } \sin \theta -\frac{i \sqrt{3} x \cos\theta }{\sqrt{5x^2+2x+2}} &~ \cos \theta -\frac{i \sqrt{3}x e^{i \psi }  \sin \theta }{\sqrt{5x^2+2x+2}} &~  \sqrt{\frac{2(x^2+x+1)}{5x^2+2x+2}} \\
  e^{-i \psi } \sin \theta -\frac{i \sqrt{3}x\cos \theta }{\sqrt{5x^2+2x+2}} &~ -\cos \theta -\frac{i \sqrt{3}x e^{i \psi}\sin\theta}{\sqrt{5x^2+2x+2}} &~ \sqrt{\frac{2(x^2+x+1)}{5x^2+2x+2}}\\
\end{pmatrix}\,,
\end{eqnarray}
The lepton mixing angles and CP rephasing invariants can be read off as
\begin{eqnarray}
\nonumber \hskip-0.3in && \sin^2\theta_{13}=\frac{3 x^2}{5 x^2+2 x+2}\,,\qquad
  \sin^2\theta_{12}=\sin ^2\theta \,,\qquad
\sin^2\theta_{23}=\frac{1}{2}\,, \\
\hskip-0.3in && J_{CP}=-\frac{\sqrt{3} x \left(x^2+x+1\right) \sin 2\theta   \sin \psi }{2(5 x^2+2 x+2)^{3/2}}\,, \qquad
 I_{1}=-\frac{\left(x^2+x+1\right)^2 \sin ^22 \theta \sin (\rho -\sigma )}{\left(5 x^2+2 x+2\right)^2}\,.
\end{eqnarray}
As same as case $\mathcal{I}_{1}$, $\theta_{23}$ is maximal and $\theta_{13}$ only depend on real parameter $x$.  For the $3\sigma$ interval $0.02068\leq\sin^2\theta_{13}\leq0.02463$,  we have $x\in[-0.123,-0.113]\cup[0.127,0.140]$. In order to know how well the predicted mixing patterns agree with the experimental data, we shall perform a $\chi^2$ analysis for $x=\frac{1}{8}$. The numerical results are
\begin{eqnarray}
\nonumber && \eta=0.993\pi, \qquad m_{a}=19 .180\,\text{meV},\qquad r=2 .890, \qquad \chi^2_{\text{min}}=32 .054, \qquad \sin^2\theta_{13}=0.0201\,, \\
\nonumber && \sin^2\theta_{12}=0.310, \qquad \sin^2\theta_{23}=0.5, \qquad \delta_{CP}=-0.876\pi, \qquad \beta=0.510\pi, \\
&&  m_1=49 .377\,\text{meV}, \qquad m_2=50 .120\,\text{meV}, \qquad m_3=0\,\text{meV}, \qquad m_{ee}=36 .250\,\text{meV}\,.
\end{eqnarray}
For $x=\frac{1}{8}$, the absolute value of the third column of PMNS is fixed to be $\left(\sqrt{\frac{3}{149}},\sqrt{\frac{73}{149}},\sqrt{\frac{73}{149}}\right)^T$. Due to the requirement of the three mixing angles and mass ratio $m^2_1/m^2_2$ in their $3\sigma$ ranges~\cite{Esteban:2018azc}, the input parameters $|\eta|$ and $r$ are restricted in the range of  $[0.9913\pi,0.9957\pi]$ and $[2.862,2.915]$, respectively. Then the values of $\delta_{CP}$ lie in the range $[-0.904\pi,-0.842\pi]\cup[-0.158\pi,-0.0962\pi]\cup[0.0666\pi,0.106\pi]\cup[0.894\pi,0.933\pi]$, and the allowed range of the absolute value of the Majorana phase is $[0.375\pi,0.406\pi]\cup[0.492\pi,0.546\pi]$. The other mixing angles can take any value in their $3\sigma$ ranges.

\item[~~($\mathcal{I}_{7}$)]{$(G_{l},G_{\text{atm}},G_{\text{sol}})=(Z_3^T,Z_2^{U},Z_2^{TU})$, $X_{\text{atm}}=\{1,U\}$, $X_{\text{sol}}=\{U,T\}$}

Here only the residual CP transformation $X_{\text{sol}}=\{U,T\}$ can give phenomenological predictions. Then the VEV alignments of the flavons $\phi_{\text{atm}}$ and $\phi_{\text{sol}}$ are
\begin{equation}\label{eq_VEV_I7}
\langle\phi_{\text{atm}}\rangle=v_{\phi_a}\left(0,1,-1\right)^T\,, \quad \langle\phi_{\text{sol}}\rangle=v_{\phi_s}\left(1,x\omega , x\omega^2 \right)^T\,.
\end{equation}
Form the general VEVs of flavon fields $\phi_{\text{atm}}$ and $\phi_{\text{sol}}$, we find the general form of the neutrino mass matrix is given by
\begin{equation}
 m_{\nu}=m_{a}\begin{pmatrix}
 0 &~ 0 &~ 0 \\
 0 &~ 1 &~ -1 \\
 0 &~ -1 &~ 1 \\
\end{pmatrix}+m_{s}e^{i\eta}
\begin{pmatrix}
  1 &~ x \omega ^2 &~ x \omega  \\
 x \omega ^2 &~ x^2 \omega  &~ x^2 \\
 x \omega  &~ x^2 &~ x^2 \omega ^2 \\
\end{pmatrix}\,,
\end{equation}
It can be block diagonalised by the unitary matrix $U_{\nu1}$, where unitary matrix $U_{\nu1}$ takes the following form
\begin{equation}
U_{\nu1}=\begin{pmatrix}
 0 &~ \frac{\sqrt{2} }{\sqrt{2+x^2}} &~ -\frac{x}{\sqrt{2+x^2}} \\
 -\frac{1}{\sqrt{2}} &~ -\frac{x}{ \sqrt{2(2+x^2)}} &~ -\frac{1}{\sqrt{2+x^2}} \\
 \frac{1}{\sqrt{2}} &~ -\frac{x}{ \sqrt{2(2+x^2)}} &~ -\frac{1}{\sqrt{2+x^2}} \\
\end{pmatrix}\,,
\end{equation}
Then the three nonzero elements $y$, $z$ and $w$ are determined to be
\begin{equation}
y=2 m_{a}-\frac{3}{2} x^2  m_{s}e^{i \eta }, \quad
z=\frac{i}{2} x  \sqrt{3(2+x^2)} m_{s}e^{i \eta },  \quad
w=\frac{1}{2}  \left(2+x^2\right) m_{s}e^{i \eta }\,.
\end{equation}
Then neutrino mass matrix $m^\prime_{\nu}$ can be diagonalized by performing the unitary transformation matrix $U_{\nu2}$ which is given in Eq.~\eqref{eq:Unu2}.  As a consequence, the PMNS matrix is
\begin{equation}\label{eq:PMNS_I7}
\hskip-0.1in U=\frac{1}{\sqrt{2}}
\begin{pmatrix}
 \frac{2 e^{-i \psi } \sin \theta }{\sqrt{2 +x^2}} &~ \frac{2 \cos \theta }{\sqrt{2 +x^2}} &~ -\frac{\sqrt{2} x}{\sqrt{2 +x^2}} \\
 -\cos \theta-\frac{x e^{-i \psi } \sin \theta }{\sqrt{2 +x^2}}  &~ e^{i \psi } \sin \theta-\frac{x \cos \theta }{\sqrt{2 +x^2}}  &~ -\frac{\sqrt{2} }{\sqrt{2 +x^2}} \\
 \cos \theta -\frac{x e^{-i \psi } \sin \theta }{\sqrt{2 +x^2}} &~ -e^{i \psi } \sin \theta -\frac{x \cos \theta }{\sqrt{2 +x^2}} &~ -\frac{\sqrt{2}}{\sqrt{2 +x^2}} \\
\end{pmatrix}\,,
\end{equation}
Its predictions for the three mixing angles and the two CP rephasing invariants are
\begin{eqnarray}
 \nonumber && \sin^2\theta_{13}=\frac{x^2}{2+x^2}\,, \qquad
  \sin^2\theta_{12}=\cos ^2\theta \,, \qquad
\sin^2\theta_{23}=\frac{1}{2}\,, \\
&& J_{CP}=-\frac{x \sin 2\theta  \sin \psi }{2\left(2 +x^2\right)^{3/2}}\,, \qquad I_{1}=-\frac{ \sin ^22 \theta \sin (\rho -\sigma )}{\left(2 +x^2\right)^2}\,.
\end{eqnarray}
It predicts a maximal atmospheric mixing angle $\theta_{23}$. The viable range of $x$ can be obtained by varying $\theta_{13}$ over its $3\sigma$ range, i.e. $|x|\in[0.206,0.225]$.  In order to see how well the lepton mixing angles can be described by this breaking pattern and its predictions for CP phases, we shall perform a numerical analysis.  When the experimentally allowed regions at $3\sigma$ confidence level of mixing parameters and mass ratio $m^2_1/m^2_2$ are considered, the viable ranges of input parameters $|\eta|$ and $r$ are $[0.0138\pi,0.0201\pi]$ and $[1.803,1.831]$, respectively. Then the Dirac CP phase and the absolute value of the Majorana CP phase are limited to  narrow ranges $[0.0151\pi,0.0357\pi]\cup[0.964\pi,0.985\pi]$ and $[0.166\pi,0.189\pi]$, respectively.

Furthermore we perform a comprehensive numerical analysis for $x=\pm2/9$ which give  relatively simple VEVs of $\phi_{\text{sol}}$. From PMNS matrix in Eq.~\eqref{eq:PMNS_I7}, we find the that fixed column of the PMNS matrix for $x=\pm2/9$ is $\left(\sqrt{\frac{2}{83}},\frac{9}{\sqrt{166}},\frac{9}{\sqrt{166}}\right)^T\simeq\left(0.155, 0.699, 0.699\right)^T$. It agrees with all measurements to date~\cite{Esteban:2018azc}. The usual $\chi^2$ analysis results for $x=\pm2/9$  are
\begin{eqnarray}
\nonumber && \eta=0.0162\pi, \qquad m_{a}=25 .829\,\text{meV},\qquad r=1 .810, \qquad \chi^2_{\text{min}}=22 .509, \qquad \sin^2\theta_{13}=0.0241\,, \\
\nonumber && \sin^2\theta_{12}=0.310, \qquad \sin^2\theta_{23}=0.5, \qquad \delta_{CP}=0.0264\pi, \qquad \beta=0.181\pi, \\
&&  m_1=49 .377\,\text{meV}, \qquad m_2=50 .120\,\text{meV}, \qquad m_3=0\,\text{meV}, \qquad m_{ee}=46 .753\,\text{meV}\,.
\end{eqnarray}

\item[~~($\mathcal{I}_{8}$)]{$(G_{l},G_{\text{atm}},G_{\text{sol}})=(Z_3^T,Z_2^{U},Z_2^{STSU})$, $X_{\text{atm}}=\{1,U\}$, $X_{\text{sol}}=\{U,STS\}$ }

In this combination of residual flavor symmetries, the residual CP transformation $X_{\text{sol}}$ is viable only when it takes the transformations $\{U,STS\}$. In this breaking pattern, the vacuum alignments of flavons $\phi_{\text{atm}}$ and $\phi_{\text{sol}}$ can be read out from table~\ref{tab:inv_VEV_CP}:
\begin{equation}
\langle\phi_{\text{atm}}\rangle=v_{\phi_a}\left(0,1,-1\right)\,, \qquad \langle\phi_{\text{sol}}\rangle=v_{\phi_s}\left(\frac{\sqrt{3} x-1}{2} , 1+i x, 1-i x \right)^T\,.
\end{equation}
The general form of the light neutrino mass matrix is determined to be
\begin{equation}
 m_{\nu}=m_{a}\begin{pmatrix}
 0 &~ 0 &~ 0 \\
 0 &~ 1 &~ -1 \\
 0 &~ -1 &~ 1 \\
\end{pmatrix}+m_{s}e^{i\eta}
\begin{pmatrix}
 \frac{1}{4} \left(1-\sqrt{3} x\right)^2 &~ \frac{1}{2} (1-i x) \left(\sqrt{3} x-1\right) &~ \frac{1}{2} (1+ix ) \left(\sqrt{3} x-1\right) \\
 \frac{1}{2} (1-i x) \left(\sqrt{3} x-1\right) &~ (1-i x)^2 &~ 1+x^2 \\
 \frac{1}{2} (1+ix) \left(\sqrt{3} x-1\right) &~ 1+x^2 &~ (1+ix )^2 \\
\end{pmatrix}\,,
\end{equation}
Before diagonalizing the neutrino mass matrix, we first perform a unitary $U_{\nu1}$ to it, where the unitary transformation matrix $U_{\nu1}$ takes the following form
\begin{equation}
U_{\nu1}=\begin{pmatrix}
 0 &~ \frac{1-\sqrt{3} x}{\sqrt{3 x^2-2 \sqrt{3} x+9}} &~ \frac{2 \sqrt{2}}{\sqrt{3 x^2-2 \sqrt{3} x+9}} \\
 -\frac{1}{\sqrt{2}} &~ -\frac{2}{\sqrt{3 x^2-2 \sqrt{3} x+9}} &~ \frac{1-\sqrt{3} x}{\sqrt{6 x^2-4 \sqrt{3} x+18}} \\
 \frac{1}{\sqrt{2}} &~ -\frac{2}{\sqrt{3 x^2-2 \sqrt{3} x+9}} &~ \frac{1-\sqrt{3} x}{\sqrt{6 x^2-4 \sqrt{3} x+18}} \\
\end{pmatrix}\,,
\end{equation}
The expressions of the parameters $y$, $z$ and $w$ are
\begin{equation}
 \hskip-0.11in y=2 m_{a}-2 x^2  m_{s}e^{i \eta }, \quad
z= -i x\sqrt{\frac{9 -2 \sqrt{3} x+3 x^2}{2}}  m_{s}e^{i \eta },  \quad
w=\frac{1}{4}  \left(9-2 \sqrt{3} x+3 x^2\right) m_{s}e^{i \eta }\,.
\end{equation}
The unitary transformation $U_{\nu2}$ diagonalizing the neutrino mass matrix $m^\prime_{\nu}$ is of the form given in Eq.~\eqref{eq:Unu2} for IO case. Then the lepton mixing matrix  is determined to be of the form
\begin{equation}\label{eq:PMNS_IO8}
  U =\frac{1}{\sqrt{2}}
\begin{pmatrix}
 \frac{\left(\sqrt{3} x-1\right) e^{-i \psi } \sin \theta }{\sqrt{9-2 \sqrt{3} x+3 x^2}} &~ \frac{\left(\sqrt{3} x-1\right) \cos \theta }{\sqrt{9-2 \sqrt{3} x+3 x^2}} &~ \frac{2 \sqrt{2} }{\sqrt{9-2 \sqrt{3} x+3 x^2}} \\
 \frac{\cos \theta }{\sqrt{2}}+\frac{2 e^{-i \psi } \sin \theta }{\sqrt{9-2 \sqrt{3} x+3 x^2}} &~ \frac{2 \cos \theta }{\sqrt{9-2 \sqrt{3} x+3 x^2}}-\frac{e^{i \psi } \sin \theta }{\sqrt{2}} &~ \frac{1-\sqrt{3} x}{\sqrt{2(9-2 \sqrt{3} x+3 x^2)}} \\
 -\frac{\cos \theta }{\sqrt{2}}+\frac{2 e^{-i \psi } \sin \theta }{\sqrt{9-2 \sqrt{3} x+3 x^2}} &~ \frac{2 \cos \theta }{\sqrt{9-2 \sqrt{3} x+3 x^2}}+\frac{e^{i \psi } \sin \theta }{\sqrt{2}} &~ \frac{1-\sqrt{3} x}{\sqrt{2(9-2 \sqrt{3} x+3 x^2)}} \\
\end{pmatrix}\,,
\end{equation}
The mixing parameters extracted from above PMNS matrix are:
\begin{eqnarray}
\nonumber && \sin^2\theta_{13}=\frac{8}{9-2 \sqrt{3} x+3 x^2}\,, \qquad
  \sin^2\theta_{12}=\cos^2\theta \,, \qquad
\sin^2\theta_{23}=\frac{1}{2}\,, \\
&& J_{CP}=-\frac{\sqrt{2} \left(1-\sqrt{3} x\right)^2 \sin 2\theta  \sin \psi }{2\left(9-2 \sqrt{3} x+3 x^2\right)^{3/2}}\,, \qquad I_{1}=-\frac{\left(1-\sqrt{3} x\right)^4 \sin ^22 \theta  \sin (\rho -\sigma )}{4 \left(9-2 \sqrt{3} x+3 x^2\right)^2}\,.
\end{eqnarray}
Inputting the $3\sigma$ ranges of the third column of the PMNS matrix, we find the parameter $x$ should vary in the interval $[-10.660,-9.699]\cup[10.854,11.815]$. As an example,  we take $x=-6\sqrt{3}$.  the fixed column of the PMNS matrix is $\frac{1}{3 \sqrt{82}}\left(4, 19,19\right)^T\simeq\left(0.147,0.699,0.699\right)^T$ which is not beyond $3\sigma$ confidence level~\cite{Esteban:2018azc}.  Furthermore, we perform a conventional $\chi^2$ analysis  and the numerical results are
\begin{eqnarray}
\nonumber && \eta=0.00227\pi, \quad m_{a}=45 .595\,\text{meV},\quad r=0.00645, \quad \chi^2_{\text{min}}=19 .755, \quad \sin^2\theta_{13}=0.0217\,, \\
\nonumber && \sin^2\theta_{12}=0.310, \quad \sin^2\theta_{23}=0.5, \quad \delta_{CP}=0.210\pi, \quad \beta=0.716\pi, \\
&&  m_1=49 .377\,\text{meV}, \quad m_2=50 .120\,\text{meV}, \quad m_3=0\,\text{meV}, \quad m_{ee}=26 .550\,\text{meV}\,.
\end{eqnarray}
In the case that all the three input parameters $x$, $\eta$ and $r$ being free parameters, we find that the three mixing angels and mass ratio $m^2_1/m^2_2$ can lie in their $3\sigma$ ranges at the same time only when $x$, $|\eta|$ and $r$  are restricted in the range of $[-10.660,-9.699]\cup[10.854,11.815]$, $[0.00121\pi,0.00295\pi]$ and $[0.00530,0.00738]$, respectively. We find that any values of $\theta_{12}$ and $\theta_{13}$ within their $3\sigma$ ranges can be achieved, and the two CP phases are predicted to be $\delta_{CP}\in[-0.827\pi,-0.681\pi]\cup[-0.319\pi,-0.173\pi]\cup[0.153\pi,0.274\pi]\cup[0.726\pi,0.847\pi]$ and $|\beta|\in[0.677\pi,0.825\pi]$.

\item[~~($\mathcal{I}_{9}$)]{$(G_{l},G_{\text{atm}},G_{\text{sol}})=(Z_3^T,Z_2^{SU},Z_2^{STSU})$, $X_{\text{atm}}=\{1,SU\}$, $X_{\text{sol}}=\{U,STS\}$}

In this combination of residual flavor symmetries, only the residual CP symmetry $X_{\text{sol}}=\{U,STS\}$ is viable. From table~\ref{tab:inv_VEV_CP}, we can know the VEVs of flavons $\phi_{\text{atm}}$ and $\phi_{\text{sol}}$. Then the light neutrino mass matrix takes the following form
\begin{equation}
 m_{\nu}=m_{a}\begin{pmatrix}
 4 &~ -2 &~ -2 \\
 -2 &~ 1 &~ 1 \\
 -2 &~ 1 &~ 1 \\
\end{pmatrix}+m_{s}e^{i\eta}
\begin{pmatrix}
 \frac{1}{4} \left(1-\sqrt{3} x\right)^2 &~ \frac{1}{2} (1-i x) \left(\sqrt{3} x-1\right) &~ \frac{1}{2} (1+ix) \left(\sqrt{3} x-1\right) \\
 \frac{1}{2} (1-i x) \left(\sqrt{3} x-1\right) &~ (1-i x)^2 &~ 1+x^2 \\
 \frac{1}{2} (1+ix) \left(\sqrt{3} x-1\right) &~ 1+x^2 &~ (1+ix)^2 \\
\end{pmatrix}\,,
\end{equation}
This neutrino mass matrix can become a block diagonal matrix when we perform a unitary transformation $U_{\nu1}$, where $U_{\nu1}$ is
\begin{equation}
U_{\nu1}=\begin{pmatrix}
 0 &~ -\frac{i \left(x \left(19 x+6 \sqrt{3}\right)+9\right)}{\sqrt{3 \left(9 x^2+2 \sqrt{3} x+3\right) \left(19 x^2+6 \sqrt{3} x+9\right)}} &~ \frac{2 \sqrt{2}i x}{\sqrt{3 \left(9 x^2+2 \sqrt{3} x+3\right)}} \\
 \frac{x \left(4 i+\sqrt{3}\right)+3}{\sqrt{2 \left(19 x^2+6 \sqrt{3} x+9\right)}} &~ \frac{2 x \left(\left(\sqrt{3}+4 i\right) x+3\right)}{\sqrt{3 \left(9 x^2+2 \sqrt{3} x+3\right) \left(19 x^2+6 \sqrt{3} x+9\right)}} &~ \frac{x \left(4 i+\sqrt{3}\right)+3}{\sqrt{6 \left(9 x^2+2 \sqrt{3} x+3\right)}} \\
 \frac{x \left(-4 i+\sqrt{3}\right)+3}{\sqrt{2 \left(19 x^2+6 \sqrt{3} x+9\right)}} &~ \frac{2 x \left(\left(-\sqrt{3}+4 i\right) x-3\right)}{\sqrt{3 \left(9 x^2+2 \sqrt{3} x+3\right) \left(19 x^2+6 \sqrt{3} x+9\right)}} &~ \frac{x \left(4 i-\sqrt{3}\right)+3}{\sqrt{6 \left(9 x^2+2 \sqrt{3} x+3\right)}} \\
\end{pmatrix}\,,
\end{equation}
The parameters $y$, $z$ and $w$ are take the following form
\begin{eqnarray}
\nonumber &&y=\frac{6\left(x^2+2 \sqrt{3} x+3\right) m_{a} +2\left(16 x^4+8 \sqrt{3} x^3+27 x^2+6 \sqrt{3} x+9\right) m_{s}e^{i \eta }  }{19 x^2+6 \sqrt{3} x+9}\,,  \\
\nonumber && z=\frac{i \sqrt{2 \left(9 x^2+2 \sqrt{3} x+3\right)} \left(12\left(x+\sqrt{3}\right) m_{a} +\left(-12 x^3+\sqrt{3} x^2-6 x+3 \sqrt{3}\right)m_{s}e^{i \eta }  \right)}{2 \left(19 x^2+6 \sqrt{3} x+9\right)}\,, \\
&& w= \frac{-48\left(9 x^2+2 \sqrt{3} x+3\right) m_{a} -3 \left(27 x^4-12 \sqrt{3} x^3+6 x^2-4 \sqrt{3} x+3\right)m_{s}e^{i \eta }  }{4 \left(19 x^2+6 \sqrt{3} x+9\right)}\,.
\end{eqnarray}
The neutrino mass matrix $m^\prime_{\nu}$ can be diagonalized by the unitary matrix $U_{\nu2}$. Then the PMNS matrix is
\begin{equation}
\hskip-0.1in U=
\begin{pmatrix}
  \sqrt{\frac{19 x^2+6 \sqrt{3} x+9}{3(9 x^2+2 \sqrt{3} x+3)}}e^{-i \psi } \sin \theta  &~ -\sqrt{\frac{19 x^2+6 \sqrt{3} x+9}{3(9 x^2+2 \sqrt{3} x+3)}} \cos \theta  &~ \frac{2 \sqrt{2} x}{\sqrt{3(9 x^2+2 \sqrt{3} x+3)}} \\
 \frac{\cos \theta }{\sqrt{2}}-\frac{2x e^{-i \psi } \sin \theta }{\sqrt{3(9 x^2+2 \sqrt{3} x+3)}} &~ \frac{2 x \cos \theta }{\sqrt{3(9 x^2+2 \sqrt{3} x+3)}}+\frac{e^{i \psi } \sin \theta }{\sqrt{2}} &~ \sqrt{\frac{19 x^2+6 \sqrt{3} x+9}{6(9 x^2+2 \sqrt{3} x+3)}} \\
 -\frac{\cos \theta }{\sqrt{2}}-\frac{2x e^{-i \psi }  \sin \theta }{\sqrt{3(9 x^2+2 \sqrt{3} x+3)}} &~ \frac{2 x \cos \theta }{\sqrt{3(9 x^2+2 \sqrt{3} x+3)}}-\frac{e^{i \psi } \sin \theta }{\sqrt{2}} &~\sqrt{\frac{19 x^2+6 \sqrt{3} x+9}{6(9 x^2+2 \sqrt{3} x+3)}} \\
\end{pmatrix}\,,
\end{equation}
The three lepton mixing angles are predicted to be
\begin{eqnarray}
 \nonumber \hskip-0.5in && \sin^2\theta_{13}=\frac{8 x^2}{9 +6 \sqrt{3} x+27 x^2},\qquad \sin^2\theta_{12}=\cos^2\theta\,,\qquad \sin^2\theta_{23}=\frac{1}{2}\,, \\
\hskip-0.5in &&J_{CP}=\frac{x \left(19 x^2+6 \sqrt{3} x+9\right) \sin 2 \theta  \sin \psi }{3 \sqrt{6} \left(9 x^2+2 \sqrt{3} x+3\right)^{3/2}}\,, \qquad
 I_{1}=\frac{\left(19 x^2+6 \sqrt{3} x+9\right)^2 \sin ^22 \theta \sin (\sigma-\rho  )}{36 \left(9 x^2+2 \sqrt{3} x+3\right)^2}\,.
\end{eqnarray}
The atmospheric mixing angle is maximal and the reactor mixing angle only depends on input parameter $x$. Inputting the $3\sigma$ ranges of the third column of the PMNS matrix, we find the parameter $x$ should vary in the interval $[-0.157, -0.144]\cup[ 0.173 , 0.192]$. As an example,  we take $x=\sqrt{3}/10$. Then the third column of the PMNS matrix is $\frac{1}{3\sqrt{86}}\left(4,\sqrt{379},\sqrt{379}\right)^T\simeq\left(0.144,0.700,0.700\right)^T$ which is agrees with all measurements to date~\cite{Esteban:2018azc}. The $\chi^2$ analysis results are
\begin{eqnarray}
\nonumber && \eta=0.996\pi, \qquad m_{a}=12 .428\,\text{meV},\qquad r=2 .760, \qquad \chi^2_{\text{min}}=26 .533, \qquad \sin^2\theta_{13}=0.0207\,, \\
\nonumber && \sin^2\theta_{12}=0.310, \qquad \sin^2\theta_{23}=0.5, \qquad \delta_{CP}=-0.954\pi, \qquad \beta=0.246\pi, \\
&&  m_1=49 .377\,\text{meV}, \qquad m_2=50 .120\,\text{meV}, \qquad m_3=0\,\text{meV}, \qquad m_{ee}=45 .510\,\text{meV}\,.
\end{eqnarray}
After calculation and analysis, we find that only when the input parameters $|\eta|$  and $r$ lie in the ranges $[0.9954\pi,0.9976\pi]$ and $[2.450,2.472]\cup[2.755,2.763]$ respectively can bring the three mixing angles and mass ratio $m^2_1/m^2_2$ in their $3\sigma$ ranges. Then the two CP phases $\delta_{CP}$ and $\beta$ are predicted to be $\delta_{CP}\in[-0.966\pi,-0.942\pi]\cup[-0.0583\pi,-0.0343\pi]\cup[0.0150\pi,0.0260\pi]\cup[0.974\pi,0.985\pi]$ and $|\beta|\in[0.115\pi,0.127\pi]\cup[0.240\pi,0.260\pi]$.

\item[~~($\mathcal{I}_{10}$)]{$(G_{l},G_{\text{atm}},G_{\text{sol}})=(Z_4^{TSU},Z_2^S,Z_2^{TU})$, $X_{\text{atm}}=\{1,S,TST^2U,T^2STU\}$}

$\bullet$ $X_{\text{sol}}=\{U,T\}$

In this combination of the remnant flavor symmetries, both the residual CP transformations $X_{\text{sol}}=\{U,T\}$ and $X_{\text{sol}}=\{STS,T^2STU\}$ are viable. For the former, the general VEVs of flavons $\phi_{\text{atm}}$ and $\phi_{\text{sol}}$ are proportional to column vectors $(1,1,1)^T$ and $(1,x\omega , x\omega^2)^T$, respectively. The most general form of the neutrino mass matrix is given by
\begin{equation}
 m_{\nu}=m_{a}\begin{pmatrix}
 1 &~ 1 &~ 1 \\
 1 &~ 1 &~ 1 \\
 1 &~ 1 &~ 1 \\
\end{pmatrix}+m_{s}e^{i\eta}
\begin{pmatrix}
1 &~ x \omega ^2 &~ x \omega  \\
 x \omega ^2 &~ x^2 \omega  &~ x^2 \\
 x \omega  &~ x^2 &~ x^2 \omega ^2 \\
\end{pmatrix}\,,
\end{equation}
This neutrino mass matrix can be diagonalized into a block diagonal form by performing a unitary transformation $U_{\nu1}$ to $m_{\nu}$, where unitary matrix $U_{\nu1}$ takes the following form
\begin{equation}
U_{\nu1}=\begin{pmatrix}
 \frac{1}{\sqrt{3}} &~ -\frac{2+x}{\sqrt{3(2 +2 x+5 x^2)}} &~ -\frac{i \sqrt{3} x}{\sqrt{2 +2 x+5 x^2}} \\
 \frac{1}{\sqrt{3}} &~ \frac{1-\left(\omega +i \sqrt{3}\right) x}{ \sqrt{3(2 +2 x+5 x^2)}} &~ \frac{\omega  x-1}{\sqrt{2 +2 x+5 x^2}} \\
 \frac{1}{\sqrt{3}} &~ \frac{1-\left(3 \omega ^2+1\right) x}{\sqrt{3(2 +2 x+5 x^2)}} &~ \frac{1-\omega ^2 x}{\sqrt{2 +2 x+5 x^2}} \\
\end{pmatrix}\,,
\end{equation}
Then the neutrino mass matrix $m^\prime_{\nu}$ is a block diagonal matrix with elements
\begin{equation}
y=3 m_{a}+\frac{(1-x)^2}{3}   m_{s}e^{i \eta }, \quad
z=\frac{x-1}{3}  \sqrt{2 +2 x+5 x^2}m_{s}e^{i \eta },  \quad
w=\frac{2 +2 x+5 x^2}{3}   m_{s}e^{i \eta }\,.
\end{equation}
The diagonalization matrix of $m^\prime_{\nu}$ can be written as the form $U_{\nu2}$ in Eq.~\eqref{eq:Unu2} for IO case. Then the PMNS matrix is determined to be
\begin{equation}\label{eq:PMNS_I10_1}
\hskip-0.1in  U
=\begin{pmatrix}
 -\frac{i (1+2 x) e^{-i \psi } \sin \theta }{\sqrt{2+2 x+5 x^2}} &~ -\frac{i (1+2 x) \cos \theta }{\sqrt{2+2 x+5 x^2}} &~ \frac{1-x}{\sqrt{2+2 x+5 x^2}} \\
 \frac{\cos \theta }{\sqrt{2}}-\frac{(1-x) e^{-i \psi } i \sin \theta }{\sqrt{2(2+2 x+5 x^2)}} &~ \frac{i (x-1) \cos \theta }{\sqrt{2(2+2 x+5 x^2)}}-\frac{e^{i \psi } \sin \theta }{\sqrt{2}} &~ \frac{-1-2x}{\sqrt{2(2+2 x+5 x^2)}} \\
 \frac{\cos \theta }{\sqrt{2}}+\frac{i (1-x) e^{-i \psi } \sin \theta }{\sqrt{2(2+2 x+5 x^2)}} &~ \frac{(1-x) i \cos \theta }{\sqrt{2(2+2 x+5 x^2)}}-\frac{e^{i \psi } \sin \theta }{\sqrt{2}} &~ \frac{1+2 x}{\sqrt{2(2+2 x+5 x^2)}} \\
\end{pmatrix}\,,
\end{equation}
Its predictions for the mixing angles and CP invariants are
\begin{eqnarray}
\nonumber &&  \sin^2\theta_{13}=\frac{(1-x)^2}{2 +2 x+5 x^2}\,, \qquad
  \sin^2\theta_{12}=\cos ^2\theta \,, \qquad
\sin^2\theta_{23}=\frac{1}{2}\,, \\
&& J_{CP}=\frac{(x-1) (1+2 x)^2 \sin 2 \theta  \cos \psi }{4 \left(2 +2x+5 x^2\right)^{3/2}}\,, \qquad I_{1}=-\frac{(1+2 x)^4 \sin ^22 \theta  \sin (\rho -\sigma )}{4 \left(2+2 x+5 x^2\right)^2}\,.
\end{eqnarray}
The atmospheric mixing angle $\theta_{23}$ is predicted to be maximal. We find that the correct value of $\theta_{13}$ can be obtained for $x$ being restricted in the range of $[0.638,0.662]\cup[1.615,1.699]$. If we require all three mixing angles and mass ratio $m^2_1/m^2_2$ in their $3\sigma$ ranges, the two CP phases $\delta_{CP}$ and $|\beta|$ are determined to take values in the intervals $[-\pi,-0.987\pi]\cup[-0.0128\pi,0.00336\pi]\cup[0.997\pi,\pi]$ and $[0.0964\pi,0.111\pi]$, respectively. Furthermore, we find that the viable ranges of $|\eta|$  and $r$ are $[0.965\pi,0.976\pi]$ and $[0.442,0.478]\cup[1.595,1.647]$.

For illustration, we shall give the $\chi^2$ results for the typical value $x=2/3$. The vacuum alignment of flavon $\phi_{\text{sol}}$ is proportional to the column vector $\left(1,\frac{2}{3} \omega,\frac{2}{3} \omega^2\right)^T$ and the third column of the PMNS matrix is $\left(\frac{1}{5 \sqrt{2}},\frac{7}{10},\frac{7}{10}\right)^T$. The $\chi^2$ analysis results are
\begin{eqnarray}
\nonumber && \eta=-0.969\pi, \qquad m_{a}=16 .794\,\text{meV},\qquad r=1 .579, \qquad \chi^2_{\text{min}}=33 .640, \qquad \sin^2\theta_{13}=0.02\,, \\
\nonumber && \sin^2\theta_{12}=0.310, \qquad \sin^2\theta_{23}= 0.5, \qquad \delta_{CP}=-0.997\pi, \qquad \beta=0.0964\pi, \\
&&  m_1=49 .377\,\text{meV}, \qquad m_2=50 .120\,\text{meV}, \qquad m_3=0\,\text{meV}, \qquad m_{ee}=48 .137\,\text{meV}\,.
\end{eqnarray}
We see that $\theta_{13}$ is rather close to its $3\sigma$ lower limit $0.2068$~\cite{Esteban:2018azc}. Hence this example should be considered as a good leading order approximation. The reason is that if subleading contributions are taken into account, accordance with experimental data are easily achieved.

$\bullet$ $X_{\text{sol}}=\{STS,T^2STU\}$

For $X_{\text{sol}}=\{STS,T^2STU\}$,  the most general vacuum alignment of flovan $\phi_{\text{sol}}$ is
\begin{equation}
\langle\phi_{\text{sol}}\rangle=v_{\phi_s}\left(1+2 i x, \omega  (1-i x), \omega ^2 (1-i x) \right)^T\,.
\end{equation}
Then the most general form of the light neutrino mass matrix can be easily obtained:
\begin{equation}
 m_{\nu}=m_{a}\begin{pmatrix}
 1 &~ 1 &~ 1 \\
 1 &~ 1 &~ 1 \\
 1 &~ 1 &~ 1 \\
\end{pmatrix}+m_{s}e^{i\eta}
\begin{pmatrix}
 (1+2i x )^2 &~ \left(1+ix+2 x^2\right) \omega ^2 &~ \left(1+ix+2 x^2\right) \omega  \\
 \left(1+ix+2 x^2\right) \omega ^2 &~ (1-i x)^2 \omega  &~ (1-i x)^2 \\
 \left(1+ix+2 x^2\right) \omega  &~ (1-i x)^2 &~ (1-i x)^2 \omega ^2 \\
\end{pmatrix}\,.
\end{equation}
After performing a $U_{\nu1}$ transformation, $m^\prime_{\nu}$ can be a block diagonal matrix, where the unitary matrix $U_{\nu1}$ takes the form
\begin{equation}
U_{\nu1}=\frac{1}{\sqrt{3}}\begin{pmatrix}
 1 &~ \frac{1-i x}{\sqrt{1+x^2}} &~ \frac{-x-i }{\sqrt{1+x^2}} \\
 1 &~ \frac{ \omega -i x \omega ^2}{\sqrt{1+x^2}} &~ \frac{-i  \omega ^2-x \omega }{\sqrt{1+x^2}} \\
 1 &~ \frac{ \omega ^2-i x \omega }{\sqrt{1+x^2}} &~ \frac{-x \omega ^2-i \omega  }{\sqrt{1+x^2}} \\
\end{pmatrix}\,.
\end{equation}
The nonzero elements $y$, $z$ and $w$ of the block diagonal $m^\prime_{\nu}$ are
\begin{equation}
y=3 m_{a}-3 x^2  m_{s}e^{i \eta }, \qquad
z= -3 i x  \sqrt{1+x^2} m_{s}e^{i \eta },  \qquad
w=3   \left(1+x^2\right)m_{s}e^{i \eta }\,.
\end{equation}
Then the neutrino mass matrix $m^\prime_{\nu}$ can be diagonalized by the unitary matrix $U_{\nu2}$ which is given in Eq.~\eqref{eq:Unu2}. As a consequence, the PMNS matrix can take the following form
\begin{equation}\label{eq:PMNS_I10_2}
\hskip-0.1in  U
=\frac{1}{\sqrt{2}}\begin{pmatrix}
 -\frac{\sqrt{2}  e^{-i \psi } \sin \theta }{\sqrt{1+x^2}} &~ -\frac{\sqrt{2} \cos \theta }{\sqrt{1+x^2}} & ~ \frac{\sqrt{2} x}{\sqrt{1+x^2}} \\
 \cos \theta -\frac{x e^{-i \psi } \sin \theta }{\sqrt{1+x^2}} &~ -e^{i \psi } \sin \theta -\frac{x \cos \theta }{\sqrt{1+x^2}} &~ -\frac{1}{\sqrt{1+x^2}} \\
 \cos \theta +\frac{x e^{-i \psi } \sin \theta }{\sqrt{1+x^2}} &~- e^{i \psi } \sin \theta+\frac{x \cos \theta }{\sqrt{1+x^2}}  &~ \frac{1}{\sqrt{1+x^2}} \\
\end{pmatrix}\,,
\end{equation}
The lepton mixing parameters are predicted to be
\begin{eqnarray}
\nonumber &&  \sin^2\theta_{13}=\frac{x^2}{1+x^2}\,, \qquad
  \sin^2\theta_{12}=\cos ^2\theta \,, \qquad
\sin^2\theta_{23}=\frac{1}{2}\,, \\
&& J_{CP}=\frac{x \sin 2 \theta  \sin \psi }{4 \left(1+x^2\right)^{3/2}}\,, \qquad
I_{1}=-\frac{ \sin ^22\theta   \sin (\rho -\sigma )}{4\left(1+x^2\right)^2} \,.
\end{eqnarray}
This mixing pattern gives a maximal $\theta_{23}$. The reactor mixing angle $\theta_{13}$ only depend on parameter $x$ which comes from the vacuum alignment of flavon $\phi_{\text{sol}}$. In order to obtain the value of $\theta_{13}$ allowed  by experiment data, the input parameter $|x|$ must be restricted in $[0.145,0.159]$.  In order to accommodate the experimentally favored $3\sigma$ ranges~\cite{Esteban:2018azc} of mixing angles and mass ratio $m^2_1/m^2_2$, we find the allowed region of the parameters $|\eta|$  and $r$ are $[0.0241\pi,0.0346\pi]$ and $[0.945,0.954]$, respectively. Any values of $\theta_{13}$ and $\theta_{23}$ in their $3\sigma$ can be obtained. The two CP phases are predicted to be $\delta_{CP}\in[-\pi,-0.988\pi]\cup[-0.0120\pi,0.00241\pi]\cup[0.997\pi,\pi]$ and $|\beta|\in[0.0973\pi,0.111\pi]$. Detailed numerical analyses show that accordance with experimental data can be achieved for $x=\frac{1}{4\sqrt{3}}$ and the best fit values of mixing parameters and neutrino masses are
\begin{eqnarray}
\nonumber && \eta=-0.0308\pi, \qquad m_{a}=16.797\,\text{meV},\qquad r=0.955, \qquad \chi^2_{\text{min}}=29.075 \qquad \sin^2\theta_{13}=0.0204\,, \\
\nonumber && \sin^2\theta_{12}=0.307, \qquad \sin^2\theta_{23}= 0.5, \qquad \delta_{CP}=-0.00289\pi, \qquad \beta=-0.0973\pi, \\
&&  m_1=49.377\,\text{meV}, \qquad m_2=50.120\,\text{meV}, \qquad m_3=0\,\text{meV}, \qquad m_{ee}=48.108\,\text{meV}\,.
\end{eqnarray}
Furthermore, the fixed column of PMNS matrix is $\frac{1}{7}\left(1,2\sqrt{6},2\sqrt{6}\right)^T$ in the case of $x=\frac{1}{4\sqrt{3}}$.

\item[~~($\mathcal{I}_{11}$)]{$(G_{l},G_{\text{atm}},G_{\text{sol}})=(Z_4^{TSU},Z_2^S,Z_2^{T^2U})$, $X_{\text{atm}}=\{1,S,TST^2U,T^2STU\}$}

$\bullet$ $X_{\text{sol}}=\{U,T^2\}$

In this combination of the flavor symmetries, both of the residual CP transformations $X_{\text{sol}}=\{U,T^2\}$ and $X_{\text{sol}}=\{ST^2S,TST^2U\}$ are viable. For the former residual CP symmetry, the invariant vacuum of $\phi_{\text{sol}}$ takes
\begin{equation}
\langle\phi_{\text{sol}}\rangle=v_{\phi_s}\left(1,x\omega^2, x\omega \right)^T\,.
\end{equation}
Then the most general form of the neutrino mass matrix is determined to be
\begin{equation}
 m_{\nu}=m_{a}\begin{pmatrix}
 1 &~ 1 &~ 1 \\
 1 &~ 1 &~ 1 \\
 1 &~ 1 &~ 1 \\
\end{pmatrix}+m_{s}e^{i\eta}
\begin{pmatrix}
 1 &~ x \omega  &~x \omega ^2 \\
 x \omega  &~ x^2 \omega ^2 &~ x^2 \\
 x \omega ^2 &~ x^2 &~x^2 \omega  \\
\end{pmatrix}\,,
\end{equation}
 Firstly, we perform a unitary $U_{\nu1}$ to $m_{\nu}$, where the unitary matrix $U_{\nu1}$ takes the following form
 \begin{equation}
U_{\nu1}=\begin{pmatrix}
 \frac{1}{\sqrt{3}} &~ -\frac{2 +x}{\sqrt{3(2+2 x+5 x^2)}} &~ \frac{i \sqrt{3} x}{\sqrt{2+2 x+5 x^2}} \\
 \frac{1}{\sqrt{3}} &~ \frac{1-\left(3 \omega ^2+1\right) x}{\sqrt{3(2+2 x+5 x^2)}} &~ \frac{\omega ^2 x-1}{\sqrt{2+2 x+5 x^2}} \\
 \frac{1}{\sqrt{3}} &~ \frac{1-\left(\omega +i \sqrt{3}\right) x}{ \sqrt{3(2+2 x+5 x^2)}} &~ \frac{1-\omega x}{\sqrt{2+2 x+5 x^2}} \\
\end{pmatrix}\,,
\end{equation}
 Then the neutrino mass matrix $m_{\nu}$ is a  block diagonal matrix and the parameters $y$, $z$ and $w$ which are introduced in Eq.~\eqref{eq:mnup} are
 \begin{equation}
y=3 m_{a}+\frac{(1-x)^2}{3}   m_{s}e^{i \eta }, \quad
z=\frac{x-1}{3}  \sqrt{2 +2 x+5 x^2} m_{s}e^{i \eta },  \quad
w=\frac{2+2 x+5 x^2}{3} m_{s}e^{i \eta }\,.
\end{equation}
The neutrino mass matrix $m^\prime_{\nu}$ can be diagonalized by unitary matrix $U_{\nu2}$ which show in Eq.~\eqref{eq:Unu2}. Then the lepton mixing matrix takes the following form
\begin{equation}\label{eq:PMNS_IO11_1}
 U
=\frac{1}{\sqrt{2}}\begin{pmatrix}
\frac{i \sqrt{2}  (x-1) e^{-i \psi }\sin \theta }{\sqrt{5 x^2+2 x+2}} &~ \frac{i \sqrt{2} (x-1) \cos\theta }{\sqrt{5 x^2+2 x+2}} &~ \frac{\sqrt{2} (2 x+1)}{\sqrt{5 x^2+2 x+2}} \\
 \cos \theta -\frac{i  (2 x+1)e^{-i \psi } \sin \theta }{\sqrt{5 x^2+2 x+2}} &~ -\frac{i (2 x+1) \cos \theta }{\sqrt{5 x^2+2 x+2}}-e^{i \psi } \sin \theta  &~ \frac{x-1}{\sqrt{5 x^2+2 x+2}} \\
 \cos \theta +\frac{ i (2 x+1) e^{-i \psi }\sin \theta }{\sqrt{5 x^2+2 x+2}} &~ \frac{i (2 x+1) \cos \theta }{\sqrt{5 x^2+2 x+2}}-e^{i \psi } \sin \theta  &~ \frac{1-x}{\sqrt{5 x^2+2 x+2}} \\
\end{pmatrix}\,,
\end{equation}
The predictions for the lepton mixing angles as well as CP invariants are
\begin{eqnarray}
 \nonumber && \sin^2\theta_{13}=\frac{(1+2 x)^2}{2 +2 x+5x^2}\,, \qquad
  \sin^2\theta_{12}=\cos ^2\theta \,, \qquad
\sin^2\theta_{23}=\frac{1}{2}\,, \\
&& J_{CP}=-\frac{(1-x)^2 (1+2 x) \sin 2 \theta  \cos \psi }{4 \left(2+2 x+5 x^2\right)^{3/2}}\,, \qquad
I_{1}=-\frac{(1-x)^4 \sin ^22 \theta  \sin (\rho -\sigma )}{4 \left(2+2 x+5 x^2\right)^2} \,.
\end{eqnarray}
The atmospheric mixing angle $\theta_{23}$ is maximal and $\theta_{13}$ only depends on $x$.  Inputting the $3\sigma$ range of $\theta_{13}$, we find the parameter $x$ should vary in the interval $[-0.629, -0.618]\cup[ -0.398 , -0.390]$. Let us give a relatively simple example which is easier to present in an explicit model, i.e. example with $x=-2/5$. Then the vacuum alignment of flavon field $\phi_{\text{sol}}$ is proportional to the column vector $\left(1,-\frac{2}{5} \omega^2,-\frac{2}{5} \omega\right)^T$ and the third column of the PMNS matrix is $\left(\frac{1}{5 \sqrt{2}},\frac{7}{10},\frac{7}{10}\right)^T$. The $\chi^2$ analysis results for this example are
\begin{eqnarray}
\nonumber && \eta=-0.996\pi, \qquad m_{a}=23 .399\,\text{meV},\qquad r=2 .260, \qquad \chi^2_{\text{min}}=33 .640, \qquad \sin^2\theta_{13}=0.02\,, \\
\nonumber && \sin^2\theta_{12}=0.310, \qquad \sin^2\theta_{23}= 0.5, \qquad \delta_{CP}=-0.872\pi, \qquad \beta=0.548\pi, \\
&&  m_1=49 .377\,\text{meV}, \qquad m_2=50 .120\,\text{meV}, \qquad m_3=0\,\text{meV}, \qquad m_{ee}=34 .550\,\text{meV}\,.
\end{eqnarray}
Furthermore, we think it is necessary to give the predictions for three mixing angles and two CP phases. We obtain the viable ranges of mixing angles and CP phases by scanning the input parameters $x$, $r$ and $\eta$ in their ranges. Then we find that mixing angles $\theta_{13}$ and $\theta_{12}$ can take any values in their $3\sigma$ ranges, the absolute values of the two CP phases are restricted in $|\delta_{CP}|\in[0.0956\pi,0.161\pi]\cup[0.839\pi,0.904\pi]$ and $|\beta|\in[0.526\pi,0.574\pi]$. Moreover, in order to above viable ranges of mixing parameters and the mass ration $m^2_1/m^2_2$, the input parameters $|\eta|$ and $r$ should take values in the range of $[0.9951\pi,0.9966\pi]$ and $[1.661,1.691]\cup[2.274,2.282]$, respectively.

$\bullet$ $X_{\text{sol}}=\{ST^2S,TST^2U\}$

For this combination of residual symmetries, the vacuum alignments of the flavon field $\phi_{\text{sol}}$ is
\begin{equation}
\langle\phi_{\text{sol}}\rangle=v_{\phi_s}\left(1+2 i x, \omega ^2 (1-i x),\omega  (1-i x) \right)^T\,.
\end{equation}
 Then the light neutrino mass matrix is given by
\begin{equation}
 m_{\nu}=m_{a}\begin{pmatrix}
 1 &~ 1 &~ 1 \\
 1 &~ 1 &~ 1 \\
 1 &~ 1 &~ 1 \\
\end{pmatrix}+m_{s}e^{i\eta}
\begin{pmatrix}
 (1+2i x )^2 &~ \left(1+ix+2 x^2\right) \omega  &~ \left(1+ix+2 x^2\right) \omega ^2 \\
 \left(1+ix+2 x^2\right) \omega  &~ (1-i x)^2 \omega ^2 &~ (1-i x)^2 \\
 \left(1+ix+2x^2\right) \omega ^2 &~ (1-i x)^2 &~ (1-i x)^2 \omega  \\
\end{pmatrix}\,,
\end{equation}
In order to diagonalize this mass matrix, we first perform a unitary $U_{\nu1}$ to it, where the unitary matrix $U_{\nu1}$ is
\begin{equation}
U_{\nu1}=\frac{1}{\sqrt{3}}\begin{pmatrix}
 1 &~ \frac{i x-1}{\sqrt{x^2+1}} &~ \frac{i+x}{\sqrt{x^2+1}} \\
 1 &~ \frac{i x \omega -\omega ^2}{\sqrt{x^2+1}} &~ \frac{x \omega ^2+i \omega }{\sqrt{x^2+1}} \\
 1 &~ \frac{i x \omega ^2-\omega }{\sqrt{x^2+1}} &~ \frac{i \omega ^2+x \omega }{\sqrt{x^2+1}} \\
\end{pmatrix}\,.
\end{equation}
Then the light neutrino mass matrix $m^\prime_{\nu}$ is a block diagonal matrix with the nonzero parameters $y$, $z$ and $w$:
\begin{equation}
y=3 m_{a}-3 x^2  m_{s}e^{i \eta }, \qquad
z=-3 i x   \sqrt{1+x^2}m_{s}e^{i \eta },  \qquad
w=3  \left(1+x^2\right) m_{s}e^{i \eta }\,.
\end{equation}
Following the procedures presented in section~\ref{sec:introduction}, we know that the neutrino mass matrix $m^\prime_{\nu}$ can be diagonalized by the unitary matrix $U_{\nu2}$. Then the PMNS matrix is given by
\begin{equation}\label{eq:PMNS_IO11_2}
\hskip-0.1in  U=\frac{1}{\sqrt{2}}\begin{pmatrix}
 -\frac{\sqrt{2}x e^{-i \psi }  \sin \theta }{\sqrt{x^2+1}} &~ \frac{\sqrt{2} x \cos \theta }{\sqrt{x^2+1}} &~ \frac{\sqrt{2}}{\sqrt{x^2+1}} \\
 i \cos \theta +\frac{e^{-i \psi } \sin \theta }{\sqrt{x^2+1}} &~ i e^{i \psi } \sin \theta -\frac{\cos \theta }{\sqrt{x^2+1}} &~ \frac{x}{\sqrt{x^2+1}} \\
 -i \cos \theta+\frac{e^{-i \psi } \sin \theta }{\sqrt{x^2+1}}  &~ -i e^{i \psi } \sin \theta-\frac{\cos \theta }{\sqrt{x^2+1}}  &~ \frac{x}{\sqrt{x^2+1}} \\
\end{pmatrix}\,,
\end{equation}
One can straightforwardly extract the lepton mixing angles and CP phases as follows
\begin{eqnarray}
\nonumber &&  \sin^2\theta_{13}=\frac{1}{1+x^2}, \qquad
  \sin^2\theta_{12}=\cos ^2\theta , \qquad
\sin^2\theta_{23}=\frac{1}{2}\,, \\
&& J_{CP}=-\frac{x^2 \sin 2 \theta  \cos \psi }{4 \left(1+x^2\right)^{3/2}}, \qquad
I_{1}=-\frac{x^4 \sin ^22\theta   \sin (\rho -\sigma )}{4\left(1+x^2\right)^2}\,.
\end{eqnarray}
$\theta_{23}$ is predicted to be maximal and $\theta_{13}$ only depends on input parameter $x$ which comes from the general VEV invariant under the action of the residual symmetry $Z_2^{T^2U}\times H^{\text{sol}}_{CP}$ with $\mathbf{3^\prime}$ representation. When we take into account the current $3\sigma$ bounds of $\theta_{13}$, we find that parameter $|x|$ is constrained to be in the range of $[6.293 , 6.882]$.

In order to give the predictions for mixing parameters,  we could focus on the admissible values of $x$, $r$ and $\eta$ in their ranges given $\mathcal{I}_1$. The admissible ranges of $x$, $r$ and $\eta$ can be obtained  from the requirement that the three mixing angles and mass ratio $m^2_1/m^2_2$ in their experimentally preferred $3\sigma$ ranges, i.e. $|x|\in[6.293 , 6.882]$, $|\eta|\in[0.0034\pi,0.0049\pi]$ and $r\in[0.0105,0.0120]$. Then the possible ranges of the absolute values of the two CP phases $\delta_{CP}$ and $\beta$ are $[0.598\pi,0.661\pi]$ and $[0.527\pi,0.574\pi]$, respectively. This mixing pattern gives not predictions for mixing angles $\theta_{12}$ and $\theta_{13}$. Here we shall give an example ($x=4\sqrt{3}$) to show how well the lepton mixing angles can be described by this mixing pattern and the predictions for CP phases. In this example, the fixed column which only depends on the VEVs of $\phi_{\text{atm}}$ and $\phi_{\text{sol}}$ is determined to be $\frac{1}{7}\left(1,2 \sqrt{6},2 \sqrt{6}\right)^T$. Then the best fit point and the predictions for  various observable quantities obtained at the best fit point are
\begin{eqnarray}
\nonumber && \eta=0.00408\pi, \qquad m_{a}=23 .396\,\text{meV},\qquad r=0.0103, \qquad \chi^2_{\text{min}}=29 .075, \qquad \sin^2\theta_{13}=0.0204\,, \\
\nonumber && \sin^2\theta_{12}=0.310, \qquad \sin^2\theta_{23}= 0.5, \qquad \delta_{CP}=-0.372\pi, \qquad \beta=0.548\pi, \\
&&  m_1=49 .377\,\text{meV}, \qquad m_2=50 .120\,\text{meV}, \qquad m_3=0\,\text{meV}, \qquad m_{ee}=34 .539\,\text{meV}\,.
\end{eqnarray}

\item[~~($\mathcal{I}_{12}$)]{$(G_{l},G_{\text{atm}},G_{\text{sol}})=(Z_4^{TSU},Z_2^U,Z_2^{TU})$, $X_{\text{atm}}=\{1,U\}$, $X_{\text{sol}}=\{U,T\}$ }

For this combination of residual flavor symmetries, the admissible residual CP symmetry in the solar neutrino sector is $X_{\text{sol}}=\{U,T\}$. From the most general invariant vacuum alignments of $\phi_{\text{atm}}$ and $\phi_{\text{sol}}$, one can obtained the most general light neutrino mass matrix with the form
\begin{equation}
 m_{\nu}=m_{a}\begin{pmatrix}
 0 &~ 0 &~ 0 \\
 0 &~ 1 &~ -1 \\
 0 &~ -1 &~ 1 \\
\end{pmatrix}+m_{s}e^{i\eta}
\begin{pmatrix}
 1 &~ x \omega ^2 &~ x \omega  \\
 x \omega ^2 &~ x^2 \omega  &~ x^2 \\
 x \omega  &~ x^2 &~ x^2 \omega ^2 \\
\end{pmatrix}\,,
\end{equation}
In order to diagonalise the neutrino mass $m_{\nu}$, we first apply the unitary transformation $U_{\nu1}$ to yield $m^\prime_{\nu}=U^T_{\nu1}m_{\nu}U_{\nu1}$ being a block diagonal matrix, where the unitary matrix $U_{\nu1}$ is
\begin{equation}
U_{\nu1}=\begin{pmatrix}
 -\frac{i \sqrt{3}}{\sqrt{2 x^2+2 x+5}} &~ \frac{x+2}{\sqrt{\left(x^2+2\right) \left(2 x^2+2 x+5\right)}} &~ -\frac{x}{\sqrt{x^2+2}} \\
 \frac{\omega  x-1}{\sqrt{2 x^2+2 x+5}} &~ \frac{i \sqrt{3}+x (\omega  x-1)}{\sqrt{\left(x^2+2\right) \left(2 x^2+2 x+5\right)}} &~ -\frac{1}{\sqrt{x^2+2}} \\
 \frac{1-\omega ^2 x}{\sqrt{2 x^2+2 x+5}} &~ \frac{-i \sqrt{3}-x \left(1-\omega ^2 x\right)}{\sqrt{\left(x^2+2\right) \left(2 x^2+2 x+5\right)}} &~ -\frac{1}{\sqrt{x^2+2}} \\
\end{pmatrix}\,,
\end{equation}
The parameters $y$, $z$ and $w$ of $m^\prime_{\nu}$ is given by
\begin{eqnarray}
\nonumber &&y=\frac{(x+2)^2m_{a} -3 (x-1)^2m_{s}e^{i \eta }  }{2 x^2+2 x+5}\,,  \\
\nonumber && z=\frac{i  \sqrt{3(x^2+2)} \left(-(x+2)m_{a} +(x-1) (2 x+1) m_{s}e^{i \eta } \right)}{2 x^2+2 x+5}\,, \\
&& w=\frac{\left(x^2+2\right) \left(-3 m_{a}+(2 x+1)^2m_{s}e^{i \eta }  \right)}{2 x^2+2 x+5}\,.
\end{eqnarray}
The block diagonal neutrino mass matrix $m^\prime_{\nu}$ can be diagonalized by the unitary transformation matrix $U_{\nu2}$. Then the following PMNS matrix can be obtained
\begin{equation}\label{eq:PMNS_I12}
\hskip-0.1in  U
=\frac{1}{\sqrt{6}}\begin{pmatrix}
 - \sqrt{\frac{2(2 x^2+2 x+5)}{x^2+2}} e^{-i \psi }\sin \theta  &~ \sqrt{\frac{2(2 x^2+2 x+5)}{x^2+2}} \cos \theta  &~ -\frac{\sqrt{2} (x-1)}{\sqrt{x^2+2}} \\
 \sqrt{3} \cos \theta -\frac{ (x-1) e^{-i \psi }\sin \theta }{\sqrt{x^2+2}} &~ \frac{(x-1) \cos \theta }{\sqrt{x^2+2}}+\sqrt{3}e^{i \psi } \sin \theta   &~ \sqrt{\frac{2 x^2+2 x+5}{x^2+2}} \\
 -\sqrt{3}\cos \theta-\frac{ (x-1) e^{-i \psi }\sin \theta }{\sqrt{x^2+2}}   &~ \frac{(x-1) \cos \theta }{\sqrt{x^2+2}}-\sqrt{3} e^{i \psi } \sin \theta  &~ \sqrt{\frac{2 x^2+2 x+5}{x^2+2}} \\
\end{pmatrix}\,,
\end{equation}
The lepton mixing angles and CP phases can be read off as
\begin{eqnarray}
\nonumber &&  \sin^2\theta_{13}=\frac{(1-x)^2}{3 \left(2+x^2\right)}\,,\qquad
 \sin^2\theta_{12}=\cos^2\theta\,,\qquad
\sin^2\theta_{23}=\frac{1}{2}\,, \\
&& J_{CP}=\frac{\left(2 x^3+3 x-5\right) \sin 2\theta   \sin \psi }{12 \sqrt{3} \left(x^2+2\right)^{3/2}}\,, \qquad
 I_{1}=-\frac{(2 x^2+2 x+5)^2 \sin ^22 \theta  \sin (\rho -\sigma )}{36 \left(x^2+2\right)^2}\,.
\end{eqnarray}
We see that this breaking pattern predicts a maximal $\theta_{23}$. The reactor mixing angle $\theta_{13}$ only depends on input parameter $x$,  while the solar mixing angle $\theta_{12}$ and two CP phases are depend on input parameters $x$, $\eta$ and $r$. Varying the mixing angle $\theta_{13}$ over its $3\sigma$ range~\cite{Esteban:2018azc}, we obtain the allowed region of $x$, i.e. $x\in[0.584,0.616]\cup[1.516,1.575]$. As an example, we take $x=\frac{3}{5}$. Then the VEV alignment of flavon $\phi_{\text{sol}}$ is proportional to $\left(1,\frac{3}{5} \omega,\frac{3}{5}\omega^2\right)^T$ and the third column of PMNS matrix is fixed to be $\left(\frac{2}{\sqrt{177}},\sqrt{\frac{173}{354}},\sqrt{\frac{173}{354}}\right)^T$. The $\chi^2$ results are given by
\begin{eqnarray}
\nonumber && \eta=0.00408\pi, \qquad m_{a}=30 .164\,\text{meV},\qquad r=1 .153, \qquad \chi^2_{\text{min}}=17 .644, \qquad \sin^2\theta_{13}=0.0226\,, \\
\nonumber && \sin^2\theta_{12}=0.310, \qquad \sin^2\theta_{23}= 0.5, \qquad \delta_{CP}=-0.228\pi, \qquad \beta=-0.729\pi, \\
&&  m_1=49 .377\,\text{meV}, \qquad m_2=50 .120\,\text{meV}, \qquad m_3=0\,\text{meV}, \qquad m_{ee}=25.931\,\text{meV}\,.
\end{eqnarray}

If the three input parameters $x$, $\eta$ and $r$ vary in their ranges. Requiring the three mixing angles are in the experimentally preferred $3\sigma$ ranges,  the allowed regions of $|\eta|$ and $r$ are  $[0.0027\pi,0.0065\pi]$ and $[0.335,0.354]\cup[1.127,1.181]$, respectively. Any values of $\theta_{13}$ and $\theta_{12}$ allowed $3\sigma$ ranges can be taken in this mixing pattern. The two CP phases are determined to take values in the intervals
\begin{eqnarray}
\nonumber && \delta_{CP}\in[-0.835,-0.695]\cup[-0.305,-0.165]\cup[0.0871,0.140]\cup[0.860,0.913]\,, \\
&& |\beta|\in[0.465,0.507]\cup[0.680,0.800]\,.
\end{eqnarray}

\item[~~($\mathcal{I}_{13}$)]{$(G_{l},G_{\text{atm}},G_{\text{sol}})=(Z_4^{TSU},Z_2^{TU},Z_2^U)$, $X_{\text{atm}}=\{U,T\}$, $X_{\text{sol}}=\{1,U\}$}

For this breaking pattern, the most general neutrino mass matrix is
\begin{equation}
 m_{\nu}=m_{a}\begin{pmatrix}
 0 &~ 0 &~ 0 \\
 0 &~ \omega  &~ -1 \\
 0 &~ -1 &~ \omega ^2 \\
\end{pmatrix}+m_{s}e^{i\eta}
\begin{pmatrix}
 1 &~ x &~ x \\
 x &~ x^2 &~ x^2 \\
 x &~ x^2 &~ x^2 \\
\end{pmatrix}\,,
\end{equation}
A unitary transformation $U_{\nu1}$ is performed on the light neutrino fields, then the neutrino mass matrix $m_{\nu}$ becomes $m^\prime_{\nu}=U^T_{\nu1}m_{\nu}U_{\nu1}$ and $m^\prime_{\nu}$ is a block diagonal form, where the unitary matrix $U_{\nu1}$ takes the following form
\begin{equation}
U_{\nu1}=\begin{pmatrix}
 0 &~ -\frac{\sqrt{2} }{\sqrt{2 +x^2}} &~ \frac{x}{\sqrt{2 +x^2}} \\
 \frac{\omega }{\sqrt{2}} &~ \frac{\omega  x}{ \sqrt{2(2 +x^2)}} &~ \frac{\omega }{\sqrt{2 +x^2}} \\
 -\frac{\omega ^2}{\sqrt{2}} &~ \frac{\omega ^2 x}{ \sqrt{2(2 +x^2)}} &~ \frac{\omega ^2}{\sqrt{2 +x^2}} \\
\end{pmatrix}\,,
\end{equation}
The nonzero parameters $y$, $z$ and $w$ in $m^\prime_{\nu}$ are
\begin{equation}
y=2 m_{a}-\frac{3}{2}x^2  m_{s}e^{i \eta }, \quad
z=-\frac{i}{2}   x   \sqrt{3(2+x^2)}m_{s}e^{i \eta },  \quad
w=\frac{1}{2}  \left(2+x^2\right) m_{s}e^{i \eta }\,.
\end{equation}
The general form of the diagonalization matrix of the neutrino mass matrix $m^\prime_{\nu}$ is the form of $U_{\nu2}$ shown in Eq.~\eqref{eq:Unu2} for IO case. Using the diagonalization matrix of the charged lepton mass matrix in Eq.~\ref{eq:ch_dia_matrix}, we find that the lepton mixing matrix is fixed to be
\begin{equation}\label{eq:PMNS_I13}
\hskip-0.01in  U
=\begin{pmatrix}
 \frac{\sqrt{\frac{2}{3}} (1-x) e^{-i \psi } \sin \theta }{\sqrt{2+x^2}} &~ \frac{\sqrt{\frac{2}{3}} (1-x) \cos \theta }{\sqrt{2+x^2}} &~ \frac{2+x}{ \sqrt{3(2+x^2)}} \\
 \frac{(2+x) e^{-i \psi } \sin \theta }{ \sqrt{6(2+x^2)}}-\frac{\cos \theta }{\sqrt{2}} &~ \frac{(2+x) \cos \theta }{\sqrt{6(2+x^2)}}+\frac{e^{i \psi } \sin \theta }{\sqrt{2}} &~ \frac{x-1}{\sqrt{3(2+x^2)}} \\
 -\frac{(2+x) e^{-i \psi } \sin \theta }{\sqrt{6(2+x^2)}}-\frac{\cos \theta }{\sqrt{2}} &~ -\frac{(2+x) \cos \theta }{\sqrt{6(2+x^2)}}+\frac{e^{i \psi } \sin \theta }{\sqrt{2}} &~ \frac{1-x}{\sqrt{3(2+x^2)}} \\
\end{pmatrix}\,,
\end{equation}
The lepton mixing parameters are predicted to be
\begin{eqnarray}
\nonumber &&  \sin^2\theta_{13}=\frac{(2+x)^2}{3 \left(2 +x^2\right)}, \qquad
  \sin^2\theta_{12}=\cos ^2\theta , \qquad
\sin^2\theta_{23}=\frac{1}{2}\,, \\
&& J_{CP}=-\frac{(1-x)^2 (2+x) \sin 2 \theta  \sin \psi }{6 \sqrt{3} \left(2+x^2\right)^{3/2}}, \qquad
I_{1}=-\frac{(1-x)^4 \sin ^22 \theta  \sin (\rho -\sigma )}{9 \left(2+x^2\right)^2}\,.
\end{eqnarray}
It predicts a maximal atmospheric mixing angles and the experimentally allowed $3\sigma$ range of $\theta_{13}$ requires input parameter $x$ in the range of $[-2.870,-2.776]\cup[-1.489,-1.450]$. For certain value of $x=- \frac{14}{5}$, accordance with experimental data can be achieved, and the corresponding $\chi^2$ results are
\begin{eqnarray}
\nonumber && \eta=0 .00223\pi, \qquad m_{a}=45.926\,\text{meV},\qquad r=0 .119, \qquad \chi^2_{\text{min}}=19 .755, \qquad \sin^2\theta_{13}=0.0217\,, \\
\nonumber && \sin^2\theta_{12}=0.310, \qquad \sin^2\theta_{23}= 0.5, \qquad \delta_{CP}=-0.787\pi, \qquad \beta=0 .721\pi, \\
&&  m_1=49 .337\,\text{meV}, \qquad m_2=50 .120\,\text{meV}, \qquad m_3=0\,\text{meV}, \qquad m_{ee}=26 .358\,\text{meV}\,.
\end{eqnarray}
In the case $x=- \frac{14}{5}$, the third of PMNS matrix is predicted to be $\frac{1}{3 \sqrt{82}}\left(4,19,19\right)^T$. If we require the three mixing angles and mass ratio $m^2_1/m^2_2$ lie in their $3\sigma$ regions ~\cite{Esteban:2018azc}. We find all the other two input parameters $|\eta|$  and $r$ are found to lie in rather narrow regions  $[0.0017\pi,0.0037\pi]$ and $[0.114,0.121]\cup[0.367,0.381]$, respectively. Then the two CP phases are predicted to be in the ranges of $\delta_{CP}\in[-0.842\pi,-0.725\pi]\cup[-0.275\pi,-0.158\pi]\cup[0.124\pi,0.211\pi]\cup[0.789\pi,0.876\pi]$ and $|\beta|\in[0.608\pi,0.674\pi]\cup[0.683\pi,0.772\pi]$, respectively.

\item[~~($\mathcal{I}_{14}$)]{$(G_{l},G_{\text{atm}},G_{\text{sol}})=(K_4^{(S,TST^2)},Z_3^T,Z_2^{SU})$, $X_{\text{atm}}=\{1,T,T^2\}$}

$\bullet$ $X_{\text{sol}}=\{1,SU\}$

In this combination of residual flavor symmetries, both the residual CP transformations $X_{\text{sol}}=\{1,SU\}$ and $X_{\text{sol}}=\{S,U\}$ are viable. These two breaking patterns are discussed in the case of $\mathcal{N}_7$ of NO case. Then the neutrino mass matrix and the lepton mixing matrix can be easy obtain from the results in the case of $\mathcal{N}_7$. For $X_{\text{sol}}=\{1,SU\}$, the lepton mixing matrix can takes the following form
\begin{equation}\label{eq:PMNS_I14_1}
\hskip-0.1in U=
\begin{pmatrix}
 -\sqrt{\frac{x^2-2 x+4}{3(x^2-2 x+2)}} \cos \theta  &~ - \sqrt{\frac{x^2-2 x+4}{3(x^2-2 x+2)}}e^{i \psi } \sin \theta &~  \frac{\sqrt{2} (x-1)}{\sqrt{3(x^2-2 x+2)}}  \\
  \frac{(x-1) \cos \theta }{ \sqrt{3(x^2-2 x+2)}}-\frac{e^{-i \psi } \sin \theta }{\sqrt{2}} &~\frac{ (x-1)e^{i \psi } \sin \theta }{\sqrt{3(x^2-2 x+2)}}+ \frac{\cos \theta }{\sqrt{2}} &~ \sqrt{\frac{x^2-2 x+4}{6(x^2-2 x+2)}}\\
  \frac{(x-1) \cos \theta }{ \sqrt{3(x^2-2 x+2)}}+\frac{e^{-i \psi } \sin \theta }{\sqrt{2}} &~ \frac{ (x-1)e^{i \psi } \sin \theta }{ \sqrt{3(x^2-2 x+2)}}-\frac{\cos \theta }{\sqrt{2}} &~ \sqrt{\frac{x^2-2 x+4}{6(x^2-2 x+2)}} \\
\end{pmatrix}\,,
\end{equation}
Here the expressions of $\theta$, $\psi$, $\rho$ and $\sigma$ which  with respect to free parameter $x$, $\eta$ and $r$ are the same as them in Eq.~\eqref{eq:PMNS_N7_1}. Furthermore, the expressions of the neutrino masses $m_{1}$ and $m_{2}$ are the same as the expressions of $m_2$ and $m_{3}$ in the case of $\mathcal{N}_7$, respectively. For the mixing matrix in Eq.~\eqref{eq:PMNS_I14_1}, we can obtain the following expressions for the mixing angles and the CP
invariants
\begin{eqnarray}
\nonumber \hskip-0.4in &&  \sin^2\theta_{13}=\frac{2 (x-1)^2}{3 \left(x^2-2 x+2\right)}\,,\qquad
  \sin^2\theta_{12}=\sin^2\theta\,,\qquad
\sin^2\theta_{23}=\frac{1}{2}\,, \\
\hskip-0.4in  && J_{CP}=\frac{(1-x) (x^2-2 x+4) \sin 2 \theta  \sin \psi }{6 \sqrt{6} (x^2-2 x+2)^{3/2}}\,, \qquad
I_{1}=-\frac{(x^2-2 x+4)^2 \sin ^22 \theta  \sin (\rho -\sigma )}{36 (x^2-2 x+2)^2}\,.
\end{eqnarray}
We find that the atmospheric mixing angle $\theta_{23}$ is $45^\circ$. The experimentally allowed region of $x$ depends of the $3\sigma$ range of $\theta_{13}$. We find the viable range of $x$ is $[0.804,0.821]\cup[1.179,1.196]$. Here we shall give the predictions for $x=\frac{4}{5}$. Firstly, it predicts the third column of PMNS matrix being $\frac{1}{ \sqrt{39}}\left(1,\sqrt{19},\sqrt{19}\right)^T$. Secondly, the results of the $\chi^2$ analysis are
\begin{eqnarray}
\nonumber &&  \eta=-0.994\pi, \qquad m_{a}=60 .715\,\text{meV},\qquad r=0 .323, \qquad \chi^2_{\text{min}}=38 .315, \qquad \sin^2\theta_{13}=0 .0256\,, \\
\nonumber && \sin^2\theta_{12}=0.310, \qquad \sin^2\theta_{23}= 0.5, \qquad \delta_{CP}=0.895\pi, \qquad \beta=-0.451\pi, \\
&&  m_1=49 .377\,\text{meV}, \qquad m_2=50 .120\,\text{meV}, \qquad m_3=0\,\text{meV}, \qquad m_{ee}=38 .556\,\text{meV}\,.
\end{eqnarray}
We note that the best fit value of $\sin^2\theta_{13}$ is rather close to its $3\sigma$ upper limit $0.02463$. Hence we think that this breaking pattern with $x=\frac{4}{5}$ is a good  leading order approximation. Furthermore we perform a comprehensive numerical analysis. When three mixing angles and mass ratio $m^2_1/m^2_2$ are restricted in their $3\sigma$ ranges, the input parameters $|\eta|$  and $r$ have to in the ranges of $[0.9929\pi,0.9950\pi]$ and $[0.322,0.325]$, respectively. The limiting of the input parameters leads to $|\delta_{CP}|\in[0.0798\pi,0.128\pi]\cup[0.872\pi,0.920\pi]$ and $|\beta|\in[0.430\pi,0.469\pi]$. The mixing angles $\theta_{12}$ and $\theta_{13}$ can take any values in their $3\sigma$ ranges.

$\bullet$ $X_{\text{sol}}=\{S,U\}$

For the residual CP transformation $X_{\text{sol}}=\{S,U\}$, we can get the corresponding results directly from the results of $\mathcal{N}_7$. From PMNS matrix in Eq.~\eqref{eq:PMNS_N7_2}, we can read the PMNS matrix in this case
\begin{equation}\label{eq:PMNS_I14_2}
\hskip-0.1in U=
\begin{pmatrix}
   -\sqrt{\frac{x^2+3}{3(x^2+1)}} \cos \theta  &~ - \sqrt{\frac{x^2+3}{3(x^2+1)}} e^{i \psi }\sin \theta &~ \frac{\sqrt{2} x}{\sqrt{3(x^2+1)}}  \\
  \frac{x \left(x +\sqrt{3}\right) \cos \theta }{ \sqrt{3(x^2+1)( x^2+3)}}+\frac{ \left(x-\sqrt{3}\right) e^{-i \psi }\sin \theta }{ \sqrt{2(x^2+3)}} &~ \frac{\left(\sqrt{3}-x\right) \cos \theta }{ \sqrt{2(x^2+3)}}+\frac{ x \left(x +\sqrt{3}\right)e^{i \psi } \sin \theta }{\sqrt{3(x^2+1)( x^2+3)}} &~  \frac{x+\sqrt{3}}{ \sqrt{6(x^2+1)}} \\
  \frac{x \left( x-\sqrt{3}\right) \cos \theta }{\sqrt{3(x^2+1)( x^2+3)}}-\frac{ \left(x+\sqrt{3}\right) e^{-i \psi }\sin \theta }{ \sqrt{2(x^2+3)}} & ~ \frac{\left(x+\sqrt{3}\right) \cos \theta }{ \sqrt{2(x^2+3)}}+\frac{ x \left( x-\sqrt{3}\right) e^{i \psi }\sin \theta }{\sqrt{3(x^2+1)( x^2+3)}} &~ \frac{x-\sqrt{3}}{ \sqrt{6(x^2+1)}} \\
\end{pmatrix}\,,
\end{equation}
Then the predictions for the three mixing angles and two CP invariants are
\begin{eqnarray}
\nonumber &&  \sin^2\theta_{13}=\frac{2 x^2}{3 \left(1+x^2\right)}\,,\qquad
 \sin^2\theta_{12}=\sin^2\theta\,,\qquad
\sin^2\theta_{23}=\frac{1}{2}+\frac{\sqrt{3} x}{3+x^2}\,, \\
&& J_{CP}=\frac{x \left(x^2-3\right) \sin 2 \theta  \sin \psi }{6 \sqrt{6} \left(x^2+1\right)^{3/2}}\,, \qquad
I_{1}=-\frac{\left(x^2+3\right)^2 \sin ^22 \theta  \sin (\rho -\sigma )}{36 \left(x^2+1\right)^2}\,.
\end{eqnarray}
From Eq.~\eqref{eq:PMNS_I14_2}, we find that the third column of PMNS matrix only depends on the parameter $x$ which dictates the vacuum alignment of flavon $\phi_{\text{sol}}$. Since both mixing angles $\theta_{13}$ and $\theta_{23}$ depend on only one input parameter $x$. Then we can obtain the following sum rule
\begin{equation}
\sin^2\theta_{23}=\frac{1}{2}\pm\frac{\tan\theta_{13}}{2}\sqrt{2-\tan^2\theta_{13}}\simeq\frac{1}{2}\pm\frac{\sqrt{2}\tan\theta_{13}}{2}\,,
\end{equation}
where ``$+$'' sign in $\pm$ is satisfied for $x>0$ and ``$-$'' for $x<0$. Given the experimental $3\sigma$ range of $\theta_{13}$, we have $0.602\leq\sin^2\theta_{23}\leq0.612$ or $0.388\leq\sin^2\theta_{23}\leq0.398$. The later range has been removed by the $3\sigma$ range of $\theta_{23}$. The experimental data of the third column of PMNS matrix at $3\sigma$ level can be accommodated for the parameter $x$ with $x\in[0.179,0.196]$. The requirement with three mixing angles and mass ratio $m^2_1/m^2_2$ in their $3\sigma$ ranges require the other two input parameters $|\eta|$ and $r$ in the ranges of $[0.9929\pi,0.9950\pi]$ and $[0.3226,0.3248]$, respectively. Then the mixing angle $\theta_{23}$ and two CP phases are predicted to be $0.602\leq\sin^2\theta_{23}\leq0.612$, $0.580\pi\leq|\delta_{CP}|\leq0.628\pi$ and $0.431\pi\leq|\beta|\leq0.469\pi$. The other two mixing angles can take any values in their $3\sigma$ ranges.

Now let us give the numerical results of a relatively simple example with $x=\frac{1}{3\sqrt{3}}$.  In this example, the third column of the PMNS matrix is $\frac{1}{\sqrt{42}}\left(1,5,4\right)^T$ which is agrees with all measurements to date~\cite{Esteban:2018azc}. When we perform a $\chi^2$ analysis, the predictions for various observable quantities are
\begin{eqnarray}
\nonumber && \eta=-0.994\pi, \qquad m_{a}=60.743\,\text{meV},\qquad r=0.323, \qquad \chi^2_{\text{min}}=5.731, \qquad \sin^2\theta_{13}=0.0238\,, \\
\nonumber && \sin^2\theta_{12}=0.310, \qquad \sin^2\theta_{23}= 0.610, \qquad \delta_{CP}=-0.604\pi, \qquad \beta=-0.449\pi, \\
&&  m_1=49.377\,\text{meV}, \qquad m_2=50.120\,\text{meV}, \qquad m_3=0\,\text{meV}, \qquad m_{ee}=38.688\,\text{meV}\,.
\end{eqnarray}

\item[~~($\mathcal{I}_{15}$)]{$(G_{l},G_{\text{atm}},G_{\text{sol}})=(K_4^{(S,TST^2)},Z_2^U,Z_2^{TU})$, $X_{\text{atm}}=\{1,U\}$, $X_{\text{sol}}=\{U,T\}$}

For this combination of residual symmetries, the vacuum alignments of flavons $\phi_{\text{atm}}$ and $\phi_{\text{sol}}$ are given in table~\ref{tab:inv_VEV_CP}. Then the most general neutrino mass matrix is given by
\begin{equation}
 m_{\nu}=m_{a}\begin{pmatrix}
 0 &~ 0 &~ 0 \\
 0 &~ 1 &~ -1 \\
 0 &~ -1 &~ 1 \\
\end{pmatrix}+m_{s}e^{i\eta}
\begin{pmatrix}
 1 &~ x \omega ^2 &~ x \omega  \\
 x \omega ^2 &~ x^2 \omega  &~ x^2 \\
x \omega  &~ x^2 &~ x^2 \omega ^2 \\
\end{pmatrix}\,,
\end{equation}
We firstly perform the following unitary transformation to light neutrino fields
\begin{equation}
U_{\nu1}=\begin{pmatrix}
 0 &~ -\frac{\sqrt{2} }{\sqrt{2 +x^2}} &~ -\frac{x}{\sqrt{2 +x^2}} \\
 -\frac{1}{\sqrt{2}} &~ \frac{x}{\sqrt{2(2 +x^2)}} &~ -\frac{1}{\sqrt{2 +x^2}} \\
 \frac{1}{\sqrt{2}} &~ \frac{x}{ \sqrt{2(2 +x^2)}} &~ -\frac{1}{\sqrt{2 +x^2}} \\
\end{pmatrix}\,,
\end{equation}
Then the neutrino mass matrix $m^\prime_{\nu}=U^T_{\nu1}m_{\nu}U_{\nu1}$ is a block diagonal matrix with nonzero elements
\begin{equation}
y=2 m_{a}-\frac{3}{2} x^2  m_{s}e^{i \eta }, \qquad
z=-\frac{i}{2}  x  \sqrt{3(2+x^2)}m_{s}e^{i \eta },  \qquad
w=\frac{1}{2}   \left(2+x^2\right)m_{s}e^{i \eta }\,.
\end{equation}
The most general diagonalization matrix of $m^\prime_{\nu}$ is the unitary matrix $U_{\nu2}$ which is given in Eq.~\eqref{eq:Unu2}. We can obtain the PMNS matrix taking the following form
\begin{equation}\label{eq:PMNS_I15}
\hskip-0.1in  U
=\begin{pmatrix}
 -\frac{\sqrt{2} (1-x) e^{-i \psi } \sin \theta }{\sqrt{3(2+x^2)}} &~ -\frac{\sqrt{2} (1-x) \cos \theta }{\sqrt{3(2+x^2)}} &~ \frac{2+x}{ \sqrt{3(2+x^2)}} \\
 \frac{i \cos \theta}{\sqrt{2}}+\frac{(2 a+b) e^{-i \psi } \sin \theta }{ \sqrt{6(2+x^2)}} &~ \frac{(2+x) \cos \theta }{\sqrt{6(2+x^2)}}-\frac{ie^{i \psi }  \sin \theta }{\sqrt{2}} &~ \frac{1-x}{\sqrt{3(2+x^2)}} \\
 -\frac{i \cos \theta }{\sqrt{2}}+\frac{(2+x) e^{-i \psi } \sin \theta }{\sqrt{6(2+x^2)}} &~ \frac{(2 +x) \cos \theta }{\sqrt{6(2+x^2)}}+\frac{i e^{i \psi } \sin \theta }{\sqrt{2}} &~ \frac{1-x}{\sqrt{3(2+x^2)}} \\
\end{pmatrix}\,,
\end{equation}
 Then we can extract the expressions for the lepton mixing angles and CP invariants as follows
 \begin{eqnarray}
\nonumber &&  \sin^2\theta_{13}=\frac{(2+x)^2}{3 \left(2+x^2\right)}, \qquad
  \sin^2\theta_{12}=\cos ^2\theta , \qquad
\sin^2\theta_{23}=\frac{1}{2}\,, \\
&& J_{CP}=-\frac{(1-x)^2 (2+x) \sin 2 \theta  \cos \psi }{6 \sqrt{3} \left(2+x^2\right)^{3/2}}, \qquad
I_{1}=-\frac{(1-x)^4 \sin ^22 \theta  \sin (\rho -\sigma )}{9 \left(2+x^2\right)^2}\,.
\end{eqnarray}
We see that the atmospheric mixing angle is maximal. Inserting the $3\sigma$ range of $\theta_{13}$, we find the parameter $x$ should vary in the interval $[-2.870,-2.776]\cup[-1.489,-1.450]$. The mixing angles $\theta_{12}$ and the mass ratio $m^2_1/m^2_2$ depend on the three input parameters $x$, $\eta$ and $r$. Hence if we require $\theta_{12}$ and $m^2_1/m^2_2$ in their $3\sigma$ ranges, we can obtain the restrictions of $|\eta|$ and $r$ are $|\eta|\in[0.0017\pi,0.0037\pi]$ and $r\in[0.114,0.121]\cup[0.367,0.381]$. Then the allowed values of the two CP phases would generically be constrained in regions $|\delta_{CP}|\in[0.225\pi,0.342\pi]\cup[0.624\pi,0.711\pi]$ and $|\beta|\in[0.608\pi,0.673\pi]\cup[0.683\pi,0.772\pi]$. As an example easily achievable in a model,  we consider the case of $x=-3/2$. Then the vacuum alignment of flavon $\phi_{\text{sol}}$ is proportional to the column vector $\left(1,-\frac{3}{2} \omega,-\frac{3}{2}\omega^2\right)^T$, and the fixed column of PMNS matrix is $\frac{1}{\sqrt{51}}\left(1,5,5\right)^T$. The best fit values of the mixing parameters read
\begin{eqnarray}
\nonumber && \eta=-0.00301\pi, \qquad m_{a}=40 .128\,\text{meV},\qquad r=0.362, \qquad \chi^2_{\text{min}}=38 .746, \qquad \sin^2\theta_{13}=0.0196\,, \\
\nonumber && \sin^2\theta_{12}=0.310, \qquad \sin^2\theta_{23}= 0.5, \qquad \delta_{CP}=-0.668\pi, \qquad \beta=-0.640\pi, \\
&&  m_1=49.377\,\text{meV}, \qquad m_2=50 .120\,\text{meV}, \qquad m_3=0\,\text{meV}, \qquad m_{ee}=30 .231\,\text{meV}\,.
\end{eqnarray}
We see that the best fit value of $\theta_{13}$ is a bit smaller than its $3\sigma$ lower limit $0.02068$~\cite{Esteban:2018azc}. We think it is a good leading order approximation.

\item[~~($\mathcal{I}_{16}$)]{$(G_{l},G_{\text{atm}},G_{\text{sol}})=(K_4^{(S,U)},Z_2^{TST^2},Z_2^{TU})$, $X_{\text{atm}}=\{SU,ST^2S,T^2,T^2STU\}$, $X_{\text{sol}}=\{STS,T^2STU\}$}

The vacuum alignments which invariant under the residual symmetries in the atmospheric neutrino sector and the solar neutrino sector are shown in table~\ref{tab:inv_VEV_CP}.  Then the neutrino mass matrix is given by
\begin{equation}
m_{\nu}=m_{a}\begin{pmatrix}
 1 &~ \omega  &~ \omega ^2 \\
 \omega  &~ \omega ^2 &~ 1 \\
 \omega ^2 &~ 1 &~ \omega  \\
\end{pmatrix}+m_{s}e^{i\eta}
\begin{pmatrix}
 (1+2 x i)^2 &~ \left(1+x i +2 x^2\right) \omega ^2 &~ (1-i x) (1+2 x i) \omega  \\
 \left(1+x i +2 x^2\right) \omega ^2 &~ (1-i x)^2 \omega  &~ (1-i x)^2 \\
 (1-i x) (1+2 x i) \omega  &~ (1-i x)^2 &~ (1-i x)^2 \omega ^2 \\
\end{pmatrix}\,,
\end{equation}
In order to diagonalize $m_{\nu}$, we first perform a unitary transformation  $U_{\nu1}$ to $m_{\nu}$, where the unitary matrix $U_{\nu1}$ is
\begin{equation}
U_{\nu1}=\begin{pmatrix}
 \frac{1}{\sqrt{3}} &~ \frac{i x-1}{ \sqrt{3 \left(1+x^2\right)}} &~ \frac{i+x}{\sqrt{3 \left(1+x^2\right)}} \\
 \frac{\omega ^2}{\sqrt{3}} &~ \frac{  i x-\omega}{\sqrt{3 \left(1+x^2\right)}} &~ \frac{i +\omega  x}{\sqrt{3 \left(1+x^2\right)}} \\
 \frac{\omega }{\sqrt{3}} &~ \frac{ i x-\omega ^2}{\sqrt{3 \left(1+x^2\right)}} &~ \frac{i+ \omega ^2x}{\sqrt{3 \left(1+x^2\right)}} \\
\end{pmatrix}
\end{equation}
When we perform the unitary transformation $U_{\nu1}$ to $m_{\nu}$, we obtain a block diagonal $m^\prime_{\nu}$ with parameters $y$, $z$ and $w$ being
\begin{equation}
y=3 m_{a}-3 x^2 m_{s}e^{i \eta } , \qquad
z= -3 i x \sqrt{1+x^2} m_{s}e^{i \eta } ,  \qquad
w=3\left(1+x^2\right)  m_{s}e^{i \eta } \,.
\end{equation}
The diagonalization matrix of block diagonal $m^\prime_{\nu}$ can take to be $U_{\nu2}$ which is given in Eq.~\eqref{eq:Unu2}. As a consequence, the PMNS matrix can take the following form
\begin{equation}\label{eq:PMNS_I16}
\hskip-0.1in  U=\frac{1}{\sqrt{2}}\begin{pmatrix}
- \frac{\sqrt{2} x e^{-i \psi } \sin \theta }{\sqrt{1+x^2}} &~ -\frac{\sqrt{2} x \cos \theta }{\sqrt{1+x^2}} &~ \frac{\sqrt{2} }{\sqrt{1+x^2}} \\
 \cos \theta +\frac{ e^{-i \psi } \sin \theta }{\sqrt{1+x^2}} &~ \frac{ \cos \theta }{\sqrt{1+x^2}}-e^{i \psi } \sin \theta  &~ \frac{x}{\sqrt{1+x^2}} \\
 -\cos \theta +\frac{ e^{-i \psi } \sin \theta }{\sqrt{1+x^2}} &~ \frac{ \cos \theta }{\sqrt{1+x^2}}+e^{i \psi } \sin \theta  &~ \frac{x}{\sqrt{1+x^2}} \\
\end{pmatrix}\,,
\end{equation}
The lepton mixing angles and CP phases are found to be of the form
\begin{eqnarray}
\nonumber &&  \sin^2\theta_{13}=\frac{1}{1+x^2}, \qquad
  \sin^2\theta_{12}=\cos ^2\theta , \qquad
\sin^2\theta_{23}=\frac{1}{2}\,, \\
&& J_{CP}=-\frac{x^2 \sin 2 \theta  \sin \psi }{4 \left(1+x^2\right)^{3/2}}, \qquad
I_{1}=-\frac{x^4 \sin ^22\theta  \sin (\rho -\sigma )}{4\left(1+x^2\right)^2}\,.
\end{eqnarray}
We see that $\theta_{23}$ is maximal. In order to obtain viable $\theta_{13}$, the absolute value of the input parameter $x$ must lie in the range of $[6.293,6.882]$. Freely varying the three mixing angles and the mass ratio $m^2_1/m^2_2$ in their $3\sigma$ ranges, we find that other input parameters $|\eta|$ and $r$ are limited in the range $[0.0033\pi,0.0049\pi]$ and $[0.0104,0.0124]$, respectively.  Moreover, we find the values of CP phases $|\delta_{CP}|$ and $|\beta|$ are in the intervals $[0.839\pi,0.905\pi]\cup[0.0955\pi,0.162\pi]$ and $[0.526\pi,0.574\pi]$, respectively.  Any value in the $3\sigma$ range of $\theta_{12}$ can be taken in this mixing pattern. For model building convenience, we shall give the analysis for $x=-7$. In this example, the third column of PMNS matrix is determined to $\left(\frac{1}{5 \sqrt{2}},\frac{7}{10},\frac{7}{10}\right)^T$. The best fit values for all parameters are given by
\begin{eqnarray}
\nonumber && \eta=0.00408\pi, \qquad m_{a}=23 .399\,\text{meV},\qquad r=0.0100, \qquad \chi^2_{\text{min}}=33 .640, \qquad \sin^2\theta_{13}=0.02\,, \\
\nonumber && \sin^2\theta_{12}=0.310, \qquad \sin^2\theta_{23}= 0.5, \qquad \delta_{CP}=-0.872\pi, \qquad \beta=0.548\pi, \\
&&  m_1=49 .377\,\text{meV}, \qquad m_2=50 .120\,\text{meV}, \qquad m_3=0\,\text{meV}, \qquad m_{ee}=34 .550\,\text{meV}\,.
\end{eqnarray}

\item[~~($\mathcal{I}_{17}$)]{$(G_{l},G_{\text{atm}},G_{\text{sol}})=(K_4^{(S,U)},Z_2^{TU},Z_2^{U})$,  $X_{\text{atm}}=\{U,T\}$, $X_{\text{sol}}=\{S,SU\}$}

For this kind of combination of residual symmetries, the light neutrino mass matrix is
\begin{equation}
m_{\nu}=m_{a}\begin{pmatrix}
 0 &~ 0 &~ 0 \\
 0 &~ \omega  &~ -1 \\
 0 &~ -1 &~ \omega ^2 \\
\end{pmatrix}+m_{s}e^{i\eta}
\begin{pmatrix}
 (1+2 x i)^2 &~ (1-i x) (1+2 x i) &~ (1-i x) (1+2 x i) \\
 (1-i x) (1+2 x i) &~ (1-i x)^2 &~ (1-i x)^2 \\
 (1-i x) (1+2 x i) &~ (1-i x)^2 &~ (1-i x)^2 \\
\end{pmatrix}\,,
\end{equation}
This neutrino mass matrix can be diagonalized to a block diagonal matrix when we perform a unitary transformation $U_{\nu1}$, where the unitary matrix $U_{\nu1}$ is
\begin{equation}
U_{\nu1}=\begin{pmatrix}
 \frac{2 i x^2+x+i}{\sqrt{3 \left(2 x^2+1\right) \left(3 x^2+1\right)}} &~ -\frac{4 x^2+1}{\sqrt{3 \left(2 x^2+1\right) \left(4 x^2+1\right)}} &~ \frac{1-i x}{\sqrt{3 \left(3 x^2+1\right)}} \\
 \frac{i \left(1-2 \omega ^2\right) x^2+\omega  x+i}{\sqrt{3 \left(2 x^2+1\right) \left(3 x^2+1\right)}} &~ \frac{2 x^2-\sqrt{3} x-\omega ^2}{\sqrt{3 \left(2 x^2+1\right) \left(4 x^2+1\right)}} &~ \frac{\omega  (2 i x+1)}{\sqrt{3 \left(3 x^2+1\right)}} \\
 \frac{i (1-2 \omega ) x^2+\omega ^2 x+i}{\sqrt{3 \left(2 x^2+1\right) \left(3 x^2+1\right)}} &~ \frac{2 x^2+\sqrt{3} x-\omega }{\sqrt{3 \left(2 x^2+1\right) \left(4 x^2+1\right)}} &~ \frac{\omega ^2 (2 i x+1)}{\sqrt{3 \left(3 x^2+1\right)}} \\
\end{pmatrix}
\end{equation}
Then we can obtain the three parameters $y$, $z$ and $w$ in $m^\prime_{\nu}$ are
\begin{eqnarray}
\nonumber &&y=\frac{\left(3 x^2+1\right) \left(m_{a}-3 m_{s} e^{i \eta }\right)}{2 x^2+1}\,,  \\
\nonumber && z=\frac{ \left(3 x^2+1\right) \left((-2ix^2+x-i)m_{a}+3x(4x^2+1)m_{s} e^{i \eta }  \right)}{\left(2 x^2+1\right) \sqrt{(3x^2+1)(4x^2+1)}}\,, \\
\label{eq:yzw_I17}&& w=-\frac{ \left((2x^2+ix+1)^2m_{a} +3x^2(4x^2+1)^2 m_{s} e^{i \eta } \right)}{(2x^2+1)(4x^2+1)}\,.
\end{eqnarray}
As a result, $m^\prime_{\nu}$ can be diagonalized by the unitary matrix $U_{\nu2}$ which is given in Eq.~\eqref{eq:Unu2}. From expressions of parameters $y$, $z$ and $w$ in Eq.~\eqref{eq:yzw_I17}, the absolute values and the phases of them can be obtained. Then from Eq.~\eqref{eq:nu_masses}, we can get the neutrino masses $m^2_1$ and $m^2_2$.
Following the procedures listed in section~\ref{sec:framework}, the expressions of $\sin2\theta$, $\cos2\theta$, $\sin\psi$, $\cos\psi$, $\sin\rho$, $\cos\rho$, $\sin\sigma$ and $\cos\sigma$  can be obtained. The lepton mixing matrix read as
\begin{equation}\label{eq:PMNS_IO17}
\hskip-0.1in U
=\frac{1}{\sqrt{2}}
\begin{pmatrix}
 - \sqrt{\frac{2(2 x^2+1)}{3 x^2+1}} \cos \theta  &~ -  \sqrt{\frac{2(2 x^2+1)}{3 x^2+1}}e^{i \psi } \sin \theta  &~ \frac{\sqrt{2} x}{\sqrt{3 x^2+1}} \\
 \frac{x \cos \theta }{\sqrt{\left(2 x^2+1\right) \left(3 x^2+1\right)}}+\sqrt{\frac{4 x^2+1}{2 x^2+1}}e^{-i \psi } \sin \theta   &~ \frac{ x e^{i \psi }\sin \theta }{\sqrt{\left(2 x^2+1\right) \left(3 x^2+1\right)}}-\sqrt{\frac{4 x^2+1}{2 x^2+1}} \cos \theta  &~ \frac{1}{\sqrt{3 x^2+1}} \\
 x \sqrt{\frac{4 x^2+1}{\left(2 x^2+1\right) \left(3 x^2+1\right)}} \cos \theta -\frac{ e^{-i \psi } \sin \theta }{\sqrt{2 x^2+1}} &~ \frac{\cos \theta  }{\sqrt{2 x^2+1}}+ x  \sqrt{\frac{4 x^2+1}{\left(2 x^2+1\right) \left(3 x^2+1\right)}}e^{i \psi } \sin \theta &~ \sqrt{\frac{4 x^2+1}{3 x^2+1}} \\
\end{pmatrix}\,,
\end{equation}
It is straightforward to extract the mixing angles and the two CP rephasing invariants
\begin{eqnarray}
\nonumber && \sin^2\theta_{13}=\frac{x^2}{1+3 x^2}\,,\qquad
  \sin^2\theta_{12}=\sin^2\theta\,,\qquad
\sin^2\theta_{23}=\frac{1}{2 +4 x^2}\,, \\
&& J_{CP}=\frac{x\sqrt{1+4x^2}\sin2\theta\sin\psi}{4(1+3x^2)^{3/2}}\,, \qquad
I_{1}=-\frac{\left(2 x^2+1\right)^2 \sin ^22 \theta  \sin (\rho -\sigma )}{4 \left(3 x^2+1\right)^2}\,.
\end{eqnarray}
We see that both the atmospheric mixing angle and the reactor mixing angle only depend on one input parameter $x$ which decides the vacuum alignment of flavon $\phi_{\text{sol}}$. Then a sum rule between mixing angle $\theta_{13}$ and $\theta_{23}$ is obtained
\begin{equation}
\sin^2\theta_{23}=\frac{1}{2}-\tan^2\theta_{13}\,.
\end{equation}
This sum rule has also been obtained in Ref.~\cite{Li:2014eia}. It implies that the atmospheric mixing angle is in the first octant, i.e. $\theta_{23}<45^\circ$. For the fitted $3\sigma$ range of $\theta_{13}$, the atmospheric mixing angles is constrained to be in the interval of $0.475\leq\sin^2\theta_{23}\leq0.479$. This can be tested in future neutrino oscillation experiments. Inserting the $3\sigma$ ranges of $\theta_{13}$, we find the viable range of $|x|$ is $[0.148,0.163]$. Detailed numerical analyses show that accordance with experimental data can be achieved for certain values of $x$, $m_{a}$, $r$ and $\eta$. As an example, the fixed column of PMNS matrix is $\left(\frac{1}{4 \sqrt{3}},\frac{3\sqrt{5}}{4 \sqrt{6}},\frac{7}{4 \sqrt{6}}\right)^T$ for $x=\frac{\sqrt{5}}{15}$. The best fit values of all parameters are
\begin{eqnarray}
\nonumber && \eta=0.0969\pi, \qquad m_{a}=34 .963\,\text{meV},\qquad r=0 .633, \qquad \chi^2_{\text{min}}=35 .201, \qquad \sin^2\theta_{13}=0.0208\,, \\
\nonumber && \sin^2\theta_{12}=0.310, \qquad \sin^2\theta_{23}= 0.479, \qquad \delta_{CP}=-0.184\pi, \qquad \beta=-0.546\pi, \\
&&  m_1=49 .377\,\text{meV}, \qquad m_2=50 .120\,\text{meV}, \qquad m_3=0\,\text{meV}, \qquad m_{ee}=34 .625\,\text{meV}\,.
\end{eqnarray}
Furthermore, it is necessary to give allowed ranges of all mixing parameters. Freely varying the input parameters we find that $\sin^2\theta_{23}$ can take any value between $0.475$ and  $0.479$. The other two mixing angles are restricted in their $3\sigma$ ranges. The two CP phases are predicted to be $|\delta_{CP}|\in[0.152\pi,0.219\pi]\cup[0.881\pi,0.949\pi]$ and $|\beta|\in[0.522\pi,0.575\pi]$. The requirement of mixing angels and mass ratio in their $3\sigma$ ranges also requires input parameters $|\eta|$ and $r$ in the ranges $[0.0888\pi,0.106\pi]$ and $[0.626,0.638]$, respectively.

\item[~~($\mathcal{I}_{18}$)]{$(G_{l},G_{\text{atm}},G_{\text{sol}})=(K_4^{(S,U)},Z_2^{TU},Z_2^{T^2U})$, $X_{\text{atm}}=\{U,T\}$, $X_{\text{sol}}=\{ST^2S,TST^2U\}$}

For this kind of residual symmetries, the light neutrino mass matrix can be easily obtained
\begin{equation}
m_{\nu}=m_{a}\begin{pmatrix}
 0 &~ 0 &~ 0 \\
 0 &~ \omega  &~ -1 \\
 0 &~ -1 &~ \omega ^2 \\
\end{pmatrix}+m_{s}e^{i\eta}
\begin{pmatrix}
 (1+2 ix)^2 &~ \left(1+ix +2 x^2\right) \omega  &~ \left(1+ix +2 x^2\right) \omega ^2 \\
 \left(1+ix +2 x^2\right) \omega  &~ (1-i x)^2 \omega ^2 &~ (1-i x)^2 \\
 \left(1+ix +2 x^2\right) \omega ^2 &~ (1-i x)^2 &~ (1-i x)^2 \omega  \\
\end{pmatrix}\,,
\end{equation}
In order to diagonalize the light neutrino mass matrix $m_{\nu}$, it is useful to first perform a unitary transformation $U_{\nu1}$. Here the unitary matrix $U_{\nu1}$ takes to be
\begin{equation}
U_{\nu1}=\begin{pmatrix}
 \frac{2 i x^2+x+i}{\sqrt{3 \left(2 x^2+1\right) \left(3 x^2+1\right)}} &~ -\frac{4 x^2+1}{\sqrt{3 \left(2 x^2+1\right) \left(4 x^2+1\right)}} &~ \frac{1-i x}{\sqrt{3 \left(3 x^2+1\right)}} \\
 \frac{i \left(1-2 \omega ^2\right) x^2+\omega  x+i}{\sqrt{3 \left(2 x^2+1\right) \left(3 x^2+1\right)}} &~ \frac{2 x^2-\sqrt{3} x-\omega ^2}{\sqrt{3 \left(2 x^2+1\right) \left(4 x^2+1\right)}} &~ \frac{\omega  (2 i x+1)}{\sqrt{3 \left(3 x^2+1\right)}} \\
 \frac{i (1-2 \omega ) x^2+\omega ^2 x+i}{\sqrt{3 \left(2 x^2+1\right) \left(3 x^2+1\right)}} &~ \frac{2 x^2+\sqrt{3} x-\omega }{\sqrt{3 \left(2 x^2+1\right) \left(4 x^2+1\right)}} &~ \frac{\omega ^2 (2 i x+1)}{\sqrt{3 \left(3 x^2+1\right)}} \\
\end{pmatrix}
\end{equation}
Then the neutrino mass matrix $m^\prime_{\nu}$ is a block diagonal matrix with elements
\begin{eqnarray}
\nonumber &&y=\frac{\left(3 x^2+1\right) \left(m_{a}+3x^2 m_{s} e^{i \eta }\right)}{2 x^2+1}\,,  \\
\nonumber && z=\frac{(2 x-i) \left(3 x^2+1\right) \left((1-ix) m_{a}+3x \left(i x^2+x+i\right) m_{s} e^{i \eta } \right)}{\left(2 x^2+1\right) \sqrt{(3x^2+1)(4x^2+1)}}\,, \\
&& w=\frac{(i-2 x) \left((x+i)^2m_{a} +3(x^2-ix+1)^2m_{s} e^{i \eta }  \right)}{(2 x+i) \left(2 x^2+1\right)}\,.
\end{eqnarray}
Then we can obtain the absolute values and the phases of parameters $y$, $z$ and $w$. As a result, $m^\prime_{\nu}$ can
be diagonalized by the unitary matrix $U_{\nu2}$ which is given in Eq.~\eqref{eq:Unu2}. It is easily to get the neutrino masses $m^2_1$ and $m^2_2$ by inserting the parameters $y$, $z$ and $w$ into Eq.~\eqref{eq:nu_masses}. Then the parameters $\theta$, $\psi$, $\rho$ and $\sigma$ can be given in dependence on the absolute values and the phases of parameters $y$, $z$ and $w$. Form the matrix form of $U_{l}$ in Eq.~\eqref{eq:ch_dia_matrix}, $U_{\nu1}$ and $U_{\nu2}$, we find that the lepton mixing matrix is given by
\begin{equation}\label{eq:PMNS_I18}
\hskip-0.1in  U
=\frac{1}{\sqrt{2}}
\begin{pmatrix}
 - \sqrt{\frac{2(2 x^2+1)}{3 x^2+1}} \cos \theta  &~ -  \sqrt{\frac{2(2 x^2+1)}{3 x^2+1}}e^{i \psi } \sin \theta  &~ \frac{\sqrt{2} x}{\sqrt{3 x^2+1}} \\
  x \sqrt{\frac{4 x^2+1}{\left(2 x^2+1\right) \left(3 x^2+1\right)}} \cos \theta -\frac{ e^{-i \psi } \sin \theta }{\sqrt{2 x^2+1}} &~ \frac{\cos \theta  }{\sqrt{2 x^2+1}}+ x  \sqrt{\frac{4 x^2+1}{\left(2 x^2+1\right) \left(3 x^2+1\right)}}e^{i \psi } \sin \theta &~ \sqrt{\frac{4 x^2+1}{3 x^2+1}} \\
 \frac{x \cos \theta }{\sqrt{\left(2 x^2+1\right) \left(3 x^2+1\right)}}+\sqrt{\frac{4 x^2+1}{2 x^2+1}}e^{-i \psi } \sin \theta   &~ \frac{ x e^{i \psi }\sin \theta }{\sqrt{\left(2 x^2+1\right) \left(3 x^2+1\right)}}-\sqrt{\frac{4 x^2+1}{2 x^2+1}} \cos \theta  &~ \frac{1}{\sqrt{3 x^2+1}} \\
\end{pmatrix}\,,
\end{equation}
Then the lepton mixing parameters are predicted to be
\begin{eqnarray}
 \nonumber && \sin^2\theta_{13}=\frac{x^2}{1+3 x^2}\,,\qquad
  \sin^2\theta_{12}=\sin^2\theta\,,\qquad
\sin^2\theta_{23}=1-\frac{1}{2 \left(1+2 x^2\right)}\,, \\
&& J_{CP}=-\frac{x\sqrt{1+4x^2}\sin2\theta\sin\psi}{4(1+3x^2)^{3/2}}\,, \qquad
I_{1}=-\frac{\left(2 x^2+1\right)^2 \sin ^22 \theta  \sin (\rho -\sigma )}{4 \left(3 x^2+1\right)^2}\,.
\end{eqnarray}
Both the atmospheric mixing angle and the reactor mixing angle depend on only one input parameter $x$ which comes from the vacuum alignment invariant under the action of the residual symmetry in the solar neutrino sector. As a consequence, the following sum rule between the reactor mixing angle and the atmospheric mixing angle is found to be satisfied
\begin{equation}
\sin^2\theta_{23}=\frac{1}{2}+\tan^2\theta_{13}\,.
\end{equation}
We note that $\theta_{23}$ is constrained to lie in the second octant. Inputting the experimentally preferred $3\sigma$ range of $\theta_{13}$, the atmospheric mixing angle is predicted to be $0.521\leq\sin^2\theta_{23}\leq0.525$.  It is remarkable that a good fit to the experimental data can always be achieved for any $|x|$. When these two mixing angles are required in their $3\sigma$, the input parameter $x$ is constricted in the range of $[0.148,0.163]$. Similar to the example in $\mathcal{I}_{17}$, we also give the example of $x=\frac{\sqrt{5}}{15}$. The $\chi^2$ analysis results are
\begin{eqnarray}
\nonumber && \eta=0.0913\pi, \qquad m_{a}=34 .659\,\text{meV},\qquad r=0 .644, \qquad \chi^2_{\text{min}}=17 .329, \qquad \sin^2\theta_{13}=0.0208\,, \\
\nonumber && \sin^2\theta_{12}=0.310, \qquad \sin^2\theta_{23}= 0.521, \qquad \delta_{CP}=-0.748\pi, \qquad \beta=0.513\pi, \\
&&  m_1=49 .377\,\text{meV}, \qquad m_2=50 .120\,\text{meV}, \qquad m_3=0\,\text{meV}, \qquad m_{ee}=36 .089\,\text{meV}\,.
\end{eqnarray}
The input parameters $x$, $r$ and $\eta$ are treated as random numbers, and mixing angles and mass ratio are required in their $3\sigma$ ranges. We find the allowed regions of the parameter $|\eta|$ and $r$ are  $[0.0905\pi,0.108\pi]$ and  $[0.6328,0.6448]$, respectively. The predicted range of the two CP phases are predicted to be $|\delta_{CP}|\in[0,0.0619\pi]\cup[0.710\pi,0.777\pi]$ and $|\beta|\in[0.485\pi,0.536\pi]$.

\end{description}

\end{appendix}

\providecommand{\href}[2]{#2}\begingroup\raggedright\endgroup

\end{document}